\documentclass[12pt]{article}
\usepackage{rmpbib,axodraw,epsf}

\newcommand{\bd}{\begin{displaymath}}
\newcommand{\ed}{\end{displaymath}}
\newcommand{\be}{\begin{equation}}
\newcommand{\ee}{\end{equation}}
\newcommand{\bda}{\begin{eqnarray*}} 
\newcommand{\eda}{\end{eqnarray*}}
\newcommand{\bea}{\begin{eqnarray}} 
\newcommand{\eea}{\end{eqnarray}}

\newcommand{\e}{\mathrm{e}}
\newcommand{\ii}{\mathrm{i}}

\newcommand{\Tr}{\mathrm{Tr}}
\newcommand{\ms}{\overline{\mathrm{MS}}}

\def\slash#1{\mkern-1.5mu\raise0.4pt\hbox{$\not$}\mkern1.2mu #1\mkern 0.7mu}
\newcommand{\Ds}{\mkern-1.5mu\raise0.4pt\hbox{$\not$}\mkern-.1mu {D}\mkern.1mu}
\newcommand{\Gs}{\mkern-1.5mu\raise0.4pt\hbox{$\not$}\mkern2.6mu {\Gamma}\mkern.7mu}

\renewcommand{\baselinestretch}{1.1}

\makeatletter
\@addtoreset{equation}{section}
\makeatother

\begin{document}

\begin{flushright}
DESY 02-185 \\
November 2002
\end{flushright}

\vspace{0.3cm}

\begin{center}
{\bf \LARGE Lattice Perturbation Theory}
\end{center}

\vspace{0.3cm}

\begin{center}
{\Large Stefano Capitani~\footnote{ \sf E-mail:~stefano@ifh.de}}
\end{center} 

\begin{center}
{\large  DESY Zeuthen \\
         John von Neumann-Institut f\"ur Computing (NIC) \\[0.2 cm]
         Platanenallee 6, 15738 Zeuthen, Germany}
\end{center}

\vspace{0.5cm}

\begin{abstract}

The consideration of quantum fields defined on a spacetime lattice provides 
computational techniques which are invaluable for studying gauge theories 
nonperturbatively from first principles. 

Perturbation theory is an essential aspect of computations on the lattice, 
especially for investigating the behavior of lattice theories near the 
continuum limit. Particularly important is its role in connecting the outcome 
of Monte Carlo simulations to continuum physical results. For these matchings 
the calculation of the renormalization factors of lattice matrix elements 
is required.

In this review we explain the main methods and techniques of lattice
perturbation theory, focusing on the cases of Wilson and Ginsparg-Wilson 
fermions. We will illustrate, among other topics, the peculiarities of 
perturbative techniques on the lattice, the use of computer codes for the 
analytic calculations and the computation of lattice integrals. Discussed 
are also methods for the computation of 1-loop integrals with very high 
precision. 

The review presents in a pedagogical fashion also some of the recent 
developments in this kind of calculations. The coordinate method of L\"uscher 
and Weisz is explained in detail. Also discussed are the novelties that 
Ginsparg-Wilson fermions have brought from the point of view of perturbation 
theory.

Particular emphasis is given throughout the paper to the role of chiral 
symmetry on the lattice and to the mixing of lattice operators under 
renormalization. The construction of chiral gauge theories regularized 
on the lattice, made possible by the recent advances in the understanding
of chiral symmetry, is also discussed.

Finally, a few detailed examples of lattice perturbative calculations 
are presented.

\end{abstract}

\newpage

\setcounter{tocdepth}{3}
\tableofcontents

\newpage

\section{Introduction}

In a lattice field theory the quantum fields are studied and computed 
using a discretized version of the spacetime. 
The lattice spacing $a$, the distance between neighboring sites, 
induces a cutoff on the momenta of the order $1/a$.

A spacetime lattice can be viewed as a regularization which is 
nonperturbative. Since the other known regularizations, like dimensional 
regularization or Pauli-Villars, can be defined only order by order in 
perturbation theory, the lattice regularization has this unique advantage over 
them. It is a regularization which is not tied to any specific approximation 
method, and which allows calculations from first principles employing various 
numerical and analytical methods, without any need to introduce models 
for the physics or additional parameters. 

In discretizing a continuum field theory one has to give up Lorentz invariance
(and in general Poincar\'e invariance), but the internal symmetries can
usually be preserved. In particular, gauge invariance can be kept as a 
symmetry of the lattice for any finite value of the lattice spacing, and this 
makes possible to define QCD as well as chiral gauge theories like the 
electroweak theory. The construction of the latter kind of theories on a 
lattice however presents additional issues due to chiral symmetry, which have 
been understood and solved only recently. The fact that one is able to 
maintain gauge invariance for any nonzero $a$ is of great help in proving the 
renormalizability of lattice gauge theories.

Lattice gauge theories constitute a convenient regularization of QCD where
its nonperturbative features, which are essential for the description of 
the strong interactions, can be systematically studied. The lattice can 
then probe the long-distance physics, which is otherwise unaccessible to 
investigations which use continuum QCD. Precisely to study low-energy 
nonperturbative phenomena the lattice was introduced by Wilson, which went on 
to prove in the strong coupling regime the confinement of quarks. Confinement 
means that quarks, the fundamental fields of the QCD Lagrangian, are not the 
states observed in experiments, where one can see only hadrons, and the free 
theory has no resemblance to the observed physical world. The quark-gluon 
structure of hadrons is hence intrinsically different from the structure of 
other composite systems. No description in terms of two-body interactions 
is possible in QCD. Lattice simulations of QCD show that a large part of the 
mass of the proton arises from the effect of strong interactions between 
gluons, that is from the pure energy associated with the dynamics of 
confinement. Only a small fraction of the proton mass is due to the quarks. 
Similarly, the lattice confirms that only about half of the momentum and a 
small part of the spin of the proton come from the momentum and spin of the 
constituent quarks. Computations coming from the lattice can then give 
important contributions to our understanding of the strong interactions.

In this review we are interested in doing lattice calculations in the weak 
coupling regime. This is the realm of perturbation theory, which is used to 
compute the renormalization of the parameters of the Lagrangian and of the 
matrix elements, and to study the approach of the lattice to the continuum 
limit. Details of the lattice formulation that are only relevant at the 
nonperturbative level will not be discussed in this review. If the reader is 
also interested in the nonperturbative aspects of lattice field theories, they
are covered at length in the books of~\cite{Creutz:mg,Montvay:cy,Rothe:kp}
and in the just appeared book of~\cite{Smit:ug}. 
The book by Rothe is the one which contains more material about lattice 
perturbation theory. Useful shorter reviews, which also cover many 
nonperturbative aspects, sometimes with a pedagogical cut, can also be found 
in~\cite{Kogut:1982ds,Sharpe:1994dc,Sharpe:wt,DeGrand:1996ri,DeGrand:1996sc,Gupta:1998hu,Sharpe:1998hh,Wittig:1999kb,Munster:2000ez,Davies:2002cx,Kronfeld:2002nm} and in the recent~\cite{Luscher:2002pz}. 
Here we would like to explain the main methods and techniques of lattice 
perturbation theory, particularly when Wilson and Ginsparg-Wilson fermions 
are used. We will discuss, among other things, Feynman rules, aspects of the 
analytic calculations, the computer codes which are often necessary to carry 
them out, the mixing properties of lattice operators, and lattice integrals.

Chiral symmetry is a recurrent topic in the treatment of fermions on the 
lattice, and we will address some issues related to it in the course of the 
review. We feel that a discussion of the problems connected with the 
realization of chiral symmetry on the lattice is needed. The reader might 
otherwise wonder why one should do such involved calculations like the ones 
required for Ginsparg-Wilson fermions. We think that it is also interesting to
see how the lattice can offer fascinating solutions to the general quantum 
theoretical problem of defining chiral gauge theories beyond tree level. 

Also discussed is an algebraic method for the reduction of any 1-loop 
lattice integral (in the Wilson case) to a linear combination of a few basic 
constants. These constants are calculable with very high precision using 
in a clever way the behavior of the position space propagators at large 
distances. The coordinate space method, which turns out to be a very powerful 
tool for the computation of lattice integrals, allows then the calculation 
of these constants with very high precision, which is also the necessary 
requirement in order to be able to compute two-loop lattice integrals 
with many significant decimal places, as we will explain in detail. 
A lot of nice and interesting work has been done using these techniques 
in the case of bosonic integrals, which thanks to them can be easily 
computed with extraordinary precision at one-loop, and with good precision 
at two loops. A nonnegligible part of this review will discuss these 
calculations in detail in the bosonic case.

The focus of this review is on methods rather than on results. In fact, 
very few numerical results will be reported. The reader, if interested, 
can consult many perturbative results in the references given. Instead, 
our objective is different. We would like to provide computational tools 
which can be useful for physicists who are interested in doing this kind 
of calculations. Technical details will be therefore explained in a 
pedagogical fashion. Particular attention will be paid to certain aspects 
that only occur in lattice computations, and that physicists expert in 
continuum perturbative calculations might find curious.
The main objective of this review is to show how perturbation theory works 
on the lattice in the more common situations. 
It is hoped that one can learn from the material presented here.

A background in continuum quantum field theory is required, and
an acquaintance with continuum perturbative calculations in gauge field 
theories, the derivation of Feynman rules in continuum QCD and the
calculation of Feynman diagrams will be assumed. 
Familiarity with the path integral formalism, with the quantization of field 
theories through the functional integral and with the renormalization
of continuum quantum field theories is also desired.
The knowledge of elementary facts, such as the renormalization group
equations, the running of the strong coupling, the $\beta$ function and 
asymptotic freedom of QCD, will also be taken for granted.

This review is not homogeneous. I have given more space to the topics that 
I believe are more interesting and more likely to be of wider use in the 
future. Many of the choices made and of the examples reported draw from the 
experience of the author in doing this kind of calculations.

To keep this review into a manageable size, not all important topics or 
contributions will be covered. One thing that will not be discussed in detail 
is perturbation theory applied to Symanzik improvement, which, although very 
interesting and useful, would probably require a review in itself, given also 
the many important result that have been produced. 
The Schr\"odinger functional is also introduced only in a very general way.
I will not be able to do justice to other topics like numerical perturbation
theory or tadpole improvement.

Many interesting subjects had to be entirely left out because of constraints 
on space. Among the topics which are not covered at all are nonrelativistic 
theories, heavy quarks, and anisotropic lattices. I have also omitted all what
concerns finite temperature perturbation theory. Many of these things are 
treated in detail in the reviews and books cited above, where several topics 
not covered here can also be found. Moreover, we will not occupy ourselves 
with phenomenological results, but only with how perturbation theory is useful
for the extraction from the lattice of that phenomenology. In any case, there 
are by now so many perturbative calculations that have been made on the 
lattice that it would be impossible to include all of them here.

The main reason for the introduction of the lattice was to study QCD 
in its nonperturbative aspects, like confinement, and we will confine this 
review to QCD. Although very interesting, spin models, the $\lambda \phi^4$ 
theory and the Higgs sector, to name a few, will be left out.
Even after this restriction is made, lattice QCD is still quite a broad 
topic by itself, and thus to contain this review into a reasonable size 
we have been compelled to discuss only the main actions that have been used 
to study QCD on the lattice. Given the great number of different actions 
that have been proposed for studying lattice QCD in the past 30 years,
it is necessary to limit ourselves to just the few of them that are more 
widely used. The Wilson and staggered formulations have been the most 
popular ones in all this time. Recently the particular kind of chiral 
fermions known as Ginsparg-Wilson (like the overlap, domain wall and 
fixed-point fermions) have also begun to find broader application, and they 
present interesting challenges for lattice physicists interested in 
perturbation theory. These are the actions that we will cover in this paper. 

The main features of the lattice construction and of lattice perturbation 
theory will be discussed in detail in the context of Wilson fermions. 
When the other actions will be introduced the discussions will be more
general, although we will try to point out the peculiarities of perturbative 
calculations in these particular cases. The explanations of the various 
lattice actions will be rather sketchy and aimed mainly at the aspects which 
are interesting from the point of view of perturbation theory.

\subsection{Outline of the paper}

The review is divided in three parts: Sections 2 and 3, which are a sort of
motivation, Sections 4 to 12, which introduce the lattice, the various actions
and their Feynman rules, and Sections 13 to 20, which in much more technical 
detail show how lattice computations are made.

We begin with two Sections which are meant to stress the importance 
of lattice perturbation theory and explain what is meant for renormalization
of operators on the lattice. After having introduced these motivations, 
we start with the first technical points, defining in Section 4 a euclidean 
lattice and showing what the discretization of a continuum theory means 
in practice. We introduce typical lattice quantities and we pass to discuss in
detail the Wilson action (which has not chiral invariance) in Section 5, 
explaining how to derive its Feynman rules in momentum space. All Feynman 
rules necessary for one-loop calculations are then explicitly given. In 
Section 6 we focus our attention on the relation between chirality and 
fermionic modes on the lattice, and the problems which arise when one tries to
define chiral fermions on a lattice. After a brief interlude which discusses 
staggered fermions, which have some chiral symmetry and have been the major 
alternative to Wilson fermions (at least in the first two decades of the 
lattice), we then introduce in Section 8 Ginsparg-Wilson fermions, the 
long-awaited reconciliation of chirality with the lattice. We give some 
details about the overlap, the domain wall and the fixed-point actions, which 
are solutions of the Ginsparg-Wilson relation.
In Section 9 we then explain how using Ginsparg-Wilson fermions it is possible
to define on the lattice chiral gauge theories, where gauge invariance and 
chiral symmetry are maintained together at every order in perturbation theory 
and for any finite value of the lattice spacing.

In Section 10 we then see the approach of coupling and masses to the 
continuum limit and talk about the $\beta$ function and the $\Lambda$
parameter of the lattice theory. In Section 17 we briefly introduce 
the Symanzik improvement, including a short discussion about improved 
pure gauge actions on the lattice. We conclude the first part of the review
with a brief introduction to the Schr\"odinger functional, which has gained
a paramount place in the lattice landscape in recent years.

In Section 13 we begin the more technical part of the review, dedicated 
to how to actually carry out the perturbative computations on the lattice. 
We introduce at this point the symmetry group of the lattice, the hypercubic 
group. Since the lattice symmetries are not as restrictive as in the continuum 
theory, more mixings arise in general under renormalization, and we discuss 
some examples of them in Section 14. How to compute Feynman diagrams on the 
lattice is explained in great detail in Section 15, where we talk about the
lattice power counting theorem of Reisz, which is useful for the computation
of divergent integrals, and we present, step by step, the complete calculation
of the 1-loop renormalization constant of the operator measuring the first 
moment of the momentum distribution of quarks in hadrons.
This example is rather simple (compared to other operators) and contains
all the main interesting features one can think of: a logarithmic divergence, 
a covariant derivative, symmetrized indices and of course the peculiar use 
of Kronecker deltas in lattice calculations. Moreover, it is an example of 
a calculation in which the various propagators and vertices need an expansion 
in the lattice spacing $a$ (in this case, at first order). Finally, it 
includes the computation of the quark self-energy, which is quite interesting 
and useful by itself. Brief discussions about overlap calculations, tadpole 
improvement and perturbation theory with fat link actions conclude Section 15. 
In Section 16 we then discuss the use of computer codes for the automated 
computations of lattice Feynman diagrams.

In Section 17 we explain some advanced techniques for the numerical evaluation
of lattice integrals coming from Feynman diagrams (using extrapolations to 
infinite volume), while in Section 18 we introduce an algebraic method for the
exact reduction of any Wilson 1-loop integral to a few basic constants. 
The bosonic case is thoroughly  explained, so that the reader can learn to use
it, and some applications to the exact calculations of operator tadpoles are 
also explicitly given. Section 18 ends with a discussion of the main points 
of the general fermionic case, and the expression of the 1-loop quark
self-energy in terms of the basic constants.

The basic constants of the algebraic method can be computed with arbitrary 
precision, as explained in detail in Section 19. The values of the fundamental
bosonic constants, $Z_0$ and $Z_1$, are given with a precision of about 400 
significant decimal places in Appendix B. In order to be able to calculate
them to this precision, we need to introduce the coordinate space method of 
L\"uscher and Weisz, which will also be used for the computation of 2-loop 
integrals. The 2-loop bosonic integrals are discussed at length, and the 
general fermionic case is also addressed. In Section 20 we then briefly 
introduce numerical perturbation theory, which is promising for doing
calculations at higher loops. Finally, conclusions are given in Section 21, 
and Appendix A summarizes some notational conventions.

\section{Why lattice perturbation theory}
\label{sec:why}

To some readers the words ``lattice'' and ``perturbation theory'' might sound 
like a contradiction in terms, but we will see that this is not the case and 
that lattice perturbation theory has grown into a large and established 
subject. Although the main reason why the lattice is introduced is because it
constitutes a nonperturbative regularization scheme and as such it allows 
nonperturbative computations, perturbative calculations on the lattice 
are rather important, and for many reasons. 

Perturbation theory of course cannot reveal the full content of the lattice
field theory, but it can still give a lot of valuable informations. In fact, 
there are many applications where lattice perturbative calculations are useful
and in some cases even necessary. Among them we can mention the determinations
of the renormalization factors of matrix elements of operators and of the 
renormalization of the bare parameters of the Lagrangian, like couplings and 
masses. The precise knowledge of the renormalization of the strong coupling is 
essential for the determination of the $\Lambda$ parameter of QCD in the 
lattice regularization (as we will see in Section~\ref{sec:approach}) and
of its relation to its continuum counterpart, $\Lambda_{QCD}$. 
In general perturbation theory is of paramount importance in order to 
establish the right connection of the lattice scheme which is used in 
practice, and of the matrix elements simulated within that scheme, to the 
physical continuum theory. Every lattice action defines a different 
regularization scheme, and thus one needs for each new action that is 
considered a complete set of these renormalization computations in order for 
the results which come out from Monte Carlo simulations to be used and 
understood properly.

Moreover, lattice perturbation theory is important for many other aspects,
among which we can mention the study of the anomalies on the lattice, the 
study of the general approach to the continuum limit, including the recovery 
of the continuum symmetries broken by the lattice regularization (like Lorentz
or chiral symmetry) in the limit $a \rightarrow 0$, and the scaling 
violations, i.e., the corrections to the continuum limit which are of order 
$a^n$. These lattice artifacts bring a systematic error in lattice results, 
which one can try to reduce by means of an ``improvement'', as we will see 
in Section~\ref{sec:improvement}.

Perturbative calculations are thus in many cases essential, and are the only 
possibility for having some analytical control over the continuum limit.
As we will see in Section~\ref{sec:approach}, the perturbative region is the 
one that must be necessarily traversed in order to reach the continuum limit. 
Lattice perturbation theory is in fact strictly connected to the continuum 
limit of the discretized versions of QCD. Because of asymptotic freedom, 
one has $g_0 \to 0$ when $\mu \to \infty$, which means $a \to 0$. We should 
also point out that one cannot underestimate the role played by perturbative 
calculations in proving the renormalizability of lattice gauge theories.

Finally, perturbation theory will also be important for defining chiral gauge 
theories on the lattice at all orders of the gauge coupling, as we will see 
in detail in Section~\ref{sec:chiralgaugetheories}. The lattice will thus be
proven to be the only regularization that can preserve chirality and gauge 
invariance at the same time (without destroying basic features like locality 
and unitarity).

We can say in a nutshell that lattice perturbation theory is important 
for both conceptual and practical reasons. The phenomenological numbers 
that are quoted from lattice computations are very often the result 
of the combined effort of numerical simulations and analytic calculations, 
usually with some input from theoretical ideas.

In principle all known perturbative results of continuum QED and QCD can also 
be reproduced using a lattice regularization instead of the more popular ones.
However, calculating in such a way the correction to the magnetic moment of 
the muon (to make an example) would be quite laborious. A lattice cutoff would 
not be the best choice in most cases, for which instead regularizations like 
Pauli-Villars or dimensional regularization are more suited and much easier to
employ. The main virtue of the lattice regularization is instead to allow 
nonperturbative investigations, which usually need some perturbative 
calculations to be interpreted properly. As we have already mentioned, the 
connection from Monte Carlo results of matrix elements to their corresponding 
physical numbers, that is the matching with the continuum physical theory, 
has to be realized by performing a lattice renormalization. It is in this 
context that lattice perturbation theory has a wide and useful range of 
applications, and we will discuss this important aspect of lattice 
computations in more detail in the next Section. In this respect, perturbative
lattice renormalization is important by itself as well as in being a hint and 
a guide for the few cases in which one can also determine the renormalization 
constants nonperturbatively according to the method proposed 
in~\cite{Martinelli:1994ty} (for a recent review see~\cite{Sommer:2002en}).
This is even more important in the cases in which operator mixings are 
present, which are generally more complex than in the continuum. In fact, 
mixing patterns on the lattice become in general more transparent when looked 
at using perturbative renormalization than nonperturbatively. We should also 
add that perturbative coefficients can be usually computed much more 
accurately than typical quantities in numerical simulations. Perturbative 
renormalization results can in any case be quite useful in checking and 
understanding results coming from nonperturbative methods (when available). 
When short-distance quantities can be calculated using such diverse techniques
as lattice perturbation theory and Monte Carlo simulations, the comparison of 
the respective results, repeated in different situations, can give significant
suggestions on the validity of perturbative and nonperturbative methods.

In some cases a nonperturbative determination of the renormalization constants
can turn out to be rather difficult to achieve. For the method to work, it is 
necessary that there is a plateau for the signal over a substantial range of 
energies so that one can numerically extract the values of the renormalization
factors. The nonperturbative renormalization methods can then fail because 
a window which is large enough cannot be found. Moreover, where mixings are 
present these methods could come out to be useless because the amount of 
mixing is too small to be disentangled with numerical techniques, although 
still not too small to be altogether ignored. It is then not clear whether all 
operators can be renormalized nonperturbatively in such a way. In these cases 
the only possibility to compute renormalization factors seems to be provided 
by the use of lattice perturbative methods. An important exception to this  
is given by the Schr\"odinger functional scheme, where using a particular 
procedure known as recursive finite-size scaling technique (which we will 
explain in Section~\ref{sec:sf}) it is possible to carry out precise 
nonperturbative determinations of renormalized couplings, masses and operators
for an extremely wide range of energies. Computations using the Schr\"odinger 
functional are however rather more involved than average and usually require 
much bigger computational resources.

We would also like to point out that in the case of Ginsparg-Wilson fermions 
the numerical computations required to extract nonperturbative renormalization 
factors (which come on top of the already substantial effort needed to 
determine the bare matrix elements) could turn out to be quite expensive, 
especially in the cases of complicated operators like the ones that measure 
moments of parton distributions.

\begin{figure}[t]
\begin{center}
\setlength{\epsfysize}{.5\textheight}
\epsffile{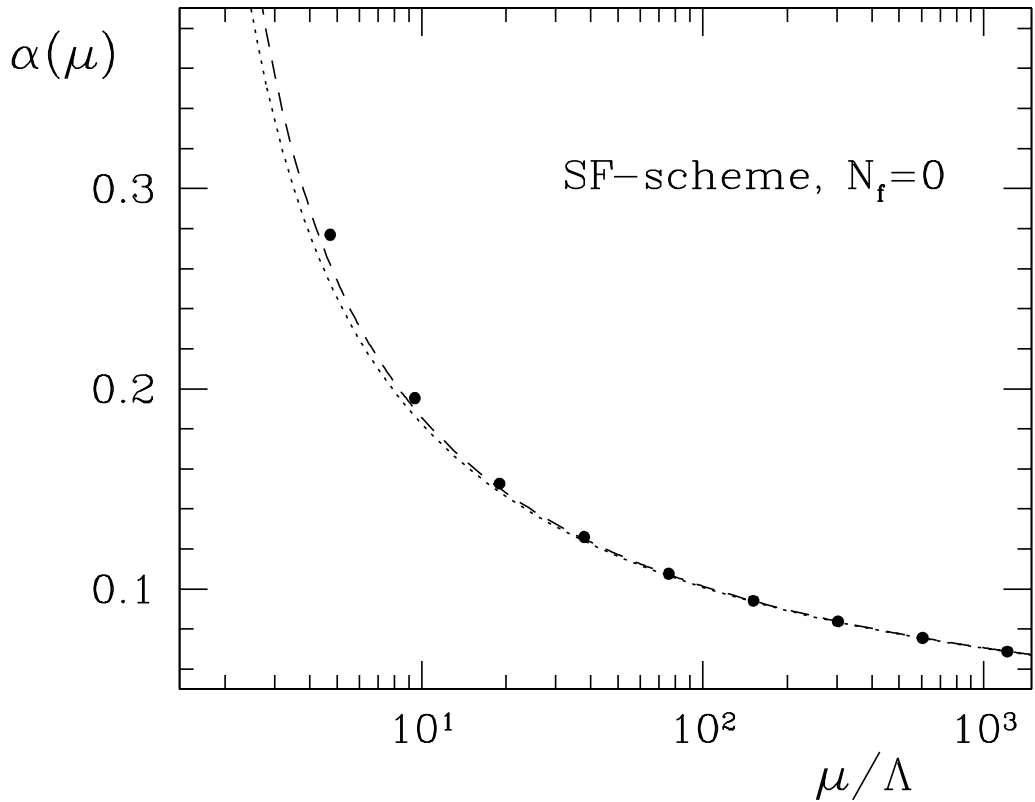}
\end{center}
\caption{\small Perturbative and nonperturbative running of the renormalized
strong coupling from the Schr\"odinger functional on the lattice, from
(Capitani {\em et al.}, 1999c). 
In this scheme $\Lambda \sim 116~{\mathrm MeV}$.}
\label{fig:alpha}
\end{figure}

\begin{figure}[t]
\begin{center}
\setlength{\epsfysize}{.5\textheight}
\epsffile{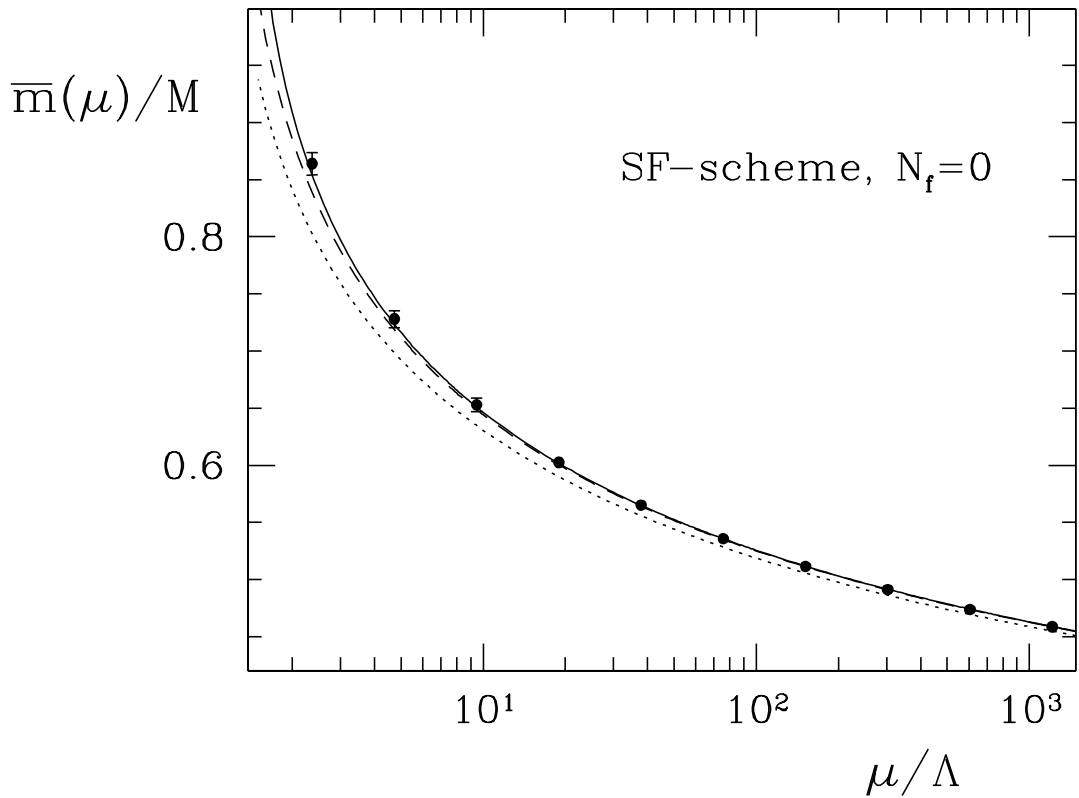}
\end{center}
\caption{\small Perturbative and nonperturbative running of the renormalized
masses from the Schr\"odinger functional on the lattice, from
(Capitani {\em et al.}, 1999c).  
In this scheme $\Lambda \sim 116~{\mathrm MeV}$.
The renormalization group invariant mass $M$ is defined in 
Eq.~(\ref{eq:rgimass}).}
\label{fig:mass}
\end{figure}

We can thus say, after having looked at all these different aspects, that 
lattice perturbation theory is important and fundamental. Of course sometimes 
there can be issues concerning its reliability when a 1-loop perturbative 
correction happens to be large, especially when the corresponding 2-loop 
calculation looks rather difficult to carry out.~\footnote{In this case 
mean-field improved perturbation theory, using Parisi's boosted bare coupling,
is known to reduce the magnitude of these corrections in many situations 
(see Section~\ref{sec:tadpoleimprovement}).}

On the other hand, there are cases in which lattice perturbation theory works 
rather well. As an example we show in Figs.~\ref{fig:alpha} and~\ref{fig:mass}
the scale evolutions of the renormalized strong coupling and masses computed 
in the Schr\"odinger functional scheme~\cite{Capitani:1998mq}.~\footnote{An 
explanation of the way these nonperturbative evolutions are obtained is given 
in Section~\ref{sec:sf}.} We can see that these scale evolutions are 
accurately described by perturbation theory for a wide range of energies. 
The perturbative and nonperturbative results are very close to each other, and 
almost identical even down to energy scales which are surprising low. The best
perturbative curves shown are obtained by including the $b_2 g_0^7$ term of 
the $\beta$ function and the $d_1 g_0^4$ term of the $\tau$ function 
(that is, the first nonuniversal coefficients). The other curves are lower 
approximations. In Section~\ref{sec:approach} more details about these 
functions can be found.

Although the coupling and masses in the figures are computed in the 
Schr\"odinger functional scheme, and their values are different from, say, 
a calculation done with standard Wilson fermions, this example shows how close 
perturbation theory can come to nonperturbative results. The case shown is 
particularly interesting, because these nonperturbative results are among 
the best that one can at present obtain. The Schr\"odinger functional coupled 
to recursive finite-size techniques allows to control the systematic errors
completely. There are fewer errors in these calculations, and they are fully 
understood. In other situations we cannot really exclude, when we see a 
discrepancy between nonperturbative and perturbative results, that at least
part of this discrepancy originates from the nonperturbative side.

\begin{figure}[t]
\begin{center}
\setlength{\epsfysize}{.6\textheight}
\epsffile{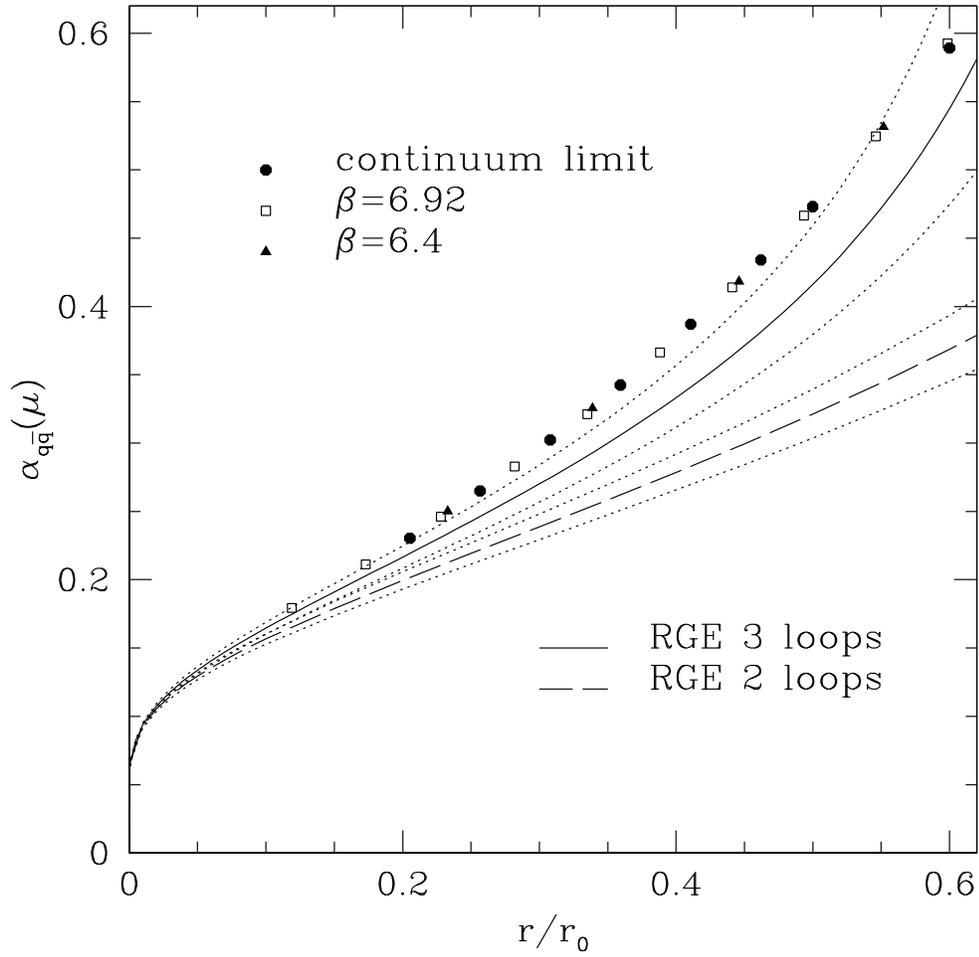}
\end{center}
\caption{\small Perturbative and nonperturbative running of the renormalized
strong coupling in the $qq$ scheme, from~(Necco and Sommer, 2001).}
\label{fig:alpha_qq}
\end{figure}

Another nice example of the good behavior of lattice perturbation theory
is given in Fig.~\ref{fig:alpha_qq}, which comes from the work
of~\cite{Necco:2001gh,Necco:2001tq}. These authors have computed the running 
coupling from the static quark force or potential in three different ways, 
corresponding to three different schemes:~\footnote{A few different ways 
of defining a strong coupling constant on the lattice are discussed 
in~\cite{Weisz:1995yz}.}
\bea
F(r) &=& \frac{dV}{dr} = C_F \, \frac{\alpha_{qq} (1/r) }{r^2} , \\
V(r) &=& - C_F \, \frac{\alpha_{\overline{V}} (1/r) }{r} , \\
\widetilde{V}(Q) &=& - 4\pi C_F \, \frac{\alpha_V (Q) }{Q^2} ,
\eea
and showed that in first case, called $qq$ scheme, the perturbative expansion 
of the coupling is rather well behaved, that is the coefficients are small 
and rapidly decreasing. This is what is shown in Fig.~\ref{fig:alpha_qq},
which indicates that the perturbative computations can be trusted to describe 
with a good approximation the nonperturbative numbers for couplings up to 
$\alpha_{qq} \approx 0.3$. 
In the other two schemes, however, the coefficients are somewhat larger
(especially in the last case), and the perturbative expansion of the coupling 
has a worse behavior. The perturbative couplings in these two schemes
have a more pronounced difference with respect to the nonperturbative results.

The three schemes above differ only by kinematics, and the results show how 
the choice of one scheme or another can have a big influence on the 
perturbative behavior of the coupling. Discussions about the validity of 
lattice perturbation theory cannot then be complete until the dependence on 
the renormalization scheme (beside the dependence on the various actions used) 
is also taken into account and investigated. Not all schemes are suitable to 
be used in lattice perturbation theory to the same extent. In particular, the 
$qq$ scheme is the best one among the three considered above. From this point 
of view the coupling in the Schr\"odinger functional scheme is even better 
behaved than $\alpha_{qq}$, in other words the coefficient $b_2$ of the 
$\beta$ function (the first nonuniversal coefficient) is the smallest in this 
scheme, and in this sense the Schr\"odinger functional scheme is the closest 
to the $\ms$ scheme. 

We would also like to mention the work of~\cite{Davies:1997mg}, where the QCD 
coupling is extracted from various Wilson loops, and which shows another 
instance of a good agreement between perturbation theory and simulations.

We can thus trust lattice perturbation theory, under the right conditions.
It is probably not worse behaved than perturbative QCD in the continuum, 
which is an asymptotic expansion, and which also gives origin to large 
corrections in some cases. We could even say that perturbation theory
can be better tested on the lattice than in the continuum, because in a lattice
scheme one has also at his disposal the nonperturbative results to compare 
perturbation theory with.
When lattice perturbation theory does not agree with nonperturbative numerical
results, perhaps a look at the systematic errors coming from the numerical 
side can sometimes be worthwhile.

\clearpage

\section{Renormalization of operators}
\label{sec:renormalizationoperators}

In general, matrix elements of operators computed on the lattice using 
numerical simulations still require a renormalization in order to be converted
into meaningful physical quantities. The Monte Carlo matrix elements can be 
considered as (regulated) bare numbers, and to get physical results one needs 
to perform a renormalization, matching the numbers to some continuum scheme, 
which is usually chosen to be the $\ms$ scheme of dimensional regularization.

In many physical problems one evaluates matrix elements of operators that 
appear in an operator product expansion. These matrix elements contain the 
long-distance physics of the system, and are computed numerically on the 
lattice, while the Wilson coefficients contain the short-distance physics, 
which can be gathered from perturbative calculations in the continuum. In this
case the operators computed on the lattice must at the end be matched to 
the same continuum scheme in which the Wilson coefficient are known. Therefore
one usually chooses to do the matching from the lattice to the $\ms$ scheme of 
dimensional regularization. A typical example is given by the moments of 
deep inelastic structure functions, and we will illustrate many features of 
perturbation theory in the course of this review using lattice operators
appearing in operator product expansions which are important for the analysis 
of structure functions.

Thus, on the lattice one has to perform a matching to a continuum scheme, that 
is one looks for numbers which connect the bare lattice results to physical 
continuum renormalized numbers. We will now discuss how a perturbative 
matching can be done at one loop. Some good introductory material on the 
matching between lattice and continuum and on the basic concepts of lattice 
perturbation theory can be found 
in~\cite{Sachrajda:1989qa,Sharpe:1994dc,Sharpe:wt}. A short review 
of the situation of perturbative calculations at around 1995 is given 
in~\cite{Morningstar:1995vd}.

It turns out that to obtain physical continuum matrix elements from Monte 
Carlo simulations one needs to compute 1-loop matrix elements on the lattice 
as well as in the continuum. Lattice operators are chosen so that at tree 
level they have the same matrix elements as the original continuum operators,
at least for momenta much lower than the lattice cutoff, $p \ll \pi/a$. 
Then at one loop one has 
\bea
\langle q | O_i^{\mathrm lat} | q \rangle &=& \sum_j \Bigg( \delta_{ij} 
+\frac{g_0^2}{16 \pi^2} \Big( -\gamma_{ij}^{(0)} \log a^2p^2 
+ R_{ij}^{\mathrm lat} \Big) \Bigg) \cdot 
\langle q | O_j^{\mathrm tree} | q \rangle  
\label{eq:1looplat} \\
\langle q | O_i^{\ms} | q \rangle &=& \sum_j \Bigg( \delta_{ij} 
+\frac{g_{\ms}^2}{16 \pi^2} \Big( -\gamma_{ij}^{(0)} \log \frac{p^2}{\mu^2} 
+ R_{ij}^{\ms} \Big) \Bigg) \cdot \langle q | O_j^{\mathrm tree} | q \rangle .
\label{eq:1loopcont}
\eea
The lattice and continuum finite constants, $R_{ij}^{\mathrm lat}$ and
$R_{ij}^{\ms}$, and therefore the lattice and continuum 1-loop renormalization
constants, do not have the same value.~\footnote{We note that while 
$R_{ij}^{\mathrm lat}$ is the whole momentum-independent 1-loop correction, 
$R_{ij}^{\ms}$ does not contain the pole in $\epsilon$ and the factors 
proportional to $\gamma_E$ and $\log 4\pi$.} This happens because lattice 
propagators and vertices, as will be seen in detail later on, are quite 
different from their continuum counterparts, especially when the loop momentum
is of order $1/a$. Therefore the 1-loop renormalization factors on the lattice
and in the continuum are in general not equal. Note however that the 1-loop 
anomalous dimensions are the same.

The connection between the original lattice numbers and the final continuum 
physical results is given, looking at Eqs.~(\ref{eq:1looplat}) and 
(\ref{eq:1loopcont}), by
\be
\langle q | O_i^{\ms} | q \rangle = \sum_j \Bigg( \delta_{ij} 
-\frac{g_0^2}{16 \pi^2} \Big( -\gamma_{ij}^{(0)} \log a^2\mu^2
+ R_{ij}^{\mathrm lat} -R_{ij}^{\ms} \Big) \Bigg) 
\cdot \langle q | O_j^{\mathrm lat} | q \rangle .
\label{eq:1loopcontlat}
\ee
The differences $\Delta R_{ij} = R_{ij}^{\mathrm lat} - R_{ij}^{\ms}$ 
enter then in the matching factors
\be
Z_{ij} (a\mu,g_0)=\delta_{ij} -\frac{g_0^2}{16 \pi^2} 
\Big( -\gamma_{ij}^{(0)} \log a^2\mu^2 + \Delta R_{ij} \Big) ,
\ee
which are the main objective when one performs computations of renormalization
constants of operators on the lattice.~\footnote{The coupling that appears in 
Eq.~(\ref{eq:1loopcontlat}) is usually chosen to be the lattice one, $g_0$, as
advocated in~\cite{Sachrajda:1989qa}. Of course choosing one coupling or the 
other makes only a 2-loop difference, but these terms could still be 
numerically important. The validity of this procedure should be checked by 
looking at the size of higher-order corrections. Unfortunately on the lattice 
these terms are known only in a very few cases and no definite conclusions can
then be reached with regard to this point.

In the work of~\cite{Ji:1995vv} a generalization to higher loops was proposed, 
which gives an exact matching condition to all orders. This is done using the 
lattice and continuum renormalization group evolutions (see 
Section~\ref{sec:approach}). Using the lattice evolution, one goes to very 
high energies, which means very small couplings because of asymptotic freedom,
and there he does the matching to $\ms$. It thus essentially becomes a 
matching at the tree level. After this, one goes back to the original scale 
$\mu$, using the continuum renormalization group evolution backwards. For the 
matching at the scale $\mu$ one then obtains a factor
\be
\exp \Bigg\{
-\int_0^{g_{\ms} (\mu)} du \, \frac{\gamma_{\ms} (u)}{\beta_{\ms} (u)} 
\Bigg\} \,  \cdot
\exp \Bigg\{
-\int^0_{g_0 (a)} dv \, \frac{\gamma_{\mathrm lat} (v)}{
\beta_{\mathrm lat} (v)} 
\Bigg\} ,
\ee
where the $\gamma$ function governs the evolution of the renormalized 
operator. Effectively this formula uses the high-order coefficients of the 
$\beta$ and $\gamma$ functions. This approach has been used 
in~\cite{Gupta:1996yt}, where a discussion on these issues is made.} Lattice 
operators have more mixing options than continuum ones, due to the lower 
symmetry of the lattice theory. There is no Lorentz invariance (as we will see
in more detail in Section 13), and in many cases other symmetries, like chiral
symmetry, are also broken. Thus, the matching factors are not in general 
square matrices, that is one has $i \le j$ in Eq.~(\ref{eq:1loopcontlat}). 
To include all relevant operators one should look at the tree-level structures 
which appear in the calculation of lattice radiative corrections.

While $R^{\mathrm lat}$ and $R^{\ms}$ depend on the state $| q \rangle$, 
$\Delta R$ is independent of it and the $Z$ depends only on $a\mu$. This is as
it should be, since the renormalization factors are a property of the 
operators and are independent of the particular states with which a given 
matrix element is constructed. This is the reason why we have not specified 
$| q \rangle$. Furthermore, the matching factors between the lattice and the 
$\ms$ scheme are gauge invariant, and this property can be exploited to make 
important checks of lattice perturbative calculations.

At the end of the process that we have just explained one has computed,
using both lattice and continuum perturbative techniques, the renormalization 
factor $Z_O (a\mu)$ which converts the lattice operator $O(a)$ into the 
physical renormalized operator $\widehat{O} (\mu)$:
\be
\widehat{O} (\mu) = Z_O (a\mu) \, O(a) .
\ee
The reader should always keep in mind that in this way one does the matching 
of the bare Monte Carlo results (obtained using a lattice regulator)
to the physical renormalized results in the $\ms$ scheme. 

As for any general quantum field theory, the process at the end of which
one obtains physical numbers is accomplished in two steps. First one 
regularizes the ultraviolet divergences, and in this case the regulator 
is given by the lattice itself. Afterwards one renormalizes the theory so 
regulated, and on the lattice this results in a matching to a continuum scheme.
In the end, after renormalization the lattice cutoff must be removed, 
which means that one has to go to the continuum limit $a \to 0$ of the lattice
theory, and that it is safe to do so at this point (if the renormalization 
procedure is consistent). What remains after all this is only the scale $\mu$ 
brought in by the renormalization. In our case the scale $\mu$ at which the 
matrix elements are renormalized should be in the range
\be
\Lambda_{QCD} < \mu < \frac{\pi}{a} .
\ee
The lower bound ensures that perturbation theory is valid, while the upper
bound ensures that the cutoff effects, proportional to positive powers
of the lattice spacing, are small. One usually sets
\be
\mu = \frac{1}{a} ,
\ee
and since the 1-loop anomalous dimensions are the same on the lattice and in 
the continuum, only a finite renormalization connects the lattice to the 
$\ms$ scheme:
\be
\langle q | O_i^{\ms} | q \rangle = \sum_j \Bigg( \delta_{ij} 
-\frac{g_0^2}{16 \pi^2} \Big( R_{ij}^{\mathrm lat} -R_{ij}^{\ms} \Big) \Bigg) 
\cdot \langle q | O_j^{\mathrm lat} | q \rangle .
\ee
Every lattice action defines a different regularization scheme, and therefore 
these finite renormalization factors are in principle different for different
actions. The bare numbers, that is the Monte Carlo results for a given matrix
element, are also different, and so everything adjusts to give the
same physical result. 

We conclude mentioning that Sharpe (1994) has observed that when the operators
come from an operator product expansion one should multiply the 1-loop matching
factors introduced above with the {\em 2-loop} Wilson coefficients, in order 
to be consistent. This can be seen by looking at the 2-loop renormalization 
group evolution for the Wilson coefficients,
\be
\frac{c (\mu_1)}{c (\mu_2)} = \Bigg( \frac{g_0^2 (\mu_1)}{g_0^2 (\mu_2)}
\Bigg)^{- \gamma^{(0)}/2\beta_0}
\Bigg[1+ \frac{g_0^2 (\mu_2) - g_0^2 (\mu_1)}{16\pi^2} \Bigg( 
\frac{\gamma^{(1)}}{2\beta_0} - \frac{\gamma^{(0)} \beta_1}{2\beta_0^2} 
\Bigg) + O(g_0^4) \Bigg] .
\ee
The term proportional to $g_0^2 (\mu_2) - g_0^2 (\mu_1)$ is analogous to a 
1-loop matching factor, but this term contains $\beta_1$ and $\gamma^{(1)}$,  
which are 2-loop coefficients. So, it is only combining 1-loop renormalization
matching with 2-loop Wilson coefficients that one is doing calculations in a 
consistent way. 

In this Section we have learned that we need perturbative lattice calculations
in order to extract a physical number from Monte Carlo simulations of matrix 
elements of operators (unless one opts for nonperturbative renormalization, 
when this is possible). We will try to explain how to do this kind of 
calculations in the rest of the review.

\section{Discretization}

Lattice calculations are done in euclidean space. A new time coordinate is 
introduced by doing a Wick rotation from Minkowski space to imaginary times:
\be
x_0^E = \ii x_0^M .
\ee
In momentum space this corresponds to $k_0^E = - \ii k_0^M$, so that the 
Fourier transforms continue in euclidean space to be defined through the 
same phase factor. The reason for working with imaginary times is that the 
imaginary factor in front of the Minkowski-space action becomes a minus sign 
in the euclidean functional integral,
\be
\e^{\ii S_M}  \longrightarrow  \e^{- S_E} ,
\ee
and the lattice field theory in euclidean space acquires many analogies with a 
statistical system. The path integral of the particular quantum field theory 
under study becomes the partition function of the corresponding statistical 
system. The transition to imaginary times brings a close connection between 
field theory and statistical physics which has many interesting facets. 
In particular, when the euclidean action is real and bounded from below 
one can see the functional integral as a probability system weighted by 
a Boltzmann distribution $\e^{- S_E}$. It is this feature that allows 
Monte Carlo methods to be used.~\footnote{However, when the action is complex,
like in the case of QCD with a finite baryon number density, this is not 
possible. It is this circumstance that has hampered progress in the lattice 
studies of finite density QCD.} 
Furthermore, on a euclidean lattice of finite volume the path integral 
is naturally well defined, since the measure contains only a finite number 
of variables and the exponential factor, which has a negative exponent, 
gives an absolutely convergent multi-dimensional integral. 
One then can generate configurations with the appropriate probability
distribution and in this way sample the functional integral with Monte Carlo 
techniques. This is the practical basis of Monte Carlo simulations.

From now on we will work in the euclidean space in four dimensions, with 
metric (1,1,1,1), and we will drop all euclidean subscripts from lattice 
quantities, so that $x_0$ is for example the time component after the Wick 
rotation. The Dirac matrices in euclidean space satisfy an anticommutation 
relation with $g_{\mu\nu}$ replaced by $\delta_{\mu\nu}$:
\be
\{\gamma_\mu, \gamma_\nu\} = 2 \delta_{\mu\nu} ,
\ee
and they are all hermitian:
\be
(\gamma_\mu)^\dagger = \gamma_\mu.
\ee
The euclidean $\gamma_5$ matrix is defined by
\be
\gamma_5 = \gamma_0 \gamma_1 \gamma_2 \gamma_3 ,
\ee
is also hermitian, and satisfies $(\gamma_5)^2=1$.
The relation between Dirac matrices in Minkowski and euclidean space is 
\be
\gamma_0^E = \gamma_0^M, \quad \gamma_i^E = -\ii \gamma_i^M .
\ee
This can be inferred from the kinetic term of
the Dirac action in the functional integral:
\be
\exp \Big\{ \ii \overline{\psi} \gamma_\mu^M \partial_\mu^M \psi \Big\}
\longrightarrow
\exp \Big\{ - \overline{\psi} \gamma_\mu^E \partial_\mu^E \psi \Big\} .
\ee
The explicit euclidean Dirac matrices in the chiral representation are 
given in Appendix A.

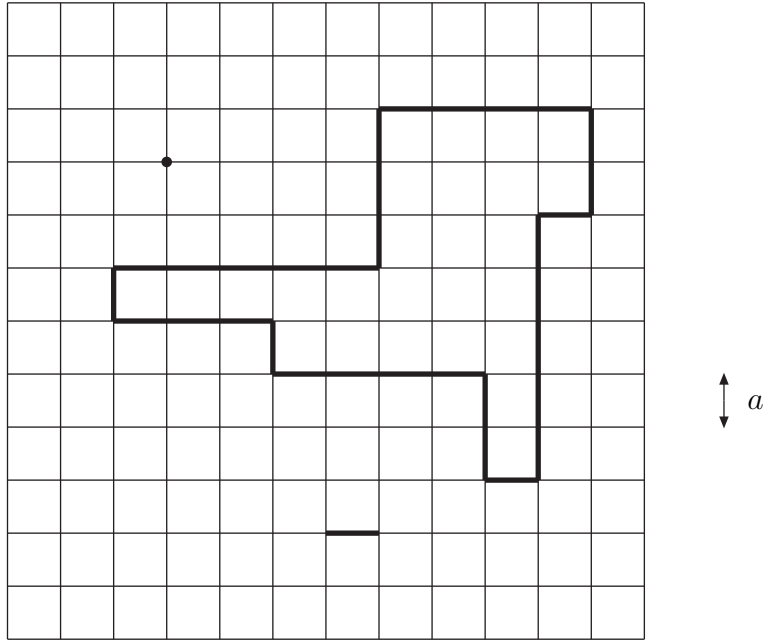
\begin{figure}[t]
\begin{center}
\begin{picture}(300,250)(35,35)
\Line(40,40)(40,280)
\Line(60,40)(60,280)
\Line(80,40)(80,280)
\Line(100,40)(100,280)
\Line(120,40)(120,280)
\Line(140,40)(140,280)
\Line(160,40)(160,280)
\Line(180,40)(180,280)
\Line(200,40)(200,280)
\Line(220,40)(220,280)
\Line(240,40)(240,280)
\Line(260,40)(260,280)
\Line(280,40)(280,280)
\Line(40,40)(280,40)
\Line(40,60)(280,60)
\Line(40,80)(280,80)
\Line(40,100)(280,100)
\Line(40,120)(280,120)
\Line(40,140)(280,140)
\Line(40,160)(280,160)
\Line(40,180)(280,180)
\Line(40,200)(280,200)
\Line(40,220)(280,220)
\Line(40,240)(280,240)
\Line(40,260)(280,260)
\Line(40,280)(280,280)
\Vertex(100,220){2}
\LongArrow(310,130)(310,121)
\LongArrow(310,130)(310,139)
\Text(320,130)[l]{$a$}
\SetWidth{2}
\Line(160,80)(180,80)
\Line(140,140)(220,140)
\Line(220,140)(220,100)
\Line(220,100)(240,100)
\Line(240,100)(240,200)
\Line(240,200)(260,200)
\Line(260,200)(260,240)
\Line(260,240)(180,240)
\Line(180,240)(180,180)
\Line(180,180)(80,180)
\Line(80,180)(80,160)
\Line(80,160)(140,160)
\Line(140,160)(140,140)
\end{picture}
\end{center}
\caption{\small A 2-dimensional projection of a lattice. A site, a link
and a closed loop are also shown.}
\label{fig:lattice}
\end{figure}

We want to construct field theories on a hypercubic lattice. This is a 
discrete subset of the euclidean spacetime, where the sites are denoted 
by $x_\mu=a n_\mu$ (with $n_\mu$ integers). 
We will work in this review only with hypercubic lattices, where the lattice
spacing is the same in all directions. A 2-dimensional projection of such
a (finite) lattice is given in Fig.~\ref{fig:lattice}. 
For convenience we will sometimes omit to indicate the lattice spacing $a$,
that is we will use $a=1$. The missing factors of $a$ can always be reinstated
by doing a naive dimensional counting.

In going from continuum to lattice actions one replaces integrals with sums, 
\be
\int d^4 x \rightarrow a^4 \sum_{x} ,
\ee
where on the right-hand side $x$ means now sites: $x=an$.~\footnote{We use
in general the same symbols for continuum and lattice quantities, hoping that 
this does not cause confusion in the reader. An exception is given by the
lattice derivatives, for which we will use special symbols.}
Lattice actions are then written in terms of sums over lattice sites.
The distance between neighboring sites is $a$, and this minimum distance 
induces a cutoff on the modes in momentum space, so that the lattice spacing 
$a$ acts as an ultraviolet regulator. The range of momenta is thus restricted 
to an interval of range $2\pi/a$, called the first Brillouin zone, and
which can be chosen to be 
\be
B_Z = \Big\{ k : -\frac{\pi}{a} < k_\mu \le \frac{\pi}{a}  \Big\} .
\ee
This is region of the allowed values of $k$, and is the domain of integration
when lattice calculations are made in momentum space. 
For a lattice of finite volume $V = L_0 L_1 L_2 L_3$ the allowed momenta
in the first Brillouin zone become a discrete set, given by 
\be
(k_n)_\mu = \frac{2\pi}{a} \, \frac{n_\mu}{L_\mu} 
\qquad n_\mu= -L_\mu/2 +1, \dots, 0, 1, \dots, L_\mu/2 ,
\ee
and so in principle one deals with sums also in momentum space.
However, as we will see later, when doing perturbation theory one assumes 
to do the calculations in infinite volume and so the sums over the modes 
of the first Brillouin zone become integrals:
\be
\frac{1}{V} \sum_k \longrightarrow 
\int^{\frac{\pi}{a}}_{-\frac{\pi}{a}} \frac{d k_0}{2\pi} \,
\int^{\frac{\pi}{a}}_{-\frac{\pi}{a}} \frac{d k_1}{2\pi} \,
\int^{\frac{\pi}{a}}_{-\frac{\pi}{a}} \frac{d k_2}{2\pi} \,
\int^{\frac{\pi}{a}}_{-\frac{\pi}{a}} \frac{d k_3}{2\pi} .
\ee

The one-sided forward and backward lattice derivatives (also known as right
and left derivatives) can be respectively written as
\bea
\nabla_\mu \psi (x) &=& \frac{\psi(x+a\hat{\mu}) - \psi(x)}{a} , \\
\nabla_\mu^\star \psi (x) &=& \frac{\psi(x) - \psi(x-a\hat{\mu})}{a} ,
\eea
where $\hat{\mu}$ denotes the unit vector in the $\mu$ direction.
It is easy to see that
\bea
(\nabla_\mu )^\dagger &=& -\nabla_\mu^\star , \\
(\nabla_\mu^\star )^\dagger &=& -\nabla_\mu  ,
\eea
that is they are (up to a sign) conjugate to each other. Therefore $\nabla_\mu$
or $\nabla_\mu^\star$ alone cannot be chosen in a lattice theory that is 
supposed to have a hermitian Hamiltonian. In this case one needs their sum,
$\nabla_\mu + \nabla_\mu^\star$, which is anti-hermitian and gives a lattice 
derivative operator extending over the length of two lattice spacings:
\be
\frac{1}{2} \, (\nabla + \nabla^\star)_\mu \psi (x) = 
\frac{\psi(x+a\hat{\mu}) - \psi(x-a\hat{\mu})}{2a} .
\label{eq:symmderiv}
\ee
Note that the second-order differential operator 
$\nabla_\mu\nabla_\mu^\star = \nabla_\mu^\star\nabla_\mu$ is hermitian, 
and when $\mu$ is summed corresponds to the 4-dimensional lattice Laplacian,
\be
\Delta \psi (x) = \sum_\mu \nabla^\star_\mu \nabla_\mu \psi (x)  
= \sum_\mu \frac{\psi(x+a\hat{\mu}) + \psi(x-a\hat{\mu}) -2 \psi(x)}{a^2} .
\ee
It is also useful to know that
\be
\sum_x (\nabla_\mu f(x)) g(x) = - \sum_x f(x) (\nabla_\mu^\star g(x)) ,  
\ee
that is,
\be
\sum_x \Big( f(x+a\hat{\mu}) g(x) - f(x) g(x) \Big) = 
\sum_x \Big( f(x) g(x-a\hat{\mu}) - f(x) g(x) \Big) ,  
\ee
which is valid for an infinite lattice, and also for a finite one if $f$ and 
$g$ are periodic (or their support is smaller than the lattice). The formula 
above corresponds to an integration by parts on the lattice, which in practical
terms amounts to a shift of the summation variable.

There is in general some freedom in the construction of lattice actions. 
For the discretization of continuum actions and operators and the practical 
setting of the corresponding lattice theory many choices are possible. Since
the lattice symmetries are less restrictive than the continuum ones, 
there is more than one possibility in formulating a gauge theory starting 
from a given continuum gauge theory. In particular, one has quite a few 
choices for the exact form of the QCD action on the lattice, depending on what
features are of interest in the studies that one wants to carry out using the 
discretized version. There is not an optimal lattice action to use in all 
cases, and each action has some advantages and disadvantages which weigh 
differently in different contexts. This means that deciding for one action 
instead of another depends on whether chiral symmetry, flavor symmetry, 
locality, or unitarity are more or less relevant to the physical system under 
study. There is a special emphasis on the symmetry properties. 
One also evaluates the convenience of lattice actions by considering 
some balance between costs and gains from the point of view of numerical 
simulations and of perturbation theory. Therefore, the final choice of a 
lattice action depends also on the problem that one wants to study. 

There are many lattice actions which fall in the same universality class,
that is they have the same naive continuum limit, and each of them constitutes 
a different regularization, for finite $a$, of the same physical theory. 
Since every lattice action defines a different regularization scheme,
one needs for each action that is used a new complete set of renormalization 
computations of the type discussed in 
Section~\ref{sec:renormalizationoperators}, 
in order for the results which come out from Monte Carlo simulations 
to be used, interpreted and understood properly. Using different actions 
leads to different numerical results for the matrix elements computed 
in Monte Carlo simulations, and also the values of the renormalization 
factors, and of the $\Lambda$ parameter, depend in general on the lattice 
action chosen. Even the number and type of counterterms required in the 
renormalization of operators can be different in each case. For example, a 
weak operator which is computed on the lattice using the Wilson action, where 
chiral symmetry is broken, needs in general more counterterms to be 
renormalized than when is computed with the overlap action, which is chiral 
invariant. Of course all the various differences among lattice actions arise 
only at the level of finite lattice spacing. The final extrapolations to the 
continuum limit must lead, within errors, to the same physical results.

In the case of QCD there seems to be a lot of room in choosing an action 
for the fermion part, although also the pure gauge action has some popular 
variants (but the plaquette, or Wilson, action has a clear predominance over 
the other ones here, except in particular situations, where for example 
improved gauge actions may be more convenient).
The main features of perturbation theory will be first introduced
in the context of Wilson fermions, which are one of the most widely used
lattice formulations. Then a few other fermion actions will be discussed
along the way, pointing out the differences with the standard perturbation 
theory made with Wilson fermions, which is generally simpler.

\section{Wilson's formulation of lattice QCD}
\label{sec:Wilsonfermions}

One of the most popular lattice formulations of QCD is the one invented by 
Wilson (1974; 1977), which was also the first formulation ever for a lattice 
gauge theory. Its remarkable feature is that it maintains exact gauge 
invariance also at any nonzero values of the lattice spacing.
The discretization of the (euclidean) QCD action for one quark flavor
\be
S = \int d^4 x \Bigg[ \overline{\psi} (x) 
\Big( \Ds + m_{f} \Big) \psi (x) 
+ \frac{1}{2} \Tr \, \Big[ F_{\mu \nu} (x) F_{\mu \nu} (x) \Big] \Bigg]
\ee
that Wilson proposed is the following:
\bea
S_W &=& 
a^4 \sum_{x} \Bigg[ -\frac{1}{2a} \sum_{\mu} \Big[ \overline{\psi} (x) 
( r - \gamma_\mu ) U_\mu (x) \psi (x + a\hat{\mu}) \nonumber \\
&& +\overline{\psi} (x + a\hat{\mu}) ( r + \gamma_\mu ) 
U_\mu^\dagger (x) \psi (x) \Big] + \overline{\psi} (x) \left( m_0 
+ \frac{4 r}{a} \right) \psi (x) \Bigg] \nonumber \\
&& + {\displaystyle\frac{1}{g_0^2}} \, a^4 \sum_{x,\mu \nu}  \Bigg[  
N_c - {\mathrm Re} \, \Tr \left[  U_\mu (x) U_\nu (x + a\hat{\mu}) 
U_\mu^\dagger (x + a\hat{\nu}) U_\nu^\dagger (x) \right] \Bigg] ,
\label{eq:wilsonaction}
\eea
where $x=an$ and $0 < r \le 1$. This action has only nearest-neighbor 
interactions.~\footnote{Other actions can have more complicated interactions, 
like for example overlap fermions 
(see Section~\ref{sec:ginspargwilsonfermions}).}
The derivative in the Dirac operator is 
the symmetric one, Eq.~(\ref{eq:symmderiv}) (with an integration by parts). 
The fields $U_\mu (x)$ live on the links which connect two neighboring lattice
sites, and these variables are naturally defined in the middle point of
a link. Each link carries a direction, so that
\be
U_\mu^{-1} (x) = U_\mu^\dagger (x) = U_{-\mu} (x+a\hat{\mu}) .
\ee
These link variables, which are unitary, are not linear in the gauge potential
$A_\mu (x)$. The fact is that they belong to the group $SU(N_c)$ rather than 
to the corresponding Lie algebra, as would be the case in the continuum. 
The relation of the $U_\mu (x)$ matrices with the gauge fields $A_\mu (x)$, 
the variables which have a direct correspondence with the continuum, 
is then given by
\be
U_\mu (x) = \e^{\displaystyle \ii g_0 a T^{a} A_\mu^a (x)} \ \ \ \ 
(a = 1 , \ldots , N_c^{2} -1 ) ,
\label{eq:ua}
\ee
where the $T^a$ are $SU(N_c)$ matrices in the fundamental representation.

The Wilson action is a gauge-invariant regularization of QCD, and it has exact 
local gauge invariance on the lattice at any finite $a$. The gauge-invariant
construction is done directly on the lattice, extending a discretized version 
of the {\em free} continuum fermionic action. It is not therefore a trivial 
straightforward discretization of the whole gauge-invariant continuum QCD
action, which would recover the gauge invariance only in the continuum limit. 
A naive lattice discretization of the minimal substitution rule 
$\partial_\mu \to D_\mu$ would in fact result in an action that violates gauge
invariance on the lattice, whereas with the choice made by Wilson gauge 
invariance is kept as a symmetry of the theory for any $a$. It is this 
requirement that causes the group variables $U_\mu$ to appear in the action 
instead of the algebra variables $A_\mu$. The lattice gauge transformations 
are given by
\bea
U_\mu (x) & \rightarrow & \Omega (x) \, U_\mu (x) \, 
\Omega^{-1} (x+a\hat{\mu})  \nonumber \\ 
\psi (x) & \rightarrow & \Omega (x) \, \psi (x) \nonumber \\ 
\overline{\psi} (x) & \rightarrow & \overline{\psi} (x) \, \Omega^{-1} (x) , 
\eea
with $\Omega \in SU(N_c)$, and it is easy to see that they leave the 
quark-gluon interaction term in the Wilson action invariant. Note that also in 
the lattice theory the local character of the invariance group is maintained.

This form of local gauge invariance imposes strong constraints on the form 
of the gauge field-strength tensor $F_{\mu\nu}$. Given the above formula for 
the lattice gauge transformations of the $U_\mu$'s, it is easy to see that the 
simplest gauge-invariant object that one can build from the link variables 
involves their path-ordered product. In particular, one obtains a 
gauge-invariant quantity by taking the trace of the product of $U_\mu$'s
on adjoining links forming a closed path, thanks to the unitarity of the 
$U_\mu$'s and the cyclic property of the trace.

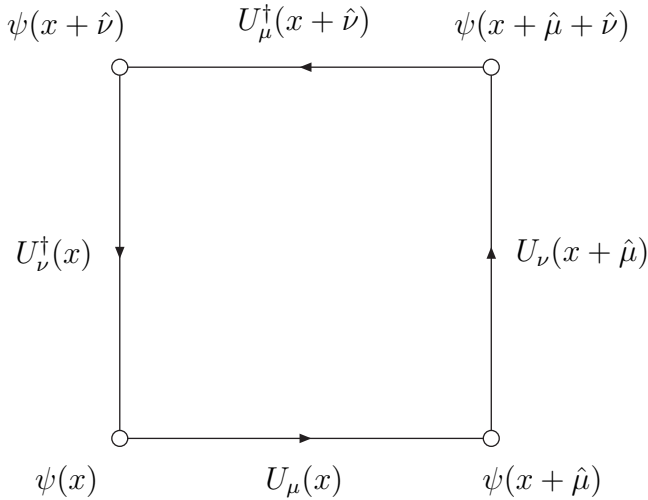
\begin{figure}[t]
\begin{center}
\begin{picture}(200,200)(0,0)
\ArrowLine(30,30)(170,30)
\ArrowLine(170,30)(170,170)
\ArrowLine(170,170)(30,170)
\ArrowLine(30,170)(30,30)
\BCirc(30,30){3}
\BCirc(170,30){3}
\BCirc(170,170){3}
\BCirc(30,170){3}
\Text(10,20)[t]{$\psi (x)$}
\Text(100,20)[t]{$U_\mu (x)$}
\Text(190,20)[t]{$\psi (x + \hat{\mu})$}
\Text(180,100)[l]{$U_\nu (x + \hat{\mu})$}
\Text(190,180)[b]{$\psi (x + \hat{\mu} + \hat{\nu})$}
\Text(100,180)[b]{$U_\mu^\dagger (x + \hat{\nu})$}
\Text(10,180)[b]{$\psi (x + \hat{\nu})$}
\Text(20,100)[r]{$U_\nu^\dagger (x)$}
\end{picture}
\end{center}
\caption{\small The plaquette.}
\label{fig:plaquette}
\end{figure}

The physical theory is a local one, and so in constructing the pure gauge 
action we should direct our attention toward small loops. The simplest lattice 
approximation of $F_{\mu\nu}$ is then the product of the links of an 
elementary square, called ``plaquette'':
\be
P_{\mu \nu}(x) = U_\mu (x) U_\nu (x + a\hat{\mu}) 
U_\mu^\dagger (x + a\hat{\nu}) U_\nu^\dagger (x) ,
\label{eq:uuuu}
\ee
which is shown in Fig.~\ref{fig:plaquette}. This form is not surprising, given
that the gauge field-strength tensor is in differential geometry the curvature
of the metric tensor. One could also take larger closed loops, but this 
minimal choice gives better signal-to-noise ratios, and for the standard 
Wilson action the trace of the plaquette is then used.~\footnote{Other 
actions which use different approximations for $F_{\mu\nu}$, and which have 
the aim of reducing the discretization errors, are discussed in 
Section~\ref{sec:improvedgluons}.} This is the expression appearing in the 
last line of Eq.~(\ref{eq:wilsonaction}).
The factor $N_c$ can be understood by looking at the expansion of the 
plaquette Eq.~(\ref{eq:uuuu}) in powers of $a$, which is 
\be
P_{\mu \nu}(x) = 1 + \ii g_0 a^2 F_{\mu \nu}(x) 
-\frac{1}{2} g_0^2 a^4 F^2_{\mu \nu}(x) 
+ \ii a^3 G_{\mu \nu}(x) + \ii a^4 H_{\mu \nu}(x) ,
\ee
with $G$ and $H$ hermitian fields.~\footnote{This expansion can be derived 
using 
\be
A_\mu (x+a\hat{\nu}) = A_\mu (x) + a \partial_\nu A_\mu (x) + \cdots .
\ee
}
We have then 
\be
{\mathrm Re} \, \Tr P_{\mu \nu}(x) = N_c 
-\frac{1}{2} g_0^2 a^4 \, \Tr F^2_{\mu \nu}(x) +O(a^6) ,
\ee
where we have used $\Tr F_{\mu\nu}=0$, because the trace of the generators 
is zero. The plaquette action then has the right continuum limit, and the 
first corrections to the continuum pure gauge action are of order $a^2$.
These are irrelevant terms, which are zero in the continuum limit, but they 
are important for determining the rate of convergence to the continuum physics.
It can also be shown that in the fermionic part of the action the corrections 
with respect to the continuum limit are of order $a$. In 
Section~\ref{sec:improvement} we will see how to modify the fermion action 
in order to decrease the error on the fermionic part to order $a^2$.

The plaquette action is also often written as
\be
\beta \cdot a^4 \sum_P \Bigg( 1 - \frac{1}{N_c} {\mathrm Re} \, \Tr U_P \Bigg) 
\label{eq:plaq_action} ,
\ee
where $U_P$ is given in Eq.~(\ref{eq:uuuu}), and in numerical simulations of 
QCD the coefficient in front of the action is
\be
\beta = 
{\displaystyle \frac{2 N_c}{g_0^2}} = {\displaystyle \frac{6}{g_0^2}} .
\ee
The factor two comes out because here one takes the sum over the oriented 
plaquettes, that is a sum over ordered indices (for example, $\mu > \nu$), 
while in Eq.~(\ref{eq:wilsonaction}) the sum over $\mu$ and $\nu$ is free.

In the weak coupling regime, where $g_0$ is small, the functional integral
is dominated by the configurations which are near the trivial field 
configuration $U_\mu (x) = 1$. Perturbation theory is then a saddle-point 
expansion around the classical vacuum configurations, and the relevant degrees
of freedom are given by the components of the gauge potential, $A^a_\mu (x)$. 
Thus, while the fundamental gauge variables for the Monte Carlo simulations 
are the $U_\mu$'s and the action is relatively simple when expressed 
in terms of these variables, in perturbation theory the true dynamical 
variables are the $A_\mu$'s. This mismatch causes a good part of the
complications of lattice perturbation theory. In fact, when the Wilson action 
is written in terms of the $A_\mu$'s, using 
$U_\mu = 1 + \ii g_0 a A_\mu - g_0^2 a^2 A_\mu^2 + \cdots$,  it becomes much 
more complicated. Moreover, it consists of an infinite number of terms, which 
give rise to an infinite number of interaction vertices. Fortunately, only a 
finite number of vertices is needed at any given order in $g_0$.

All vertices except a few of them are ``irrelevant'', that is they are 
proportional to some positive power of the lattice spacing $a$ and so they
vanish in the naive continuum limit. However, this does not mean that they
can be thrown away when doing perturbation theory. Quite on the contrary, they 
usually contribute to correlation functions in the continuum limit through 
divergent ($\sim 1/a^n$) loop corrections. These irrelevant vertices are 
indeed important in many cases, contributing to the mass, coupling and 
wave-function renormalizations~\cite{Sharatchandra:af}. All these vertices are
in fact necessary to ensure the gauge invariance of the physical amplitudes. 
Only when they are included can gauge-invariant Ward Identities be 
constructed, and the renormalizability of the lattice theory can be proven.

\begin{figure}[t]
\begin{center}
\begin{picture}(400,150)(0,0)
\Gluon(0,100)(20,100){4}{2}
\Gluon(60,100)(80,100){4}{2}
\GlueArc(40,100)(20,0,360){4}{16}
\Gluon(100,100)(120,100){4}{2}
\Gluon(160,100)(180,100){4}{2}
\ArrowArc(140,100)(20,-90,270)
\Gluon(200,100)(220,100){4}{2}
\Gluon(260,100)(280,100){4}{2}
\DashArrowArc(240,100)(20,-90,270){4}
\Gluon(0,10)(80,10){4}{8}
\GlueArc(40,34)(20,-90,270){4}{16}
\Gluon(100,10)(180,10){4}{8}
\ArrowArc(140,33)(20,-90,270)
\Gluon(200,10)(280,10){4}{8}
\DashArrowArc(240,33)(20,-90,270){4}
\Gluon(300,10)(380,10){4}{8}
\SetWidth{1}
\Line(330,0)(350,20)
\Line(330,20)(350,0)
\end{picture}
\end{center}
\caption{\small Diagrams for the self-energy of the gluon on the lattice. 
The diagrams on the upper row have a continuum analog, while the diagrams 
on the lower row are a pure lattice artifact. They are however necessary 
to maintain the gauge invariance of the lattice theory, and are important 
for its renormalizability.}
\label{fig:gluon_self}
\end{figure}
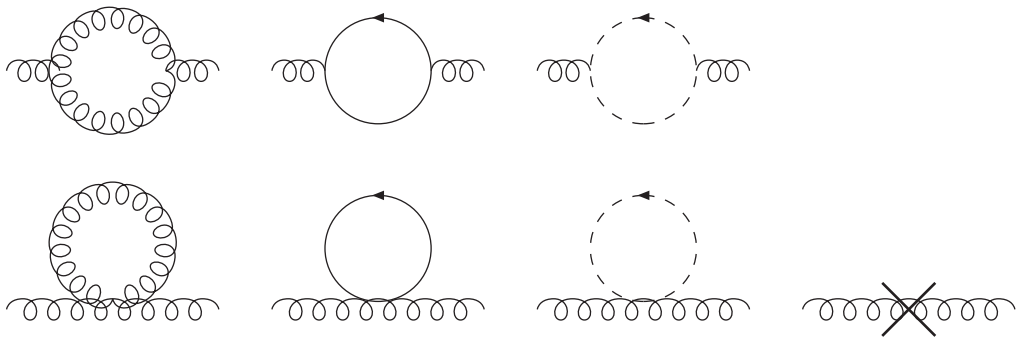

An example of this fact is given by the diagrams contributing to the
gluon self-energy at one loop (Fig.~\ref{fig:gluon_self}). 
If one would only consider the diagrams on the upper row, that is the ones 
that would also exist in the continuum, the lattice results would contain 
an unphysical $1/(am)^2$ divergence. This divergence is canceled away only 
when the results of the diagrams on the lower row are added, that is only 
when gauge invariance is fully restored. Notice that for this to happen
also the measure counterterm is needed (see Section~\ref{sec:measure}).
In a similar way, terms of the type $p_\mu^2 \delta_{\mu\nu}$, which are not 
Lorentz covariant and are often present in the individual diagrams, disappear 
only after all diagrams have been considered and summed.

From what we have seen so far, we can understand that a lattice regularization
does not just amount to introducing in the theory a momentum cutoff. In fact, 
it is a more complicated regularization than just setting a nonzero lattice 
spacing, because one has also to provide a lattice action. Different actions 
define different lattice regularizations. Because of the particular form
of lattice actions, the Feynman rules are much more complicated that 
in the continuum, and in the case of gauge theories new interaction vertices
appear which have no analog in the continuum.
The structure of lattice integrals is also completely different, due to the 
overall periodicity which causes the appearance of trigonometric functions. 
The lattice integrands are then given by rational functions of trigonometric 
expressions. 

At the end of the day, lattice perturbation theory is much more complicated 
than continuum perturbation theory: there are more fundamental vertices and 
more diagrams, and these propagators and vertices, with which one builds the 
Feynman diagrams, are more complicated on the lattice than they are in the 
continuum, which can lead to expressions containing a huge number of terms. 
Finally, one has also to evaluate more complicated integrals. Lattice 
perturbative calculations are thus rather involved. As a consequence, for the 
calculation of all but the simplest matrix elements computer codes have to be 
used (see Section~\ref{sec:computercodes}).

Matrix elements computed in euclidean space do not always correspond
to the analytic continuation of matrix elements of a physical theory 
in Minkowski space. For this to happen, the lattice action has to satisfy 
a property known as reflection positivity, which involves time reflections 
and complex conjugations (roughly speaking is the analog of hermitian 
conjugation in Minkowski space).
In this case the reconstruction theorem of Osterwalder and Schrader 
(1973; 1975) says that it is possible to reconstruct a Hilbert space 
in Minkowski space in the usual way starting from the lattice theory.

The Wilson action with $r=1$ is reflection positive, and therefore 
corresponds to a well-defined physical theory in Minkowski 
space~\cite{Luscher:1976ms,Creutz:1986ky}.
For $r \neq 1$ instead the lattice theory contains additional time doublers, 
which disappear in the continuum limit.~\footnote{This is at variance with the 
doublers which appear for $r=0$ (naive fermion action), which do not disappear
in the continuum limit, as we will see in Section~\ref{sec:aspects}.} 
In the following we will only work with $r=1$. 
This is what is usually meant for Wilson action.

\subsection{Fourier transforms}

To perform calculations of Feynman diagrams in momentum space (the main topic 
of this review) we need to define the Fourier transforms on the lattice. 
They are given in infinite volume (which is the standard setting in 
perturbation theory) by
\bea
\psi (x) &=& \int^{\frac{\pi}{a}}_{-\frac{\pi}{a}} \frac{d^4 p}{(2\pi)^4} \,
\e^{\ii xp} \, \psi (p) , \nonumber \\
\overline{\psi} (x) &=& 
\int^{\frac{\pi}{a}}_{-\frac{\pi}{a}} \frac{d^4 p}{(2\pi)^4} \,
\e^{-\ii xp} \, \overline{\psi} (p) , \nonumber  \\
A_\mu (x) &=& \int^{\frac{\pi}{a}}_{-\frac{\pi}{a}} \frac{d^4 k}{(2\pi)^4} \,
\e^{\ii (x+a{\hat \mu}/2)k} \, A_\mu (k) , 
\label{eq:fourier}
\eea
where with abuse of notation we indicate $\psi (x)$ and its Fourier 
transformed function with the same symbol. The inverse Fourier transforms 
are given by
\bea
\psi (p) &=& a^4 \, \sum_x \, \e^{-\ii xp} \, \psi (x) \nonumber \\
\overline{\psi} (p) &=& a^4 \, \sum_x \, \e^{\ii xp} \, \overline{\psi} (x) 
\nonumber \\
A_\mu (k) &=& a^4 \, \sum_x \, \e^{-\ii (x+a{\hat \mu}/2)k} \, A_\mu (x) ,
\eea
and this means that
\be
\delta^{(4)} (p) =\frac{a^4}{(2\pi)^4} \, \sum_x \e^{-\ii xp} .
\ee
This lattice delta function is zero except at $p=2\pi n$. 
The Kronecker delta in position space is 
\be
\delta_{xy} = a^4 \int_{-\frac{\pi}{a}}^{\frac{\pi}{a}}
\frac{d^4 p}{(2\pi)^4} \, \e^{\ii (x-y) p} .
\ee
Of course on a lattice of finite volume the allowed momenta are a discrete
set. However, in perturbation theory we will always consider the limit of 
infinite volume.

Notice that the Fourier transform of $A_\mu(x)$ is taken at the point 
$x+a{\hat \mu}/2$, that is in the middle of the link. This comes out naturally
from its definition. This choice turns out to be quite important also for the 
general economy of the calculations, as we can see from the following example.

Let us consider the gauge interaction of the quarks in the Wilson 
action at first order in the gauge coupling, which gives rise to the 
quark-quark-gluon vertex. Going to momentum space we have
\bea
S_{qqg} &=&-\frac{\ii  g_0}{2} a^4 \sum_{x,\mu} \Big( \overline{\psi} (x) 
( r - \gamma_\mu ) A_\mu (x) \psi (x + a\hat{\mu}) 
-\overline{\psi} (x + a\hat{\mu}) ( r + \gamma_\mu ) 
A_\mu (x) \psi (x) \Big)  \nonumber \\
&=& -\frac{\ii  g_0}{2} a^4 \sum_{x,\mu}
\int^{\frac{\pi}{a}}_{-\frac{\pi}{a}} \frac{d^4 p}{(2\pi)^4} \,
\int^{\frac{\pi}{a}}_{-\frac{\pi}{a}} \frac{d^4 k}{(2\pi)^4} \, 
\int^{\frac{\pi}{a}}_{-\frac{\pi}{a}} \frac{d^4 p'}{(2\pi)^4} \,
\e^{\ii x(p+k-p')} \,
\e^{\ii ak_\mu/2} \\
&& \times \Big( \overline{\psi} (p') 
( r - \gamma_\mu ) A_\mu (k) \psi (p) \e^{\ii ap_\mu} 
-\overline{\psi} (p') \e^{-\ii ap'_\mu} ( r + \gamma_\mu ) 
A_\mu (k) \psi (p) \Big) \nonumber \\
&=& \frac{\ii  g_0}{2} \sum_\mu
\int^{\frac{\pi}{a}}_{-\frac{\pi}{a}} \frac{d^4 p}{(2\pi)^4} \,
\int^{\frac{\pi}{a}}_{-\frac{\pi}{a}} \frac{d^4 k}{(2\pi)^4} \, 
\int^{\frac{\pi}{a}}_{-\frac{\pi}{a}} \frac{d^4 p'}{(2\pi)^4} \,
(2\pi)^4 \delta^{(4)} (p+k-p') \, \e^{\ii ak_\mu/2} \nonumber \\
&& \times \Big( 
\overline{\psi} (p') \gamma_\mu A_\mu (k) \psi (p)
(\e^{\ii a p_\mu} + \e^{-\ii a p'_\mu} )
+ r \overline{\psi} (p') A_\mu (k) \psi (p)
(-\e^{\ii a p_\mu} + \e^{-\ii a p'_\mu} ) \Big) \nonumber \\
&=& \frac{\ii  g_0}{2} \sum_\mu
\int^{\frac{\pi}{a}}_{-\frac{\pi}{a}} \frac{d^4 p}{(2\pi)^4} \,
\int^{\frac{\pi}{a}}_{-\frac{\pi}{a}} \frac{d^4 k}{(2\pi)^4} \, 
\int^{\frac{\pi}{a}}_{-\frac{\pi}{a}} \frac{d^4 p'}{(2\pi)^4} \,
(2\pi)^4 \delta^{(4)} (p+k-p') \, \e^{\ii ak_\mu/2} \nonumber \\
&& \times \Bigg( 
\overline{\psi} (p') \gamma_\mu A_\mu (k) \psi (p) \,
\e^{\ii ap_\mu/2} \e^{-\ii ap'_\mu/2} \cdot
2 \cos \frac{a (p+p')_\mu}{2} 
\nonumber \\ && \quad
+ r \overline{\psi} (p') A_\mu (k) \psi (p) \,
\e^{\ii ap_\mu/2} \e^{-\ii ap'_\mu/2} \cdot 
( -2 \ii ) \sin \frac{a (p+p')_\mu}{2} \Bigg) .
\eea
We can notice at this point that all exponential phases cancel with each 
other, because of the $\delta $ function expressing the momentum conservation 
at the vertex (where $p'=p+k$). We are then left with
\bea
S_{qqg} &=& 
\int^{\frac{\pi}{a}}_{-\frac{\pi}{a}} \frac{d^4 p}{(2\pi)^4} \,
\int^{\frac{\pi}{a}}_{-\frac{\pi}{a}} \frac{d^4 k}{(2\pi)^4} \, 
\int^{\frac{\pi}{a}}_{-\frac{\pi}{a}} \frac{d^4 p'}{(2\pi)^4} \, 
(2\pi)^4 \delta^{(4)} (p+k-p') \,  
\nonumber \\
&& \times \, \ii g_0 \sum_\mu \overline{\psi} (p') \Bigg(
\gamma_\mu \cos \frac{a (p+p')_\mu}{2} - \ii  r \sin \frac{a (p+p')_\mu}{2}
\Bigg) A_\mu (k) \psi (p) \label{eq:qqgft},
\eea
which gives us the lattice Feynman rule for this vertex. It is easy to see
that in the continuum limit this Wilson vertex reduces to the familiar vertex 
of QCD,
\be
\int_{-\infty}^\infty \frac{d^4 p}{(2\pi)^4} \, 
\int_{-\infty}^\infty \frac{d^4 k}{(2\pi)^4} \,
\int_{-\infty}^\infty \frac{d^4 p'}{(2\pi)^4} \,
(2\pi)^4 \delta^{(4)} (p+k-p') \,  
\cdot \ii g_0 \sum_\mu \overline{\psi} (p') \gamma_\mu A_\mu (k) \psi (p).
\ee
Notice that had we chosen for the Fourier transform of the gauge potential
the expression
\be
A_\mu (x) = \int^{\frac{\pi}{a}}_{-\frac{\pi}{a}} \frac{d^4 k}{(2\pi)^4} \,
\e^{\ii xk} \, A_\mu (k)
\ee
the exponential phases would not have canceled, and the
$\e^{\ii a p_\mu/2} \e^{-\ii a p'_\mu/2}$ terms would still be present 
in the final expression of the vertex. This is a general feature of 
lattice perturbation theory: if one uses the Fourier transforms 
in Eq.~(\ref{eq:fourier}), all terms of the type $\e^{\ii a k_\mu/2}$ coming 
from the various gluons exactly combine to cancel all other phases floating 
around, and then only sine and cosine functions remain in the Feynman 
rules in momentum space. This cancellation becomes rather convenient in the 
case of complicated vertices containing a large number of gluons.

We are now going to give the explicit expressions for the propagators and 
for the vertices of order $g_0$ and $g_0^2$ of the Wilson action, which is 
all what is needed for 1-loop calculations. In the following we will not 
explicitly write the $\delta$ function of the momenta present in each vertex 
and propagator. In our conventions for the vertices, all gluon lines are 
entering, and when there are two quarks or ghosts one of them is entering 
and the other one is exiting.

\subsection{Pure gauge action}

As we have seen, in the Wilson formulation gauge invariance is imposed 
directly on the free fermion lattice action, so that the group elements 
$U_\mu (x)$ appear instead of the algebra elements $A_\mu (x)$, which are
the fundamental perturbative variables. To derive the gluon vertices from 
the pure gauge action, one has then to expand the $U_\mu$'s in the plaquette
in terms of the $A_\mu$'s. As a consequence, an infinite number of interaction
vertices are generated, which express the self-interaction of $n$ gluons, 
with any $n$. Since the power of the coupling which appears in these vertices 
grows with the number of gluons, only a finite number of them is needed at any
given order in $g_0$.

In lattice QCD the $A_\mu$'s are also nontrivial color matrices, and
therefore they do not commute with each other. The expansion of the plaquette 
in terms of the $A_\mu$'s can be carried out using the 
Baker-Campbell-Hausdorff formula
\be
\e^A \e^B  = \exp \Bigg\{A+B+\frac{1}{2} \Big[A,B\Big] 
    + \frac{1}{12} \Big[A-B,\Big[A,B\Big]\Big]
    + \frac{1}{24} \Big[\Big[A,\Big[A,B\Big],B\Big]\Big] + \cdots \Bigg\} .
\ee
Since the color matrices $T^a$ are traceless and are closed under commutation,
the exponent in the expansion of the plaquette obtained using the 
Baker-Campbell-Hausdorff formula is also traceless, so that the knowledge of
the terms of this expansion which are cubic in $A_\mu^a$ is sufficient to 
calculate all vertices with a maximum of four gluons, that is to order 
$g_0^2$, which is all that is needed for 1-loop calculations.~\footnote{For 
computing vertices of higher order it is useful to know that the 
Baker-Campbell-Hausdorff formula can be written as
\be
\e^A \e^B  = \exp \Big\{\sum_{n=1}^\infty C_n(A,B) \Big\} ,
\ee
where the $C_n$'s can be determined recursively:
\bea
C_{n+1} (A,B) &=& \frac{1}{2(n+1)} \Big[A-B,C_n(A,B)\Big] \\ 
&& +\sum_{\stackrel{\scriptstyle p\ge 1}{2p \le n}} \frac{B_{2p}}{(2p)!(n+1)}
 \sum_{\stackrel{\scriptstyle m_1, \dots, m_{2p} > 0}{m_1+ \dots + m_{2p} = n}}
 \Big[C_{m_1} (A,B), \Big[\dots, \Big[C_{m_{2p}} (A,B), A+B\Big] 
 \cdots \Big]\Big], \nonumber
\eea
with $C_1 (A,B) = A+B$, and $B_{2p}$ a Bernoulli number.
The Bernoulli numbers $B_i$ are defined by 
\be
\frac{x}{\e^x-1} = 1 - \frac{x}{2} + \frac{B_1x^2}{2!} - \frac{B_2x^4}{4!}
 + \frac{B_3x^6}{6!} -\cdots, \qquad \qquad |x| < 2\pi .
\ee
The first few Bernoulli numbers are: $B_1=1/6$, $B_2=1/30$, $B_3=1/42$, 
$B_4=1/30$, $B_5=5/66$, $B_6=691/2730$, $B_7=7/6$.}

\begin{figure}[t]
\begin{center}
\begin{picture}(470,390)(0,0)
\Text(10,80)[l]{\Large $O(g_0^2):$}

\Gluon(80,115)(115,80){-4}{6}
\Gluon(115,80)(150,115){-4}{6}
\Gluon(80,45)(115,80){4}{6}
\Gluon(115,80)(150,45){4}{6}
\Text(80,125)[t]{$p$}
\Text(150,125)[t]{$q$}
\Text(150,35)[b]{$r$}
\Text(80,35)[b]{$s$}
\Text(80,97)[t]{$a$}
\Text(150,100)[t]{$b$}
\Text(150,60)[b]{$c$}
\Text(80,60)[b]{$d$}
\Text(104,115)[t]{$\mu$}
\Text(130,115)[t]{$\nu$}
\Text(130,45)[b]{$\lambda$}
\Text(100,43)[b]{$\rho$}

\Gluon(185,80)(255,80){4}{8}
\Text(235,95)[b]{$k$}
\Text(175,70)[l]{$a$}
\Text(265,70)[r]{$b$}
\Text(175,90)[l]{$\mu$}
\Text(265,90)[r]{$\nu$}

\DashArrowLine(290,45)(325,80){4}
\DashArrowLine(325,80)(360,45){4}
\Gluon(290,115)(325,80){-4}{6}
\Gluon(325,80)(360,115){-4}{6}
\Text(290,35)[b]{$p_1$}
\Text(360,35)[b]{$p_2$}
\Text(290,97)[t]{$a$}
\Text(360,100)[t]{$b$}
\Text(360,60)[b]{$c$}
\Text(290,60)[b]{$d$}
\Text(314,115)[t]{$\mu$}
\Text(340,115)[t]{$\nu$}

\ArrowLine(395,45)(430,80)
\ArrowLine(430,80)(465,45)
\Gluon(395,115)(430,80){-4}{6}
\Gluon(430,80)(465,115){-4}{6}
\Text(395,35)[b]{$p_1$}
\Text(465,35)[b]{$p_2$}
\Text(395,97)[t]{$a$}
\Text(465,100)[t]{$b$}
\Text(465,60)[b]{$c$}
\Text(395,60)[b]{$d$}
\Text(419,115)[t]{$\mu$}
\Text(445,115)[t]{$\nu$}

\Text(10,240)[l]{\Large $O(g_0):$}

\Gluon(80,205)(115,240){4}{6}
\Gluon(115,240)(150,205){4}{6}
\Gluon(115,240)(115,275){4}{5}
\Text(115,290)[t]{$p$}
\Text(150,195)[b]{$q$}
\Text(80,195)[b]{$r$}
\Text(100,265)[t]{$a$}
\Text(150,220)[b]{$b$}
\Text(80,220)[b]{$c$}
\Text(130,265)[t]{$\mu$}
\Text(130,205)[b]{$\nu$}
\Text(100,205)[b]{$\lambda$}

\DashArrowLine(290,205)(325,240){4}
\DashArrowLine(325,240)(360,205){4}
\Gluon(325,240)(325,275){4}{5}
\Text(290,195)[b]{$p_1$}
\Text(360,195)[b]{$p_2$}
\Text(310,265)[t]{$a$}
\Text(360,220)[b]{$b$}
\Text(290,220)[b]{$c$}
\Text(340,265)[t]{$\mu$}

\ArrowLine(395,205)(430,240)
\ArrowLine(430,240)(465,205)
\Gluon(430,240)(430,275){4}{5}
\Text(395,195)[b]{$p_1$}
\Text(465,195)[b]{$p_2$}
\Text(415,265)[t]{$a$}
\Text(465,220)[b]{$b$}
\Text(395,220)[b]{$c$}
\Text(445,265)[t]{$\mu$}

\Text(10,365)[l]{\Large $O(1):$}

\Gluon(80,365)(150,365){4}{10}
\Text(130,380)[b]{$k$}
\Text(70,355)[l]{$a$}
\Text(160,355)[r]{$b$}
\Text(70,375)[l]{$\mu$}
\Text(160,375)[r]{$\nu$}

\DashArrowLine(290,365)(360,365){4}
\Text(340,380)[b]{$k$}
\Text(280,355)[l]{$a$}
\Text(370,355)[r]{$b$}

\ArrowLine(395,365)(465,365)
\Text(445,380)[b]{$k$}
\Text(385,355)[l]{$a$}
\Text(475,355)[r]{$b$}

\Text(115,0)[b]{Eq.~(\ref{eq:gggg})}
\Text(220,0)[b]{Eq.~(\ref{eq:meas})}
\Text(325,0)[b]{Eq.~(\ref{eq:ccgg})}
\Text(430,0)[b]{Eq.~(\ref{eq:qqgg})}
\Text(115,160)[b]{Eq.~(\ref{eq:ggg})}
\Text(325,160)[b]{Eq.~(\ref{eq:ccg})}
\Text(430,160)[b]{Eq.~(\ref{eq:qqg})}
\Text(115,320)[b]{Eq.~(\ref{eq:gg})}
\Text(325,320)[b]{Eq.~(\ref{eq:cc})}
\Text(430,320)[b]{Eq.~(\ref{eq:qq})}
\SetWidth{1}
\Line(211,71)(229,89)
\Line(229,71)(211,89)
\end{picture}
\end{center}
\caption{\small Propagators and vertices sufficient for 1-loop calculations 
in lattice QCD.}
\label{fig:prop_vert}
\end{figure}

Looking at Eq.~(\ref{eq:ua}), we see that the entries of the matrices 
$ag_0A_\mu (x)$ are angular variables, which thus assume values between zero 
and $2\pi$. In the functional integral the range of integration of the fields 
$A_\mu^a(x)$ is extended to infinity, so that it is possible to work with 
Gaussian integrals in the zeroth-order computations. It is only after 
$A_\mu^a(x)$ has been decompactified that the propagators and the correlation 
functions of operators can then be computed.

The propagators come as usual from the inverse of the quadratic part of 
the action. For the full expression of the gluon propagator we have to wait 
until gauge fixing is implemented, and we will report it later. 
The 3-gluon vertex is: 
\bea
W^{abc}_{\mu\nu\lambda} (p,q,r) &=& - \ii  g_0 \, f^{abc} \, \frac{2}{a} 
    \Bigg\{\delta_{\mu\nu} \, \sin \frac{a (p-q)_\lambda}{2} 
    \, \cos \frac{a r_\mu}{2} 
\label{eq:ggg} \\ &&
   +\delta_{\nu\lambda} \, \sin \frac{a (q-r)_\mu}{2} 
    \, \cos \frac{a p_\nu}{2}
   +\delta_{\lambda\mu} \, \sin \frac{a (r-p)_\nu}{2} 
    \, \cos \frac{a q_\lambda}{2} \Bigg\} \nonumber ,
\eea
where $p+q+r=0$, all momenta are entering and are assigned clockwise
(see Fig.~\ref{fig:prop_vert}).
This lattice vertex for $a \to 0$ reduces to the continuum vertex 
\be
- \ii  g_0 \, f^{abc} \, \Big\{
    \delta_{\mu\nu} \, (p-q)_\lambda
   +\delta_{\nu\lambda} \, (q-r)_\mu
   +\delta_{\lambda\mu} \, (r-p)_\nu \Big\} .
\ee

It is useful from now on to introduce the shorthand notation
\be
\widehat{a k}_\mu = \frac{2}{a} \sin \frac{a k_\mu}{2}, 
\ee
especially for writing the 4-gluon vertex, which is quite complicated.
It is given by~\cite{Rothe:kp}:
\bea
W^{abcd}_{\mu\nu\lambda\rho} (p,q,r,s) &=& -g_0^2 
\Bigg\{ \sum_e f_{abe} f_{cde} 
\Bigg[ \delta_{\mu\lambda} \delta_{\nu\rho} \Bigg( 
\cos \frac{a (q-s)_\mu}{2} \cos \frac{a (k-r)_\nu}{2}
-\frac{a^4}{12} \, 
\widehat{k}_\nu \widehat{q}_\mu \widehat{r}_\nu \widehat{s}_\mu \Bigg)
\nonumber \\ &&
-\delta_{\mu\rho} \delta_{\nu\lambda} \Bigg(
\cos \frac{a (q-r)_\mu}{2} \cos \frac{a (k-s)_\nu}{2}
-\frac{a^4}{12} \, 
\widehat{k}_\nu \widehat{q}_\mu \widehat{r}_\mu \widehat{s}_\nu \Bigg)
\nonumber \\ &&
+\frac{1}{6} 
\delta_{\nu\lambda} \delta_{\nu\rho} a^2 \, 
(\widehat{s-r})_\mu \, \widehat{k}_\nu \, \cos \frac{a q_\mu}{2}
-\frac{1}{6}
\delta_{\mu\lambda} \delta_{\mu\rho} a^2 \, 
(\widehat{s-r})_\nu \, \widehat{q}_\mu \, \cos \frac{a k_\nu}{2}
\nonumber \\ &&
+\frac{1}{6}
\delta_{\mu\nu} \delta_{\mu\rho} a^2 \, 
(\widehat{q-k})_\lambda \, \widehat{r}_\rho \, \cos \frac{a s_\lambda}{2}
-\frac{1}{6}
\delta_{\mu\nu} \delta_{\mu\lambda} a^2 \, 
(\widehat{q-k})_\rho \, \widehat{s}_\lambda \, \cos \frac{a r_\rho}{2}
\nonumber \\ &&
+\frac{1}{12}
\delta_{\mu\nu} \delta_{\mu\lambda} \delta_{\mu\rho} 
a^2 \sum_\sigma (\widehat{q-k})_\sigma (\widehat{s-r})_\sigma
\Bigg]
\nonumber \\ &&
+( b \leftrightarrow c , \nu \leftrightarrow \lambda , q \leftrightarrow r)
+( b \leftrightarrow d , \nu \leftrightarrow \rho , q \leftrightarrow s)
\Bigg\}
\nonumber \\ &&
+\frac{g_0^2}{12} \, a^4 \Bigg\{
\frac{2}{3} ( \delta_{ab} \delta_{cd} + \delta_{ac} \delta_{bd} 
  + \delta_{ad} \delta_{bc})
+\sum_e ( \delta_{abe} \delta_{cde} + \delta_{ace} \delta_{bde} 
  + \delta_{ade} \delta_{bce}) \Bigg\}
\nonumber \\ &&
\times
\Bigg\{ \delta_{\mu\nu} \delta_{\mu\lambda} \delta_{\mu\rho} \sum_\sigma
\widehat{k}_\sigma \widehat{q}_\sigma \widehat{r}_\sigma \widehat{s}_\sigma
-\delta_{\mu\nu} \delta_{\mu\lambda} \, 
\widehat{k}_\rho \widehat{q}_\rho \widehat{r}_\rho \widehat{s}_\mu
\nonumber \\ &&
-\delta_{\mu\nu} \delta_{\mu\rho} \,
\widehat{k}_\lambda \widehat{q}_\lambda \widehat{s}_\lambda \widehat{r}_\mu
-\delta_{\mu\lambda} \delta_{\mu\rho} \,
\widehat{k}_\nu \widehat{r}_\nu \widehat{s}_\nu \widehat{q}_\mu
-\delta_{\nu\lambda} \delta_{\nu\rho} \,
\widehat{q}_\mu \widehat{r}_\mu \widehat{s}_\mu \widehat{k}_\nu
\nonumber \\ &&
+\delta_{\mu\nu} \delta_{\lambda\rho} \,
\widehat{k}_\lambda \widehat{q}_\lambda \widehat{r}_\mu \widehat{s}_\mu
+\delta_{\mu\lambda} \delta_{\nu\rho} \,
\widehat{k}_\nu \widehat{r}_\nu \widehat{q}_\mu \widehat{s}_\mu
+\delta_{\mu\rho} \delta_{\nu\lambda} \,
\widehat{k}_\nu \widehat{s}_\nu \widehat{q}_\mu \widehat{r}_\mu
\Bigg\} .
\label{eq:gggg} 
\eea
In the continuum limit this expression becomes the four-gluon vertex
of continuum QCD.~\footnote{It is interesting to note that also in lattice QED
there are vertices in which the gauge particles are self-interacting, but 
of course they have to be irrelevant in the continuum limit. The lowest-order 
vertex in pure gauge contains four lattice photons coming from the 
$a^4 g_0^2 F_{\mu\nu}^4$ term of the expansion of the lattice QED action
\be
\frac{1}{2 a^4 g_0^2} \cdot a^4 
\sum_x \sum_{\mu\nu} \Big( 1 -\cos (a^2 g_0 F_{\mu\nu} (x)) \Big) ,
\ee
where
\be
F_{\mu\nu} (x) = \nabla_\mu A_\nu (x) - \nabla_\nu A_\mu (x) .
\ee
This vertex contains four derivatives, it is of order $a^4$ and it disappears 
in the continuum limit.} 

The vertices containing five or more gluons are at least of order $g_0^3$,
and thus they are not necessary for calculations at one loop. To my knowledge,
an explicit expression for the five-gluon vertex has not yet been given in the 
literature. General algorithms for the automated calculation of higher-order 
vertices (for a given configuration of external momenta) have been reported 
in~\cite{Luscher:1985wf}.

For nonabelian gauge theories, the calculation of the pure gauge part of 
course does not end here. One has still to consider the gauge integration 
measure, which gives a $1/a^2$ mass counterterm at order $g_0^2$ and an 
infinite number of measure vertices of order $g_0^3$ and higher, as well as  
the Faddeev-Popov procedure on the lattice, with which the Feynman rules for 
the ghost propagator and the various ghost vertices can be derived. We 
anticipate that the effective ghost-gauge field interaction, at variance with 
the continuum, is not linear in the gauge potential $A_\mu$, and thus also in 
this sector we find an infinite number of new vertices that have no continuum 
analog, like for example the vertex involving two ghosts and two gluons.

\subsubsection{Measure}
\label{sec:measure}

The definition of the gauge-invariant integration measure on the lattice 
turns out for nonabelian gauge groups to be nontrivial, and again 
generates an infinite number of vertices.

Let us at first consider only the gauge potential, $A_\mu$, at a certain point.
We start with the 2-form
\be
d^2s = \Tr  (dU_\mu^\dagger dU_\mu),
\ee
where $dU_\mu=U_\mu(A_\mu + dA_\mu)-U_\mu(A_\mu)$, which is invariant under 
left or right multiplication of $U_\mu$ with an $SU(3)$ matrix. When rewritten 
in terms of $A_\mu$, this 2-form defines a metric $g$ on the space of the 
$A_\mu$'s, 
\be
d^2s = g_{ab} (A) \, dA_\mu^a dA_\mu^b ,
\ee
which gives the gauge-invariant measure (the Haar measure)~\footnote{
The gauge-invariant Haar measure on the group has the properties:
\bea
\int [dU] &=& 1 , \\
\int [dU] f(U) &=& \int [dU] f(U_0U)  , \label{eq:haar}
\eea
for any arbitrary but sensible function $f$, and for any arbitrary element
$U_0$ of the gauge group.}
\be
d\mu (A) = \sqrt{\det g(A)} \, \prod_{a,\mu} dA_\mu^a .
\ee
The calculation of $g(A)$ can be done using the properties of the $SU(3)$ 
group. It turns out that the infinitesimal transformation can be written 
as~\cite{Boulware}
\be
U_\mu(A_\mu + dA_\mu) = U_\mu(A_\mu) 
(1+\ii a g_0 \, dA_\mu^a \, M_{ab} (A_\mu) \, T^b) ,
\ee
where the matrix $M$ is given by 
\be
M (A_\mu) = \Bigg( \frac{\e^{\ii ag_0 \widetilde{A}_\mu}-1}{\ii ag_0 
\widetilde{A}_\mu} \Bigg) ,
\label{eq:m_matrix}
\ee
and $\widetilde{A}$ denotes the gauge potential in the {\em adjoint} 
representation:
\be
\widetilde{A}_\mu = A_\mu^a t^a, 
\qquad  (t^a)^{bc} = -\ii f^{abc}, \qquad \Tr  (t^a t^b) = 3\delta^{ab} .
\ee
One can then write~\cite{Kawai:1980ja}
\bea
g_{ab} d A_\mu^a d A_\mu^b &=& \Tr (dU_\mu^\dagger dU_\mu) \nonumber \\
&=& \Tr \Big\{ T^c \, M_{ca}^\dagger (A_\mu) \, (-\ii) d(a g_0 A_\mu^a) \, 
U_\mu^\dagger (A_\mu) \cdot 
U_\mu (A_\mu) \ii d(a g_0 A_\mu^b) \, M_{bd} (A_\mu) \, T^d \Big\}  
\nonumber \\
&=& a^2 g_0^2 \, \cdot \frac{1}{2} \, M_{ca}^\dagger (A_\mu) M^{bc} (A_\mu)  
\, d A_\mu^a d A_\mu^b . 
\eea
The metric is thus given, leaving aside a factor that will simplify 
in the ratios expressing the expectation values of operators, by
\be
g(A) = \frac{1}{2} \Big( M^\dagger (A) M (A) \Big),
\ee
which explicitly is
\be
g(A) = \frac{1-\cos (ag_0 A_\mu^a t^a)}{(ag_0 A_\mu^a t^a)^2}
= \frac{1}{2} + \sum_{l=1}^\infty \frac{(-1)^l}{(2l+2)!}
\Big( \ii  ag_0 A_\mu^a t^a \Big)^{2l} .
\label{expansmeas}
\ee
The measure for the Wilson action can then be written as the product over 
all sites of the above expression, and is given by
\be
{\cal D} U = \prod_{x,\mu} 
\sqrt{\det \Bigg( \frac{1}{2} M^\dagger (A_\mu (x)) M (A_\mu (x) ) \Bigg)} 
\, {\cal D} A , \qquad {\cal D} A = \prod_{x,\mu,a} dA_\mu^a (x).
\ee
It is convenient to write this measure term in the form
\be
{\cal D}U = \e^{-S_{\mathrm meas} [A]} \, {\cal D}A .
\ee
This can be done by using the identity $\det g= \exp(\Tr \log g)$, so that 
at the end we obtain, using Eq.~(\ref{expansmeas}),
\be
S_{\mathrm meas} [A] = -\frac{1}{2} \sum_{x,\mu} \Tr  \, \log
\frac{2(1-\cos (ag_0 A_\mu^a t^a))}{(ag_0 A_\mu^a t^a)^2} 
= -\frac{1}{2} \sum_{x,\mu} \Tr  \, \log \Bigg[ 1 
+ 2 \sum_{l=1}^\infty \frac{(-1)^l}{(2l+2)!}
\Big( ag_0 A_\mu^a t^a \Big)^{2l} \Bigg].
\label{eq:sermeas}
\ee
This formula contains the infinite number of vertices coming from the
measure term. These measure vertices do not contribute at the tree level.
The lowest order is just
\be
S_{\mathrm meas} [A] = \frac{g_0^2}{8a^2} \sum_{x,a,\mu} (A_\mu^a)^2 ,
\ee
and this term, which is quadratic in $A_\mu$, is nonetheless part of the 
interaction and not a kinetic term. It acts like a mass counterterm 
of order $g_0^2$, and is needed to restore gauge invariance in lattice 
Feynman amplitudes. It cancels, for example, the quadratic divergence 
in the 1-loop gluon self-energy (see Fig.~\ref{fig:gluon_self}).
In momentum space this mass counterterm is (see Fig.~\ref{fig:prop_vert}) 
\be
- \frac{g_0^2}{4a^2} \delta_{\mu\nu} \delta_{ab} .
\label{eq:meas} 
\ee
The higher orders in Eq.~(\ref{eq:sermeas}), which give self-interaction
vertices of the gluons, are at least of order $g_0^3$ and thus only relevant 
for calculations with two or more loops.

As a last comment, we mention that in lattice QED things are much simpler: 
the abelian measure is just given by
\be
{\cal D} U = \prod_{x,\mu,a} dA_\mu^a (x),
\ee
and thus there are no measure counterterms.

\subsubsection{Gauge fixing and the Faddeev-Popov procedure}

Although in some situations can be convenient, gauge fixing is not 
necessary on the lattice when one works with actions which are expressed 
in terms of the $U_\mu$'s, as is done in Monte Carlo calculations.
The reason is that the volume of the gauge group is finite, and it can be 
factored out and simplified when one normalizes the path integral with the 
partition function for the calculation of expectation values of operators.
In perturbation theory, where one makes a saddle-point approximation 
of the functional integral around $U_\mu=1$ and the $A_\mu$'s become 
the real variables (which moreover are decompactified), gauge fixing 
is instead necessary~\cite{Baaquie:1977hz,Stehr:en}. 
A gauge has then to be fixed in order to avoid the zero modes in the 
quadratic part of the action (expressed in terms of the $A_\mu$'s). 
This degeneracy is due to gauge invariance. 

We can see why it is necessary to fix a gauge in perturbative lattice QCD also 
from the following argument. Strictly speaking, perturbation theory arises as 
an expansion around the minimum of the action, that is the plaquette. 
We can see from looking at the Wilson action that the value of the plaquette
$P_{\mu \nu}(x) = 1$ minimizes the pure gauge action, but this does not yet 
imply $U_{\mu}(x) = 1$. Quite on the contrary, even if one fixes 
$U_{\mu}(x) = 1$ for each link from the beginning, a gauge transformation will
lead to $1 \to \Omega (x) \, \Omega^{-1} (x+a\hat{\mu})$, which can assume any
value. In order for perturbation theory to be a weak coupling expansion around
the configuration $U_{\mu}(x) = 1$, one must then fix the gauge.

Gauge fixing is thus an essential step in the perturbative calculations made 
on the lattice, and can be implemented by using a lattice Faddeev-Popov 
procedure, which goes along lines similar to the continuum. The final result, 
however, will be rather different. In fact, as another consequence of the 
gauge invariance on the lattice, one obtains from the Faddeev-Popov procedure 
an infinite number of vertices. Although cumbersome, this procedure is 
consistent and gives a meaning to the lattice functional integral.

We illustrate the lattice Faddeev-Popov method for a covariant gauge, 
which is the most commonly used, implementing the gauge-fixing condition
\be
{\cal F}^a_x [A,\chi] = \nabla_\mu^\star A_\mu^a (x) - \chi^a (x) = 0,
\ee
with $\chi$ some arbitrary fields. One chooses the backward lattice derivative
$\nabla_\mu^\star$ here because only in this way the gluon propagator can be 
expressed in a simple form, as we will see below.
The Faddeev-Popov determinant is defined by
\be
1 = \Delta_{FP} [A,\chi] \cdot \int {\cal D} g \prod_{x,a} \delta
({\cal F}^a_x [A_g,\chi]),
\label{fp}
\ee
where $g$ is a gauge transformation and $A_g$ is the result of this gauge
transformation applied to $A$. The measure ${\cal D} g = \prod_x d\mu (g_x)$ 
is given by the product of the Haar measures over the lattice sites. From the 
property~(\ref{eq:haar}) of an Haar measure it is easy to see that the 
Faddeev-Popov determinant so constructed is gauge invariant: 
$\Delta_{FP} [A_g,\chi] = \Delta_{FP} [A,\chi]$.

One makes now the same steps as in the continuum, that is Eq.~(\ref{fp}) 
is inserted into the partition function, and, as in the continuum, the
gauge invariance of the Faddeev-Popov determinant is exploited so that 
at the end one can factorize the ${\cal D} g$ integration and drop it out.
Finally, after adding a gauge-fixing term to the action like is done
in the continuum and integrating in $\chi$, one obtains for the expectation 
value of a generic gauge-invariant operator~\footnote{This formula is not
generally valid in a nonperturbative context. It can happen that the partition 
function of a gauge-fixed BRS-invariant theory reduces to zero on the lattice,
because of the contribution of Gribov copies~\cite{Neuberger:vv,Neuberger:xz}.
One of the possible ways to overcome this is proposed in~\cite{Testa:1998az}.}
\be
\langle O \rangle =  
\frac{{\displaystyle \int} {\cal D} \psi {\cal D} \overline{\psi} {\cal D} A 
{\cal D} \chi
\Delta_{FP} [A,\chi] \prod_{x,a} \delta ({\cal F}^a_x [A,\chi]) 
\cdot O \cdot
\e^{\displaystyle -S_{QCD}} \e^{\displaystyle -S_{meas}} 
\e^{\displaystyle -\frac{1}{2\alpha} \, a^4 \sum_{x,a} \chi^a (x) \chi^a (x)}}{
{\displaystyle \int} {\cal D} \psi {\cal D} \overline{\psi} {\cal D} 
A {\cal D} \chi
\Delta_{FP} [A,\chi] \prod_{x,a} \delta ({\cal F}^a_x [A,\chi]) 
\cdot
\e^{\displaystyle -S_{QCD}} \e^{\displaystyle -S_{meas}} 
\e^{\displaystyle -\frac{1}{2\alpha} \, a^4 \sum_{x,a} \chi^a (x) \chi^a (x)}},
\ee
where $\alpha$ is the gauge parameter (particular cases are the 
Feynman gauge $\alpha=1$ and the Landau gauge $\alpha=0$). 
What is left at this point is the computation of the Faddeev-Popov 
determinant, which turns out to be independent of $\chi$. To do this, 
one only needs to know $A_g$ in the infinitesimal neighborhood of the identity 
transformation $g=1$. The infinitesimal transformation with parameter 
$\epsilon^a (x)$ gives
\be
U_\mu (x) \rightarrow 
\e^{\ii \epsilon (x)} U_\mu (x) \e^{-\ii \epsilon (x+a\hat{\mu})} 
= \e^{\ii a g_0 \Big(A_\mu (x) + \delta_{(\epsilon)} A_\mu (x) \Big)} ,
\ee
where
\be
g_0 \delta_{(\epsilon)} A_\mu^a (x) = - \sum_b \widehat{D}_\mu [A]_{ab} 
\epsilon^b (x) ,
\label{eq:fp1}
\ee
with
\be
\widehat{D}_\mu [A] = (M^\dagger)^{-1} (A_\mu(x)) \cdot \nabla_\mu 
+ \ii g_0 A_\mu^a (x) t^a .
\label{eq:fp2}
\ee
We remark that $M$ is the same matrix of Eq.~(\ref{eq:m_matrix}) and $t^a$ 
is a matrix in the {\em adjoint} representation of $SU(3)$. This is similar
to the continuum case. In fact $\widehat{D}_\mu [A]$ is a discretized form
of the covariant derivative acting on fields in the adjoint representation.
The result for the Faddeev-Popov determinant is indeed still reminiscent 
of the continuum:~\footnote{For the detailed derivation, see~\cite{Rothe:kp}.} 
\be
\Delta_{FP} [A] = \det \, (-\nabla_\mu^\star \widehat{D}_\mu [A]) .
\ee
However, the important difference with the continuum case is that the lattice
operator $\widehat{D}_\mu [A]$ is not linear in $A$, because of the expansion
\be
(M^\dagger)^{-1} (A_\mu(x)) = 1 +\frac{\ii}{2} a g_0 A_\mu (x) 
-\frac{1}{12} (a g_0 A_\mu (x))^2 + \cdots ,
\ee
and an infinite number of ghost vertices are thus generated.~\footnote{The part
of the lattice Faddeev-Popov determinant that gives a direct correspondence 
with the continuum ghosts can be obtained putting $M=1$, as one can easily see 
by combining the formulae (\ref{eq:fp1}) and (\ref{eq:fp2}) to reconstruct the 
continuum formulae
\be
g_0 \delta_{(\epsilon)} A_\mu^a (x) 
= - ( \partial_\mu + \ii g_0 A_\mu^c (x) t^c )^{ab} \, \epsilon^b (x)
= - D_\mu [A] (x) \cdot \epsilon (x) 
\ee
and
\be
\Delta_{FP} [A] = \det \, (-\partial_\mu D_\mu [A]) .
\ee
}
In fact, using the well-known formula for Grassmann variables $c$ and 
$\overline{c}$,
\be
\int {\cal D} (\overline{c}c) \, 
\e^{\displaystyle - a^4 \sum_{ij} \overline{c}_i Q_{ij} c_j} = \det Q ,
\label{eq:detgrassmann}
\ee
we can write the Faddeev-Popov determinant in terms of an action
involving ghosts,
\be
\Delta_{FP} [A] = \int \Bigg( \prod_{a,x} d\overline{c}^a(x)c^a(x) \Bigg)
\e^{\displaystyle a^4 \sum_x  \overline{c}^a(x) \nabla_\mu^\star 
\widehat{D}^{ab}_\mu [A] c^b(x)} ,
\label{eq:fpghosts}
\ee
so that the expectation value of a generic gauge-invariant operator is
\be
\langle O \rangle =  
\frac{{\displaystyle \int} {\cal D} \psi {\cal D} \overline{\psi} {\cal D} A 
{\cal D} \overline{c} {\cal D} c 
\cdot O \cdot
\e^{\displaystyle -S_{QCD}} 
\e^{\displaystyle a^4 \sum_x \overline{c}^a(x) \nabla_\mu^\star 
\widehat{D}^{ab}_\mu [A] c^b(x)} 
\e^{\displaystyle -S_{meas}} 
\e^{\displaystyle -S_{gf}}}{
{\displaystyle \int} {\cal D} \psi {\cal D} \overline{\psi} {\cal D} A 
{\cal D} \overline{c} {\cal D} c 
\cdot 
\e^{\displaystyle -S_{QCD}} 
\e^{\displaystyle a^4 \sum_x \overline{c}^a(x) \nabla_\mu^\star 
\widehat{D}^{ab}_\mu [A] c^b(x)} 
\e^{\displaystyle -S_{meas}} 
\e^{\displaystyle -S_{gf}}} ,
\ee
where the gauge-fixing term has been written, thanks to the delta function
$\delta ({\cal F}^a_x [A,\chi])$, as
\be
S_{gf} = \frac{a^4}{2\alpha} \sum_x \Bigg( \sum_\mu  
\nabla^\star_\mu A_\mu (x) \Bigg)^2 
= \frac{a^2}{2\alpha} \sum_x \Bigg( \sum_\mu \Big( 
A_\mu (x) - A_\mu (x -a\hat{\mu} \Big) \Bigg)^2 .
\ee
The ghosts are Grassmann variables which carry a color index, but they have no 
spin index. They transform according to the adjoint representation of $SU(3)$.

We are now at last ready to compute the gluon propagator in the covariant 
gauge $\partial_\mu A_\mu=0$, which is:
\be
G^{ab}_{\mu\nu} (k) = \delta^{ab} \, \frac{1}{{\displaystyle \frac{4}{a^2}} 
\sum_\lambda \sin^2 {\displaystyle \frac{a k_\lambda}{2}} } \,
\Bigg\{\delta_{\mu\nu} 
-(1-\alpha)  \frac{ \sin {\displaystyle \frac{a k_\mu}{2}} \, 
\sin {\displaystyle \frac{a k_\nu}{2}}    }{
\sum_\lambda \sin^2 {\displaystyle \frac{a k_\lambda}{2}} } \Bigg\} . 
\label{eq:gg} 
\ee
This expression is the result of adding the free part of the gluon action,
\bea
S_g &=& \frac{1}{4} \Big(\nabla_\mu A_\nu - \nabla_\nu A_\mu \Big)^2  
=\frac{1}{2} \Big(\nabla_\mu A_\nu \nabla_\mu A_\nu -
\nabla_\mu A_\nu \nabla_\nu A_\mu \Big) 
=- \frac{1}{2} \Big(A_\nu \nabla_\mu^\star \nabla_\mu A_\nu -
A_\nu \nabla_\mu^\star \nabla_\nu A_\mu \Big) \nonumber \\
&=&  - \frac{1}{2} \Big(A_\nu ( \Delta 
\delta_{\mu\nu} - \nabla_\mu^\star \nabla_\nu ) A_\mu \Big) ,
\eea
to the gauge-fixing term,
\be
S_{gf} = \frac{1}{2\alpha} (\nabla_\nu^\star A_\nu ) 
(\nabla_\mu^\star A_\mu ) = 
- \frac{1}{2\alpha} A_\nu \nabla_\nu \nabla_\mu^\star A_\mu , 
\ee
where we have integrated by parts. We can see that one is forced to choose 
the backward lattice derivative in the gauge-fixing term because in the pure 
gauge action the forward derivative is present. Only in this way the 
longitudinal components of the lattice propagator have a simple form. 
In the limit $a \to 0$ the lattice gluon propagator of course reduces to the 
well-known expression
\be
\delta^{ab} \cdot \frac{1}{k^2} \, \Bigg[\delta_{\mu\nu} 
-(1-\alpha)  \frac{k_\mu  k_\nu}{k^2} \Bigg] . 
\ee

From Eqs.~(\ref{eq:fpghosts}) and (\ref{eq:fp2}) one can derive the
Feynman rules for the ghosts on the lattice. The ghost propagator is
\be
\delta^{ab} \cdot \frac{1}{{\displaystyle \frac{4}{a^2}} 
\sum_\lambda \sin^2 {\displaystyle \frac{a k_\lambda}{2}} } ,
\label{eq:cc} 
\ee
and the ghost-gluon-gluon vertex (see Fig.~\ref{fig:prop_vert}) is
\be
\ii  g_0 f_{abc} (\widehat{p}_2)_\mu \cos \frac{(a p_1)_\mu}{2} ,
\label{eq:ccg} 
\ee
while the ghost-ghost-gluon-gluon vertex,
\be
\frac{1}{12} g_0^2 a^2 \{t^a,t^b\}_{cd} \, \delta_{\mu\nu} \, 
(\widehat{p}_1)_\mu (\widehat{p}_2)_\mu ,
\label{eq:ccgg} 
\ee
is the first ghost vertex which is a lattice artifact, that is it vanishes
in the continuum limit. Vertices containing three or more ghosts are at least 
of order $g_0^3$ and do not enter in 1-loop calculations.

The lattice theory so defined after gauge fixing has an exact BRS 
symmetry~\cite{Baaquie:1977hz,Kawai:1980ja}, which is given by
\bea
\delta A_\mu^a (x) &=& \frac{1}{g_0 a} \, \Bigg[ 
   M_{ab}^{-1} (A_\mu (x)) \, c^b (x) 
 - M_{ba}^{-1} (A_\mu (x)) \, c^b (x+a\hat{\mu}) \Bigg] \\
\delta c^a (x) &=& -\frac{1}{2} \, f_{abc} \, c^b (x) c^c (x) \\
\delta \overline{c}^a (x) &=& -\frac{1}{\alpha g_0 a} \,  
    \sum_\nu \Big( A_\nu^a (x) - A_\nu^a (x -a\hat{\nu}) \Big) \\
\delta \psi (x) &=& \ii \, c^a (x) \, T^a \, \psi (x)   \\
\delta \overline{\psi} (x) &=& \ii \, \overline{\psi} (x) \, T^a \, c^a (x) .
\eea
The BRS variation of the gauge-fixing term in the action is the opposite of 
the BRS variation of the Faddeev-Popov term, and they cancel. This can be seen
from the nilpotence of the BRS transformation, $\delta^2 A_\mu^a (x) = 0$.
The transformations of quarks, antiquarks and ghosts are the same as
in the continuum BRS. The transformation of the antighosts, which contains 
the gauge-fixing term, replaces the continuum derivative with the lattice 
backward derivative, and the difference with the continuum is then of $O(a)$. 
The transformation of the gauge potential is instead quite different from the 
continuum, and because of $M$ is nonlinear in the gauge potential $A$. 
For $a \to 0$ it reduces to the continuum BRS transformation. 

In the abelian case the Faddeev-Popov determinant reduces to a trivial factor 
that can be eliminated from the path integral when computing expectation 
values, analogously to the continuum QED theory.

The derivation of the Faddeev-Popov determinant for arbitrary linear 
gauge-fixing conditions and for lattices with a large class of boundaries 
can be found in~\cite{Luscher:1985wf}.

\subsection{Fermion action}

We now discuss the Feynman rules coming from the fermion part of the action.
The quark propagator can be computed by inverting the lattice Dirac operator
in momentum space, and is given by
\be
S^{ab} (k, m_0) = \delta^{ab} \cdot a \, 
\frac{- \ii  \sum_\mu \gamma_\mu \sin a k_\mu + a m_0 + 2 r 
\sum_\mu \sin^2 {\displaystyle \frac{a k_\mu}{2}} }{ \sum_\mu \sin^2 a k_\mu +
\Big( 2 r \sum_\mu \sin^2 {\displaystyle \frac{a k_\mu}{2}} + a m_0 \Big)^2} .
\label{eq:qq} 
\ee
In the continuum limit this is the well-known expression
\be
\delta^{ab} \cdot \frac{-\ii  \sum_\mu \gamma_\mu k_\mu + m_0}{\sum_\mu k_\mu^2
+ m_0^2} .
\ee
The vertices are obtained from the expansion of the fermionic part of the 
action in powers of $g_0A$, with a Fourier transform. We get again an
infinite tower of vertices, which involve two (anti-)quarks and $n$ gluons.  
Fortunately, only a finite number of them is needed at any given 
order in $g_0$. Here we only give the explicit expressions for the vertices 
which are sufficient to do 1-loop calculations.

We have already derived the quark-quark-gluon vertex (see Eq.~(\ref{eq:qqgft}),
and is 
\be
(V^a_1)_\mu^{bc} (p_1,p_2) = - g_0 (T^a)^{bc} \,
\Bigg( \ii  \gamma_\mu \cos \frac{a(p_1+p_2)_\mu}{2}
+ r \sin \frac{a(p_1+p_2)_\mu}{2} \Bigg) ,
\label{eq:qqg} 
\ee
where $p_1$ and $p_2$ are the quark momenta flowing in and out of the vertices
(see Fig.~\ref{fig:prop_vert}). For $a \to 0$ this is the familiar continuum 
QCD vertex
\be
- g_0 (T^a)^{bc} \, \ii  \gamma_\mu .
\ee
Doing a similar calculation, one can derive that the quark-quark-gluon-gluon 
vertex is given by 
\be
(V^{ab}_2 )_{\mu_1\mu_2}^{cd}(p_1,p_2) = 
- \frac{1}{2} a g_0^2 \, \delta_{\mu_1 \mu_2} 
\Big( \frac{1}{N_c} \delta^{ab} + d^{abe} T^e \Big)^{cd}
\, \Bigg( -\ii  \gamma_\mu \sin \frac{a(p_1+p_2)_\mu}{2}
              + r \cos \frac{a(p_1+p_2)_\mu}{2} \Bigg) .
\label{eq:qqgg} 
\ee
This vertex is zero in the continuum limit and thus has no continuum analog. 
However it can still give nonvanishing contributions to Feynman diagrams, 
like in divergent loops or in tadpoles, which have a factor $1/a^2$ coming
from the gluon propagator. Vertices of two quarks and $n$ gluons are also 
irrelevant and are associated with a factor $a^{n-1} g_0^n$.

In spite of the complexities brought from the fact that the lattice gauge 
theory has an infinite number of vertices, it turns out that the superficial 
degree of divergence of a Feynman diagram depends only on the number of its 
external lines~\cite{Kawai:1980ja}:
\be
D = 4 - E_G - E_g - \frac{3}{2} E_q .
\ee
The type of vertex functions which give rise to divergences is the same as in 
the continuum, and this is very important for the renormalizability of the 
theory. Thus, vertices which are of higher order in $a$ and $g_0$ do not 
modify this continuum picture.

It is customary to define and use the vector and axial currents
on the lattice as follows:
\bea
V_\mu (x) &=& \overline{\psi} (x) \, \gamma_\mu \, \psi (x) \\
A_\mu (x) &=& \overline{\psi} (x) \, \gamma_\mu \gamma_5 \, \psi (x) .
\eea
However, these currents are not conserved in the Wilson formulation (in fact,
they are not equal to the currents appearing in the action), and therefore 
they are not protected from renormalization. One then has
\bea
Z_V & \neq & 1 , \\ 
Z_A & \neq & 1 ,
\eea
at variance with the continuum results. The Noether current corresponding to 
the vector transformation that leaves the Wilson action invariant is instead 
given by
\be
V_\mu^{\mathrm cons} (x) = \frac{1}{2} \Bigg(
   \overline{\psi} (x) \, (\gamma_\mu-r) \, U_\mu (x) \, \psi (x+a\hat{\mu}) 
 + \overline{\psi} (x+a\hat{\mu}) \, (\gamma_\mu+r) \, U_\mu^\dagger (x) 
   \, \psi (x) \Bigg) ,
\label{eq:extcurr}
\ee
and it is an extended operator, which needs for its definition two lattice 
sites. The renormalization constant of this conserved current is one.
It is not possible to make a corresponding construction which leads to a
conserved axial current because, as we will see in detail in the next Section,
the Wilson action breaks chiral symmetry.

\begin{figure}[t]
\begin{center}
\begin{picture}(450,160)(0,0)
\Line(0,0)(100,150)
\Line(100,150)(200,0)
\Gluon(20,30)(180,30){5}{20}
\BCirc(115,70){20}
\Gluon(80,120)(110,90){-5}{6}
\Gluon(125,36)(110,50){5}{2}
\Gluon(60,35)(90,105){5}{9}
\Line(250,0)(350,150)
\Line(350,150)(450,0)
\Gluon(270,30)(430,30){5}{20}
\GlueArc(365,70)(20,-100,260){5}{15}
\Gluon(330,120)(360,90){-5}{6}
\Gluon(375,36)(360,50){5}{2}
\Gluon(310,35)(340,105){5}{9}
\end{picture}
\end{center}
\caption{\small The diagram on the left is zero in the quenched approximation,
while the diagram on the right does not contain internal quark loops and 
has to be included also in quenched calculations.}
\label{fig:quenching}
\end{figure}
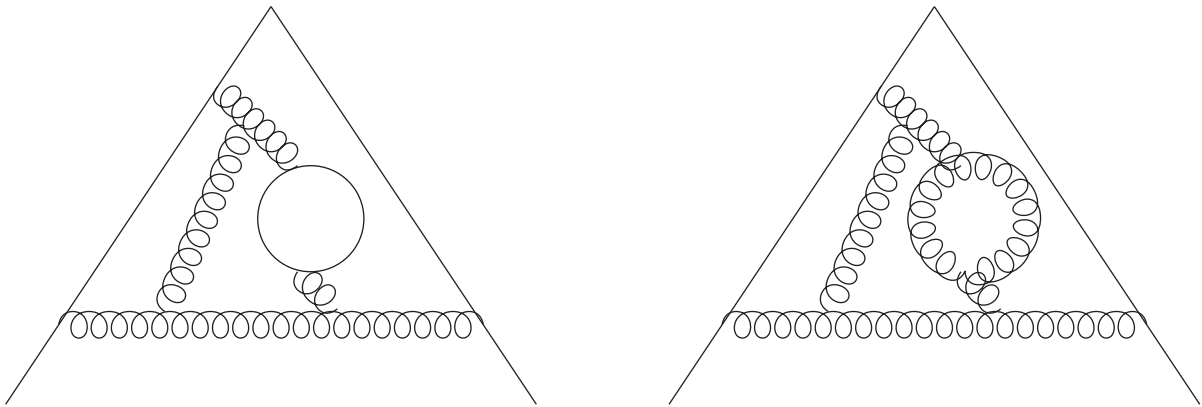

We want to conclude this Section by mentioning that, for reasons of computing 
power, the so-called quenched approximation is often used in Monte Carlo 
simulations. In order to perform a statistical sampling of the functional
integral, the fermion variables (which are Grassmann numbers) are analytically
integrated (using Eq.~(\ref{eq:detgrassmann})) and one does the numerical 
simulations using the partition function
\be
Z = \int {\cal D} U \, \, \det (\Ds [U]+m_0) \, \, \e^{-S_g[U]} .
\ee
This is equivalent to use the effective action
\be
S_{\mathrm eff}[U] = S_g[U] - \log \, \det (\Ds [U]+m_0) 
= S_g[U] -\Tr \, \log (\Ds [U]+m_0) .
\ee
Simulations in full QCD have to include the full contribution of the 
determinant, while quenching amounts to doing simulations where one puts
$\det (\Ds [U]+m_0) = 1$, which saves a couple of orders of magnitudes of 
computer time. In physical terms, this means that there are no sea quarks 
in the calculations: the internal quark loops are neglected 
(see Fig.~\ref{fig:quenching}). Quenching is often summarized by saying that 
$N_f=0$. Although it looks quite drastic, in many cases this does not turn out
to be a bad approximation.

This approximation has been introduced only for reasons having to do with the 
simulations. In perturbation theory, quenching means dropping all diagrams 
which contain an internal quark loop, but the inclusion of these diagrams 
is usually not so challenging as it is in the simulations. 
For consistency one can exclude the diagrams containing internal quark loops 
when the perturbative numbers have to be used in connection with quenched 
simulation results.

\section{Aspects of chiral symmetry on the lattice}
\label{sec:aspects}

Due to the presence of the Wilson term (the part of the action proportional 
to the parameter $r$), the Wilson action breaks chiral symmetry, and thus 
Wilson fermions do not possess chiral invariance even when the bare mass of 
the quark is zero. This additional symmetry-spoiling term turns out however to
be necessary in order to get rid of the extra fermions (also called doublers) 
which are unavoidable in the naive lattice discretization of the QCD action.

Let us see what would happen putting $r=0$ in the Wilson action 
(which corresponds to naive fermions). 
In this case, the free fermion propagator is just 
\be
S^{ab} (k, m_0) = \delta^{ab} \cdot a \, 
\frac{- \ii  \sum_\mu \gamma_\mu \sin a k_\mu + a m_0}{ 
\sum_\mu \sin^2 a k_\mu + (a m_0 )^2} .
\label{eq:naiveprop}
\ee
Let us for simplicity consider the massless naive propagator, setting $m_0=0$ 
in the above equation. This propagator has a pole at $ak=(0,0,0,0)$, as 
expected. However, there are also poles at $ak=(\pi,0,0,0)$, $ak=(0,\pi,0,0)$, 
\dots, $ak=(\pi,\pi,0,0)$, \dots, $ak=(\pi,\pi,\pi,\pi)$, that is at all 
points at the edges of the first Brillouin zone, because 
$\sum_\mu \sin^2 a k_\mu$ vanishes at these 16 points, at which any component 
is either $k_\mu =0$ or $k_\mu = \pi/a$. This propagator describes a fermion 
mode at each of these poles. All these fermions are acceptable and legitimate 
particles, even if they correspond to poles at the edges of the Brillouin 
zone. In fact, we can always think of shifting the integration in momentum 
space, thanks to the periodicity of the lattice; for example, we could shift 
it from $\int_{-\pi/a}^{\pi/a}$ to $\int_{-\pi/2a}^{3\pi/2a}$ and so no pole 
would then be found at the edges. At these poles we have then low-energy 
excitations as legitimate as the fermion at $ak=(0,0,0,0)$. 

For $r=0$ we would then have to take into account all these 16 Dirac 
particles when doing computations with the lattice theory. Although they are 
a lattice artifact, they would be pair produced as soon as the interactions 
are switched on, and would appear in internal loops and contribute to 
intermediate processes. 

Let us consider in more detail one of these poles, for example the one at 
$ak=(\pi,0,0,0)$. To see things more clearly, we make the change of variables
\be
k_0' = \frac{\pi}{a} - k_0 , \quad k_i' = k_i ,
\ee
and correspondingly~\footnote{The Dirac matrices that one has to use in the 
transformed variables are equivalent to the standard set: 
$\gamma_\mu'= (\gamma_0 \gamma_5) \gamma_\mu (\gamma_0 \gamma_5)^\dagger$.}
\be
\gamma_0' = - \gamma_0 , \quad \gamma_i' = \gamma_i ,
\ee
so that $\gamma_5' = - \gamma_5$. The propagator in the new variables assumes 
near $ak'=(0,0,0,0)$ the same form as the original propagator 
Eq.~(\ref{eq:naiveprop}) near $ak=(0,0,0,0)$. This means that there is a 
fermion mode also at $ak=(\pi,0,0,0)$, and moreover the chirality of this new 
particle is opposite to the chirality of the mode at $ak=(0,0,0,0)$, because 
$\gamma_5'=-\gamma_5$, so that $(1+\gamma_5)/2=(1-\gamma_5')/2$. It can be 
easily seen that the 16 doublers split in 8 particles of one chirality and 8 
particles of the opposite chirality, so that even if the massless continuum 
theory had chiral symmetry and the physical particle (the pole at the origin) 
was a chiral mode we end up with a vector theory on the lattice. The main 
problem then is not that there are more species of fermions than expected, but
that the doublers destroy the chirality properties of the continuum theory.

The Wilson term does precisely the work of suppressing these 15 unwanted 
additional fermions in the continuum limit.
In fact, the Wilson action in short form can be written as
\be
D_W = \frac{1}{2} \Big( \gamma_\mu (\widetilde{\nabla}^\star_\mu 
+ \widetilde{\nabla}_\mu)
- a r \widetilde{\nabla}^\star_\mu \widetilde{\nabla}_\mu \Big) , 
\ee
where the gauge covariant forward derivative is given by
\be
\widetilde{\nabla}_\mu \psi (x) = 
\frac{1}{a} \Big( U(x,\mu)\psi(x+a\hat{\mu}) - \psi (x) \Big) ,  
\ee
and the Wilson term is the Laplacian part of this action,
\be
- \frac{1}{2} a r \widetilde{\nabla}^\star_\mu \widetilde{\nabla}_\mu .
\ee
This term is irrelevant in the continuum limit but modifies the lattice 
dispersion relation for finite lattice spacing, so that this ``bosonic'' 
energy term gives the extra doublers at the edges of the Brillouin zone 
a mass of the order of the cutoff. This mass then becomes large in the 
continuum limit and decouples the doublers from the ``right'' physical 
fermion. Of course, being a generalized (momentum-dependent) mass term, 
it automatically breaks chiral symmetry. This connection of the doublers 
with chiral symmetry is a deep one, and what we have shown is a particular 
case of a general phenomenon, as we will shortly see.

Thus, if one uses Wilson fermions chiral symmetry can only be recovered in 
the continuum limit. The breaking of chiral symmetry at finite lattice spacing
has serious consequences on the Wilson theory, among them the appearance of an
additive renormalization to the quark mass. The nonzero value of the bare 
quark mass which corresponds to a vanishing renormalized quark mass is called 
critical mass. Its value depends on the strength of the interaction.
There is then a critical line in the plane of bare parameters,
\be
m_0 = m_c (g_0) , \qquad m_c (0) = 0 ,
\ee
where the physical quark mass vanishes. It is only the subtracted mass 
$m_0 - m_c$ that is multiplicatively renormalized:
\be
m_R = Z_m (m_0 - m_c) .
\ee
Thus, the renormalized quark mass has no protection from chiral symmetry 
and acquires a nonzero value even when the bare quark mass is zero. 
The bare and the renormalized quark mass cannot vanish at the same time,
and the bare mass has to be carefully tuned in order to extract 
physical information from the simulations. 
Since numerical simulations of QCD are performed with lattice spacings 
that are not small, the violation of chiral symmetry by these lattice effects 
is rather pronounced, and this mass renormalization is quite large.

The quark self-energy at one loop has a correction proportional to $1/a$, 
which corresponds to the mass counterterm which has to be introduced due to 
the breaking of chirality, and we will show in detail how to compute it in 
Section~\ref{sec:analyticcomputations}. This critical mass $m_c$ is one of the
best known quantities in lattice perturbation theory, and its high-precision 
determination at 1-loop, as well as its value calculated to two loop, are 
given in Section~\ref{sec:fermioniccase}.

Another bad consequence of the loss of chirality is the appearance of more 
mixings under renormalization, because the mixing among operators of different
chirality is not a priori forbidden for Wilson fermions. One example of this 
kind of mixings will be given in Section~\ref{sec:amixing}. Moreover, it is 
not possible to define a conserved axial current.

The impossibility of removing the doublers in the naive fermion formulation 
without breaking at the same time chiral symmetry (as it happens in the Wilson
action) or some other important symmetry is a special case of a very important
no-go theorem, established by Nielsen and Ninomiya many years 
ago~(Nielsen and Ninomiya, 1981a; 1981b; 1981c; Friedan, 1982).~\footnote{An
alternative proof of the theorem which makes use of the Poincar\'e-Hopf
theorem has been given in~\cite{Karsten:1981gd}.}
In short, the theorem says that a lattice fermion formulation without species 
doubling and with an explicit continuous chiral symmetry is impossible, 
unless one is prepared to give up some other fundamental properties like 
locality, or unitarity.~\footnote{This statement only applies to the chiral
symmetry which acts on the spinor fields like 
\bea
\psi &\rightarrow& \psi + \epsilon \cdot \gamma_5 \, \psi , \\
\overline{\psi} &\rightarrow& \overline{\psi} +  
\epsilon \cdot \overline{\psi} \, \gamma_5 .
\eea
As we will shortly see, one of the major theoretical advances of the last 
years has been the understanding that there are other kinds of transformations
that can define a lattice chiral symmetry and which do not necessarily 
imply a fermion doubling.}

It is quite common in quantum field theory that the introduction of an 
ultraviolet regulator brings some unphysical features, and this chiral 
symmetry issue is perhaps the most serious and unpleasant drawback of the 
lattice regularization. The difficulties regarding chiral symmetry with 
lattice fermions arise already at the level of free fields, and they are 
present even when one does not discretize space, but only the time direction. 
We can see why this happens from general topological considerations on the 
free fermion propagator, where the close interplay between chirality and 
doublers follows from the continuity of the energy-momentum relation in the 
Brillouin zone~\cite{Karsten:1980wd} (see also the lectures 
of~\cite{Smit:1985nu}). The general form of a massless fermion propagator on 
the lattice which is compatible with continuous chiral invariance is
\be
\frac{1}{\ii \sum_\mu \gamma_\mu P_\mu (k)} ,
\label{eq:genprop}
\ee
where the function $P_\mu (k)$ is real. This function has to cross each 
$P_\mu=0$ axis only once if this propagator is meant to describe for 
$a \to 0$ only a single fermion.

\begin{figure}[t]
\begin{center}
\begin{picture}(450,270)(0,0)

\LongArrow(0,200)(200,200)
\LongArrow(100,150)(100,250)
\Text(195,205)[b]{$k_\mu$}
\Text(100,255)[b]{$P_\mu (k)$}
\Text(40,195)[t]{-$\frac{\pi}{a}$}
\Text(145,195)[t]{$\frac{\pi}{a}$}

\LongArrow(250,200)(450,200)
\LongArrow(350,150)(350,250)
\Text(445,205)[b]{$k_\mu$}
\Text(350,255)[b]{$P_\mu (k)$}
\Text(290,195)[t]{-$\frac{\pi}{a}$}
\Text(395,195)[t]{$\frac{\pi}{a}$}

\LongArrow(0,50)(200,50)
\LongArrow(100,0)(100,100)
\Text(195,55)[b]{$k_\mu$}
\Text(100,105)[b]{$P_\mu (k)$}
\Text(40,45)[t]{-$\frac{\pi}{a}$}
\Text(145,45)[t]{$\frac{\pi}{a}$}

\LongArrow(250,50)(450,50)
\LongArrow(350,0)(350,100)
\Text(445,55)[b]{$k_\mu$}
\Text(350,105)[b]{$(4 S (k))^{-1}$}
\Text(290,45)[t]{-$\frac{\pi}{a}$}
\Text(395,45)[t]{$\frac{\pi}{a}$}

\SetWidth{1}
\Curve{(50,200)
(52,215)
(53,220)
(60,220)
(62,215)
(65,206)
(70,184)
(75,170)
(80,174)
(85,169)
(90,173)
(95,192)
(100,200)
(105,208)
(110,227)
(115,231)
(120,226)
(125,230)
(130,216)
(135,194)
(138,185)
(140,180)
(147,180)
(148,185)
(150,200)}

\DashCurve{(0,200)
(5,208)
(10,227)
(15,231)
(20,226)
(25,230)
(30,216)
(35,194)
(38,185)
(40,180)
(47,180)
(48,185)
(50,200)}{3}

\DashCurve{(150,200)
(152,215)
(153,220)
(160,220)
(162,215)
(165,206)
(170,184)
(175,170)
(180,174)
(185,169)
(190,173)
(195,192)
(200,200)}{3}

\Curve{(300,200)
(302.5,197.510268)
(305,195.081842)
(307.5,192.774517)
(310,190.645107)
(312.5,188.746046)
(315,187.124095)
(317.5,185.819191)
(320,184.863466)
(322.5,184.280452)
(325,184.084506)
(327.5,184.280452)
(330,184.863466)
(332.5,185.819191)
(335,187.124095)
(337.5,188.746046)
(340,190.645107)
(342.5,192.774517)
(345,195.081842)
(347.5,197.510268)
(350,200)
(352.5,202.489732)
(355,204.918158)
(357.5,207.225483)
(360,209.354893)
(362.5,211.253954)
(365,212.875906)
(367.5,214.180809)
(370,215.136535)
(372.5,215.719548)
(375,215.915494)
(377.5,215.719548)
(380,215.136535)
(382.5,214.180809)
(385,212.875906)
(387.5,211.253954)
(390,209.354893)
(392.5,207.225483)
(395,204.918158)
(397.5,202.489732)
(400,200)}

\DashCurve{(250,200)
(252.5,202.489732)
(255,204.918158)
(257.5,207.225483)
(260,209.354893)
(262.5,211.253954)
(265,212.875906)
(267.5,214.180809)
(270,215.136535)
(272.5,215.719548)
(275,215.915494)
(277.5,215.719548)
(280,215.136535)
(282.5,214.180809)
(285,212.875906)
(287.5,211.253954)
(290,209.354893)
(292.5,207.225483)
(295,204.918158)
(297.5,202.489732)
(300,200)}{3}

\DashCurve{(400,200)
(402.5,197.510268)
(405,195.081842)
(407.5,192.774517)
(410,190.645107)
(412.5,188.746046)
(415,187.124095)
(417.5,185.819191)
(420,184.863466)
(422.5,184.280452)
(425,184.084506)
(427.5,184.280452)
(430,184.863466)
(432.5,185.819191)
(435,187.124095)
(437.5,188.746046)
(440,190.645107)
(442.5,192.774517)
(445,195.081842)
(447.5,197.510268)
(450,200)}{3}

\Line(50,0)(150,100)
\DashLine(0,50)(50,100){3}
\DashLine(150,0)(200,50){3}

\Curve{(300,65.915494)
(302.5,65.817521)
(305,65.526014)
(307.5,65.048152)
(310,64.395700)
(312.5,63.584724)
(315,62.635194)
(317.5,61.570489)
(320,60.416827)
(322.5,59.202613)
(325,57.957747)
(327.5,56.712881)
(330,55.498668)
(332.5,54.345006)
(335,53.280301)
(337.5,52.330770)
(340,51.519794)
(342.5,50.867343)
(345,50.389480)
(347.5,50.097973)
(350,50)
(352.5,50.097973)
(355,50.389480)
(357.5,50.867343)
(360,51.519794)
(362.5,52.330770)
(365,53.280301)
(367.5,54.345006)
(370,55.498668)
(372.5,56.712881)
(375,57.957747)
(377.5,59.202613)
(380,60.416827)
(382.5,61.570489)
(385,62.635194)
(387.5,63.584724)
(390,64.395700)
(392.5,65.048152)
(395,65.526014)
(397.5,65.817521)
(400,65.915494)}

\DashCurve{(250,50)
(252.5,50.097973)
(255,50.389480)
(257.5,50.867343)
(260,51.519794)
(262.5,52.330770)
(265,53.280301)
(267.5,54.345006)
(270,55.498668)
(272.5,56.712881)
(275,57.957747)
(277.5,59.202613)
(280,60.416827)
(282.5,61.570489)
(285,62.635194)
(287.5,63.584724)
(290,64.395700)
(292.5,65.048152)
(295,65.526014)
(297.5,65.817521)
(300,65.915494)}{3}

\DashCurve{(400,65.915494)
(402.5,65.817521)
(405,65.526014)
(407.5,65.048152)
(410,64.395700)
(412.5,63.584724)
(415,62.635194)
(417.5,61.570489)
(420,60.416827)
(422.5,59.202613)
(425,57.957747)
(427.5,56.712881)
(430,55.498668)
(432.5,54.345006)
(435,53.280301)
(437.5,52.330770)
(440,51.519794)
(442.5,50.867343)
(445,50.389480)
(447.5,50.097973)
(450,50)}{3}

\end{picture}
\end{center}
\caption{\small Examples of inverse propagator functions: a smooth function 
$P_\mu (k)$ which gives a particle and its three doublers (top left), 
$P_\mu (k)$ for the naive fermion propagator (top right) and the SLAC 
propagator (bottom left), and the inverse of the bosonic propagator 
(bottom right). The dashed parts of the curves lie outside the first 
Brillouin zone.}  
\label{fig:doublers}
\end{figure}
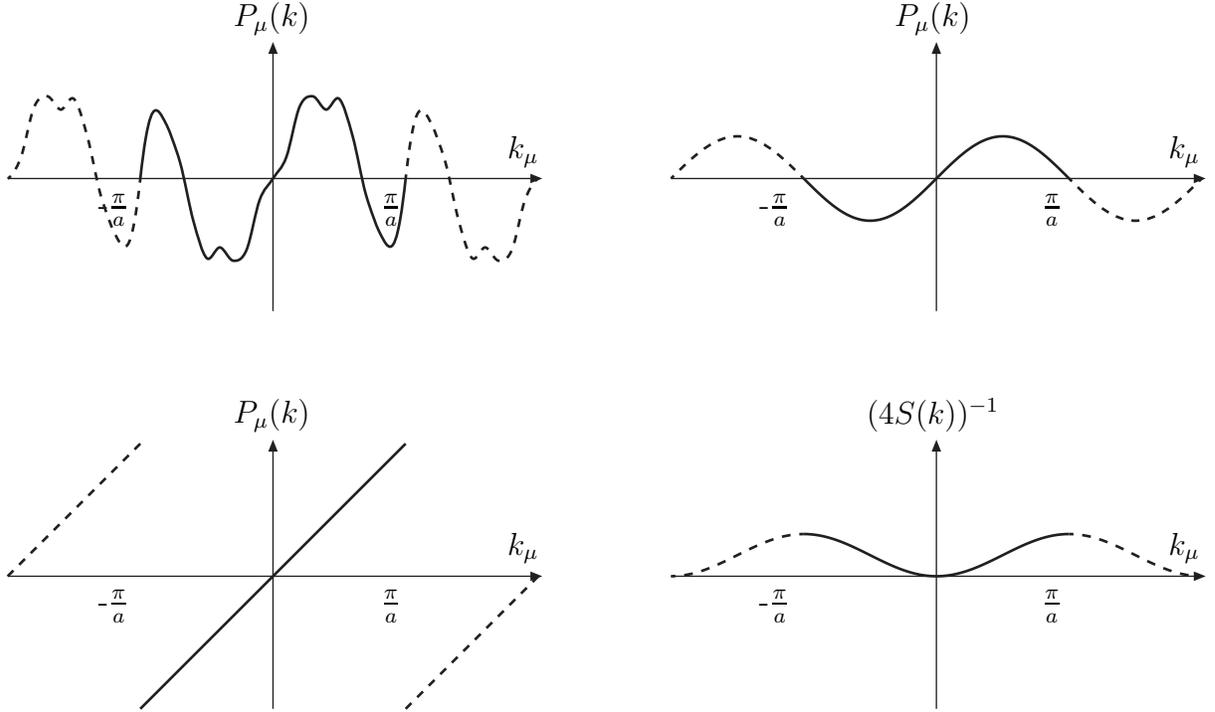

Let us assume at first that $P_\mu (k)$ is a continuous function. If the 
derivative in the Dirac operator in the action is anti-hermitian (like in 
Wilson), the theory is unitary and the function $P_\mu (k)$ is periodic in 
$2\pi$. Since it has a first order zero at $k_\mu=0$ (that is at 
$k_\mu = 2 \pi n_\mu$), it has to cross the $P_\mu=0$ axis one other time 
somewhere else in the first Brillouin zone, and with the opposite derivative, 
which corresponds to the opposite chirality (see Fig.~\ref{fig:doublers}, top 
left). This crossing is then precisely the doubler, and therefore it is 
unavoidable to have these extra particles in the theory. This reasoning only 
comes from global features, and is independent of the shape of the particular 
function, as long as it is continuous. Another heuristic proof coming from 
topological arguments, as well as a more general explanation of why half of 
the doublers have opposite chirality, has been given in~\cite{Wilczek:kw}. 
In this work a particular choice of the function $P_\mu (k)$ which minimizes 
the numbers of doublers is also proposed, which has the drawback of the loss 
of the equivalence of all four directions under discrete permutations, and the 
need of new operators (but no $1/a$ counterterm as in Wilson).

Let us come back to the naive propagator in Eq.~(\ref{eq:naiveprop}), i.e., 
$P_\mu (k) = \sin ak_\mu$. One crucial feature here is that the argument 
of the sine function is $k$, which is a consequence of the fact that one uses 
the anti-hermitian derivative $\nabla_\mu + \nabla_\mu^\star$. This choice 
causes all our problems. If the argument of the sine function were $k/2$, 
things would be completely different, since $\sin ak/2$ is antiperiodic, that 
is it has a period of $4\pi$ in $ak$, and there would be no need to cross the 
$P_\mu=0$ axis a second time {\em inside} the first Brillouin zone. Then no 
doublers would appear. However, $\sin ak/2$ can only appear in the propagator 
if one uses either the forward or the backward derivative, but not their sum. 
These derivatives are not anti-hermitian, and unfortunately in this case
the theory turns out not to be unitary, which causes other kinds of serious
problems. The fact is that $\nabla$ and $\nabla^\star$ are unphysical, and 
they propagate the fermion only in the forward or backward direction, and so 
using only one of them cannot give a Lorentz invariant theory in Minkowski 
space. It has also been shown, in lattice QED, that if one uses only the 
forward or backward derivative, then the interactions generate noncovariant 
contributions to the quark self-energy and vertex function, and the theory is 
nonrenormalizable~\cite{Sadooghi:1996ip}.

Note that a boson propagator does not have this problem, as it is the 
solution of a second order differential equation. Therefore the function 
$P_\mu (k)$ is in this case quadratic (and $\gamma_\mu$ disappears), 
and crossings are replaced by second-order zeros. Therefore, even if doublers 
were present, there would be no problems regarding the chiral properties
of the lattice theory. In any case, the Dirac operator can be chosen to be 
discretized using $\nabla^\star \nabla$, which is hermitian and produces a 
$\sin^2 k/2$ function in the propagator, as is well known.
Therefore in this case the function has only a minimum
at the origin without further crossings in the first Brillouin zone 
(see Fig.~\ref{fig:doublers}, bottom right), and there are no doublers. 

In the fermionic case, the only other way to avoid the second crossing would be
to consider a discontinuous function $P_\mu (k)$. The most famous example 
of this is given by the SLAC 
propagator~(Drell, Weinstein and Yankielowicz, 1976a; 1976b), for which 
$P_\mu (k) = k_\mu$ throughout the whole Brillouin
zone (see Fig.~\ref{fig:doublers}, bottom left). SLAC fermions have been 
studied in perturbation theory in~\cite{Karsten:1978nb,Karsten:1979wh}. 
However, this implies a nonlocality in the lattice action (it corresponds to 
a nonlocal lattice derivative), which brings many problems in the continuum 
limit. The locality assumption is important in order to avoid a disaster in 
the straightforward weak coupling expansion, otherwise propagators and vertex 
functions would not be analytic in the momenta. Another chiral action which 
is nonlocal has been proposed in~\cite{Rebbi:1986ra}.

At the end of the day the origin of the fermion doubling lies in the fact that
the Dirac equation is of first order. Doublers in the naive fermion action are
a necessary feature, and we can understand why also by looking at quantum 
anomalies. We know that in continuum quantum field theory the quantum 
corrections, and precisely the process of regularization, break chiral 
symmetry. A mass scale appears in the renormalized theory,~\footnote{For 
example, in dimensional regularization (which preserves the gauge symmetry) 
one needs to introduce a scale $\mu$ to define the coupling in noninteger 
dimensions, and in Pauli-Villars one introduces a heavy mass in the 
propagators.} which gives rise to the phenomenon of dimensional transmutation 
and causes the appearance of quantum anomalies. So even in theories that are 
chirally symmetric the axial current acquires an anomalous divergence through 
quantum effects.

Naive lattice fermions can be thought as a regularization that does not break 
chiral symmetry for any finite $a$. The lattice theory has no anomalies, and 
this implies that one needs some extra particles in order to cancel the 
continuum anomaly which is associated with the ``right'' fermion. When one 
tries to remove only the extra doublers from the game then the anomaly 
becomes again nonzero (as in the continuum), which must then correspond to a 
regularization which somehow has to break chiral symmetry, and in fact we end 
up with the lattice Wilson action. So, everything fits in the general picture.

After these general considerations we are now ready to state the 
Nielsen-Ninomiya theorem. In one of its formulations it says that the lattice 
massless Dirac operator $D = \gamma_\mu D_\mu$ in the fermionic action
\be
S_F = a^4 \sum_{x,y} \overline{\psi} (x) D(x-y) \psi (y)
\ee
cannot satisfy all of the following four properties at the same time:
\begin{description}
\item{(a)} $D(x)$ is local (in the sense that is bounded by 
           $C \e^{-\gamma |x|}$);
\item{(b)} its Fourier transform has the right continuum limit:
           $D(p) = \ii \gamma_\mu p_\mu +O(ap^2) $ for small $p$;
\item{(c)} $D(p)$ is invertible for $p \neq 0$ (and hence there are no 
           massless doublers);
\item{(d)} $\gamma_5 D + D \gamma_5 =0$ (it has chiral symmetry).
\end{description}
Therefore, for any given lattice action at least one of the these conditions 
has to fail. In particular, naive fermions have doublers and therefore do not 
satisfy (c), Wilson fermions break chiral symmetry and therefore do not 
satisfy (d), and SLAC fermions are not local and therefore do not satisfy (a).
The case of staggered fermions, another widely used (and old) fermion 
formulation which has been useful for studying problems in which chiral 
symmetry is relevant, and which will be discussed in the next Section, 
is more complicated from this point of view: only a $U(1) \otimes U(1)$ 
subgroup of the $SU(N_f) \otimes SU(N_f)$ chiral group remains unbroken, 
and the doublers are removed only partially.

Contrary to what one would naively expect after looking at the 
Nielsen-Ninomiya theorem, it is still possible to construct a Dirac operator 
which satisfies (a), (b) and (c) and is also chiral invariant. The solution to
this apparent paradox is that the corresponding chiral symmetry cannot be the 
one associated with a Dirac operator which anticommutes with $\gamma_5$, and 
the condition (d) has instead to be replaced by the Ginsparg-Wilson relation: 
$\gamma_5 D + D \gamma_5$ is not zero, but is proportional to $a D \gamma_5 D$.
Thus, the real lattice chiral symmetry turns out not to be what one would 
naively expect, the Nielsen-Ninomiya theorem is still valid and one can have a
nonpathological formulation of chiral fermions with no doublers.~\footnote{When
the condition that the Dirac operator anticommutes with $\gamma_5$ is 
released (at $a \neq 0$), the lattice quark propagator is not restricted to 
the form in Eq.~(\ref{eq:genprop}) and the considerations about the presence
of the doublers deriving from it are not anymore valid. In fact, one finds
more general propagator functions which in particular do not have the simple 
Dirac structure of Eq.~(\ref{eq:genprop}), as one can see by looking 
at the overlap propagator in Eqs.~(\ref{eq:ovprop}) and~(\ref{eq:omega}).}

This kind of lattice chiral fermions will be discussed in detail in 
Section~\ref{sec:ginspargwilsonfermions}. Before that, we will shortly present 
staggered fermions. If one wants to maintain some form of chiral symmetry, but
is prepared to give up flavor symmetry, then staggered fermions are the ideal 
fermions to work with. Otherwise, the only way to maintain chiral symmetry and
flavor symmetry at the same time (and of course all other fundamental 
properties like locality, unitarity etc.) leads again to the Ginsparg-Wilson 
relation.

\section{Staggered fermions}
\label{sec:staggeredfermions}

Another formulation of fermions on the lattice which is quite popular can be 
obtained by keeping part of the doublers of the naive fermion action and 
interpreting them as extra flavors. One remains with 4 fermion flavors whose 
16 components are split over a unit hypercube by assigning only a single 
fermion field component to each lattice site. This construction gives 
the staggered, or Kogut-Susskind, 
fermions~\cite{Kogut:ag,Banks:1975gq,Susskind:1976jm}, and can only be 
carried out in an even number of spacetime dimensions. It turns out that a 
continuous subgroup of the original chiral transformations remains a symmetry 
of this lattice action even at finite lattice spacing, and thus no mass 
counterterms are needed for vanishing bare quark masses. All this is achieved 
at the expense of flavor and translational symmetry, which become in fact all 
mixed together. 

The idea is to use only one spinor component for each Dirac spinor at each 
site. One has to decouple this component from the remaining three, which 
can then be kept out of the dynamics. This is accomplished doing a 
change of variables called spin-diagonalization~\cite{Kawamoto:1981hw}
\bea
\psi (x) &=& \gamma (x) \chi (x), \\
\overline{\psi} (x) &=& \overline{\chi} (x) \gamma^\dagger (x),
\eea
where
\be
\gamma (x=an) = \gamma_0^{n_0} \gamma_1^{n_1} \gamma_2^{n_2} \gamma_3^{n_3} 
\ee
depends only on $\bmod_2 (n_\mu)$ (because $\gamma_\mu^2=1$).
In these new spinor variables $\chi (x)$ and $\overline{\chi} (x)$ 
the naive fermion action
\be
S^f = a^4 \sum_{x} \sum_{\mu} \frac{1}{2a} \, \overline{\psi} (x) 
\gamma_\mu \Big[ U_\mu (x) \psi (x + a \hat{\mu})
- U_\mu^\dagger (x - a \hat{\mu}) \psi (x - a \hat{\mu}) \Big] 
+ a^4 \sum_{x} m_ f \overline{\psi} (x) \psi (x) 
\ee
becomes
\be
S^f_{stagg} = a^4 \sum_{x} \sum_{\mu} \frac{1}{2a} \, \overline{\chi} (x) 
\eta_\mu (x) \Big[ U_\mu (x) \chi (x + a \hat{\mu})
- U_\mu^\dagger (x - a \hat{\mu}) \chi (x - a \hat{\mu}) \Big] 
+ a^4 \sum_{x} m_ f \overline{\chi} (x) \chi (x) .
\label{eq:staggered}
\ee
Up to now, what we have done is just a rewriting of the usual naive action 
in terms of new variables. The crucial thing now is that the Dirac matrices 
have disappeared, and they have been replaced by the phase factors
\be
\eta_\mu (x=an)  = (-1)^{\sum_{\mu < \nu} n_\nu} .
\ee
Thus in the action (\ref{eq:staggered}) the 4 components of the spinor 
$\chi (x)$ are decoupled from each other. We are then allowed to keep 
only one spinor component out of four, and to forget about the others.
Since we started with the action of naive fermions, which presents 16
doublers, after the spin-diagonalization we end up only with 4 doublers.
This theory is a theory of 4 flavors.

The phase factors $\eta_\mu (x)$ bring a minus sign in the action for every 
translation of one lattice spacing $a$, and divide the lattice in even and odd
sites, so that they form two independent sublattices. This makes the action 
invariant only for translations of {\em two} lattice spacings. Translations of
an even number of lattice spacings correspond to ordinary continuum 
translations, whereas a translation of an odd number of lattice spacings 
interchanges the chiral components (i.e., is a chiral rotation of $\pi/2$).

It is then natural to take as fundamental objects hypercubes of linear
size $a$ (which contain 16 sites) rather than single sites. The 16 spinor 
components of the 4 flavors of the theory at each site $x$ can be assigned at 
the vertices of such a $2^4$-hypercube whose $(0,0,0,0)$ vertex is the point 
$x$. In the continuum limit each $2^4$-hypercube will then be mapped to a 
single physical point. In this way one distributes the fermionic degrees of 
freedom over the lattice, leaving only one degree of freedom per lattice site.
Beside the 4 components of each of the 4 continuum quarks, the components of 
the continuum $\gamma$ matrices are also spread over these $2^4$-hypercubes.  
With this construction there is a doubling of the effective lattice spacing, 
which becomes $2a$, and the staggered formulation effectively halves the size 
of the Brillouin zone. This is the way it (partially) solves the problem of 
the doublers. One works with an action in which there is only one independent 
Grassmann variable $\chi (x)$ per site. Only the color degrees of freedom 
remain at each site.

Since in $d$ dimensions the unit hypercube has $2^d$ sites, and a Dirac spinor
has $2^{d/2}$ components (for even $d$), one needs $2^{d/2}$ fermion fields 
to carry out this construction. In four dimensions this corresponds to 4 
flavors, and everything fits together. This also shows that it is not possible
to remove all the doublers in this way, because 16 is the minimal number of
sites of a four-dimensional unit hypercube, and not 4.

With the construction above, the 16-fold original degeneracy has thus been 
reduced to a 4-fold degeneracy. Actually, strictly speaking the 4 flavors are 
degenerate only in the continuum limit, while at finite lattice spacing the 
$SU(4)$ symmetry is broken. For nonzero $a$ the action maintains nevertheless
an exact $U(1)_V$ symmetry,
\be
\chi (x) \rightarrow \e^{\ii  \theta_V} \chi (x) , \quad
\overline{\chi} (x) \rightarrow \overline{\chi} (x) 
\e^{-\ii  \theta_V}  , 
\ee
corresponding to fermion number conservation, which in the case of vanishing 
bare masses is supplemented by a flavor nonsinglet axial $U(1)_A$ symmetry,
\be
\chi (x) \rightarrow \e^{\ii \theta_A \cdot (-1)^{n_0+n_1+n_2+n_3}} 
\chi (x) , \quad
\overline{\chi} (x) \rightarrow \overline{\chi} (x) 
\e^{\ii \theta_A \cdot (-1)^{n_0+n_1+n_2+n_3}} . 
\ee
This chiral $U(1)_L \otimes U(1)_R$ group is all what remains of the original 
$SU(4)_L \otimes SU(4)_R$ symmetry, but it is enough to guarantee that quark
masses are not additively renormalized. Thus, no mass counterterm 
is required if one starts with a zero bare mass, and this is a great advantage
of staggered over Wilson fermions.

The intertwining of spin and flavor is a major disadvantage of staggered 
fermions, and the correct spin-flavor structure is only recovered in the 
continuum limit. Much of the work in staggered calculations goes in the 
reconstruction of the continuum quark fields and operators from the 16 
one-component fields (carrying flavor and spin components) spread over 
the hypercubes. This can be complicated, and calculations with staggered 
fermions can thus become rather involved. For details on the way staggered
perturbation theory is set up, the reader can turn to the works 
of~\cite{Sharatchandra:1981si,vandenDoel:mf,Golterman:1984ds,Golterman:1984cy,Gockeler:1984rq}
and~\cite{Daniel:1987aa,Patel:1992vu,Sharpe:1993ur,Ishizuka:1993fs}. 
The difficult part of dealing with staggered fermions is in writing the 
lattice discretization of a continuum operator which has certain given 
symmetry properties. In fact, the construction of staggered lattice 
operators and the interpretation of their components in terms of spin 
and flavor turns out to be quite complicated. Moreover, there are various 
possibilities for the assignments of the various spin and  flavor components. 
One can choose to make these assignments in momentum space as well.

This is the price to pay for the fact that staggered fermions are numerically 
quite cheap. The mixing of operators under renormalization when staggered 
fermions are used can also become quite complicated. Flavor symmetry breaking 
can generate mixings with operators which were not present in the original 
continuum theory. If for example we consider a four-fermion operator with a 
certain flavor structure, then using staggered fermions this operator will mix
already at one loop with many other four-fermion operators which have 
different flavor components. This quickly renders these calculations 
technically involved, and complicates a lot perturbative calculations with 
staggered fermions. Gamma matrices are split as well, and also the presence of
color matrices (if one has $U$ fields in the operators) or derivatives 
contributes to entangle things more. To my knowledge no one has done the 
calculation of the renormalization of an operator like 
$\overline{\psi} \gamma_\mu D_\nu \psi$, which involves a covariant 
derivative, with staggered fermions. We will present the calculation of this 
operator with Wilson fermions in Section~\ref{sec:examplewilson}. On the other
hand, the perturbative calculations for $\epsilon'/\epsilon$ are at a rather 
advanced stage (for a recent work see~\cite{Lee:2001hc}, and other recent 
perturbative calculations with staggered fermions can be found 
in~\cite{Hein:2001kw,Nobes:2001tf} and~\cite{Lee:2002ui,Lee:2002bf}). 
Of course it makes much more sense to spend a lot of effort for weak 
operators, given the good chiral properties of staggered fermions. But 
recently the full understanding of the implications of the Ginsparg-Wilson 
relation has opened more ways for the investigation of these matrix elements 
on the lattice.

\section{Ginsparg-Wilson fermions}
\label{sec:ginspargwilsonfermions}

\subsection{The Ginsparg-Wilson relation}

For many years it was believed that the Nielsen-Ninomiya theorem had said
the final word with regard to the possibility of having reasonable chiral 
fermions on the lattice. 
The (now) fundamental paper by Ginsparg and Wilson (1982), in which the 
mildest breaking of chiral symmetry was introduced in a study of the 
block-spin renormalization group in lattice QCD, appeared shortly after 
the work of Nielsen and Ninomiya, but remained almost unnoticed.
Its importance was only recognized fifteen years later.

The Ginsparg-Wilson relation lay indeed dormant for all this time, until it 
was rediscovered in the context of perfect actions~\cite{Hasenfratz:1997ft}.
Shortly after, also overlap fermions and domain wall fermions, which had been
formulated some time before following ingenious ideas, and well before anyone
were thinking about the Ginsparg-Wilson relation, were also recognized to be 
a solution of this relation. Many interesting developments have then come out 
after the rediscovery of the paper of Ginsparg and Wilson, and for a general 
overview of these developments the reader can study the excellent reviews 
of~\cite{Niedermayer:1998bi},~\cite{Creutz:2000bs},~\cite{Luscher:2000hn},
and~\cite{Neuberger:2001nb}, which cover different aspects of them, as
well as the shorter and original discussion of the main ideas given 
in~\cite{Hernandez:2002hu}.
An up-to-date discussion of the numerical results which have been obtained 
using Ginsparg-Wilson fermions can be found in~\cite{Giusti:2002rx}.

A Dirac operator $D$ which satisfies the Ginsparg-Wilson relation
\be
\gamma_5 D + D \gamma_5 = a \, \frac{1}{\rho} \, D \gamma_5 D
\label{eq:gw}
\ee
and the hermiticity condition $D^\dagger = \gamma_5 D \gamma_5$ 
defines fermions which have exact chiral symmetry, present no 
doublers~\footnote{Although in general the Ginsparg-Wilson relation does not 
guarantee the absence of doublers, we are of course only interested in actions
which have no doublers and the solutions that we will discuss are all 
of this kind.} and which also possess all other fundamental principles 
like flavor symmetry, locality~\cite{Hernandez:1998et},~\footnote{Locality 
in this context does not have the meaning of strict locality, but it is to be 
understood in the larger sense that the strength of the interaction decays 
exponentially with the distance in lattice units. It becomes microscopically 
small when one considers the continuum limit.}  unitarity and gauge invariance.

L\"uscher has shown that fermions obeying the Ginsparg-Wilson 
relation possess an exact chiral symmetry at finite lattice spacing,
which is of the form~\cite{Luscher:1998pq}
\bea
\psi &\rightarrow& \psi + \epsilon \cdot \gamma_5 \, \Big(1-\frac{a}{\rho} \, 
D \Big) \, \psi , \\
\overline{\psi} &\rightarrow& \overline{\psi} +  
\epsilon \cdot \overline{\psi} \, \gamma_5 .
\eea
Note the asymmetric way in which $\psi$ and $\overline{\psi}$ are treated.
The global anomaly of the original continuum fermions is also reproduced at 
nonzero lattice spacing (as was already noticed in the case of domain wall 
fermions by~\cite{Jansen:1992yj}). In terms of the quark propagator the 
Ginsparg-Wilson relation reads 
\be
S(x,y) \gamma_5 + \gamma_5 S(x,y) = a \, \frac{1}{\rho} \, 
\gamma_5 \, \delta(x-y) ,
\ee
which implies that the propagator is chirally invariant at all nonzero 
distances, i.e., on the mass shell.

This surprising result is a new form of chiral symmetry that can coexist with 
a momentum cutoff. The fact is that chiral symmetry can be realized on the 
lattice in different ways other than the naive expectation, and this does not 
constitute an exception to the Nielsen-Ninomiya theorem. In this case, the 
condition of the theorem that is not fulfilled is the anticommutation of the 
Dirac operator with $\gamma_5$, which can only be recovered in the continuum 
limit. Chirality (in this new formulation) remains a symmetry of the lattice 
theory also for nonzero lattice spacing, as are flavor symmetry and the other 
fundamental symmetries.

The chiral symmetry associated with the Ginsparg-Wilson relation can be
used to define left- and right-handed fermions. We first note that the
operator
\be
\widehat{\gamma}_5 = \gamma_5 \, \Big( 1 -\frac{a}{\rho} \, D \Big) 
\ee
satisfies 
\be
(\widehat{\gamma}_5)^\dagger = \widehat{\gamma}_5, \qquad 
(\widehat{\gamma}_5)^2 =1, \qquad
D \widehat{\gamma}_5 = -\gamma_5 D .
\ee
The projectors 
\bea
\widehat{P}_\pm &=& \frac{1}{2} (1 \pm \widehat{\gamma}_5) 
\label{eq:chiralproj}\\
P_\pm &=& \frac{1}{2} (1 \pm \gamma_5) \nonumber
\eea
can then separate the two chiral sectors. In particular,
the constraints 
\bea
\widehat{P}_- \psi &=& \psi \\ 
\overline{\psi} P_+ &=& \overline{\psi} 
\eea
define the left-handed fermions. L\"uscher's chiral symmetry 
can then be rewritten in the appealing form
\bea
\psi &\rightarrow& \psi + \epsilon \cdot \widehat{\gamma}_5 \, \psi \\
\overline{\psi} &\rightarrow& \overline{\psi} +  
\epsilon \cdot \overline{\psi} \, \gamma_5 .
\eea
It is to be remarked here that the definition of left-handed fields is not 
independent of the gauge fields $U$. The full implications of this fact will 
be seen when one wants to construct chiral gauge theories on the lattice,
and we will discuss them in Section~\ref{sec:chiralgaugetheories}.
One of the most remarkable developments has indeed been the discovery that 
Ginsparg-Wilson fermions constitute a nonperturbative regularization of gauge 
theories with an explicit chiral symmetry.

The known solutions of the Ginsparg-Wilson relation are given by overlap, 
domain wall and fixed-point (``classically perfect'') fermions. We will now 
have a look at these three cases in some detail. A few other actions which are 
approximate solutions of the Ginsparg-Wilson relation have sprung up in recent
years. We will not discuss them here. They are much more complicated than 
these three cases, especially from the point of view of perturbation theory. 
We will see that overlap, domain wall and fixed-point fermions bring already 
many complications into the perturbative features of the theory.

We mention at last that for Ginsparg-Wilson fermions it is also possible to 
define a conserved axial current, and of course a conserved vector current 
(which was possible already in the Wilson case). The form of these Noether 
currents is rather complicated. They extend over all lattice sites, and the 
kernels decay exponentially with the distance from the ``physical'' point 
(in which the corresponding continuum current is defined). These are then 
still local currents in the sense explained above. They are given explicitly 
in~\cite{Kikukawa:1998py} for overlap fermions.

\subsection{Overlap fermions}

Almost ten years ago Narayanan and Neuberger (1993a; 1993b; 1994; 1995),
motivated from mathematical insights and previous theoretical 
developments~\cite{Callan:sa,Kaplan:1992bt,Frolov:ck}, devised an ingenious 
construction with which it was shown how to define chiral fermions on the 
lattice. The chiral mode resulted from the overlap between two infinite towers
of chiral fermion fields, and the infinite number of fermions for each lattice
site was the crucial new feature which was understood as necessary in order to
construct chiral fermions on the lattice. The Dirac operator coming from this 
overlap formalism was later recognized by Neuberger (1998a; 1998b; 1998c) to 
be a solution of the Ginsparg-Wilson relation, and its action was given a 
simple form.

In the massless case~\footnote{When the quarks have a nonzero bare mass $m_0$ 
then the overlap-Dirac operator is given by
\be
D_N^{(m_0)} = \Big( 1 - \frac{1}{2\rho} \, a m_0 \Big) D_N + m_0 .
\ee
} 
the overlap-Dirac operator is~\footnote{We note that any $X$ which satisfies 
\be
\gamma_5 X^\dagger = X \gamma_5
\ee
makes $D_N$ a solution of the Ginsparg-Wilson relation. Using the Wilson 
action is the standard option, although other actions have been sometimes 
chosen for $X$, which are supposed to improve things. Such generalized overlap
fermions have been proposed by~\cite{Bietenholz:1998ut,Bietenholz:1999km,Bietenholz:2000iy,Bietenholz:2002ks}. They showed that when $X$ is a truncated 
perfect action (see later) the convergence properties of these fermions are 
improved, together with their locality and other symmetries. The presence in 
$X$ of fat link actions (Section~\ref{sec:fatlinks}) can further improve 
things~\cite{Bietenholz:2000iy,DeGrand:2000tf}. Perturbation theory however 
becomes much more complicated in all these cases.}
\be
D_N = \frac{1}{a} \, \rho \, \Bigg( 1+ \frac{X}{\sqrt{X^\dagger X}} \Bigg) ,
\qquad X=D_W -\frac{1}{a} \, \rho ,
\ee
where $D_W$ is the usual Wilson-Dirac operator:
\bea
D_W &=& \frac{1}{2} \Big( \gamma_\mu (\widetilde{\nabla}^\star_\mu 
+ \widetilde{\nabla}_\mu) - a r \widetilde{\nabla}^\star_\mu 
\widetilde{\nabla}_\mu \Big) \\
\widetilde{\nabla}_\mu \psi (x) &=& 
\frac{1}{a} \Big( U(x,\mu)\psi(x+a\hat{\mu}) - \psi (x) \Big) . \nonumber 
\eea
In the range $0 <\rho <2r$ (at tree level) the right spectrum of massless 
fermions is obtained. For the pure gauge part one usually uses the Wilson 
plaquette action. Sometimes improved pure gauge actions are also used.

Since additive mass renormalization is forbidden by chiral symmetry, when 
using overlap fermions one avoids altogether a source of systematic errors 
that is always present with Wilson fermions.

Although the interactions in the overlap action are not limited to 
nearest-neighbor sites, and not even to next-to-nearest-neighbor sites, 
but in fact involve all sites, the strength of these interactions
falls off exponentially with the distance, given in lattice units,
and in this sense the theory is still local.

Let us now have a look at perturbation theory with overlap fermions.
The interaction vertices and the quark propagator are much more complicated 
than the ones in the Wilson formulation. This causes the perturbative 
computations to be rather cumbersome, and one then needs computer programs, 
even in the simplest cases. The calculations in~\cite{Capitani:2000wi,Capitani:2000aq,Capitani:2000da,Capitani:2000bm,Capitani:2001yq} have been carried out 
using an ensemble of routines written in the symbolic manipulation language 
FORM. These routines are an extension of the ones used to do calculations with
the Wilson action in several cases.

In deriving the Feynman rules we start by noticing that in momentum space
\be
X_0 (k) = \frac{1}{a} \Bigg( \ii  \sum_\mu \gamma_\mu \sin ak_\mu
+ 2r \sum_\mu \sin^2 \frac{ak_\mu}{2} -\rho \Bigg) .
\label{eq:xzero}
\ee
The massless quark propagator, which is computed inverting 
\be
\frac{1}{a} \, \rho \, \Bigg( 
1+ \frac{X_0 (k)}{\sqrt{X^\dagger_0 (k) X_0 (k)}} \Bigg) ,
\ee
is then given by
\be
S^{ab} (k) = \delta^{ab} \cdot \Bigg(
\frac{-\ii  \sum_\mu \gamma_\mu \sin ak_\mu}{2 \rho
\left( \omega(k)+b(k) \right) } + \frac{a}{2 \rho} \Bigg) ,
\label{eq:ovprop}
\ee
where
\bea
\omega(k) &=& \Big(\sqrt{X^\dagger X} \Big)_0 (k) = 
\frac{1}{a} \sqrt{ \sum_\mu \sin^2 ak_\mu + \Bigg( 2r \sum_\mu
\sin^2 \frac{ak_\mu}{2} -\rho \Bigg)^2 } , 
\label{eq:omega} \\
b(k) &=& \frac{1}{a} \Bigg( 2r \sum_\mu
\sin^2 \frac{ak_\mu}{2} -\rho \Bigg) . \nonumber
\eea

To compute the interaction vertices one first expands $X$ order by order:
\be
X(p_1,p_2) = X_0 (p_1) (2\pi)^4 \delta^{(4)} (p_1 - p_2) 
+ X_1 (p_1,p_2) + X_2 (p_1,p_2) + O(g_0^3),
\ee
where $X_0$ is given in Eq.~(\ref{eq:xzero}), and the $X_i$'s are the same
interaction vertices of the Wilson action. For one-loop calculations one needs
to compute only the vertices of order $g_0$ and $g_0^2$.
Let us now write the expansion of the inverse of the square root as follows:
\be
\frac{1}{\sqrt{X^\dagger X}} (p_1,p_2) = \Bigg( \frac{1}{\sqrt{X^\dagger X}} 
\Bigg)_0 (p_1,p_2) + Y_1 (p_1,p_2) + Y_2 (p_1,p_2) + O(g_0^3).
\ee
$Y_1$ and $Y_2$ can be obtained imposing the identity~\cite{Kikukawa:1998pd}
\be
\int_{-\frac{\pi}{a}}^{\frac{\pi}{a}} \frac{d^4 q}{(2\pi)^4} \,  
\int_{-\frac{\pi}{a}}^{\frac{\pi}{a}} \frac{d^4 r}{(2\pi)^4} \,  
(X^\dagger X)(p_1,q) \, \Bigg( \frac{1}{\sqrt{X^\dagger X}} 
\Bigg) (q,r) \, \Bigg( \frac{1}{\sqrt{X^\dagger X}} \Bigg) (r,p_2)
= (2\pi)^4 \delta^{(4)} (p_1 - p_2)
\ee 
order by order in the coupling. For example, at first order one has
\bea
&& (X^\dagger X)_1 (p_1,p_2) \,
\Bigg( \frac{1}{\sqrt{X^\dagger X}} \Bigg)_0^2 (p_2) + 
(X^\dagger X)_0 (p_1) \, Y_1 (p_1, p_2) \, \Bigg( \frac{1}{\sqrt{X^\dagger X}} 
\Bigg)_0 (p_2) \nonumber \\
&& \qquad \qquad + (\sqrt{X^\dagger X})_0 (p_1) \, Y_1 (p_1,p_2) = 0,
\eea
which solving for $Y_1$ gives
\be
Y_1 (p_1,p_2) = 
- \frac{1}{(\omega(p_1) + \omega(p_2)) \omega(p_1) \omega(p_2)} \,
(X^\dagger X)_1 (p_1,p_2).
\ee
One can easily compute $(X^\dagger X)_1$ and thus obtain an explicit 
expression for $Y_1$. With somewhat more manipulations one also arrives at 
an expression for $Y_2$. The overlap vertices can be finally computed using 
\bea
\Bigg( \frac{X}{\sqrt{X^\dagger X}} \Bigg)_1 &=& X_0 Y_1 + X_1 
\Bigg( \frac{1}{\sqrt{X^\dagger X}} \Bigg)_0 \\
\Bigg( \frac{X}{\sqrt{X^\dagger X}} \Bigg)_2 &=& X_0 Y_2 + X_1 Y_1 + X_2 
\Bigg( \frac{1}{\sqrt{X^\dagger X}} \Bigg)_0 .
\eea
It is clear that the interaction vertices can be entirely given in terms 
of the vertices of the QED Wilson action,
\bea
W_{1\mu} (p_1,p_2) &=&
         -g_0\Bigg( \ii  \gamma_\mu \cos \frac{a(p_1+p_2)_\mu}{2}
              + r \sin \frac{a(p_1+p_2)_\mu}{2} \Bigg) \\
W_{2\mu} (p_1,p_2) &=& 
- \frac{1}{2} a g_0^2 \Bigg( -\ii  \gamma_\mu \sin \frac{a(p_1+p_2)_\mu}{2}
              + r \cos \frac{a(p_1+p_2)_\mu}{2} \Bigg) \nonumber
\eea
(where $p_1$ and $p_2$ are the quark momenta flowing 
in and out of the vertices), and of $X_0(p)$.
The quark-quark-gluon vertex in the overlap theory has then the 
expression
\be
(V^a_1)^{bc}_\mu (p_1,p_2) = (T^a)^{bc} \, \cdot 
\rho \, \frac{1}{\omega(p_1) + \omega(p_2)} 
\Bigg[ W_{1\mu} (p_1,p_2) -\frac{1}{\omega(p_1)\omega(p_2)} X_0(p_2) 
W_{1\mu}^\dagger (p_1,p_2) X_0(p_1) \Bigg] ,
\ee
which in the continuum limit is the usual QCD vertex,
$ - g_0 (T^a)^{bc} \, \ii  \gamma_\mu $.
The quark-quark-gluon-gluon vertex in the overlap theory is
\bea
&& (V^{ab}_2)^{cd}_{\mu\nu} (p_1,p_2) = 
\delta_{\mu\nu} \, \Big( \frac{1}{N_c} \delta^{ab} + d^{abe} T^e \Big)^{cd} 
\nonumber \\  
&&\qquad \times \Bigg\{ \rho \, \frac{1}{\omega(p_1) + \omega(p_2)}   
\Bigg[ W_{2\mu} (p_1,p_2) -\frac{1}{\omega(p_1)\omega(p_2)} X_0(p_2) 
W_{2\mu}^\dagger (p_1,p_2) X_0(p_1) \Bigg] \nonumber \\
&& +\frac{1}{2}\rho \, \frac{1}{\omega(p_1) +\omega(p_2)} 
\frac{1}{\omega(p_1) + \omega(k)} \frac{1}{\omega(k) + \omega(p_2)} 
\label{eq:overlapqqgg} \\
&& \times \Bigg[ X_0(p_2) W_{1\mu}^\dagger (p_2,k) W_{1\nu} (k,p_1)
+W_{1\mu} (p_2,k) X_0^\dagger (k) W_{1\nu} (k,p_1) 
+W_{1\mu} (p_2,k) W_{1\nu}^\dagger (k,p_1) X_0 (p_1) \nonumber \\
&& \qquad 
-\frac{\omega(p_1)+\omega(k)+\omega(p_2)}{\omega(p_1)\omega(k)\omega(p_2)} 
X_0 (p_2) W_{1\mu}^\dagger (p_2,k) X_0 (k) W_{1\nu}^\dagger (k,p_1) X_0 (p_1)
\Bigg] \Bigg\} . \nonumber
\eea
This is an irrelevant operator and then, although it is not obvious
by its visual appearance, it vanishes in the continuum limit.

Beyond the lowest orders the calculation of the vertices becomes increasingly 
complicated, but in principle it can be carried through. Perhaps in this
case it is more convenient to use an alternative method to derive the Feynman 
rules, which involves less algebra but more integrals.
It makes use of the integral representation 
\be
\frac{1}{\sqrt{X^\dagger X}} = \int_{-\infty}^\infty \frac{dt}{\pi} \,
\frac{1}{t^2 + X^\dagger X}.
\ee
The expansion of the square root to the second order in this representation
is given by
\bea
\frac{1}{\sqrt{X^\dagger X}} &=& \int_{-\infty}^\infty \frac{dt}{\pi} \, 
\frac{1}{t^2+X^\dagger_0 X_0} - \int_{-\infty}^\infty \frac{dt}{\pi} \, 
\frac{1}{t^2+X^\dagger_0 X_0} (X^\dagger_0 X_1 + X^\dagger_1 X_0) 
\frac{1}{t^2+X^\dagger_0 X_0} \\
&& - \int_{-\infty}^\infty \frac{dt}{\pi} \, \frac{1}{t^2+X^\dagger_0 X_0} 
(X^\dagger_0 X_2 + X^\dagger_1 X_1 + X^\dagger_2 X_0) 
\frac{1}{t^2+X^\dagger_0 X_0} \nonumber \\
&& + \int_{-\infty}^\infty \frac{dt}{\pi} \, \frac{1}{t^2+X^\dagger_0 X_0} 
(X^\dagger_0 X_1 + X^\dagger_1 X_0) \frac{1}{t^2+X^\dagger_0 X_0} 
(X^\dagger_0 X_1 + X^\dagger_1 X_0) \frac{1}{t^2+X^\dagger_0 X_0}, \nonumber 
\eea
so that, using the fact that $1/(t^2+X^\dagger_0 X_0)$ commutes with $X_0$, 
we have that
\bea
aD &=& aD_0 + \int_{-\infty}^\infty \frac{dt}{\pi} \, 
\frac{1}{t^2+X^\dagger_0 X_0} (t^2 X_1 - X_0 X^\dagger_1 X_0) 
\frac{1}{t^2+X^\dagger_0 X_0} \\
&& + \int_{-\infty}^\infty \frac{dt}{\pi} \, \frac{1}{t^2+X^\dagger_0 X_0} 
(t^2 X_2 - X_0 X^\dagger_2 X_0) \frac{1}{t^2+X^\dagger_0 X_0} \nonumber \\
&& - \int_{-\infty}^\infty \frac{dt}{\pi} \, t^2 \frac{1}{t^2+X^\dagger_0 X_0}
X_1 \frac{1}{t^2+X^\dagger_0 X_0} (X^\dagger_0 X_1 + X^\dagger_1 X_0) 
\frac{1}{t^2+X^\dagger_0 X_0} \nonumber \\
&& - \int_{-\infty}^\infty \frac{dt}{\pi} \, t^2 
\frac{1}{t^2+X^\dagger_0 X_0} X_0 X^\dagger_1 
\frac{1}{t^2+X^\dagger_0 X_0} X_1 \frac{1}{t^2+X^\dagger_0 X_0} \nonumber \\
&& + \int_{-\infty}^\infty \frac{dt}{\pi} \, X_0 \frac{1}{t^2+X^\dagger_0 X_0}
X^\dagger_1 X_0 \frac{1}{t^2+X^\dagger_0 X_0} 
X^\dagger_1 X_0 \frac{1}{t^2+X^\dagger_0 X_0} + \cdots \nonumber .
\eea
The $t$ integration can now be performed using the residue theorem, and at 
the end one obtain the various vertices.

For more details on one-loop perturbative calculations with overlap fermions
the works of~\cite{Chiu:1998aa,Ishibashi:1999ik} and~\cite{Yamada:1998se} are 
useful. One can also see~\cite{Fujikawa:2002ei}, which contains many technical
details about perturbation theory, although for generalized Ginsparg-Wilson 
relations. 

Many 1-loop calculations with overlap fermions have been already 
carried out, including the relation between the $\Lambda$ parameter in the 
lattice scheme defined by the overlap operator and in the 
$\ms$ scheme~\cite{Alexandrou:1999wr}, and the renormalization factors of the 
quark bilinears $\overline{\psi} \Gamma \psi$~\cite{Alexandrou:2000kj}, 
of the lowest moments of all structure 
functions~\cite{Capitani:2000wi,Capitani:2000aq,Capitani:2001yq}, 
and of the $\Delta S=2$ and $\Delta S=1$ effective weak Hamiltonians, which 
are important for the calculation of $\Delta I=1/2$ amplitudes (see 
Section~\ref{sec:amixing}) and of the parameter $\epsilon'/\epsilon$ on the 
lattice~\cite{Capitani:2000da,Capitani:2000bm}.

A calculation of the axial anomaly to order $g_0^4$ has been presented 
in~\cite{Chiu:2001ja}.

\subsection{Domain wall fermions}

\begin{figure}[t]
\begin{center}
\begin{picture}(320,220)(-20,-20)
\LongArrow(0,100)(300,100)
\LongArrow(150,0)(150,200)
\Text(300,105)[b]{$s$}
\Text(150,205)[b]{$\Phi (s)$}
\Line(140,180)(160,180)
\Text(165,185)[b]{$M$}
\Line(140,20)(160,20)
\Text(165,25)[b]{$-M$}

\SetWidth{1}
\Curve{(10,20.01414706)
(20,20.02755242)
(30,20.05365602)
(40,20.10447446)
(50,20.20336260)
(60,20.39561970)
(70,20.76876046)
(75,21.07085614)
(80,21.49055350)
(85,22.07259640)
(90,22.87779359)
(95,23.98790826)
(100,25.51123130)
(105,27.58813971)
(110,30.39506706)
(115,34.14394834)
(120,39.07246752)
(125,45.41905678)
(130,53.37736437)
(135,63.03062742)
(140,74.27898100)
(145,86.78876696)
(150,100)
(155,113.2112330)
(160,125.7210190)
(165,136.9693726)
(170,146.6226356)
(175,154.5809432)
(180,160.9275325)
(185,165.8560517)
(190,169.6049329)
(195,172.4118603)
(200,174.4887687)
(205,176.0120917)
(210,177.1222064)
(215,177.9274036)
(220,178.5094465)
(225,178.9291439)
(230,179.2312395)
(240,179.6043803)
(250,179.7966374)
(260,179.8955255)
(270,179.9463440)
(280,179.9724476)
(290,179.9858529)}
\end{picture}
\end{center}
\caption{\small A background field for the domain wall.}
\label{fig:dw}
\end{figure}
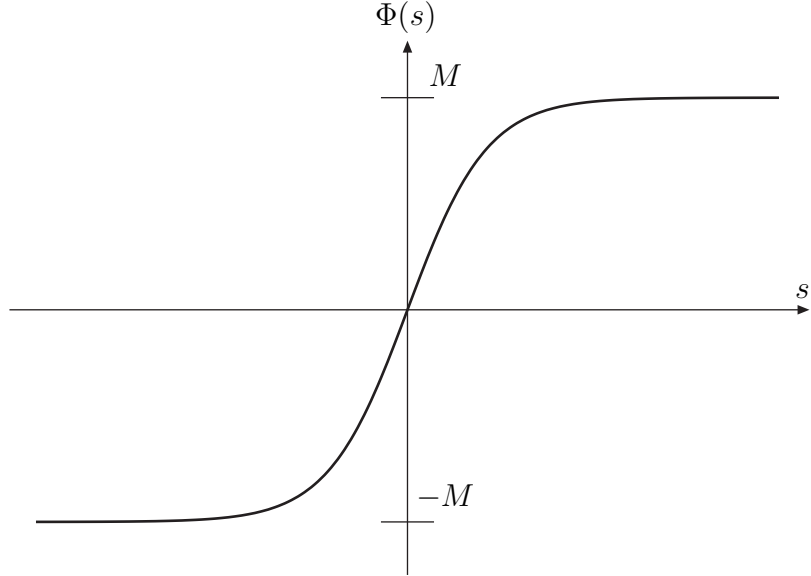

Another solution of the Ginsparg-Wilson relation is given by domain wall
fermions. The construction of domain wall fermions is not peculiar to the 
lattice, and we begin by discussing a simple continuum case. The main point, 
which traces back to ideas from~\cite{Callan:sa}, is to work in a 5-dimensional
spacetime where in the fifth dimension, denoted by $s$, there is a domain wall
separating the region $s>0$ from the region $s<0$. This domain wall can be 
described by a background field $\Phi(s)$ with a behavior like in 
Fig.~\ref{fig:dw}. An example is $\Phi(s)=M \tanh (Ms)$, but the precise form 
is not important. The essential thing is that it behaves like a step function 
with a height of order $M$ and a width of order $1/M$.

The 5-dimensional free Dirac operator in this theory is formed adding 
to the usual term a derivative term in the fifth dimension which is 
proportional to $\gamma_5$, and the background field: 
\be
D_5 = \gamma_\mu \partial_\mu + \gamma_5 \partial_s - \Phi(s).
\ee
The eigenvectors of this operator can be written in the form 
\be
\chi (x,s) = \e^{\ii px} u(s),
\ee
where the 4-dimensional plane waves are solution of the usual 4-dimensional 
Dirac equation, while $u(s)$ satisfies
\be
\Big( \gamma_5 \partial_s - \Phi(s) \Big) u(s) = -\ii \gamma_\mu p_\mu u(s).
\ee
All solutions $\chi (x,s)$ have a definite chirality and describe fermions 
which have a mass of order $M$ (the only available scale), except a mode
which is massless and which obeys the equations
\be
(\gamma_5 \partial_s - \Phi(s) ) u(s) = 0,  \qquad \gamma_\mu p_\mu u(s) =0 .
\ee
The massless chiral solutions are then given by
\be
u(s) = \exp \Big( \pm \int_0^s dt \, \Phi(t) \Big) v, 
\qquad \gamma_\mu p_\mu v = 0,
\qquad P_\pm v = v,
\ee
and we see that the solution with positive chirality is nonnormalizable for 
$|s| \to \infty$, while the interesting result is the solution with negative 
chirality. This is the massless chiral mode, the aim of the domain wall 
construction, which from the above equation can be seen to fall off 
exponentially in the fifth dimension. This mode is thus confined near the 
domain wall $s=0$, and at energies which are far below $M$ this is the only 
particle left in the theory, which has then effectively become, after this 
dimensional reduction, a chiral theory in 4 dimensions.~\footnote{For more 
details on this derivation see also~\cite{Luscher:2000hn}.}

This chiral mode is present also when the theory is formulated on a lattice 
(Kaplan 1992; 1993). The scale $M$ is now meant to be of order $1/a$, 
so that for any finite lattice spacing only the massless chiral mode can 
propagate, while the massive particles decouple from the theory.

As already noted, the precise form of the background field is not important. 
Let us consider the half space $s \ge 0$, with $P_+ \chi (x,s) |_{s=0} = 0$,
in which we can take $\Phi (s) = M$. The Dirac operator is now just 
\be
D_5 = D_4 + \gamma_5 \partial_s -M ,
\ee
with $D_4 = \gamma_\mu \partial_\mu$. It can be shown that the 5-dimensional
propagator on the domain wall is then given by
\be
G(x,s;y,t) \Big|_{s=t=0} = 2M \cdot P_- S(x,y) P_+,
\ee
where the propagator $S(x,y)$ is the inverse of the 4-dimensional operator
\be
D = M +\frac{D_4-M}{\sqrt{1-\Bigg({\displaystyle \frac{D_4}{M}}\Bigg)^2}},
\ee
which satisfies the relation
\be
\gamma_5 D + D \gamma_5 = \frac{1}{M} D \gamma_5 D.
\ee
If we make at this point the identification $M = \rho/a$, the above equations,
which we have obtained after doing the dimensional reduction, are just the 
overlap solution (when $D_4=D_W$) and the Ginsparg-Wilson relation. 
In this sense, overlap and domain wall fermions are equivalent, and one can 
consider the overlap case like a domain wall seen from the 4-dimensional world.

The lattice formulation of domain wall fermions that is commonly used in 
simulations has been proposed in~\cite{Shamir:1993zy,Furman:ky}. For a review 
see also~\cite{Jansen:1994ym}. The 5-dimensional action is constructed 
starting from the 4-dimensional Wilson action (otherwise some doublers mode 
would still be present), and is given by~\footnote{In the following we put 
$a=1$.}
\bea
S_{DW} &=& \sum_x \sum_{s=1}^{N_s} \Bigg[ 
\frac{1}{2} \sum_\mu \Big( \overline{\psi}_s(x)
(-r + \gamma_\mu) U_\mu (x) \psi_s(x+\hat{\mu}) + \overline{\psi}_s(x)
(-r -\gamma_\mu) U^\dagger_\mu (x-\hat{\mu}) \psi_s(x-\hat{\mu}) \Big) 
\nonumber \\
&& +\frac{1}{2} \Big( \overline{\psi}_s(x) (1+\gamma_5) \psi_{s+1}(x)
+ \overline{\psi}_s(x) (1-\gamma_5) \psi_{s-1}(x) \Big)
+(\rho-1+4r) \overline{\psi}_s(x) \psi_s(x) \Bigg] \nonumber \\
&& +m \sum_x \Big( \overline{\psi}_{N_s}(x)  P_+ \psi_1(x) 
+ \overline{\psi}_1(x)  P_+ \psi_{N_s}(x) \Big) .
\label{eq:dwaction}
\eea
There is a parameter $\rho$, which corresponds to the one already seen in the
overlap case, and has to take values between zero and two (at tree level). 
The Wilson parameter is set to $r=-1$, and the sign of the Wilson term is 
different. This domain wall action can be thought as a Wilson action with an 
infinite number of flavors (labeled by the index $s$) and a special mass 
matrix (Eqs.~(\ref{eq:massmatrix1}) and~(\ref{eq:massmatrix2}) below). 
The infinite number of flavors corresponds to the infinite tower of fermions 
for each lattice site of the original overlap formulation.

The pure gauge part is the usual Wilson plaquette action. The gauge fields are 
4-dimensional, and therefore there is no gauge interaction along the fifth 
dimension. The measure term, the gauge-fixing term and the Faddeev-Popov term
are also the same as in Wilson. This means that the gluon propagator and the 
quark-gluon vertices are the same as in the Wilson action. The quark 
propagator is instead different, and much more complicated. In practical 
terms this lattice action can be looked at as describing a theory of $N_s$ 
fermion flavors which have a complicated propagator.

As the reader will have noticed, the range of $s$ in Eq.~(\ref{eq:dwaction}) 
is finite, $1 \le s \le N_s$, because this is how the simulations have to be 
made. All modes have a mass of order $\rho/a$ except two modes which are 
nearly massless (that is, their masses are exponentially small in $N_s$) and 
which are localized near the boundaries $s=1$ and $s=N_s$. These two modes 
have opposite chirality. As long as $N_s$ is finite there is a small residual 
interaction (exponentially small in $N_s$) between these two chiral 
modes.~\footnote{The overlap between the chiral modes on the two walls depends
also on the strength of the gauge coupling, and for strong couplings these 
chiral modes tend to acquire some nonnegligible overlap.} It is only in the 
limit in which the number of points in the fifth dimension goes to infinity 
that the chiral mode at $s=1$ becomes massless and decouples entirely from the
other massless mode, giving an exact chiral theory. In the perturbative 
calculations made so far~\cite{Aoki:1996zz,Aoki:1997xg,Aoki:1998vv,Aoki:1998hi,Aoki:1999ky,Aoki:2000ee,Aoki:2002iq} it has been assumed that one can take the 
limit $N_s \to \infty$ before doing the Feynman integrals. The quark propagator
which one uses in perturbation theory is then the one for $N_s = \infty$. 

The domain wall Dirac operator in momentum space is
\be
D_{st} (p) = \delta_{s,t} \, \sum_\mu \ii \gamma_\mu \sin p_\mu + 
(W^+_{st} (p) + mM^+_{st}) P_+ + (W^-_{st} (p) +mM^-_{st}) P_- ,
\ee
with
\bea
W^+_{st}(p) &=& -W(p) \, \delta_{s,t} + \delta_{s+1,t} , \\
W^-_{st}(p) &=& -W(p) \, \delta_{s,t} + \delta_{s-1,t} , \\
W(p) &=& 1 - \rho -2r \sum_\mu \sin^2 \frac{p_\mu}{2} ,
\eea
and the mass matrix is
\bea
M^+_{st} &=& \delta_{s,N_s} \, \delta_{t,1} , \label{eq:massmatrix1} \\
M^-_{st} &=& \delta_{s,1} \, \delta_{t,N_s} , \label{eq:massmatrix2} .
\eea
By inverting it one gets the quark propagator 
\bea
S_{st} (p) = \langle \psi_s (-p) \overline{\psi}_t (p) \rangle
&=& (-\ii \gamma_\mu \sin p_\mu \, \delta_{s,u} +W^-_{su} (p) +mM^-_{su} (p) ) 
\, G^R_{ut} (p) P_+ \\
&& +(-\ii \gamma_\mu \sin p_\mu \, \delta_{s,u} +W^+_{su} (p) +mM^+_{su} (p) ) 
\, G^L_{ut} (p) P_-, \nonumber
\eea
where 
\bea
G^R_{st} (p) &=& 
      \frac{A}{F} \Big[ -(1-m^2) (1-W\e^{-\alpha}) \e^{\alpha(-2N_s+s+t)}
           -(1-m^2) (1-W\e^\alpha) \e^{-\alpha(s+t)} \nonumber \\
  && - 2W m (\e^{\alpha(-N+s-t)}+\e^{\alpha(-N_s-s+t)}) \sinh \alpha \Big] 
      +A\e^{-\alpha |s-t|} , \\
G^L_{st} (p) &=& 
     \frac{A}{F} \Big[ -(1-m^2) (1-W\e^\alpha) \e^{\alpha(-2N_s+s+t-2)}
           -(1-m^2) (1-W\e^{-\alpha}) \e^{-\alpha(s+t-2)} \nonumber \\
  && - 2W m (\e^{\alpha(-N+s-t)}+\e^{\alpha(-N_s-s+t)}) \sinh \alpha \Big] 
      +A\e^{-\alpha |s-t|} , \\
\cosh (\alpha) &=& \frac{1+W^2 + \sum_\mu \sin^2 p_\mu}{2|W|} , \\
A &=& \frac{1}{2W\sinh \alpha} , \\
F &=& 1 -\e^\alpha W - m^2(1-W\e^{-\alpha}) .
\eea
These formulae are valid only for positive W. For $1 < \rho \le 2$ and small 
momenta $W$ can be negative, and in this case the propagator is given by the 
above equations with the replacements
\bea
W &\rightarrow& - |W|, \\
\e^{\pm \alpha} &\rightarrow& -\e^{\pm \alpha} ,
\eea
which also imply $\sinh \alpha \rightarrow -\sinh \alpha$. 

\begin{figure}[t]
\begin{center}
\begin{picture}(400,150)(-20,-20)
\ArrowLine(0,0)(80,0)
\ArrowLine(80,0)(220,0)
\ArrowLine(220,0)(300,0)
\GlueArc(150,0)(70,0,180){6}{20}
\Text(0,-10)[l]{$\overline{q} (p)$}
\Text(80,-10)[r]{$\psi_s (-p) $}
\Text(90,-10)[l]{$\overline{\psi}_s (k) $}
\Text(220,-10)[r]{$\psi_t (-k)$ }
\Text(230,-10)[l]{$\overline{\psi}_t (p)$ }
\Text(300,-10)[r]{$q (-p)$}
\end{picture}
\end{center}
\caption{\small A typical correction to the physical quark propagator 
showing the various fermionic fields which form the various propagators.}
\label{fig:dw_props}
\end{figure}
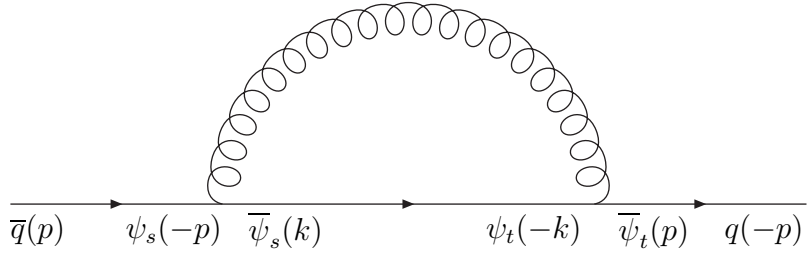

The massless fermion is given at tree level by 
\be
\chi_0 = \sqrt{1-w_0^2} \, (P_+ w_0^{s-1} \psi_s + P_- w_0^{N_s-s} \psi_s ) ,
\ee
with $w_0 = 1 -\rho$. We can see that this damping factor confines the two 
chiralities on the two different domain walls. Since this factor is 
renormalized by the interactions due to the additive renormalization of $\rho$
(this is a mass term not protected by chiral symmetry),~\footnote{The
renormalizations of $w_0$ and of the wave function turn out to be quite large.
The former in particular is of order $O(10^2)$. This is claimed to be cured 
by tadpole improvement. The wave function at one loop has been also computed
in~\cite{Shamir:2000cf}.} it is more convenient to work instead with 
the ``physical'' field 
\bea
q(x) &=& P_+ \psi_1 (x) + P_- \psi_{N_s} (x) \\
\overline{q} (x) &=& \overline{\psi}_{N_s} (x) P_+ 
+ \overline{\psi}_1 (x) P_- ,
\eea
whose renormalization is simple. The corresponding propagator is given by
\be
S_q (p) = \langle q(-p) \overline{q} (p) \rangle = 
\frac{- \ii \gamma_\mu \sin p_\mu + (1-W\e^{-\alpha}) m}{-(1-\e^\alpha W) 
+ m^2(1- W \e^{-\alpha})} .
\ee
While the 1-loop propagator $\langle \psi (-p) \overline{\psi} (p) \rangle$ 
gets an additive mass correction, this 1-loop physical propagator is protected 
from such terms, as it should be thanks to chiral symmetry. 
Composite operators in the theory are then constructed using the physical
field. The bilinears are for example given by 
\be
O (x) = \overline{q} (x) \Gamma q (x) .
\ee
One sees that, when computing Feynman diagrams in which the physical field 
is present as an external state, the propagators of the external lines are
also needed,
\bea
\langle q(-p) \overline{\psi}_s  (p) \rangle &=& \frac{1}{F} 
(\ii \gamma_\mu \sin p_\mu - m(1-W\e^{-\alpha})) 
(\e^{-\alpha(N_s-s)} P_+ + \e^{-\alpha(s-1)} P_-)\\
&& +\frac{1}{F} [m(\ii \gamma_\mu \sin p_\mu 
- m(1-W\e^{-\alpha}))-F]\e^{-\alpha}(\e^{-\alpha(s-1)} P_+ 
+ \e^{-\alpha(N_s-s)} P_-) , \nonumber \\
\langle \psi_s (-p) \overline{q} (p) \rangle &=& \frac{1}{F} 
(\e^{-\alpha(N_s-s)} P_- + \e^{-\alpha(s-1)} P_+)
(\ii \gamma_\mu \sin p_\mu - m(1-W\e^{-\alpha})) \\
&& +\frac{1}{F} (\e^{-\alpha(s-1)} P_- + \e^{-\alpha(N_s-s)} P_+) \e^{-\alpha}
[m(\ii \gamma_\mu \sin p_\mu - m(1-W\e^{-\alpha}))-F] . \nonumber
\eea
A typical situation in which these propagators are required is depicted in
Fig.~\ref{fig:dw_props}, a self-energy diagram. It is interesting to note
that logarithmic divergences, for example in the self-energy, are localized 
at the boundaries $s=1$ and $s=N_s$, because they arise when the masses are 
going to zero.

Given the expressions of the quark propagators, perturbative calculations
with domain wall fermions tend to be rather cumbersome, even in the limit 
$N_s=\infty$. Many quantities have been calculated so far, using gluon 
propagators in Feynman gauge, but no operators with covariant derivatives have
been considered, like for instance the ones measuring moments of structure 
functions.

\subsection{Fixed-point fermions}

Historically the fixed-point action, developed in the context of perfect 
actions~\cite{Hasenfratz:1993sp,Wiese:1993cb,Bietenholz:qy,DeGrand:1995ji,DeGrand:1995ab,DeGrand:1996nc},
was the actual chiral battleground where the Ginsparg-Wilson relation was 
rediscovered~\cite{Hasenfratz:1997ft}.
The key idea here goes under the name of classically perfect actions, 
that is actions for which their classical predictions (and in particular 
the properties of the chiral modes) agree with those of the continuum. 
This is true even when the lattice spacing is not small.~\footnote{For a nice 
pedagogical review of these ideas and techniques, see~\cite{Hasenfratz:1998bb};
a short review can be found in~\cite{Hasenfratz:1997ft}.}

The construction of perfect actions employs ideas which strongly resemble 
Wilson's renormalization group in statistical 
mechanics~\cite{Wilson:1973jj,Wilson:1974mb}. 
A renormalization group step in this context, also called a block 
transformation, consists in doubling the lattice spacing, which means that 
in lattice units the correlation length is halved after each step. 
In this way the short-scale fluctuations in the functional integral can be 
eliminated step by step. The fixed points of these renormalization group 
transformations, which is reached after they are iterated an infinite number 
of times, 
\be
{\cal A} \longrightarrow {\cal A}' \longrightarrow {\cal A}'' \longrightarrow
{\cal A}''' \longrightarrow \cdots \longrightarrow {\cal A}^{FP} ,
\ee
are the fixed-point actions. Their properties can be investigated using 
classical equations.

A block variable of the lattice with double lattice spacing can be constructed
from the fine variable as follows:
\be
\chi (x_B) = b \sum_x \omega (2x_B -x) \phi (x), 
\qquad \Big(\sum_x \omega (2x_B -x) = 1\Big).
\ee
One integrates out the original variables keeping the new block averages 
fixed. Blocking is then equivalent to going to a coarse lattice. The action 
that one obtains after the blocking transformations will in general contain 
all kinds of interactions, even when one starts with a very simple action. 
The fixed-point action to which one at the end arrives contains an infinite 
number of interactions, with an infinite number of corresponding couplings.

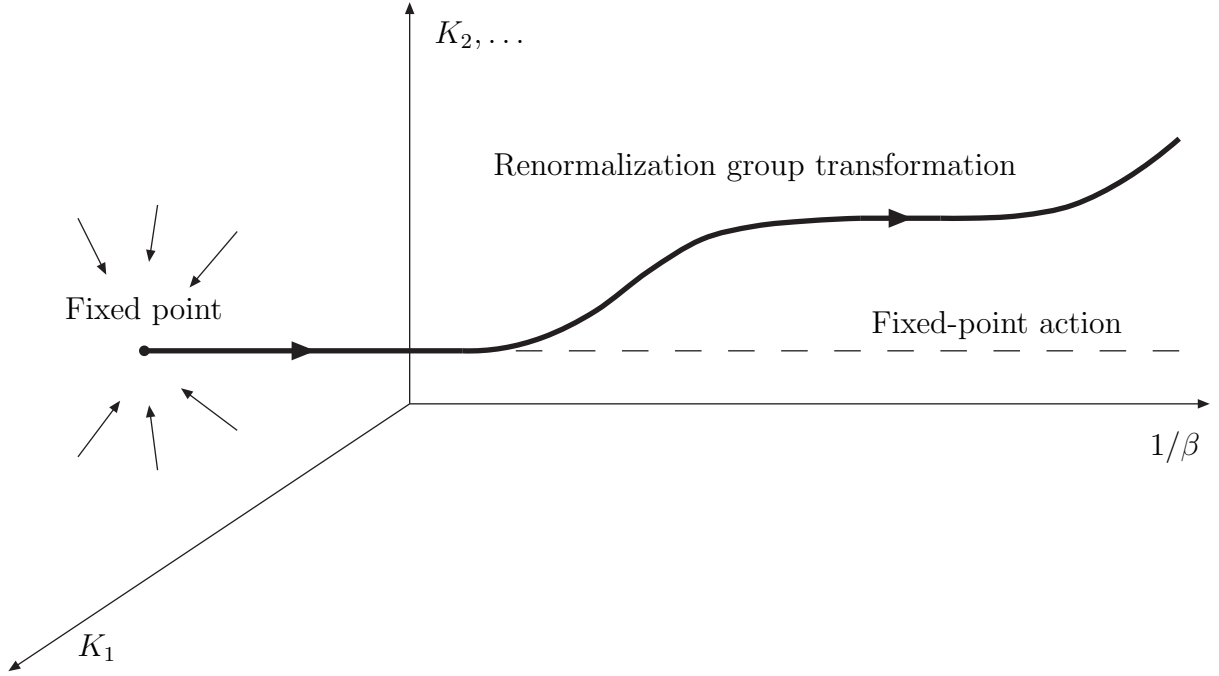
\begin{figure}[t]
\begin{center}
\begin{picture}(450,250)(0,0)
\LongArrow(150,100)(450,100)
\LongArrow(150,100)(0,0)
\LongArrow(150,100)(150,250)
\Text(440,90)[t]{$1/\beta$}
\Text(160,240)[l]{$K_2, \dots$}
\Text(25,8)[l]{$K_1$}
\DashLine(170,120)(440,120){10}
\Text(420,130)[r]{Fixed-point action}
\Text(380,190)[r]{Renormalization group transformation}
\Vertex(50,120){2}
\Text(50,130)[b]{Fixed point}
\LongArrow(25,80)(40,100)
\LongArrow(55,75)(52,98)
\LongArrow(85,90)(65,105)
\LongArrow(25,170)(35,150)
\LongArrow(55,175)(52,155)
\LongArrow(85,165)(68,145)
\SetWidth{2}
\ArrowLine(50,120)(170,120)
\Curve{(170,120)(200,125)(220,135)(240,150)
       (260,162)(280,167)(300,169)(320,170)}
\ArrowLine(320,170)(350,170)
\Curve{(350,170)(360,170)(380,171)(400,175)(420,185)(440,200)}
\end{picture}
\end{center}
\caption{\small The renormalization group flow and the classically perfect
fixed-point trajectory. The directions $K_1, K_2, \dots$ represent the 
parameter space.}  
\label{fig:rg}
\end{figure}

A fixed point is by definition an action that is invariant under further block 
transformations. One is then on a critical surface and the correlation length 
is infinite. This means that in asymptotically free theories, and in 
particular in QCD, the fixed-point action can only be reached for $g_0 \to 0$,
that is $\beta \to \infty$. 
Strictly speaking a fixed-point action ${\cal A}^{FP}$ could then only be 
simulated at $\beta = \infty$. However, if one considers an action with the
same form as the fixed-point action but for finite $\beta$, this action 
will be quite close, for large $\beta$, to the true renormalization group 
trajectory (see Fig.~\ref{fig:rg}). The lattice artifacts will then be quite 
small. In this way one has linearized the fixed-point transformation and in 
the only relevant direction this action $\beta {\cal A}^{FP}$ is classically 
perfect.~\footnote{However, it is not quantum perfect, as 1-loop calculations 
have shown~\cite{Hasenfratz:1997gs}.}

Doing a good averaging using appropriate block transformations is a kind of 
an art. A bad choice can lead to a fixed-point action with bad locality 
properties or even to no fixed point at all. A good choice can instead lead 
to a good parameterization of the fixed-point action which is convenient for 
simulations. The task is then to find a good blocking transformation which 
leads to a fixed-point action that can be well approximated by taking a small 
number of terms which are not difficult to simulate.

Let us now consider the Wilson action
\bea
\beta {\cal A}_g (U) + {\cal A}_f (\overline{\psi},\psi,U) 
&=& \beta \sum_P \Big( 1 - \frac{1}{N_c} {\mathrm Re} \, \Tr U_P \Big) \\
&&+ \frac{1}{2} \sum_{x,\mu} \Big( 
 \overline{\psi} (x) (\gamma_\mu -1) U_\mu (x) \psi (x + \hat{\mu})
-\overline{\psi} (x + \hat{\mu}) (\gamma_\mu +1) U^\dagger_\mu (x) \psi (x)
\Big) \nonumber \\
&& + \sum_x \overline{\psi} (x) (m+4) \psi (x) \nonumber . 
\eea
The renormalization group transformed action is given, after one step, by 
\be
\e^{- \left( \beta' {\cal A}'_g (V) 
+ {\cal A}_f' (\overline{\chi},\chi,V) \right)} 
= \int {\cal D} \overline{\psi} {\cal D} \psi {\cal D} U \,
\e^{- \left( \beta ( {\cal A}_g (U) + T_g (V,U) )
+ {\cal A}_f (\overline{\psi},\psi,U)  
+ T_f (\overline{\chi},\chi,\overline{\psi},\psi,U) \right)} ,
\ee
where the blocking kernels are
\be
T_g (V,U) = \sum_{x_B,\mu} \Bigg[ -\frac{\kappa_g}{N_c} {\mathrm Re} \, \Tr 
(V_\mu (x_B) Q^\dagger_\mu (x_B)) + {\cal N} (Q_\mu (x_B))
\Bigg]
\ee
for the pure gauge part (where ${\cal N}$ is needed to normalize correctly
the block variables), and
\be
T_f (\overline{\chi},\chi,\overline{\psi},\psi,U) = 
\kappa_f \sum_{x_B} \Big( \overline{\chi} (x_B) - b_f \sum_x \overline{\psi} 
(x) \omega^\dagger_{x,x_B} (U) \Big) \Big( \chi (x_B) - b_f \sum_x 
\omega_{x_B,x} (U) \psi (x) \Big) 
\ee
for the fermion part. $Q_\mu (x_B)$ is an average of the fine links near $x_B$,
while $\omega$ is an averaging function for the quark fields.
The details of these function are complicated and will not interest us here.
A simple case is given by $\omega (2x_B-x) = 2^{-d}$ for the points $x$
in the hypercube containing $x_B$.

Note that for finite $\kappa_g,\kappa_f$ the average of the fine variables 
in a block is allowed to fluctuate around the block variables. When instead 
$\kappa_g,\kappa_f \to \infty$ the blocking kernels become delta functions. 
Thus, the parameters $\kappa_g$ and $\kappa_f$ determine the stiffness of the 
averaging. 

In QCD the fixed point lies at $\beta=\infty$. In this limit one can perform 
the computations using a saddle-point approximation. Hence for the pure gauge 
part we have
\be
{\cal A}'_g (V) = \min_{\{U\}} \Big[ {\cal A}_g (U) + T_g^\infty (V,U) \Big],
\label{eq:rgt}
\ee
where $T_g^\infty$ is the $\beta=\infty$ limit of the blocking kernel, 
and the fixed-point action satisfies
\be
{\cal A}_g^{FP} (V) = \min_{\{U\}} \Big[ {\cal A}_g^{FP} (U) 
+ T_g^\infty (V,U) \Big] .
\ee
Substituting the $U_{\mathrm min}$ that minimizes Eq.~(\ref{eq:rgt}) in the 
fermion part one obtains a recursion relation and finally arrives at
the equation that determines the fixed-point fermion action,
\be
h^{FP}_{x_B,x'_B} (V)^{-1} = \frac{1}{\kappa_f} \, \delta_{x_B,x'_B}
+ b_f^2 \sum_{x,x'} \omega_{x_B,x} (U_{min}) \, h^{FP}_{x,x'} \,
(U_{\mathrm min})^{-1} \omega_{x',x'_B}^\dagger (U_{\mathrm min}) ,
\ee  
where
\be
{\cal A}_f (\overline{\psi},\psi,U) = \sum_{x,x'} \overline{\psi} (x)
h_{x,x'} (U) \psi (x') .
\ee
So far no approximations have been made. The blocking kernels are quadratic, 
and if the actions are also quadratic the renormalization group 
transformations can be computed using Gaussian integrals. The above is then 
an exact formula, and Gaussian integrals are equivalent to minimization. The 
solutions are in general obtained doing some approximations and truncations, 
but in the case of the free theory they can be still computed exactly. 

The fixed-point propagator which is obtained starting from massless Wilson 
fermions and using $\omega (2x_B-x) = 2^{-d}$ and $U_{\mathrm min}=1$ is given
by
\be
h^{-1} (q) = \sum_{l \in Z^d} 
\frac{\gamma_\mu (q_\mu + 2\pi l_\mu)}{(q + 2\pi l)^2} 
\, \prod_\nu \frac{\sin^2 {\displaystyle\frac{q_\nu}{2}}}{\Big(
{\displaystyle\frac{q_\nu}{2}}+\pi l_\nu\Big)^2 } + \frac{2}{\kappa_f} . 
\label{eq:solfp}
\ee
This is an analytic function and corresponds to a local action which has no 
doublers. It is then not surprising to learn that this action breaks chiral 
symmetry. This breaking is due to the term $2/\kappa_f$, which comes entirely 
from the fermionic blocking kernel $T_f$. The remarkable point is that this 
fixed-point propagator satisfies
\be
\{ h^{-1} , \gamma_5 \} = \frac{4}{\kappa_f} \, \gamma_5 ,
\ee
or equivalently the fixed-point action obeys
\be
\{ h , \gamma_5 \} = \frac{4}{\kappa_f} \, h \gamma_5 h ,
\ee
which we can recognize to be a form of the Ginsparg-Wilson relation. Thus this
action, although it breaks the naive chiral symmetry, has a remnant of chiral 
symmetry (from the point of view of the continuum) which is given by 
L\"uscher's symmetry, and which is a good chiral symmetry for any finite
value of the lattice spacing. It follows that the pion mass is zero when the 
bare quark mass in the action is zero, that the index theorem is satisfied,
that the theory reproduces the correct global anomalies, and so on.

We would like to point out that in the limit $\kappa_f=\infty$ this action 
becomes chirally symmetric in the naive sense, $\{ h, \gamma_5 \} =0$, 
but then it becomes also nonlocal, and so it does not contradict the 
Nielsen-Ninomiya theorem.

We have thus learned that the fixed-point action satisfies the Ginsparg-Wilson
relation. As we observed at the beginning, it is actually when studying the 
solution in Eq.~(\ref{eq:solfp}) that the Ginsparg-Wilson was rediscovered in 
all its strength after fifteen years of neglect.

In Monte Carlo computations a truncation of the fixed-point action is
necessary. It is not possible to simulate more than the first few terms. 
It is however hoped that the use of the truncated action still brings small 
errors with respect to the results that would come out using the true 
fixed-point action.

Perturbative calculations, even for a truncated action, are more cumbersome 
than average. Even when the action contains just a few terms, their 
complexity can increase swiftly, because they contain higher powers of the 
fundamental fields and of their derivatives. The propagator and vertices are 
obtained from summing the contributions of all these terms, which in general 
will give some complicated expressions. The coefficients of these various 
terms can only be determined numerically, and this constitutes another 
limitation on the accuracy of fixed-point results. When one wants to make 
simulations using a truncated action that has a small residual symmetry 
breaking and is not too far from the true fixed-point action, a relatively 
large number of terms is needed, and in this case perturbation theory looks 
quite demanding. In fact not many perturbative calculations with fixed-point 
actions have appeared in the literature. Apart from~\cite{Hasenfratz:1997gs}, 
examples of this kind of perturbative calculations can be found 
in~\cite{Bietenholz:1997rt}, where a perfect action for the anharmonic 
oscillator was perturbatively constructed, and~\cite{Farchioni:1995ne}, where 
the mass gap of the nonlinear $\sigma$-model was computed in perturbation 
theory. However, no renormalization of operators has been attempted.

Finally, we would like to point out that, although perturbation theory is 
more complicated here compared to overlap and domain wall fermions, the 
fixed-point action possesses some better physical properties than them. 
The fixed-point action is in fact classically perfect, while the overlap and
domain wall actions are only aimed to recover the classical predictions 
for chiral symmetry, and do not care to what happens to other physical 
quantities. Moreover, the strength of the interaction decays slower for
overlap and domain wall fermions, that is the effective range is greater,
and in this sense the fixed-point action is more local. Of course, the overlap
action can be given explicitly in a simple form, while to obtain the 
fixed-point action, even in an approximate form, one needs to solve 
complicated equations.

At the end of the day domain wall and overlap fermions are equivalent
formulations, while fixed-point fermions can be considered to be something
different because they possess better ``continuum'' properties than the other 
two cases.

For recent developments on the subject of fixed-point actions, 
see~\cite{Hasenfratz:2002rp}. Many technical details can also be found 
in~\cite{Jorg:2002nm}.

\subsection{Concluding remarks}

Among the various formulations of Ginsparg-Wilson fermions, we have discussed 
in some detail the perturbation theory of the overlap solution and also given 
an elementary introduction to perturbative calculations for the domain wall 
and, for what is possible without being too technical, the perfect action 
solutions.

The practical aspects of the implementation of these various fermions, and 
of the computations made with them, are quite different. In particular, the 
control over the numerical chiral symmetry breaking which necessarily occurs
in actual simulations seems to be easier for overlap 
fermions~\cite{Hernandez:2000iw}.

In the domain wall formulation, where chirality appears in the reduction of a 
five-dimensional theory to our four-dimensional world, one has to remember 
that the exact chiral symmetry is attained only when there is an infinite
number of sites in the fifth dimension, which is never the case in Monte Carlo
simulations. Reducing the amount of chiral breaking in this case means doing 
new simulations using larger lattices. 

In the fixed point action it is the necessary truncation which breaks chiral 
symmetry, and one has to add more and more terms in order to decrease 
the amount of breaking.

For overlap fermions, on the other hand, the numerical chiral symmetry 
breaking can be reduced by using more and more refined methods to compute 
the square root in the action. Techniques which use polynomial approximations
of the action and an exact evaluation of the lowest eigenvalues of the Dirac 
operator are widely employed. One does not need to repeat the simulations on 
new and larger lattices, or to add more terms to the action. Here the road to 
smaller chiral symmetry breaking goes via the extraction of more eigenvalues 
and the refinement of the polynomials, which can be easier to accomplish. 
It looks instead much more expensive and complicated to make simulations on 
lattices which are longer and longer in the domain wall fifth dimension, or to
include into the perfect action calculations more and more terms of the 
expansion, with all that this implies in terms of costs and complexities.

We mention in passing that also perturbation theory seems to be more difficult
to carry out in the case of domain wall and fixed-point actions.

\section{Perturbation theory of lattice regularized chiral gauge theories}
\label{sec:chiralgaugetheories}

The recent developments in the understanding of chiral symmetry on the lattice
have also led to very interesting insights into the subject of chiral gauge 
theories, in which left- and right-handed fermions do not couple to the gauge 
fields in the same way. This is the case of neutrino interactions, in which 
as is well known no right-handed neutrino components couple to the 
electroweak gauge field, and in general of the electroweak theory. 

We have learned in the previous Section that a Dirac operator satisfying 
the Ginsparg-Wilson relation describes fermions that have an exact chiral 
symmetry at finite lattice spacing. In this Section we discuss the fact that 
when this symmetry is gauged, that is when these fermions become chiral gauge 
fermions, the lattice regularization can at the same time preserve the gauge 
invariance at all orders in perturbation theory, and at any finite lattice 
spacing. It has been shown by L\"uscher (2000c) that using Ginsparg-Wilson 
fermions it is indeed possible to regularize chiral gauge theories in such a 
way that the regularization does not break gauge invariance and at the same 
time maintains at all orders chiral symmetry, locality and all other 
fundamental principles. We think that this has been one the major theoretical 
advances in the theory of quantum fields on the lattice in recent times, and 
we would like to give a short account of it.

Ginsparg-Wilson lattice fermions provide in this way the only known 
regularization of chiral gauge theories which is consistent at the 
nonperturbative level and which does not violate the gauge symmetry or other 
fundamental principles. For all other widely used regularization methods 
(and also the BPHZ finite-part prescription) this is impossible, because 
chirality and gauge invariance cannot be maintained together as symmetries 
beyond the classical level.~\footnote{A well-known 1-loop example in the 
electroweak theory is given by the divergent triangle diagrams 
containing $\gamma_5$, which give rise to the chiral 
anomaly~\cite{Adler:gk,Bell:ts,Bardeen:md}. The recent lectures 
of~\cite{Zinn-Justin:2002vj} contain much material about chiral anomalies 
in various regularizations and their connection with topology.} 
If one wants to keep chiral invariance unbroken, one is compelled to introduce
new counterterms order by order in perturbation theory to maintain the gauge 
invariance, making these regularizations much less appealing. Even on the 
lattice itself new counterterms need to be introduced when one just uses 
Wilson fermions, and a delicate fine tuning of the corresponding coefficients 
is required in order to recover chiral invariance. This is what was done
in the Rome approach~(Borrelli {\em et al.}, 1989; 1990; Testa, 1998a). The 
problem here is not only that one has to introduce counterterms that are not 
gauge invariant, but that the fermion modes couple to the longitudinal modes 
of the gauge field (because of the lack of gauge invariance), and the effect 
of these unwanted gauge modes on the fermions, which can lead to modifications 
of the spectrum and in particular to the appearance of doublers, has to be 
controlled and removed in some way. While in the rest of this Section we are 
only interested in chiral gauge theories which employ Ginsparg-Wilson fermions
and which maintain an exact gauge invariance, there is another nonperturbative 
approach which is worth mentioning where the coupling of the fermions to the 
unphysical degrees of freedom of the gauge field is controlled by means of 
a suitable gauge fixing~\cite{Bock:1997fu,Bock:1997vc,Bock:1999qa}. 
This gauge fixing approach uses Wilson fermions and the naive notion of 
chirality on the lattice, and skirts the consequences of the Nielsen-Ninomiya 
theorem because it has no explicit gauge invariance, which can in fact only 
be recovered in the continuum limit. It thus needs counterterms which are not 
gauge covariant (and which have to be appropriately tuned), but it still 
ensures renormalizability and at the end it achieves the decoupling of the 
longitudinal degrees of freedom while keeping the fermion spectrum intact, 
so that one can define a lattice chiral gauge theory where the fermions have 
no doublers. Some nonperturbative considerations are also needed in order to 
define a valid perturbation theory around $A_\mu=0$ which is still a reliable 
approximation of the full lattice theory at weak coupling~\cite{Bock:1997ks}. 
A naive gauge fixing does not do the job, and an appropriate classical 
potential containing higher powers of the gauge potential has to be introduced.
A gauge fixing action which leads to a chiral gauge theory has been so 
constructed in the abelian case~\cite{Golterman:1996ph,Shamir:1995gd}, 
and it needs only one counterterm to be tuned nonperturbatively. For a recent 
presentation of this method see~\cite{Golterman:2002pq}. Reviews have also 
been given in~\cite{Shamir:1995zx} and~\cite{Golterman:2000hr}.~\footnote{I 
thank M.~Golterman and Y.~Shamir for correspondence.}

It thus seems that the lattice, when Ginsparg-Wilson fermions are used,
is an exception to this vicious pattern which is common to all other 
regularizations. The lattice acquires in this sense a predominance over other 
regularizations, as it is the only nonperturbative technique which makes 
possible to regularize theories which are chiral and gauge invariant, and it 
does a much better job than to introduce counterterms. No noninvariant 
counterterms in the action are needed. One can thus regularize chiral gauge 
theories without breaking the gauge invariance and using a cutoff, which was 
thought impossible.

Such a powerful chiral regularization, which can only be realized using 
Ginsparg-Wilson fermions on the lattice, works because the gauge anomaly 
cancels when radiative corrections are included. Simple schemes which do not 
take this fact into account and try to construct chiral fermions only
at the tree level cannot work out. Many years have been indeed necessary
to understand the structure of chiral fermions on the lattice and 
go beyond the apparent impossibility given by the Nielsen-Ninomiya theorem. 
Of course at the end one does not contradict the Nielsen-Ninomiya theorem 
at all, because it is the condition (d) (see Section~\ref{sec:aspects}) 
which is not fulfilled. This was a nontrivial conceptual advance.

We know from general results in algebraic renormalization theory that if 
chiral gauge anomalies can be shown to be canceled at one loop, then they are
absent at all orders because no further anomalies can be generated 
at higher orders~\cite{Adler:er}. A feature of Ginsparg-Wilson fermions is 
that the gauge anomaly descends from a topological field in 4+2 dimensions,
and that the anomaly cancellation at one loop can be formulated in terms of 
a local cohomology problem in $4+2$ dimensions, which has been studied 
and solved in (L\"uscher, 1999a; 1999b; 2000a; 2000b; Suzuki, 1999), 
for abelian gauge theories. In general one needs a classification 
of the topological fields in 6 dimensions, and the local cohomology problem
is then solved.

Let us now explain some of the details of the construction of chiral gauge 
theories using Ginsparg-Wilson fermions. In chiral theories the fermion 
integral in the path-integral formulation is restricted to left-handed fields, 
and the propagator involves a chiral projector (defined in 
Eq.~(\ref{eq:chiralproj})):
\be
\{ \psi (x) \overline{\psi} (y) \}_F = \langle 1 \rangle_F \times 
\widehat{P}_- S(x,y) P_+ , 
\ee 
where $S(x,y)$ is the inverse of the Dirac operator: 
$\sum_z D(x,z) S(z,y) = a^{-4} \delta_{xy}$. Thus, only the left-handed 
components of the fermion field propagate. 

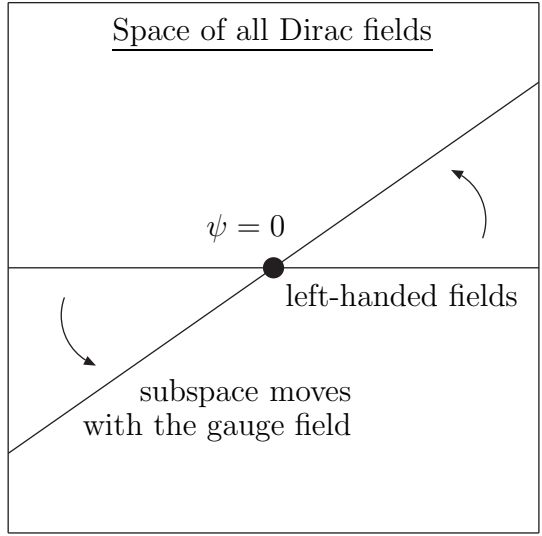
\begin{figure}[t]
\begin{center}
\begin{picture}(200,200)(0,0)
\Line(0,0)(200,0)
\Line(200,0)(200,200)
\Line(0,200)(200,200)
\Line(0,0)(0,200)
\Line(0,100)(200,100)
\Line(0,30)(200,170)
\LongArrowArc(160,118)(20,-20,70)
\LongArrowArc(40,82)(20,160,250)
\Vertex(100,100){4}
\Text(90,110)[b]{$\psi=0$}
\Text(130,53)[r]{subspace moves}
\Text(130,40)[r]{with the gauge field}
\Text(105,90)[l]{left-handed fields}
\Text(100,195)[t]{\underline{Space of all Dirac fields}}
\end{picture}
\end{center}
\caption{\small Dependence of the phase of the fermion measure on the
gauge fields.}  
\label{fig:phase}
\end{figure}

The fundamental point is that the definition of the measure for left-handed 
fermions, which is needed to construct the quantum theory, is highly 
nontrivial, because the projectors $\widehat{P}$ contain a 
$\widehat{\gamma}_5= \gamma_5 (1 -a/\rho D)$ matrix, and thus depend on the 
gauge fields. This produces an ambiguity in the measure which can be expressed
in terms of a pure phase factor, and can be seen in the following way. 
The measure can be defined by giving a complete basis of complex-valued 
fermionic fields $v_j$ which are eigenstates of the projection operators. 
The quantum field can then be expanded like
\be
\psi (x) = \sum_j c_j v_j(x) ,
\ee
and changing the basis means that the measure gets multiplied by the 
determinant of a unitary transformation matrix, which is a phase factor. 
Since the projectors depend on the gauge fields, the basis and the 
corresponding phase factor also depend on the gauge fields and cannot be fixed
independently of them (see Fig.~\ref{fig:phase}). This is the source of the 
phase ambiguity.

Thus, in chiral gauge theories the fermion integration has a nontrivial phase
ambiguity and the fermion measure is not a simple product of local measures.
Since this phase ambiguity depends on the gauge fields, it does not cancel in 
ratios of expectation values when one normalizes the path integral with the 
partition function. The right-handed projectors on the other hand give a 
constant phase factor, which can be factored out and does cancel in such 
ratios.

The phase problem, that is the definition of the measure, is the key issue 
in the construction of chiral gauge theories. The phase ambiguity for the 
left-handed components can be consistently removed only if the fermion 
multiplet is nonanomalous, that is when 
\be
d_R^{abc} = 2 \ii \, \Tr \, 
\Big\{ R(T^a) \Big[ R(T^b) R(T^c) + R(T^c) R(T^b) \Big] \Big\} 
\ee
is zero, where $R(T^a)$ are the anti-hermitian generators of the fermion 
representation of the gauge group.~\footnote{In this Section 
we use, as in the original papers, anti-hermitian color matrices, that is they 
satisfy $\Tr \{T^a T^b\} = -1/2 \, \delta^{ab}$ and $[T^a,T^b]=f^{abc} T^c$. 
This means that $U_\mu = \exp\{a g_0 A_\mu^a T^a \}$. The interested reader 
can then turn for more details to these papers without problems.}
For $U(1)$ gauge theories coupled to $N$ left-handed Weyl fermions of charges 
$e_\alpha$ this condition becomes $\sum_{\alpha=1}^N e_\alpha^3 =0$.

The key step forward is that the phase problem can be equivalently formulated 
in terms of a local current which is gauge covariant. To make this step let us 
consider the effective action when the fermionic degrees of freedom are
integrated out,
\be
\e^{-S_{\mathrm eff} [U]} = \int {\cal D} \psi {\cal D} \overline{\psi} 
\e^{-S_F [U,\overline{\psi},\psi]} .
\ee
An infinitesimal deformation of the gauge field
\be
\delta_\eta U_\mu (x) = a \eta_\mu (x) U_\mu (x) 
\ee
(where $\eta_\mu (x) = \eta_\mu^a (x) T^a$, and $\mu$ is not summed) 
induces a variation of the effective action
\be
\delta_\eta S_{\mathrm eff} = -\Tr \Big\{ \delta_\eta D \widehat{P}_-
D^{-1} P_+ \Big\} + \ii {\cal L}_\eta .
\ee
The first term in the above formula is the naive expression which one would 
normally obtain, while the second term
\be
{\cal L}_\eta  = a^4 \sum_x \eta_\mu^a (x) j_\mu^a (x) ,
\ee
which is linear in the field variation, is the new term proper of 
Ginsparg-Wilson fermions which arises because the fermion measure 
depends on the gauge fields, and in fact this term can also be written as 
${\cal L}_\eta = \ii \sum_j (v_j,\delta_\eta v_j)$. 
The axial current $j_\mu (x)$ contains all the information about the phase of 
the fermion measure, provided it is given by a gauge-covariant local field 
which satisfies the {\em integrability condition}, which in its differential 
form reads
\be
\delta_\eta {\cal L}_\zeta - \delta_\zeta {\cal L}_\eta
+ a {\cal L}_{[\eta,\zeta]} = \ii \, \Tr \Big\{ \widehat{P}_- 
\Big[ \delta_\eta \widehat{P}_- , \delta_\zeta \widehat{P}_- \Big] \Big\} ,
\label{eq:intgrabcond}
\ee
for all field variations $\eta_\mu (x)$ and $\zeta_\mu (x)$ that do not depend
on the gauge field. The reconstruction theorem then says that a given current 
with these requirements fixes the phase of the measure, and knowing the 
current is then equivalent to knowing the fermion measure. Once the measure
is fixed, the functional integral is well defined and the effective action 
is gauge invariant. We have then constructed a chiral gauge theory which 
is not spoiled by radiative corrections.
 
There is thus a one-to-one correspondence between the current and the measure, 
up to a constant phase factor in each topological sector which is not 
relevant. The current $j_\mu (x)$ defines the chiral gauge theory, and the 
problem of constructing the fermion measure is reduced to the problem of 
constructing this current. The measure does not need to be explicitly 
specified.

Let us now make an expansion in the coupling constant of the various 
quantities introduced above. The exact cancellation of the gauge anomaly can 
then be proven recursively, order by order in $g_0$. We think it is 
interesting to sketch how this construction is carried out, skipping the points
that are technically more involved, for which the reader is referred to the 
original papers. A more detailed treatment is beyond the scope of this review.

To write an expansion in the coupling, which we also need in order to be able
to derive the additional interaction vertices coming from the measure term 
${\cal L}_\eta$, we use the variation with respect to the gauge potential $A$ 
rather than to the link $U$, which is
\be
\overline{\delta}_\eta A_\mu (x) = \eta_\mu (x) .
\ee
Because
\be
\delta_\eta = g_0^{-1} \overline{\delta}_\eta + O(1) ,
\ee
the corresponding variation of the effective action contains an explicit
dependence on the coupling,
\be
\overline{\delta}_\eta S_{\mathrm eff} = 
-\Tr \Big\{ \overline{\delta}_\eta D \widehat{P}_-
D^{-1} P_+ \Big\} + \ii g_0 {\cal L}_{\overline{\eta}} ,
\ee
with
\be
\overline{\eta}_\mu (x)  = \Bigg\{ 1 + \sum_{k=1}^\infty \frac{1}{(k+1)!} 
\, \Big( g_0 a \, \widetilde{A}_\mu (x) \Big)^k \Bigg\} \eta_\mu (x) ,
\ee
where the gauge potential entering in $\overline{\eta}$ is in the adjoint 
representation. To construct the current, one starts from the curvature term 
in the right-hand side of the integrability condition 
Eq.~(\ref{eq:intgrabcond}), 
\be
{\cal F}_{\eta\zeta} = \ii \, \Tr \Big\{ \widehat{P}_- 
\Big[ \delta_\eta \widehat{P}_- , \delta_\zeta \widehat{P}_- \Big] \Big\} ,
\ee
which has an expansion in the coupling constant whose leading term is of 
order $g_0^3$. The reason why the terms of order $g_0^2$ and lower are zero 
is the anomaly cancellation condition $d_R^{abc} = 0$, and the fact that 
these lowest-order contributions are zero will be crucial in the following.
The lowest-order part of the curvature is then
\be
\widetilde{\cal F}_{\eta\zeta} = \frac{1}{3!} 
\frac{\partial^3}{\partial g_0^3} {\cal F}_{\eta\zeta}
\Bigg|_{g_0=0} ,
\ee
and the lowest-order part of the measure term is
\be
\widetilde{\cal L}_\eta = \frac{1}{4!} 
\frac{\partial^4}{\partial g_0^4} {\cal L}_\eta \Bigg|_{g_0=0} ,
\ee
where $\widetilde{\cal L}$ has to be invariant under the linearized gauge 
transformations~\footnote{One has 
\be
\eta_\mu (x) = - \widetilde{D}_\mu \omega (x) ,
\ee
with 
\be
\widetilde{D}_\mu \omega (x) = \frac{1}{a} \Big[ 
U_\mu (x) \omega (x + a \hat{\mu}) U^\dagger_\mu (x) - \omega (x) \Big] .
\ee
}
\be
A_\mu (x) \rightarrow A_\mu (x) + \nabla_\mu \omega (x) .
\label{eq:lingautra}
\ee
In terms of these quantities the lowest-order form of the integrability 
condition reads
\be
\overline{\delta}_\eta \widetilde{\cal L}_\zeta
- \overline{\delta}_\zeta \widetilde{\cal L}_\eta
= \widetilde{\cal F}_{\eta\zeta} .
\ee
From the lowest-order term of the curvature $\widetilde{\cal F}$ we now 
define the functional
\be
{\cal H}_\eta = -\frac{1}{5} \widetilde{\cal F}_{\eta\lambda} 
\Big|_{\lambda_\mu=A_\mu} , 
\ee
which satisfies the lowest-oder form of the integrability 
condition~\footnote{To prove that
\be
\overline{\delta}_\eta {\cal H}_\zeta
- \overline{\delta}_\zeta {\cal H}_\eta
= \widetilde{\cal F}_{\eta\zeta} 
\ee
one has to make use of the Bianchi identity
\be
 \overline{\delta}_\eta \widetilde{\cal F}_{\zeta\lambda}
+\overline{\delta}_\zeta \widetilde{\cal F}_{\lambda\eta}
+\overline{\delta}_\lambda \widetilde{\cal F}_{\eta\zeta} = 0 .
\ee
} 
and is linear in $\eta$, which allows to define the current $h_\mu (x)$:
\be
{\cal H}_\eta = a^4 \sum_x \eta^a (x) h_\mu^a (x) .
\ee
This current in general is not gauge invariant, but from it we can construct  
\be
q(x) = \nabla^\star_\mu h_\mu (x) ,
\ee
which is invariant under the linearized gauge transformations 
(\ref{eq:lingautra}). It can be shown that $q(x)$ is a topological field, 
that is it satisfies
\be
a^4 \sum_x \overline{\delta}_\lambda q(x) = 0
\ee
for all variations $\lambda_\mu(x)$ of the gauge potential. Not surprisingly,
this topological field turns out to correspond to the anomaly 
$-1/2 \, \Tr \{\gamma_5 T D(x,x) \}$. We stress that $q(x)$ has been derived
in a unique way from the current $j_\mu (x)$ (which also determines the 
fermion measure), via the curvature term of the integrability condition, 
$\widetilde{\cal F}$. 

Topological fields like $q(x)$ have been classified in the U(1) lattice gauge 
theory, and in absence of matter they are equal to a sum of Chern polynomials 
plus a divergence term which is topologically trivial. The field $q(x)$ that 
we have just obtained is homogeneous of degree 4 in the gauge potential, 
and since Chern polynomials in four dimensions have degree 2, they cannot 
contribute in this case. 
All that is left is thus the topologically trivial term, so that we have
\be
q(x) = \partial^\star_\mu k_\mu (x) ,
\ee
where $k_\mu (x)$ turns out to be a local current invariant under linearized 
gauge transformations. The lowest-order part of the measure term is now given 
by
\be
\widetilde{\cal L}_\eta = {\cal H}_\eta + \overline{\delta}_\eta \cdot
\frac{1}{4} \, a^4 \sum_x A_\mu^a (x) k_\mu^a (x) ,
\ee
which, since the last term has zero curvature, satisfies the lowest-oder form 
of the integrability condition. 

This completes the construction at leading order. The higher-order terms of 
${\cal L}$ can be then computed recursively according to the following
procedure. If one has already calculated the $O(g_0^n)$ term in the expansion 
of ${\cal L}$, then he subtracts to ${\cal L}$ another function, 
${\cal L}^{(n)}$, whose first $n$ terms in the coupling expansion are the same
as ${\cal L}$. Applying then the above construction (with some slight 
changes) to the difference ${\cal L} - {\cal L}^{(n)}$, one can compute 
its leading-order term, which determines ${\cal L}$ at $O(g_0^{n+1})$,
and repeating this procedure one can obtain ${\cal L}$ to the desired order.

This construction is unique up to terms that are of higher orders in the 
lattice spacing and that therefore are irrelevant in the continuum limit 
(apart for a finite renormalization). As we have already mentioned, it can 
be carried out only if the fermion multiplet is anomaly-free. If this is not
instead the case, i.e., if $d_R^{abc} \neq 0$, the lowest term of the 
curvature turns out to be of order $g_0$ instead of $g_0^3$, and the lowest 
term of the measure is of order $g_0^2$ instead of $g_0^4$. The topological  
field obtained along the lines explained above is now homogeneous of degree 2
in the gauge potential, and then this time the Chern polynomials do 
contribute, and $q(x)$ is topologically nontrivial. It can be shown that this 
leads to the presence of a lattice field corresponding to 
$F_{\mu\nu} \widetilde{F}_{\mu\nu}$:
\be
q(x) = -\frac{1}{192\pi^2} \, d_R^{abc} \epsilon_{\mu\nu\rho\sigma}
T^a F^b_{\mu\nu} (x) F^c_{\rho\sigma} (x +a\hat{\mu} +a\hat{\nu})
+ \nabla^\star_\mu k_\mu (x) ,
\ee
where $F_{\mu\nu} = \nabla_\mu A_\nu (x) - \nabla_\nu A_\mu (x)$ 
is the linearized gauge field-strength tensor. This is the well-known
covariant anomaly. Thus, for $d_R^{abc} \neq 0$, the theory is not
chiral invariant after quantum corrections, which corresponds to the
fact that the construction of a $\widetilde{\cal L}$ which satisfies the 
integrability condition, and hence of the measure term, cannot be accomplished.

For nonabelian gauge groups the cohomology problem has not yet been solved, 
that is a classification of the topological fields in $4+2$ dimensions has 
not yet been done. Thus, the general structure of the nonabelian anomaly
on the lattice is currently not known. However, although we cannot yet prove 
that $q(x)$ is topologically trivial for nonabelian gauge groups, there is 
no reason to suspect that the theorem could not be valid after all in that 
case. The cancellation of the anomalies can in fact be proven for topological 
fields in the continuum limit, and if the topological structure at finite $a$ 
matches with the one in the continuum limit (which would not constitute a 
surprising result) then the theorem can be proven also at finite lattice 
spacing.~\footnote{Although in this case one should also show that there are 
no global topological obstructions at the nonperturbative level.} Thus,
the remaining open issues are mostly technical and not of principle and one 
should expect to overcome these difficulties at some point. There is little
doubt that the construction is valid also in this case.~\footnote{The only 
nonabelian chiral gauge theory for which the solution of the cohomology 
problem is known is the $SU(2)_L \otimes U(1)_Y$ electroweak theory, which is 
easier to deal with because the representation of $SU(2)$ are 
pseudo-real~\cite{Kikukawa:2000kd}. Theoretical advances have also been 
reported in~\cite{Suzuki:2000ii,Igarashi:2000zi}, where the cancellation of 
the lattice gauge anomaly to all orders in powers of the gauge potential for 
any compact group was established, and in~\cite{Kikukawa:2001mw}, where an 
explicit construction of the chiral measure which employs domain wall fermions
was given. In the abelian case a nonperturbative construction has been so far 
carried out for weak fields satisfying $| F_{\mu\nu} | < \epsilon$ (with 
$\epsilon < 1/30$), however one can make them the only statistically 
relevant fields in the functional integral by using a modified version of 
the plaquette action which lies in the same universality class of the standard
plaquette action~\cite{Luscher:1998du}.}
 
We conclude this Section discussing the implications of this remarkable
construction for perturbative calculations. The measure term 
\be
{\cal L}_\eta = \sum_{k=4}^\infty \frac{g_0^k}{k!} \, a^{4k+4} 
\sum_{x,\dots,z_k} 
L^{(k)} (x,z_1,\dots,z_k)_{\mu\mu_1\dots\mu_k}^{aa_1\dots a_k} \,
\eta_\mu^a (x) A_{\mu_1}^{a_1} (z_1) \cdots A_{\mu_k}^{a_k} (z_k)
\ee
can be considered as a local counterterm to the action, and the implicit 
dependence of the measure on the gauge fields generates additional gauge-field
vertices in the effective action. Operating with 
$\overline{\delta}_\eta A_\mu (x)$ on $\overline{\cal L}_\eta$ an appropriate 
number of times one can obtain all these additional vertices which derive from 
the measure term. These vertices only appear at the one-loop level, and 
the $k$-th order vertex is given by
\be
\ii g_0 \, \overline{\delta}_\eta \cdots  \overline{\delta}_\eta \, 
{\cal L}_{\overline{\eta}} \Big|_{A_\mu=0} = 
g_0^k \, a^{4k} \sum_{z_1,\dots,z_k}
V_M^{(k)} (z_1,\dots,z_k)_{\mu_1\dots\mu_k}^{a_1\dots a_k} \,
\, \eta_{\mu_1}^{a_1} (z_1) \cdots \eta_{\mu_k}^{a_k} (z_k) .
\ee
In this formula the operator $\overline{\delta}_\eta$ has been applied $k-1$ 
times. The expansion of ${\cal L}_\eta$ as a power series in the coupling
only begins with the $g_0^4$ term, and these vertices are only of fifth and 
higher orders in the gauge coupling. They can be determined, using a recursive
procedure, once a local gauge-covariant current is given that satisfies the 
integrability condition. 

These interaction vertices are not explicitly known at present. Luckily, 
they are not needed in most cases of interest. In fact, since they are only of 
order $g_0^5$ and higher, and moreover they are proportional to positive 
powers of the lattice spacing (as can be seen by naive dimensional 
counting), they do not contribute to the continuum limit of one-loop diagrams.
At the two-loop level the vertices coming from the measure term can come into 
play only in some very special circumstances. In fact, the lowest-order 
vertex, which is a five-point vertex, is totally symmetric in the gauge group 
indices, so that in two-loop calculations this vertex can contribute only if 
it gives rise to more than three external lines.

All propagators and vertices of these chiral gauge theories satisfy the 
conditions for the validity of the Reisz power counting theorem (see 
Section~\ref{sec:analyticcomputations}). Although their renormalizability 
has not yet been proven, it seems unlikely that these theories are 
nonrenormalizable.

In this Section we have discussed only the theory of left-handed fermions.
The introduction of Higgs fields does not affect the measure term, 
because the chiral projectors do not refer to the Higgs sector.
Higgs field or other fields that couple vectorially can then be easily 
incorporated in this theory.

We have seen that lattice Feynman diagrams for Ginsparg-Wilson fermions are 
more complicated to calculate than continuum ones, and the vertices coming
from the fermionic measure are also quite involved. On the other hand, having 
a nonperturbative regularization that is both chiral and gauge invariant after 
quantum corrections are included can be a big advantage, especially when one 
makes calculations in the electroweak theory. It can then be worth to pay 
this price.

We have thus shown that, although certainly not simple, there is now a 
construction of chiral gauge theories, at least in the abelian case, which is 
suitable for quantum calculations, and which makes uses of the lattice.
A consistent formulation of the standard model now exists beyond 
perturbation theory.

\section{The approach to the continuum limit}
\label{sec:approach}

The range of couplings for which perturbation theory aspires to be a 
reasonable expansion, that is when $g_0$ is small, is closely related to the 
approach to the continuum limit of lattice QCD. This approach can be described
using Callan-Symanzik renormalization group equations which are similar to the 
continuum ones. 
Since the first two coefficients of the $\beta$ function, which determines 
the perturbative running of the coupling, are independent of the 
scheme~\cite{Caswell:cj,Espriu:1981eh}, they have the same values also for 
lattice QCD, which is therefore asymptotically free. This implies that the 
lattice QCD coupling goes to zero in the limit in which the momenta go to 
infinity, which when translated in the lattice language means the limit of 
vanishing lattice spacing, or in other words the continuum limit.

Let us consider, in a massless theory, a physical quantity $P$ of mass 
dimension $n$ which is computed on a lattice of spacing $a$. The product 
$a^nP$ is a dimensionless quantity, which can only depend on the bare 
coupling, which in turn depends on the scale, that on the lattice is determined
by the spacing $a$. We thus have $a^nP = f(g_0(a))$. Since the physical 
quantity $P$ has some well-defined value in the continuum limit, 
for $a \to 0$ it must happen that
\be
\lim_{a \to 0} \frac{f(g_0(a))}{a^n}
\ee
is a constant. Near the continuum limit $g_0$ is a smooth function of $a$ 
which satisfies
\be
\lim_{a \to 0} g_0(a) = g_c .
\ee
This critical point for QCD turns out to be $g_c=0$. This is a stable 
ultraviolet fixed point of the theory.

In the continuum limit the correlation length, the rate of the exponential 
falloff of the two-point correlation functions in position space, goes 
to infinity. This can be understood by thinking that the correlation length 
is fixed in physical units, and thus when $a$ decreases it just becomes 
larger and larger when instead is measured in units of the lattice spacing. 
For $a \to 0$ the correlation length must diverge in lattice units, and 
the discretization effects disappear. The continuum limit is thus a 
critical point of the theory.

What shown above is a completely nonperturbative argument. 
This is just a particular case of the well-known fact that the bare parameters
depend in general on the cutoff, which in our case is the lattice spacing 
$a$. Therefore there is always a well-defined relation between $g_0$ and $a$.
For couplings small enough for the perturbative $\beta$ function to be 
a reasonable approximation to the real running of the coupling, the approach 
to the continuum limit can be studied in perturbation theory, where $g_0$ 
and $a$ tend together to zero along a trajectory determined by 
renormalization group equations. The $\beta$ function appearing in the 
renormalization group equation for lattice QCD is defined by
\be
a \frac{dg_0}{da} = - \beta (g_0) .
\label{eq:betaf}
\ee
This $\beta$ function has for small couplings the expansion
\be
\beta(g_0) = -g_0^3 \Big[ b_0 + b_1 g_0^2 + b_2 g_0^4 + \dots  \Big] .
\label{eq:expbeta}
\ee
We see that things are similar to the continuum, with $\mu$ replaced by $1/a$.
The first two coefficients of the $\beta$ function are universal, and thus 
$b_0$ and $b_1$ in Eq.~(\ref{eq:expbeta}) are the same as in the continuum, 
were they were computed by~\cite{Gross:1973id,Politzer:fx} 
and by~\cite{Jones:mm,Caswell:gg}:
\bea
b_0 &=& \frac{1}{(4\pi)^2} \Big(11 - \frac{2}{3} N_f\Big) , \\
b_1 &=& \frac{1}{(4\pi)^4} \Big(102 - \frac{38}{3} N_f\Big) .
\eea
The coefficient $b_2$ depends on the scheme. The first calculations of this
coefficient for the Wilson action were attempted 
in~\cite{Ellis:1983af,Ellis:1984ag,Ellis:1984qt}. The coefficient $b_2$ was 
then fully computed 
in~\cite{Luscher:1995nr,Luscher:1995np,Alles:1996cy,Christou:1998ws}, 
with the result
\bea
b_2 &=& -0.00159983232(13) +0.0000799(4)\, N_f -0.00000605(2)\, N_f^2 
\quad (c_{sw}=0), \\
b_2 &=& -0.00159983232(13) -0.0009449(4)\, N_f +0.00006251(2)\, N_f^2 , 
\quad (c_{sw}=1), 
\eea
were in the last line we have also given its value in the tree-level improved 
theory (which we will introduce in the next Section), which has been computed 
by~\cite{Bode:2001uz} for general $c_{sw}$. 
The value of $b_2$ in the Schr\"odinger functional has been computed 
in~\cite{Bode:1997uj,Bode:1998hd,Bode:1999dn,Bode:1999sm}, and
for $T=L$, $\theta=\pi/5$ and $N_f=0$ has the value
\be 
b_2 = \frac{1}{(4\pi)^3} \, 0.482(7),
\ee
while for $N_f=2$ is 
\be 
b_2 = \frac{1}{(4\pi)^3} \, 0.064(10).
\ee
For comparison, $b_2$ is given in the continuum $\ms$ scheme  
by~\cite{Tarasov:au}
\be 
b_2 = \frac{1}{(4\pi)^6} \, \Bigg(
\frac{2857}{2} -\frac{5033}{18} \, N_f +\frac{325}{54} \, N_f^2 
\Bigg) .
\ee
In this scheme some higher-order coefficients are also 
known~\cite{vanRitbergen:1997va}.

Solving Eqs.~(\ref{eq:betaf}) and~(\ref{eq:expbeta}) at lowest order gives 
the solution
\be
g_0^2 \sim - \frac{1}{b_0 \log a^2 \Lambda_{\mathrm lat}^2} + O(1/\log^2 a^2).
\label{eq:rgsol}
\ee
The evolution of the bare lattice coupling with the scale $a$ defines,
like in the continuum, a $\Lambda$ parameter. The value of the $\Lambda$ 
parameter cannot be determined using the lowest-order solution, since a 
rescaling of $\Lambda$ is of the same order as terms which have been dropped.
A reasonable definition of the $\Lambda$ parameter has to take into account 
higher orders:
\be
\Lambda = a^{-1} \cdot (b_0 g_0^2)^{\displaystyle -\frac{b_1}{2b_0^2}} 
\cdot \e^{\displaystyle -\frac{1}{2b_0 g_0^2}} \cdot
\exp{\Bigg\{ -\int_0^g dt \Bigg( \frac{1}{\beta(t)} + \frac{1}{b_0t^3} -
\frac{b_1}{b_0^2t} \Bigg) \Bigg\}}.
\label{eq:lambdapar}
\ee
This definition has been used in the calculations of~\cite{Capitani:1998mq} 
reported in Section~\ref{sec:why}, which use the Schr\"odinger functional.
A $\Lambda$ parameter in a given scheme specifies the value of the coupling
constant in that scheme for any given scale $\mu$, and all dimensionful 
quantities are proportional to $\Lambda$. Since the $\Lambda$ parameter 
depends on the scheme, its values on the lattice will not be equal to 
$\Lambda_{QCD}$ as is known from the continuum. It happens indeed that the 
values of $\Lambda$ on the lattice are in general quite different from the 
continuum. Their ratio can be computed using lattice (and continuum) 
perturbation 
theory~\cite{Hasenfratz:1980kn,Dashen:vm,Weisz:1980pu,Hasenfratz:1981tw}. 
For example, for the pure gauge Wilson action one has
\be
\frac{\displaystyle 
\Lambda_{\ms}}{\displaystyle \Lambda_{\mathrm lat}} = 28.80934(1) .
\ee
The huge change in scales between the lattice theories and the continuum 
is a common phenomenon, and is more pronounced for some lattice actions
than for others. 

The two scales $\Lambda_{\mathrm lat}$ and $\Lambda_{QCD}$ can thus be 
related. Once the relation between the lattice and continuum coupling 
constants is known, combining the results of Monte Carlo simulations with 
the knowledge of $\Lambda_{\mathrm lat}$ and the quark masses allows 
in principle to predict all physical quantities.
In this way one can predict for example the value of $\alpha_S (M_Z)$ 
using only nonperturbative lattice calculations of the spectrum.

We have seen that the bare coupling $g_0$ automatically determines the size 
of the lattice spacing $a$, and computing lattice quantities near the 
continuum limit means taking both of them to zero in such a way that 
Eq.~(\ref{eq:rgsol}) (or even better Eq.~(\ref{eq:lambdapar})) is satisfied. 
The renormalized physical quantities should remain constant along this 
trajectory, and this defines the ``scaling region'' near the continuum limit. 
If when doing some lattice simulations we can show that $g_0$ and $a$ follow 
this relation, then we now that we are extracting continuum physics from the 
Monte Carlo results.

For sufficiently large scales one can also expand the running couplings
in different schemes directly in powers of each other. We can then relate 
the couplings by doing a matching at a finite scale $p$~(Celmaster and 
Gonsalves, 1979a; 1979b). On the lattice the relation between the bare 
and the renormalized coupling (defined as the three-point function 
at a certain momentum $p$) is 
\be
g_R (p) = g_0 \Big[1+g_0^2 \Big(-b_0 \log ap + C^L + O(a^2p^2 \log ap)
\Big) + O(g_0^4) \Big] ,
\ee
while for the continuum coupling one has
\be
g_R (p) = g_{\ms} \Big[1+g_{\ms}^2 \Big( -b_0 \log \frac{p}{\mu} 
+ C^{\ms} \Big) + O(g_{\ms}^4) \Big].
\ee
Combining these two equations we have then that
\be
g_0 = g_{\ms} \Big[ 1+g_0^2 \Big( C^{\ms} - C^L +b_0 \log a\mu 
\Big) + O(g_0^4) + O(a^2) \Big] .
\ee

The most effective way of computing the matching between different couplings 
is provided by background field 
techniques~\cite{DeWit:1967a,DeWit:1967a,Abbott:1980hw}. For most of the 
calculations cited above a background field method has been indeed employed. 
In the calculations of~\cite{Luscher:1995nr,Luscher:1995np} this was 
supplemented by evaluations of the Feynman diagrams using the coordinate space
method. We will discuss this method in detail in 
Section~\ref{sec:coordinatespacemethods}, and we will show there some 
integrals that occur in the calculation of $b_2$ using the background field.

A gauge theory on the lattice with a background gauge field is renormalizable 
to all orders of perturbation theory. A nice thing is that no new 
counterterms are needed, that is the ones already required in the lattice 
theory without background fields are sufficient~\cite{Luscher:1995vs}, as it 
happens in the continuum case~\cite{Kluberg-Stern:1974xv,Kluberg-Stern:1975hc}.
Thus, introducing a background field does not affect the renormalization of 
the lattice theory. The renormalization of gauge theories on the lattice 
without background fields has been proven in~\cite{Reisz:1988kk}.

The propagator of the background field is proportional to $1/g^2$, and for the
renormalization of the coupling and the perturbative determination of the 
lowest coefficients of the $\beta$ function the computation of the self-energy
of the background field is then sufficient. This is the main advantage of 
using a background field method. It is much easier to compute diagrams with 
two legs instead of three, as would be the case without background field, and 
the number of diagrams to be computed is also smaller. It is true that to 
control the renormalization of the gauge-fixing parameter the corrections to 
the gauge field propagator must also be calculated, but this is only needed 
to lower orders.

In the continuum the fields $A_\mu$ are decomposed as follows: 
\be
A_\mu (x) = B_\mu (x) + g_0 q_\mu (x) ,
\ee
where the background field $B_\mu$ is a smooth external source field 
(which is not required to satisfy the Yang-Mills equations), while
$q_\mu$ is the quantum field.
To the gauge action reexpressed in terms of $B_\mu$ and $q_\mu$
one has to add the gauge-fixing term 
\be
\frac{1}{\alpha} \int dx \, \Tr \Big( D_\mu q_\mu (x) D_\nu q_\nu (x) \Big) 
\ee
and the ghost action (coming from the Faddeev-Popov procedure)
\be
2 \int dx \, \Tr \Big( D_\mu \overline{c} (x) \, 
\Big( D_\mu + \ii g_0 \widetilde{q}_\mu (x) \Big) \, c (x) \Big) ,
\ee
where the covariant derivative is 
\be
D_\mu = \partial_\mu + \ii \widetilde{B}_\mu ,
\ee
and $\widetilde{q}$ and $\widetilde{B}$ denote the quantum field and 
the background field in the adjoint representation:
$\widetilde{B}_\mu = B_\mu^a t^a$ (like in Section~\ref{sec:measure}).

On the lattice, as expected, there is more than one choice for extending
the theory with the introduction of a nonzero background field. A convenient 
decomposition of the gauge links is
\be
U_\mu (x) = \e^{\ii a B_\mu (x)} \e^{\ii a g_0 q_\mu (x)} .
\ee
In this case the background gauge transformations take a simple form,
\bea
B_\mu^\Omega &=& - \frac{\ii}{a} \log \Big( \Omega (x) \, 
\e^{\ii a B_\mu (x)} \, \Omega^{-1} (x+a\hat{\mu}) \Big) \\
q_\mu^\Omega &=& \Omega (x) \, q_\mu (x) \, \Omega^{-1} (x) ,
\eea
with $\Omega$ a classical field. The gauge-fixing term is
\be
\frac{1}{\alpha} \cdot a^4 \sum_x \, 
\Tr \Big( D_\mu^\star q_\mu (x) D_\nu^\star q_\nu (x) \Big) 
\ee
and the ghost action is
\be
2 a^4 \sum_x \, \Tr \Big( D_\mu \overline{c} (x) \,
\Big( (M^\dagger)^{-1} (q_\mu (x)) \cdot D_\mu 
+ \ii g_0 \widetilde{q}_\mu (x) \Big) \, c (x) \Big) ,
\ee
where the forward and backward lattice covariant derivatives in 
this case act as follows:
\bea
D_\mu f(x) &=& \frac{1}{a} \Big( \e^{\ii a B_\mu (x)} \, f(x+a\hat{\mu}) \,
\e^{-\ii a B_\mu (x)} - f(x) \Big) , \\
D_\mu^\star f(x) &=& \frac{1}{a} \Big( f(x) - \e^{-\ii a B_\mu (x-a\hat{\mu})}
\, f(x-a\hat{\mu}) \, \e^{\ii a B_\mu (x-a\hat{\mu})} \Big) ,
\eea
and $M$ is the same matrix defined in Eq.~(\ref{eq:m_matrix}).

The theory so defined is gauge invariant, and has a BRS symmetry and a shift 
transformation symmetry~\cite{Luscher:1995vs}. The Feynman rules of pure QCD 
in the presence of a background field are given in~\cite{Luscher:1995np}.
The usual lattice gauge theory with the standard covariant gauge-fixing term 
can be recovered in the limit of zero background field.

The $\beta$ function can also be defined for describing the scale evolution 
of the renormalized coupling. In this form it has been used to determine the 
perturbative running of the strong coupling in the Schr\"odinger functional 
scheme of which we have discussed in Section~\ref{sec:why}. There we showed 
also the scale evolution of the renormalized masses of the quarks, which is 
described in lattice QCD by the $\tau$ function:
\be
a \frac{dm}{da} = - \tau(g) m .
\label{eq:tauf}
\ee
The expansion of the $\tau$ function for small $g$ is given by
\be
\tau(g) = -g^2 \Big[ d_0 + d_1 g^2 + d_2 g^4 + \dots  \Big] .
\ee
The leading-order coefficient coefficient $d_0$ does not depend on the 
scheme, and has been computed in~\cite{Nanopoulos:1975kd}
\be
d_0 = \frac{8}{(4\pi)^2} .
\ee
The coefficient $d_1$ is different in the various schemes. 
In the Schr\"odinger functional $d_1$ has been computed by~\cite{Sint:1998iq},
and is given (for $T=L$ and $\theta=0.5$) by
\be
d_1 = \frac{1}{(4\pi)^2} \Big( 0.217(1) + 0.084(1)\, N_f \Big) . 
\ee
In continuum $\ms$ is instead given 
by~\cite{Nanopoulos:1978hh,Tarrach:1980up,Espriu:1981eh}
\be
d_1 = \frac{1}{(4\pi)^4} \Bigg( \frac{404}{3} - \frac{40}{9}\, N_f \Bigg) .
\ee
In the $\ms$ scheme some higher-order coefficients are also 
known~\cite{Chetyrkin:1997dh,Vermaseren:1997fq}.

A very useful quantity is the renormalization group invariant mass, defined 
by~\cite{Gasser:1982ap,Gasser:1983yg,Gasser:1984gg}
\be
M = m \cdot (2b_0 g^2)^{\displaystyle -\frac{d_0}{2b_0}} \cdot
\exp{\Bigg( -\int_0^g dt \Big[ \frac{\tau(t)}{\beta(t)} - \frac{d_0}{b_0 t} 
\Big] \Bigg) } .
\label{eq:rgimass}
\ee
This quantity, at variance with the $\Lambda$ parameter, is even scheme 
independent. Moreover, renormalization group masses present the convenience
that they are nonperturbatively defined, whereas the masses renormalized at 
a certain scale in the $\ms$ scheme are not.

\section{Improvement}
\label{sec:improvement}

The results of lattice simulations are affected by statistical errors 
(which arise because only a finite number of configurations can be generated) 
and systematic errors.

The systematic errors are of various nature. Rather important are those coming 
from the finiteness of the lattice spacing, and often also finite volume 
effects, quenching effects (i.e., when internal quark loops are neglected) 
and extrapolations to the chiral limit constitute significant systematic
errors.

The techniques which go under the name of ``improvement'' aim at removing 
the systematic error due to the finiteness of the lattice spacing, which are 
generally of order $a$ with respect to the continuum limit, as in the formal 
expansion
\be
\left< p \left| \widehat{\cal O}_{L} \right| p' \right>_{\mathrm Monte~Carlo}=
a^{d} \left[ \left< p \left| \widehat{\cal O} \right| p' 
\right>_{\mathrm phys} + O(a) \right] .
\ee 
We know that at the values of the coupling which are presently simulated
these discretization errors can often have a significant effect 
on the results.

It is very expensive to reduce these cutoff effects by ``brute force'',
that is simply decreasing the lattice spacing $a$, since the computing time
required for this would grow with the fifth power of the inverse lattice 
spacing in the quenched approximation, and even faster in full QCD. 
This means for example that halving the discretization errors by brute force 
would require a calculational effort at least thirty times bigger, 
all other things being equal. A better way of reducing these errors can be 
achieved by improving the theory. These unphysical terms are then 
systematically removed by adding irrelevant terms to actions and operators
(that is, they vanish in the continuum limit and do not change the continuum 
theory). We will now discuss what this implies in terms of perturbation theory.
For a pedagogical introduction on these topics, which we will not here discuss
much in detail, the reader can turn to the reviews of~\cite{Luscher:zf} 
and~\cite{Luscher:1998pe}.

\subsection{Improved quarks}

A systematic improvement program to reduce the cutoff errors order by order
in the lattice spacing $a$ for Wilson fermions was first proposed by 
Symanzik (1979; 1981; 1983a; 1983b) and then further developed, and applied 
to on-shell matrix elements,
in~\cite{Luscher:1984xn,Luscher:1985zq,Sheikholeslami:1985ij,Heatlie:1990kg}.

In this formulation, an irrelevant operator is added to the Wilson action in 
order to cancel, in on-shell matrix elements, all terms that in the continuum 
limit are effectively of order $a$.~\footnote{This means that at $n$ loops 
these terms, because of Eq.~(\ref{eq:rgsol}), have the form 
$g_0^{2n} a \log^n a$.} In this way one can reduce the cutoff errors coming 
from the discretization of the action from $O(a)$ to $O(a^2)$:
\be
\left< p \left| \widehat{\cal O}_{L} \right| p' \right>_{\mathrm Monte~Carlo}=
a^{d} \left[ \left< p \left| \widehat{\cal O} \right| p' 
\right>_{\mathrm phys} + O(a^2) \right] .
\ee 
This is a remarkable decrease of the systematic error due to the finiteness of
the lattice spacing. The continuum limit is reached much faster, with a rate 
proportional to $a^2$.~\footnote{A similar thing can be found for example in
numerical integration methods, where the trapezoidal rule reduces the 
discretization error to the cubic power of the integration step, and
Simpson's rule further reduces it to the fifth power.}

\begin{figure}[t]
\begin{center}
\begin{picture}(220,220)(0,0)
\LongArrow(0,100)(200,100)
\LongArrow(100,0)(100,200)
\Line(95,105)(35,105)
\Line(35,105)(35,165)
\Line(35,165)(95,165)
\LongArrow(95,165)(95,110)
\Line(105,105)(105,165)
\Line(105,165)(165,165)
\Line(165,165)(165,105)
\LongArrow(165,105)(110,105)
\Line(95,95)(95,35)
\Line(95,35)(35,35)
\Line(35,35)(35,95)
\LongArrow(35,95)(90,95)
\Line(105,95)(165,95)
\Line(165,95)(165,35)
\Line(165,35)(105,35)
\LongArrow(105,35)(105,90)
\Text(210,90)[r]{$\mu$}
\Text(110,210)[r]{$\nu$}
\end{picture}
\end{center}
\caption{\small The combination of the four plaquettes which builds the clover
lattice approximation of the $F_{\mu\nu}$ tensor at the point $x$.}
\label{fig:clover}
\end{figure}
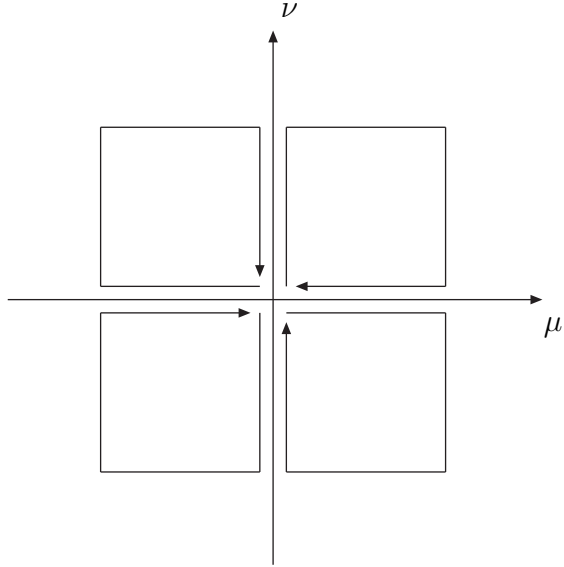

For this purpose one has to introduce the improved ``clover'' fermion action,
first proposed in~\cite{Sheikholeslami:1985ij}:
\be
\Delta S^f_I = c_{sw}\cdot \ii  g_0 a^{4} \sum_{x,\mu \nu} \frac{r}{4 a} \: 
\overline{\psi} (x)
\sigma_{\mu \nu} F_{\mu \nu}^{\mathrm clover}(x) \psi (x) ,
\ee
which vanishes in the continuum limit and has the same symmetries of the 
original unimproved Wilson action. This counterterm is the only one required 
after exploiting all the symmetries and the equations of motion.~\footnote{The
other dimension-five terms that are gauge invariant and compatible with the 
symmetries of the Wilson action are
\bea
&& \overline{\psi} \stackrel{\rightarrow}{D}_\mu \stackrel{\rightarrow}{D}_\mu
\psi +
\overline{\psi} \stackrel{\leftarrow}{D}_\mu \stackrel{\leftarrow}{D}_\mu
\psi , \\
&& m \, \Big( \overline{\psi} \gamma_\mu \stackrel{\rightarrow}{D}_\mu \psi 
- \overline{\psi} \stackrel{\leftarrow}{D}_\mu \gamma_\mu \psi \Big) ,
\eea
which can be eliminated in on-shell matrix elements using
two equations of motion, and
\bea
&& m \, \Tr \Big( F_{\mu\nu} F_{\mu\nu} \Big) , \\
&& m^2 \, \overline{\psi} \psi ,
\eea
which can be reabsorbed into a rescaling of the coupling and mass.}
The ``clover'' gluon field is here defined as 
\be
F_{\mu \nu}^{\mathrm clover}(x) = \frac{1}{4} \frac{1}{2 \ii g_0a^2}
\sum_{\mu \nu = \pm} (P_{\mu \nu} (x) - P^{\dagger}_{\mu \nu} (x)) ,
\ee
that is the average of the four plaquettes lying in the plane $\mu\nu$
stemming from the point $x$ (see Fig.~\ref{fig:clover}). This definition 
maximizes the symmetry of the field-strength tensor on the 
lattice~\cite{Mandula:1982us}.

Improved actions like this allow one to extract continuum physics already for
lattice couplings which are not too small. One just exploits the fact that 
more than one lattice action corresponds to a given continuum action, and 
looks for lattice actions with smaller discretization errors. This of course 
implies a change in the Feynman rules (and usually not for the best).
Adding the Sheikholeslami-Wohlert term to the Wilson Lagrangian means 
that we have to add to the Wilson quark-quark-gluon interaction vertex
\be
(V^a)^{bc}_{\rho} (p_1,p_2) = -g_0 (T^a)^{bc} 
\left[ \ii  \gamma_{\rho} \cos \frac{a(p_1 + p_2)_{\rho}}{2} 
   + r \sin \frac{a(p_1 + p_2)_{\rho}}{2} \right] 
\ee
the improved quark-quark-gluon interaction vertex
\be
(V^a_{\mathrm IMP})^{bc}_{\rho} (p_1,p_2) = - c_{sw} \cdot g_0 \,
\frac{r}{2} \, (T^a)^{bc} 
\cos \frac{a(p_1 - p_2)_{\rho}}{2} \sum_{\lambda}
\sigma_{\rho \lambda} \sin a(p_1 - p_2)_{\lambda} .
\label{eq:impvertex}
\ee
The fermion propagator and the vertices with an even number of gluons
are instead not modified by the improved action.

Neither is the gluon propagator. In fact, as we have seen in 
Section~\ref{sec:Wilsonfermions}, the first corrections to the pure gauge term
in the Wilson action are already of order $a^2$, and thus there is no need of 
improvement to this order, although sometimes it is useful to improve the 
gluon action reducing the discretization errors from $O(a^2)$ to $O(a^3)$
(in fact $O(a^4)$), as we will see shortly.

Only with the appropriate value of the improvement coefficient $c_{sw}$ 
(for a given value of the coupling) the cancellation of $O(a)$ effects 
can be achieved. At lowest order in perturbation theory one just needs 
the tree-level value $c_{sw}=1$, and in the work of~\cite{Heatlie:1990kg} 
it has been explicitly shown that all terms that are effectively of order $a$ 
are absent in the 1-loop matrix elements of the quark currents for $c_{sw}=1$.
Perturbative determinations of $c_{sw}$ at order $g_0^2$ have been made 
in~\cite{Wohlert:1987,Naik:1993ux,Luscher:1996vw}.
Nonperturbative determinations of $c_{sw}$ have been carried out by the 
ALPHA Collaboration, using the Schr\"odinger functional and imposing the
cancellation of all $O(a^2)$ effects in the PCAC relation.
The formula that summarizes the ALPHA Collaboration quenched results 
when $0 \le g_0 \le 1$ is~\cite{Luscher:1996ug}:
\be
c_{sw} = \frac{1-0.656 \, g_0^2 -0.152 \, g_0^4 -0.054 \, g_0^6}{1
-0.922 \, g_0^2} ,
\ee
while in the $N_f=2$ case one has~\cite{Jansen:1998mx}:
\be
c_{sw} = \frac{1-0.454 \, g_0^2 -0.175 \, g_0^4 +0.012 \, g_0^6
+0.045 \, g_0^8}{1-0.720 \, g_0^2} .
\ee
The $O(g_0^2)$ terms in these formulae correspond to the 1-loop perturbative
result.

All the above machinery is not enough to improve completely a lattice theory. 
In addition to improving the action, one also has to improve each operator 
that happens to be studied. This means that one must add a basis of 
higher-dimensional irrelevant operators with the same symmetry properties 
as the original unimproved operator. Bases of improved operators have been 
constructed for quark currents~\cite{Sint:1995ch,Sint:1997jx} and operators 
measuring unpolarized structure functions~\cite{Capitani:2000xi}, and improved 
renormalization factors have been computed 
in~\cite{Gabrielli:1990us,Frezzotti:1991pe,Borrelli:1992vf}
and~\cite{Capitani:1997nj,Capitani:2000xi}. 
We will not deal with these calculations here.

Of course one can also attempt to improve the theory to the next order,
canceling all contributions which are effectively of order $a^2$. 
For the fermion part this is quite complicated. In this case four-quark 
operators are also necessary, besides a certain number of two-quark operators 
of dimension six.

We would like briefly to mention what improving the theory looks like when 
overlap fermions are 
used~\cite{Capitani:1999uz,Capitani:1999ay,Capitani:2000xi}.
One of the many qualities of overlap fermions is that the overlap action 
is already improved to $O(a)$, and thus one needs only to improve operators.
An operator $O = \overline{\psi} \widetilde{O} \psi$ is improved,
to all orders in perturbation theory, by taking the expression
\be
O^{\mathrm imp}= \overline{\psi} \Big(1-\frac{1}{2\rho} a D_N\Big) \, \widetilde{O}
\, \Big(1-\frac{1}{2\rho} a D_N\Big) \psi .
\ee
Moreover, $O^{\mathrm imp}$ and $\widetilde{O}$ have the same 
renormalization constants.

We remind, for comparison, that for Wilson fermions an additional interaction 
vertex has instead to be introduced, the Sheikholeslami-Wohlert clover term
(proportional to $c_{sw}$), and furthermore for each operator one needs to 
construct a complete basis of operators which are one dimension higher and 
whose coefficients have then to be tuned to cancel the residual $O(a)$ 
contributions not coming from the action.

For example, for the first moment of the quark momentum distribution
the relevant operator is 
\be
O_{\{\mu\nu\}} = \overline{\psi} \gamma_{\{\mu} D_{\nu\}} \psi,
\ee 
symmetrized in $\mu$ and $\nu$, and one basis for the improvement is given by 
\be
 \overline{\psi} \gamma_{\{\mu} D_{\nu\}} \psi
-\frac{1}{4} a \ii  c_1 \, \sum_\lambda 
   \overline{\psi}\sigma_{\lambda\{\mu} \Big[ D_{\nu\}} , D_\lambda \Big] \psi
  -\frac{1}{4} a c_2 \, \overline{\psi} \Big\{ D_\mu , D_\nu 
   \Big\} \psi .
\label{eq:imprfirstmom}
\ee
Only for some particular values of the improvement coefficients
\bea
c_1 (g_0^2) &=& 1 + g_0^2 c_1^{(1)} + O(g_0^4) \\
c_2 (g_0^2) &=& 1 + g_0^2 c_2^{(1)} + O(g_0^4) \nonumber
\eea
the operator if effectively improved, i.e., all $O(a)$ corrections are
exactly canceled. Even for this simple case, these two coefficients are not 
known, because at present one knows only a relation between them.
One of the coefficients then remains unknown~\cite{Capitani:2000xi}.

One could determine all remaining coefficients using some Ward Identities or 
physical conditions, but this involves a lot of effort. Moreover, for higher 
moments, which contain more covariant derivatives, the number of operator
counterterms becomes larger and larger. This means that many improvement 
coefficients have to be determined, and they need as many conditions to be set.
Moreover, one has also to compute the contribution of each one of these 
operator counterterms to the total renormalization constant. This looks a 
formidable task.

Thus, improving the theory is much simpler for overlap fermions.

\subsection{Improved gluons}
\label{sec:improvedgluons}

\begin{figure}[t]
\begin{center}
\begin{picture}(400,140)(0,40)
\DashLine(60,80)(60,140){4}
\DashLine(180,80)(180,140){4}
\DashLine(180,140)(240,140){4}
\DashLine(200,160)(180,140){4}
\DashLine(240,80)(260,100){4}
\DashLine(260,100)(200,100){4}
\DashLine(260,160)(260,100){4}
\DashLine(300,80)(300,140){4}
\DashLine(320,160)(300,140){4}
\DashLine(360,80)(380,100){4}
\DashLine(380,160)(320,160){4}
\DashLine(320,160)(320,100){4}
\DashLine(300,140)(360,140){4}
\Text(60,60)[t]{${\cal L}_1^{(6)}$}
\Text(220,60)[t]{${\cal L}_2^{(6)}$}
\Text(340,60)[t]{${\cal L}_3^{(6)}$}
\SetWidth{2}
\Line(0,80)(120,80)
\Line(120,80)(120,140)
\Line(120,140)(0,140)
\Line(0,140)(0,80)
\Line(180,80)(240,80)
\Line(240,80)(240,140)
\Line(240,140)(260,160)
\Line(260,160)(200,160)
\Line(200,160)(200,100)
\Line(200,100)(180,80)
\Line(300,80)(360,80)
\Line(360,80)(360,140)
\Line(360,140)(380,160)
\Line(380,160)(380,100)
\Line(380,100)(320,100)
\Line(320,100)(300,80)
\end{picture}
\end{center}
\caption{\small The planar, twisted and L-shaped six-link loops.}
\label{fig:iga}
\end{figure}
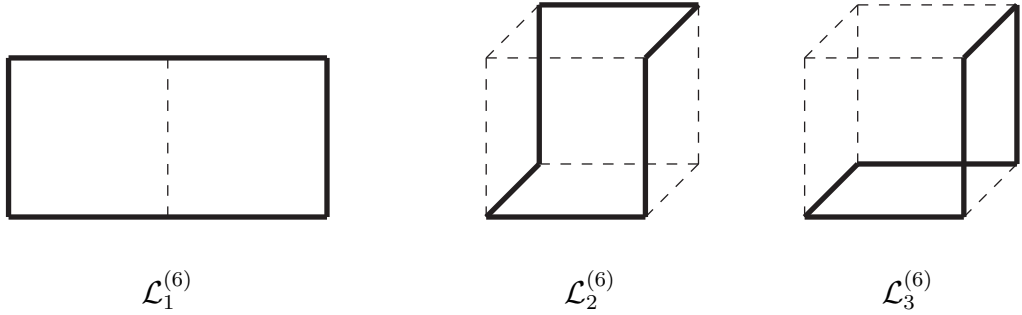

The plaquette is not the only possibility for the construction of the 
discretized version of the gauge field strength. One can also consider 
larger closed loops. As we noted in Section~\ref{sec:Wilsonfermions}, 
the corrections to the pure gauge part of the Wilson action with respect to 
the continuum are of order $a^2$. The gluon part of the action is then already 
$O(a)$ improved. The next step consists in implementing the improvement to 
$O(a^2)$, that is adding to the Wilson action some counterterms of dimension 6
that (with the appropriate values of their coefficients) can cancel all 
$O(a^2)$ effects, so that the first corrections that are left are then at 
least of order $a^3$.~\footnote{Actually, due to symmetry considerations they 
are of order $a^4$. The fact that the corrections to the pure gauge action are
of order $a^2$ or $a^4$ essentially comes from the fact that one can construct 
gauge-invariant terms of dimension 6 and 8, but not of dimension 5 and 7.}
There are 3 terms of dimension 6 with the right quantum numbers, and the 
improved gauge action can be written as
\be
S_g = \frac{6}{g_0^2} \Bigg[ c_0 (g_0^2) \, {\cal L}^{(4)} 
+ a^2 \sum_{i=1}^3 c_i (g_0^2) \, {\cal L}_i^{(6)}\Bigg],
\ee
where ${\cal L}^{(4)}$ is the usual Wilson plaquette action, and the 
dimension-6 terms are six-link closed loops which are called planar, 
twisted and L-shaped respectively (see Fig.~\ref{fig:iga}). 
Each of these terms contains the 
$\sum_{\mu\nu} \Tr \Big(F_{\mu\nu} F_{\mu\nu} \Big)$ operator plus
a linear combination of the following operators:
\be
\sum_{\mu\nu} \Tr \Big( D_\mu F_{\mu\nu} D_\mu F_{\mu\nu} \Big) ,\qquad
\sum_{\mu\nu\rho} \Tr \Big( D_\mu F_{\nu\rho} D_\mu F_{\nu\rho} \Big) ,\qquad
\sum_{\mu\nu\rho} \Tr \Big( D_\mu F_{\mu\rho} D_\nu F_{\nu\rho} \Big) .
\ee
L\"uscher and Weisz have computed the coefficients of these linear 
combinations and determined, imposing improvement conditions on the large 
Wilson loops, the values of the improvement coefficients that accomplish the 
cancellation of the $O(a^2)$ corrections~\cite{Luscher:1984xn,Luscher:1985zq}. 
At tree level one has
\be
c_0 = \frac{5}{3}, \qquad  c_1 = -\frac{1}{12} , 
\qquad c_2 = 0, \qquad c_3 = 0 ,
\ee
so that only the ${\cal L}_1^{(6)}$ counterterm (the planar loop) is needed. 
We have then
\be
S_g = \frac{6}{g_0^2} \Bigg[ \frac{5}{3} \, {\cal L}^{(4)} 
-\frac{1}{12} a^2 \, {\cal L}_1^{(6)}\Bigg]
= \frac{1}{2} \int \Tr \, F_{\mu\nu}^2 (x) + O(a^4) .
\ee
This action is a discretization of the continuum QCD pure gauge action in 
which the discretization errors have been reduced to order $a^4$.
Although both ${\cal L}^{(4)}$ and ${\cal L}_1^{(6)}$ have discretization 
errors of order $a^2$, the above combination cancels these contributions
and the first correction to the continuum becomes of order $a^4$.
Of course this is only a tree-level cancellation, and the coefficients
get corrected by quantum effects.
The values that improve the pure gauge action to 1-loop are
\bea
c_0 (g_0^2) &=& \frac{5}{3} +0.2370 g_0^2 ,     \nonumber \\ 
c_1 (g_0^2) &=& -\frac{1}{12} -0.02521 g_0^2 ,  \nonumber \\ 
c_2 (g_0^2) &=& -0.00441 g_0^2 ,    \nonumber \\ 
c_3 (g_0^2) &=&  0 . 
\eea
They define the L\"uscher-Weisz action. These coefficients satisfy 
the normalization condition 
\be
c_0 + 8 c_1 + 8 c_2 + 16 c_3 = 1,
\ee
which is valid to all orders of perturbation theory.
Since at one loop $c_2$ is small one usually drops ${\cal L}_2^{(6)}$, and 
${\cal L}_1^{(6)}$ remains the only counterterm (as it was at tree level). In 
this case the normalization condition is used to write the action in the form
\bea
S_g &=& \frac{6}{g_0^2} \Bigg[ (1 - 8 c_1) \, {\cal L}^{(4)} 
+ a^2 c_1 \, {\cal L}_1^{(6)} \Bigg] \label{eq:igaction} \\ 
&=& \frac{6}{g_0^2} \Bigg[ (1 - 8 c_1) \, 
\sum_x \sum_{\mu < \nu} P^{1\times 1}_{\mu\nu}
+ a^2 c_1 \, \sum_x \sum_{\mu < \nu} 
\Big( P^{1\times 2}_{\mu\mu,\nu} + P^{1\times 2}_{\nu\nu,\mu} \Big) \Bigg] 
\nonumber ,
\eea
where $P^{1\times 2}_{\mu\mu,\nu}$ denotes the rectangle which is two lattice 
spacings long in the $\mu$ direction.
These actions has been extensively investigated also 
in~\cite{Weisz:1982zw,Weisz:1983bn,Wohlert:1984hk,Curci:1983an,Bernreuther:wx}.

There are also other actions which go under the name of improved gauge 
actions where improvement is not done \`a la Symanzik, but instead following
renormalization group arguments (and then they still have $O(a^4)$ correction
terms). In this case one looks for actions which are close to what one obtains
after doing blocking transformations, that is renormalization group 
transformations in which the lattice spacing is doubled at each step. A 
perturbative calculation gives the action proposed by Iwasaki (1983a; 1983b), 
which is Eq.~(\ref{eq:igaction}) with
\be
c_1 = -0.331 ,
\ee
while nonperturbative calculations which use Schwinger-Dyson equations
lead to the DBW2 action~\cite{Takaishi:1996xj,deForcrand:1999bi}, 
which is Eq.~(\ref{eq:igaction}) with~\footnote{The acronym DBW2 stands for 
doubly-blocked Wilson $1\times2$ plaquette. We remark that in this case 
the relation between the coefficients is not linear, and these are only rather
approximate values.}
\be
c_1 \simeq -1.40686 .
\ee
There are other proposals in which $c_2$ and $c_3$ are nonzero, which we 
will not consider here.

All these improved gauge actions have the aim to achieve a better convergence 
to the continuum limit. Their main drawback is that they have no reflection 
positivity~\cite{Luscher:1984is}, and thus there is no corresponding theory 
in the continuum, that is it is not possible to construct a Hilbert space 
in the usual way. This also causes problems with numerical simulations.
The violation of physical positivity in fact leads to unphysical poles in the 
propagators, which correspond to unphysical states that create a sizeable
disturb in the extraction of physical observables~\cite{Necco:2002zy}.

All these actions are quite complicated to use in perturbation theory. 
The gluon propagator in a covariant gauge for generic $c_1$ is given by
\be
G_{\mu\nu} (k) = \frac{1}{(\widehat{k}^2)^2} \,
\Bigg( \alpha \widehat{k}_\mu \widehat{k}_\nu
+\sum_\sigma (\widehat{k}_\sigma \delta_{\mu\nu} - \widehat{k}_\nu 
\delta_{\mu\sigma} ) \widehat{k}_\sigma A_{\sigma\nu} (k) \Bigg) ,
\ee
with
\bea
A_{\mu\nu} (k) &=& A_{\nu\mu} (k) = (1-\delta_{\mu\nu}) \, \Delta (k)^{-1} \,
\Bigg[ (\widehat{k}^2)^2 
-c_1 \widehat{k}^2 \Bigg( 2\sum_\rho \widehat{k}_\rho^4
+\widehat{k}^2 \sum_{\rho \neq \mu,\nu} \widehat{k}_\rho ^2 \Bigg) \nonumber \\
&& \qquad \qquad 
+ c_1^2 \Bigg( \Big( \sum_\rho \widehat{k}_\rho^4 \Big)^2 
+\widehat{k}^2 \sum_\rho \widehat{k}_\rho^4 \sum_{\tau \neq \mu,\nu} 
\widehat{k}_\tau^2 + (\widehat{k}^2)^2 \prod_{\rho \neq \mu,\nu} 
\widehat{k}_\rho^2 \Bigg) \Bigg] ,
\eea
where
\bea
\Delta (k) &=& \Big( \widehat{k}^2 -c_1 \sum_\rho \widehat{k}_\rho^4 \Big)
\Bigg[ \widehat{k}^2 
-c_1 \Bigg( (\widehat{k}^2)^2 + \sum_\tau \widehat{k}_\tau^4 \Bigg) 
+\frac{1}{2} c_1^2 \Bigg( (\widehat{k}^2)^3 + 2 \sum_\tau \widehat{k}_\tau^6 
- \widehat{k}^2 \sum_\tau \widehat{k}_\tau^4 \Bigg) \Bigg] \nonumber \\
&& -4 c_1^3 \sum_\rho \widehat{k}_\rho^4 \sum_{\tau \neq \rho} 
\widehat{k}_\tau^2 .
\eea
The gluon vertices are quite complicated, and we will not report them 
here. They can be found in~\cite{Weisz:1983bn}. The quark-gluon vertices 
are of course not modified.

Perturbative calculations using improved gauge actions have been recently 
presented in~\cite{Aoki:2000ps} for three-quark operators and 
in~\cite{DeGrand:2002vu} for two- and four-quark operators, and they have 
even been employed in domain wall calculations~\cite{Aoki:2002iq}, where the
use of improved gauge actions is thought to diminish the residual chiral 
symmetry breaking which one has when working at finite $N_s$.

\section{The Schr\"odinger functional}
\label{sec:sf}

We present here a short introduction to a powerful framework for lattice
calculations that goes under the name of Schr\"odinger functional. 
This is a field of research that has grown very much in recent years and would
need a separate review in itself, given its peculiarities and technical 
complexities as well as the number and importance of the results that has 
produced. For an introductory review the lectures of~\cite{Luscher:1998pe}
are recommended.

The Schr\"odinger functional was extensively investigated on the lattice
in~\cite{Symanzik:1981wd,Luscher:iu,Luscher:1992an,Sint:1993un} 
and used in various physical situations.~\footnote{The Schr\"odinger 
functional had also been studied 
in~\cite{Rossi:1979jf,Rossi:1980pg,Rossi:1982ag,Rossi:wn,Leroy:eh},
using the temporal gauge and boundary conditions different from the ones
that we are going to introduce in the following.} 
It has been essential for the calculation of $c_{sw}$ perturbatively 
and nonperturbatively~\cite{Jansen:1995ck,Luscher:1996sc,Luscher:1996vw,Luscher:1996ug,Jansen:1998mx} and for
the nonperturbative computation of the running coupling in QCD
(which has allowed a quite precise determination of the $\Lambda$ parameter)
and of the masses of the quarks and their scale
evolution~\cite{Luscher:1991wu,Luscher:1992zx,Luscher:1993gh,Capitani:1997mw,Capitani:1998mq,Bode:2001jv,Garden:1999fg,Knechtli:2002vp}.

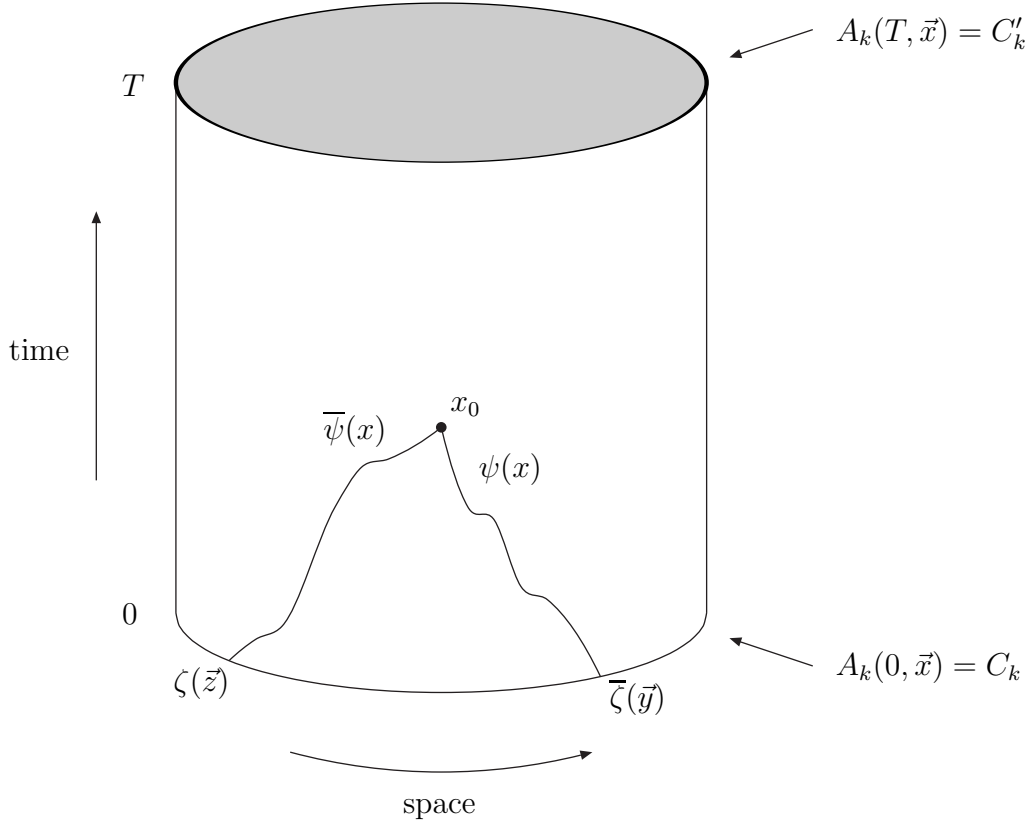
\begin{figure}[t]
\begin{center}
\begin{picture}(400,350)(0,0)
\Line(100,100)(100,300)
\Line(300,100)(300,300)
\GOval(200,300)(30,100)(0){0.8}
\Curve{(100,100)
(101,95.76797921)
(102,94.03007538)
(105,90.63250300)
(110,86.92330317)
(120,82)
(130,78.57571471)
(140,76)
(150,74.01923788)
(170,71.38182396)
(200,70)
(230,71.38182396)
(250,74.01923788)
(260,76)
(270,78.57571471)
(280,82)
(290,86.92330317)
(295,90.63250300)
(298,94.03007538)
(299,95.76797921)
(300,100)
}
\Text(80,300)[l]{$T$}
\Text(80,100)[l]{$0$}
\LongArrow(70,150)(70,250)
\Text(60,200)[r]{time}
\LongArrowArc(200,260)(220,255,285)
\Text(200,30)[t]{space}
\LongArrow(340,320)(310,310)
\Text(350,320)[l]{$A_k(T,\vec{x}) = C'_k$}
\LongArrow(340,80)(310,90)
\Text(350,80)[l]{$A_k(0,\vec{x}) = C_k$}
\Vertex(200,170){2}
\Text(210,175)[b]{$x_0$}
\Curve{(200,170)(180,158)(170,155)(160,140)(140,95)(130,90)(120,82)}
\Curve{(200,170)(210,140)(220,135)(230,110)(240,105)(260,76)}
\Text(180,170)[r]{$\overline{\psi} (x)$}
\Text(215,155)[l]{$\psi (x)$}
\Text(110,80)[t]{$\zeta (\vec{z})$}
\Text(275,75)[t]{$\overline{\zeta} (\vec{y})$}
\end{picture}
\end{center}
\caption{\small The Schr\"odinger functional, with its time boundaries.
Shown is also a correlation function involving the boundary fields.}
\label{fig:sf}
\end{figure}

The Schr\"odinger functional constitutes a finite volume renormalization
scheme. It is a standard functional integral in which fixed boundary 
conditions are imposed, and where the time direction assumes a special 
meaning. In fact, on the space directions there are generalized periodic 
conditions, while on the time direction one puts Dirichlet boundary 
conditions (see Fig.~\ref{fig:sf}).

This means that at $x_0=0$ and $x_0=T$ one fixes the spatial components of the
links $U$ to some particular values (usually constant abelian fields), 
while the temporal components, $U_0 (x)$, remain unconstrained, and they are 
defined only for $0 \le x_0 \le T-1$. The pure gauge action is given by the 
sum of the Wilson plaquettes which are fully contained between the timeslices
at $x_0=0$ and $x_0=T$. The spatial plaquettes at the boundaries $x_0=0$ and 
$x_0=T$ contribute to the action only with a weight $1/2$ to avoid double 
counting. The gauge group is local in the bulk but global on the boundaries.

The fermion fields are dynamical only for $1 \le x_0 \le T-1$, while 
at the two temporal boundaries only half their components are defined,
and they are fixed to some particular values:
\bea
P_+ \psi (x) \Big|_{x_0=0} = \rho  (\vec{x}), && \qquad
P_- \psi (x) \Big|_{x_0=T} = \rho' (\vec{x}),     \label{eq:sfbound} \\
\overline{\psi} (x) P_- \Big|_{x_0=0} = \overline{\rho}  (\vec{x}), && \qquad
\overline{\psi} (x) P_+ \Big|_{x_0=T} = \overline{\rho}' (\vec{x}) \nonumber ,
\eea
where the projectors are
\be
P_\pm = \frac{1 \pm \gamma_0}{2} .
\ee
The complementary components ($P_- \psi (x) \Big|_{x_0=0}$ etc.) 
vanish for consistency. In the spatial directions the quark fields are 
periodic up to a phase, and the generalized periodic conditions can be 
written as
\be
\psi (x + L \hat{k}) = \e^{\ii \theta_k} \psi (x), \qquad 
\overline{\psi} (x + L \hat{k}) = \overline{\psi} (x) \e^{-\ii \theta_k} , 
\qquad k=1,2,3 .
\ee
It turns out to be more convenient to work with the equivalent setting in which
the fermion fields have strictly periodic boundary conditions in the spatial 
directions and this phase $\theta$ is moved inside the definition of the 
covariant derivative,
\bea
D_\mu \psi (x) &=& \frac{1}{a} 
\Big[ \lambda_\mu U_\mu (x) \psi (x + a \hat{\mu}) - \psi(x) \Big] 
\label{eq:sf_covder}\\
D_\mu^\star \psi (x) &=& \frac{1}{a} 
\Big[ \psi(x) - \lambda_\mu^{-1} U_\mu^{-1} (x - a \hat{\mu}) 
\psi (x - a \hat{\mu}) \Big] \nonumber ,
\eea
with
\be
\lambda_\mu = \e^{\ii a \theta_\mu / L}, \quad \theta_0 = 0, \quad
-\pi < \theta_k \le \pi .
\ee
This phase $\theta$ is called a finite-size momentum, and it is not quantized 
(although we are by definition in a finite volume). It can then be chosen to 
be smaller than the minimal quantized momentum, $p_{\mathrm min} = 2\pi/L$, 
thereby reducing this kind of lattice artifacts. $\theta$ is a free parameter,
and can be tuned in such a way that one obtains the best numerical signals 
or the best perturbative expansions (or both, if one is lucky).

The boundaries do not influence the integration measure of the functional 
integral. Correlation functions in the path integral formulation can be 
computed in the usual way, the only difference being that the action has 
a more complicate form at the boundaries. The correlation functions can
then involve functional derivatives of the Boltzmann factor acting also 
on the boundary values of the quark fields,
\bea
\zeta (\vec{x}) = \frac{\delta}{\delta \overline{\rho} (\vec{x})}, &&\qquad
\overline{\zeta} (\vec{x}) = - \frac{\delta}{\delta \rho (\vec{x})}, \\
\zeta' (\vec{x}) = \frac{\delta}{\delta \overline{\rho}' (\vec{x})}, 
&&\qquad
\overline{\zeta}' (\vec{x}) = - \frac{\delta}{\delta \rho' (\vec{x})} ,
\nonumber 
\eea
if the operators under consideration contain fermions fields close to the 
boundaries. After taking the appropriate number of derivatives to build the 
desired operator, one puts 
$\rho = \rho' = \overline{\rho} = \overline{\rho}' = 0$ as usual.

The Schr\"odinger functional is not linked to a particular regularization. 
In fact, it is originally a continuum construction, and corresponds to the 
amplitude for the quantum evolution from a field configuration at the time 
zero to a field configuration at the time $T$. A lattice regularization is
of course useful for studying the nonperturbative aspects of QCD, from 
first principles. When one chooses to set the Schr\"odinger functional 
on the lattice,~\footnote{Usually one takes as lattice action in the
interior the Wilson action, which is probably the simplest setting for a 
lattice Schr\"odinger functional.} the fact that it is defined in a finite 
volume brings interesting features.

The choice of temporal boundary conditions specified above implies that zero 
modes are absent at the lowest oder of perturbation theory. For zero quark 
masses the lowest eigenvalue of the Dirac operator is then of order $1/L$. 
The frequency gap on the quark and gluon fields remains of order $1/L$ even 
in the interacting theory, and this means that one can perform simulations 
with quark masses close to zero without encountering singularities, 
because an infrared cutoff is provided by the lattice size, $L$.

A lattice of finite volume usually gives rise to systematic errors, but here 
the situation is completely different. The Schr\"odinger functional is used as
a finite volume renormalization scheme, where the renormalized quantities are 
specified at the scale $\mu = 1/L$ and for vanishing quark 
masses.~\footnote{Usually one chooses $T=2L$ or $T=L$, and everything in the 
theory is then referred to the scale $L$. It seems that some systematic 
errors are more pronounced in the case $T=2L$, and $T=L$ is then usually
preferred~\cite{Sint:1998iq}.} 
This is something quite different from the usual lattice calculations, where 
the renormalization scale is independent of the lattice size, and is instead  
determined by the lattice spacing. Thus, there cannot be by definition any 
finite volume effects in the Schr\"odinger functional. The finite volume is 
instead used to probe the theory and specify the renormalization prescriptions.

We stress again that the Schr\"odinger functional is a continuum scheme, 
because the scale at which the theory is renormalized is not proportional to 
the inverse of the lattice spacing, and so it can also be defined for a theory 
with zero lattice spacing. In practice however everything, from Monte Carlo 
simulations to perturbation theory, is carried out using lattice techniques. 
For instance, the perturbative calculations used for the running coupling in 
the Schr\"odinger functional that we mentioned in Section~\ref{sec:why} were
all done on the lattice.~\footnote{A one-loop calculation of the 
Schr\"odinger functional in dimensional regularization can be found 
in~\cite{Sint:1995rb}.} At the end one can extrapolate the results obtained at
different lattice spacings to the continuum limit, where the Schr\"odinger 
functional is well defined. Then it helps that the $\Lambda$ parameter is much
closer to the continuum that usual $\Lambda$ parameters defined on the 
lattice. For the theory with zero flavors one has
\be
\frac{\displaystyle 
\Lambda_{\mathrm SF}}{\displaystyle \Lambda_{\ms}} = 0.48811(1) .
\ee
This is related to the fact that this scheme seems ``closer'' to the $\ms$ 
scheme, as we have also discussed at the end of Section~\ref{sec:why}.

The Schr\"odinger functional scheme can also be supplemented with powerful 
finite-size recursive techniques, which allow to perform renormalization 
calculations which span a wide range of energies. Wilson suggested to 
introduce a renormalization group transformation so that one can cover large 
scale differences in a recursive manner~\cite{Wilson:1979wp}, and these ideas 
were then developed in~\cite{Luscher:1983},~\cite{Luscher:1992an} 
and~\cite{Jansen:1995ck,Luscher:1996sc}. Nice summaries of these techniques 
are also given in~\cite{Luscher:1997sv}, \cite{Sommer:1997xw} 
and~\cite{Luscher:2002pz}
The evolution of renormalized coupling and masses can then be studied from 
rather low to rather high energies. This allows nonperturbative studies 
to be carried out with a good control over the systematic errors.

We would like to sketch how this nonperturbative renormalization over 
many scales is accomplished. One takes a sequence of pairs of lattices,
where in each pair a lattice has size $L$ and the other one has size $L'=2L$, 
and computes the renormalized parameters at the new scale $\mu'=\mu/2$ 
knowing them at the scale $\mu=1/L$, maintaining in the process the bare 
parameters fixed.~\footnote{The renormalized strong coupling is defined 
as the response of the functional to the variation of the gauge fields 
at the boundaries.} This defines step scaling functions $\sigma$, which for 
the renormalized coupling and masses are given by~\footnote{In practice 
the PCAC relation
\be
\partial_\mu A_\mu (x) = 2m \cdot P(x) 
\ee
is used to define the masses, so that the renormalization of the mass is 
proportional to $Z_A/Z_P$, and since $Z_A$ does not evolve with the scale the 
scale evolution of the renormalized mass is proportional to the inverse of the
renormalization of the pseudoscalar density:
\be
m (\mu) = \frac{Z_A}{Z_P (L)} \, m_0, \qquad \mu = \frac{1}{L} .
\ee
We remind that the local currents are not conserved on the lattice, and
they are renormalized by the strong forces, so that their $Z$'s are
different from one.}
\bea
g^2 (2L) &=& \sigma ( g^2(L) )   \\
Z_P (2L) &=& \sigma_P ( Z_P(L) ) \cdot Z_P (L) .
\eea
These are a kind of integrated form of the $\beta$ and $\tau$ functions 
(see Eqs.~(\ref{eq:betaf}) and~(\ref{eq:tauf})). The errors on these step 
scaling functions can be rendered rather small when for each $L$ one repeats 
the lattice computations for a few different values of $a$ and then 
extrapolates them to the continuum limit in the $O(a)$ improved theory. 
One does not need large lattices at all, and in fact lattices as small as 
$L/a=5$ have been used. For the coupling and mass calculations it has never
been necessary to consider lattices larger than $L'/a=32$.

The step scaling functions $\sigma$ and $\sigma_P$ are used recursively 
to go from very high to very low scales. Once the renormalized quantities 
are computed at the new scale $L'=2L$, this is taken as the starting scale 
for the computation of the renormalized quantities at $L'' = 2L' = 4L$, 
and this process is repeated until after $n$ steps one reaches the scale 
$L^{(n)} = 2^n L$. One has then completed the evolution of the renormalized 
parameters from the energy $\mu$ to the energy $\mu/2^n$. The low-energy end 
of this evolution should correspond to a scale at which one can safely match 
the renormalized quantities to some low-energy hadronic quantities. In this 
way one fixes the relations between the bare coupling and masses and the 
renormalized coupling and masses at that scale.
This matching can again be done without the use of large lattices.
At the high-energy end one uses perturbation theory, which is completely
safe there (see Figs.~\ref{fig:alpha} and~\ref{fig:mass}), to compute 
the $\Lambda$ parameter and the renormalization group invariant masses 
(Eqs.~(\ref{eq:lambdapar}) and~(\ref{eq:rgimass}) respectively). 
Once these numbers are known, one can easily carry out the matching of the 
$\Lambda$ parameter to the $\ms$ scheme or to other continuum schemes, 
via continuum calculations only. We remind that the renormalization group 
invariant masses do not depend on the scheme, and thus is only 
the $\Lambda$ parameter which needs an additional conversion to $\ms$.

We have thus been able to connect the nonperturbative infrared sector 
of the theory with the high-energy perturbative regime. This is the only 
method which allows such a remarkable thing. With the Schr\"odinger functional
coupled with the recursive finite-size scaling techniques one can cover large 
scale differences, which can be of more than two orders of magnitude. 
One can arrive at energies of more than 100 GeV (as in Figs.~\ref{fig:alpha}
and~\ref{fig:mass}). In order to achieve this with conventional 
methods it would be necessary to contain all relevant scales in a single 
lattice, which is impossible. The Schr\"odinger functional acts in this process
only as an intermediate renormalization scheme. Everything can be computed 
nonperturbatively and when improvement is also implemented the systematic
errors are then completely controlled.

The methods that we have just described have allowed to establish the best 
lattice result that we currently have for the $\Lambda$ parameter of QCD, 
\be
\Lambda_{\ms} = 238 \pm 19~{\mathrm MeV} ,
\ee
in the theory with zero flavors~\cite{Capitani:1998mq}. In this work also the 
nonperturbative relation between the bare masses and the renormalization group
invariant masses has been determined with a rather small error. A spinoff of 
the latter calculation has been the computation of the nonperturbative 
renormalization of the scalar quark condensate using overlap 
fermions~\cite{Hernandez:2001yn,{Hernandez:2001hq}}. This is possible for one 
thing because in a regularization which respects chiral symmetry one has
\be
Z_S = Z_P = \frac{1}{Z_M} ,
\label{eq:ovcond}
\ee
and therefore the knowledge of the renormalization of the mass translates 
immediately in the determination of the renormalization of the scalar current,
and furthermore because the renormalization group invariant masses are 
independent of the scheme, and thus they are the same whichever lattice
action is used. This means that one can then compute the renormalization 
of the mass in the overlap formulation by comparing 
\bea
M_{RGI} &=& Z_M^{\mathrm ov} (g_0) \, m_{\mathrm ov} (g_0) \\
M_{RGI} &=& Z_M^{\mathrm W} (g_0') \, m_{\mathrm W} (g_0') ,
\eea
where $Z_{\mathrm M}^{\mathrm W}$, the relation between the bare mass in the 
Wilson formulation and the renormalization group invariant mass, has been 
nonperturbatively determined using the finite-size scaling techniques and 
the Schr\"odinger functional as an intermediate scheme~\cite{Capitani:1998mq}. 
Thus, $Z_{\mathrm M}^{\mathrm ov}$ can be computed by evaluating the ratio 
of the bare masses in the two schemes, which can be done imposing that some 
renormalized quantity at a certain scale assumes the same value in both 
schemes. Finally, the knowledge of $Z_{\mathrm M}^{\mathrm ov}$, because of
Eq.~(\ref{eq:ovcond}), allows the determination of the renormalization factor 
connecting the bare condensate to the renormalization group invariant 
condensate. 

This kind of reasoning can be applied to other physical situations and other 
lattice actions. It shows how important the calculations made using the 
Schr\"odinger functional can be in practice, and how the results obtained 
using the finite-size scaling techniques can have implications beyond the
Schr\"odinger functional and the Wilson action.

Let us now turn our attention to perturbation theory, which in the 
Schr\"odinger functional has its own peculiarities, and is more complicated 
than average. For details on the perturbative setting the reader can consult 
the article of~\cite{Luscher:1996vw}. There is an asymmetry between spatial 
indices and the temporal index, and the perturbative calculations are not 
visually 4-dimensional covariant. This is due to the fact that the 
Schr\"odinger functional has no periodic boundary conditions in the time 
direction, and so translational invariance is lost and Fourier transforms can 
be done only in the spatial directions. Thus one works in the time-momentum 
representation, and uses quantities like
\be
q(x_0, \vec{p}) ,
\ee
the three-dimensional Fourier transform of a function $q(x)$ of coordinate 
space. The presence of the boundaries also renders the form of the propagators
quite complicated. Moreover, usually the improved theory is considered and 
this causes the calculations to be even more involved. In fact, to improve 
the Schr\"odinger functional to order $a$, in addition to the usual 
Sheikholeslami-Wohlert term one also needs some $O(a)$ boundary counterterms, 
both for the gluon part and for the quark part of the action.

Let us then consider the $O(a)$ improved theory. The quark propagator obeys the
equation
\be 
\Big( D +\delta D_v +\delta D_b +m_0 \Big) S(x,y) = a^{-4} \delta_{xy}, 
\qquad 0 < x_0 < T ,
\ee
with the boundary conditions
\be
P_+ S(x,y) \Big|_{x_0=0} = P_- S(x,y) \Big|_{x_0=T} = 0, 
\ee
where the improvement counterterm to the Dirac operator in the interior is 
the same as for the Wilson action, namely the Sheikholeslami-Wohlert term
\be
\delta D_v \psi (x) = c_{sw} \frac{\ii}{4} a \sigma_{\mu\nu}
F_{\mu \nu}^{\mathrm clover}(x) \psi (x) ,
\ee
but there are also additional counterterms at the boundaries,
\be
\delta D_b \psi (x) = (\widetilde{c}_t -1) \frac{1}{a} \Bigg[ 
\delta_{x_0,a} \Big[ \psi (x) - U_0^\dagger (x-a\hat{0}) P_+ \psi (x-a\hat{0})
\Big] +\delta_{x_0,T-a} \Big[ \psi (x) - U_0 (x) P_- \psi (x+a\hat{0}) \Big] 
\Bigg] ,
\ee
where $\widetilde{c}_t$ is an improvement coefficient corresponding 
to these fermion boundary counterterms. In perturbation theory it is 
convenient to make the decomposition
\be
\psi (x) = \psi_{cl} (x) + \chi (x), \qquad
\overline{\psi} (x) = \overline{\psi}_{cl} (x) + \overline{\chi} (x) ,
\ee
where the classical field satisfies the Dirac equation
\be
\Big( D +\delta D_v +\delta D_b +m_0 \Big) \psi_{cl} (x) = 0 ,
\qquad 0 < x_0 < T ,
\ee
and has boundary values 
\be
P_+ \psi_{cl} (x) |_{x=0} = \rho (\vec{x}) , \qquad
P_- \psi_{cl} (x) |_{x=T} = \rho' (\vec{x}) 
\ee
(see Eq.~(\ref{eq:sfbound})). A similar decomposition holds for 
$\overline{\psi}_{cl}$. 
In terms of the boundaries the classical field has the expression
\be
\psi_{cl} (x) = a^3 \sum_{\vec{y}} \widetilde{c}_t \Bigg[ 
S(x,y) U_0^\dagger (y-a\hat{0}) \rho (\vec{y}) \Big|_{y_0=a} 
+S(x,y) U_0 (y) \rho' (\vec{y}) \Big|_{y_0=T-a} \Bigg] .
\ee
A useful property of this decomposition is that the quantum components 
$\chi (x)$ have vanishing boundary values. The fermionic action splits as
\be
S_f^{imp} [U,\overline{\psi},\psi] =
S_f^{imp} [U,\overline{\psi}_{cl},\psi_{cl}] 
+ S_f^{imp} [U,\overline{\chi},\chi] , 
\ee
and the quantum and classical components are then completely independent.
The generating functional with fermionic sources $\eta, \overline{\eta}$
is given by
\bea
\log Z_f &=& \log Z_f \Big|_{\overline{\rho}' = \ldots = \eta =0} \nonumber \\
&& - a^3 \sum_{\vec{x}} \Bigg[ \frac{1}{2} a \widetilde{c}_s \Big[ 
\overline{\rho} (\vec{x})\gamma_k (\nabla_k^\star + \nabla_k) \rho (\vec{x}) + 
\overline{\rho}' (\vec{x})\gamma_k (\nabla_k^\star + \nabla_k) \rho' (\vec{x}) 
\Big] \nonumber \\ 
&& -\widetilde{c}_t 
\Big[ \overline{\rho} (\vec{x}) U_0 (x-a\hat{0}) \psi_{cl} (x) \Big|_{x_0=a} 
+ \overline{\rho}' (\vec{x}) U_0^\dagger (x) \psi_{cl} (x) \Big|_{x_0=T-a} 
\Big] \Bigg] \nonumber \\
&& +a^8 \sum_{x,y} \overline{\eta} (x) S(x,y) \eta (y) 
+a^4 \sum_x \Big[ \overline{\eta} (x) \psi_{cl} (x) 
+ \overline{\psi}_{cl} (x) \eta (x) \Big] ,
\eea
where $\widetilde{c}_s$ is an improvement coefficient corresponding 
to this other kind of fermion counterterms at the boundaries
(entirely contained in the timeslices at $x_0=0$ and $x_0=T$). 
This now allows upon differentiation to derive the basic contractions,
among which we find~\footnote{The square brackets denote fermion correlations 
in a given gauge field.}
\bea
\Big[ \psi (x) \overline{\psi} (y) \Big] &=& S(x,y) , \\ 
\Big[ \psi (x) \overline{\zeta} (\vec{y}) \Big] &=& 
\frac{\delta \psi_{cl} (x)}{\delta \rho (\vec{y})} =
\widetilde{c}_t S(x,y) U_0^\dagger (y-a\hat{0}) P_+ \Big|_{y_0=a} , \\
\Big[ \zeta (\vec{z}) \overline{\psi} (x) \Big] &=& 
\frac{\delta \overline{\psi}_{cl} (x)}{\delta \overline{\rho} (\vec{z})} =
\widetilde{c}_t P_- U_0 (z-a\hat{0}) S(z,x) \Big|_{z_0=a} .
\eea
Using these contractions one can construct for example the correlation 
function of the axial current, inserted at the time $x_0$, which is
(see Fig.~\ref{fig:sf})
\be
f_A (x_0) = a^6 \sum_{\vec{y},\vec{z}} \frac{1}{2} \, \Bigg\langle \Tr
\Big\{ \, \Big[ \zeta (\vec{z}) \overline{\psi} (x) \Big] \, \gamma_0 \gamma_5
  \, \Big[ \psi (x) \overline{\zeta} (\vec{y}) \Big] \, \gamma_5 \, \Big\} 
  \Bigg\rangle,
\ee
where $\langle \cdots \rangle$ denotes gluon averages.
The complete list of contractions is given in~\cite{Luscher:1996vw}.

We give here the explicit expression of the quark propagator, which 
has a quite complicated form, even in the unimproved theory 
($\widetilde{c}_t=\widetilde{c}_s=1$). One has
\be
S(x,y) = \Big( D^\dagger + m_0 \Big) G(x,y), \qquad 0 < x_0, y_0  < T , 
\ee
where
\bea
&& G(x,y) = \frac{1}{L^3} \sum_{\vec{p}} \frac{\e^{\ii \vec{p} 
( \vec{x} - \vec{y} )}}{-2\ii a^{-1}\sin ap_0^+A(\vec{p}^+)R(p^+)} \\
&& \quad \times \Bigg\{ 
(M(p^+) -\ii a^{-1}\sin ap_0^+) \e^{-\omega(\vec{p}^+) |x_0-y_0|}
+(M(p^+) +\ii a^{-1}\sin ap_0^+) \e^{-\omega(\vec{p}^+) (2T-|x_0-y_0|)} 
\nonumber \\
&& \quad 
-(M(p^+) +\ii \gamma_0 a^{-1}\sin ap_0^+) \e^{-\omega(\vec{p}^+) (x_0+y_0)}
-(M(p^+) -\ii \gamma_0 a^{-1}\sin ap_0^+) \e^{-\omega(\vec{p}^+) (2T-x_0-y_0)} 
\Bigg\} \nonumber,
\eea
for $0 \le x_0 \le T$, with
\bea
p^+_\mu &=& p_\mu +\frac{\theta_\mu}{L}   \\
A(\vec{q}) &=& 1 + a \Bigg(
m_0 + \frac{2}{a} \sum_{k=1}^3 \sin^2 \frac{aq_k}{2} \Bigg) \\
R(q) &=& M(q) \Big[ 1-\e^{-2\omega(\vec{q})T} \Big] 
-\frac{\ii}{a} \sin aq_0  \Big[ 1+\e^{-2\omega(\vec{q})T} \Big] \\
M(q) &=& m_0 + \frac{2}{a} \sum_\mu \sin^2 \frac{aq_\mu}{2}
\eea
and
\be
p_0 = p_0^+ = \ii \omega(\vec{p}^+) \, \mathrm{mod} \frac{2\pi}{a} ,
\ee
where $\omega(\vec{q})$ is given by
\be
\sinh \Big[ \frac{a}{2} \omega(\vec{q}) \Big] = \frac{a}{2} 
\sqrt{\frac{ \frac{1}{a^2}\sum_{k=1}^3 \sin^2 aq_k 
    +\frac{1}{a^2}  (A(\vec{q}) -1)^2}{A(\vec{q})}} .
\ee
The angle $\theta$ in the formulae above is the finite-size momentum which 
comes from the boundary conditions in the spatial directions (or rather
from the modified covariant derivative, Eq.~(\ref{eq:sf_covder})). We can see 
that things easily become rather technical. We want here to report also the 
expression of the gluon propagator, which in the Feynman gauge is given by
\be
D_{\mu\nu}(x,y) = \frac{1}{L^3} \sum_{\vec{p}} 
\e^{\ii \vec{p} ( \vec{x} - \vec{y} )} d_{\mu\nu}(x_0,y_0;\vec{p}) ,
\ee
with
\bea
d_{00}(x_0,y_0;\vec{p}) &=& \frac{a}{\sinh(\varepsilon a) \sinh(\varepsilon T)}
\cosh\Bigg[\varepsilon \Big(T-x_0-\frac{1}{2}a \Big) \Bigg]
\cosh\Bigg[\varepsilon \Big(y_0+\frac{1}{2}a \Big) \Bigg]  \\
d_{kj} \, (x_0,y_0;\vec{p}) &=& \delta_{kj} \,
\frac{a}{\sinh(\varepsilon a) \sinh(\varepsilon T)}
\sinh\Big[\varepsilon (T-x_0) \Big]
\sinh\Big(\varepsilon y_0 \Big) 
\eea
for nonzero $\vec{p}$, and
\bea
d_{00}(x_0,y_0;\vec{0}) &=& y_0 + a  \\ 
d_{kj}(x_0,y_0;\vec{0}) &=& \delta_{kj} \, (T-x_0) \frac{y_0}{T} 
\eea 
for $\vec{p}=0$. The ``energy'' $\varepsilon$ is here given by
\be
\cosh (a\varepsilon) = 1 + 2 \sum_{k=1}^3 \sin^2 \frac{ap_k}{2} .
\ee
The mixed components $d_{0k}$ and $d_{k0}$ vanish. The above expression
is valid for $x_0 \ge y_0$, and for $y_0 > x_0$ one uses the symmetry
\be
d_{\mu\nu}(y_0,x_0;\vec{p}) = d_{\nu\mu}(x_0,y_0;\vec{p}) .
\ee
For more technical details, and for the remaining parts of the perturbative 
setting like the gauge fixing (which is quite complicated), 
see~\cite{Luscher:1992an,Luscher:1996vw} and also~\cite{Kurth:2002rz}, where
the renormalization of the quark mass has been computed at one loop. 
The Feynman rules for the gluon vertices are given in~\cite{Palombi:2002gw}.

The Schr\"odinger functional has been very useful also for the computation of 
many improvement coefficients. In particular the $O(a)$ corrections to the 
PCAC relation were used to fix $c_{sw}$, and because a mass term is present in
this relation the Schr\"odinger functional is much more appropriate for this 
determination than, say, usual Wilson fermions. Also calculations of moments 
of structure functions have been done using the Schr\"odinger functional, 
coupled to the recursive finite-size scaling technique with which to define 
appropriate step scaling functions for the relevant operators. 
These calculations are reported in~\cite{Bucarelli:1998mu}, 
in~(Guagnelli, Jansen and Petronzio, 1999a; 1999b; 1999c; 2000) and 
in~\cite{Jansen:2000xm,Palombi:2002gw}. Perturbation theory is in this case 
quite cumbersome. The covariant derivatives make everything more complicated, 
because of the phase factors and the boundary fields.

We conclude mentioning that also a few two-loop calculations have been 
completed using the Schr\"odinger functional~\cite{Wolff:1994ab,Narayanan:1995ex,Bode:1998hd,Bode:1999dn,Bode:1999sm}.

\section{The hypercubic group}
\label{sec:hypercubic}

With this Section we begin to explain in a more detailed way how lattice 
perturbative calculations are actually done. As a first thing, it is useful 
to discuss the symmetry group of the lattice and see what are the consequences
of the breaking of Lorentz invariance.

On the lattice one inevitably ends up with a discrete group. The symmetry 
group of the discrete rotations of a four-dimensional hypercubic lattice onto 
itself is a crystallographic group, denoted by $W_4$ and called the hypercubic
group. This is the group of $\pi/2$ rotations on the six lattice planes with 
the addition of the reflections (so that parity transformations are also 
included). It consists of 384 elements and 20 irreducible 
representations~\cite{Baake:1981qe,Baake:1982ah,Mandula:ut}.

The hypercubic group is a subgroup of the orthogonal group $O(4)$, which is 
the Lorentz group analytically continued to euclidean space. One of the main 
difficulties in doing perturbative calculations on the lattice arises from the
fact that the (euclidean) Lorentz symmetry breaks down to the hypercubic $W_4$
symmetry. The lattice has then a reduced symmetry with respect to the 
continuum, and many more mixings arise, as we will see in the next Section.

Let us first consider the special hypercubic group, $SW_4$, consisting of
proper rotations without reflections, which has 192 elements and 13 
irreducible representations. Five of these representations (of dimensions 
1, 1, 2, 3 and 3) are connected to the 4-dimensional group of permutations, 
$S_4$. There are then four representations of $O(4)$ which remain irreducible 
under the special hypercubic group:
\be
{\bf (1,0)} , {\bf (0,1)}, {\bf (\frac{1}{2},\frac{1}{2})},
{\bf (\frac{3}{2},\frac{1}{2})} . 
\ee
The product of the first three of these representations with the completely 
antisymmetric representation of the permutation group $S_4$ gives three other 
representations of $SW_4$ (which have the same dimensionality), while 
${\bf (\frac{3}{2},\frac{1}{2})}$ is invariant under this operation and
the hypercubic representations ${\bf (\frac{1}{2},\frac{3}{2})}$ and
${\bf (\frac{3}{2},\frac{1}{2})}$ turn out to be the same. 

So far we have then been able to identify 12 representations. These is still
another representation, which has dimension 6. The complete list of the 
representations of the special hypercubic group $SW_4$ is thus given by:
\be
{\bf 1_1}, {\bf 1_2}, {\bf 2}, {\bf 3_1}, {\bf 3_2}, {\bf 3_3}, {\bf 3_4}, 
{\bf 3_5}, {\bf 3_6}, {\bf 4_1}, {\bf 4_2}, {\bf 6}, {\bf 8}, 
\ee
where subscripts label different representations with the same dimensionality.
This group is a subgroup of $SO(4)$, the special orthogonal group.

We now discuss the representations of the group $W_4$.
Including the reflections doubles the number of group elements, but not
the number of the representations. This comes out from the fact that, 
contrary to what happens in three dimensions for the case of the cubic group, 
the hypercubic group is not the direct product of the rotation group and the 
reflection group. The reason is that the reflection of all four axes is still 
a rotation, which is not true for the reflection of three axes in three 
dimensions. Therefore, the representations do not double, and from 13 
they only become 20. What happens is that 9 of these 13 just 
double (generating the representations with opposite parity), while the 
remaining 4, all of dimension 3, merge into two 6-dimensional 
representations which are reflection invariant. In particular, 
the ${\bf 3_3}$ and ${\bf 3_4}$ of $SW_4$ merge
into the ${\bf 6_1}$ of $W_4$, and the ${\bf 3_5}$ and ${\bf 3_6}$ of $SW_4$ 
merge into the ${\bf 6_2}$ of $W_4$. The other representations of $SW_4$ 
just double. We can then give the complete list of the representations 
of $W_4$:~\footnote{It could be useful to note that our notation ${\bf N_m}$ 
corresponds to the representation $\tau^{(N)}_m$ (or $^s\tau^{(N)}_m$ for 
$SW_4$) in~\cite{Baake:1981qe}.}
\be
{\bf 1_1}, {\bf 1_2}, {\bf 1_3}, {\bf 1_4}, {\bf 2_1}, {\bf 2_2}, 
{\bf 3_1}, {\bf 3_2}, {\bf 3_3}, {\bf 3_4}, 
{\bf 4_1}, {\bf 4_2}, {\bf 4_3}, {\bf 4_4}, 
{\bf 6_1}, {\bf 6_2}, {\bf 6_3}, {\bf 6_4}, {\bf 8_1}, {\bf 8_2} .
\ee

The representation ${\bf 4_1}$ is the canonical one, which corresponds 
to an object with a Lorentz index, like ${\bf (\frac{1}{2},\frac{1}{2})}$
is in the continuum. When we are interested in the behavior of lattice 
operators which have more than one Lorentz index, we have to look 
at which representations of the hypercubic group are contained in the 
tensor products of the ${\bf 4_1}$ with itself, and compare the result
with what happens in the continuum, where one has to consider
the tensor products of the ${\bf (\frac{1}{2},\frac{1}{2})}$ with itself.
The relation between these two expansions determines what kind of mixings 
arise when one computes radiative corrections of lattice matrix elements
(apart from additional mixings due to the breaking of chiral symmetry 
or of other symmetries).

\section{Operator mixing on the lattice}
\label{sec:operatormixing}

Since $W_4$ is a subgroup of $O(4)$, a continuum operator which belongs 
to a given irreducible representation the (euclidean) Lorentz group 
becomes in general a sum of irreducible representations of the part of 
that group that still remains a symmetry on the lattice, the hypercubic group.
The continuum operator can then belong to various distinct lattice 
representations, according to the way in which its indices are chosen.
This implies than on the lattice the possibilities for mixings under 
renormalization are larger than in the continuum, and mixings can arise 
which are pure lattice artifacts and which have to be carefully treated. 
The number of independent renormalization factors in a lattice calculation is 
then in general larger than in the continuum. In particular, operators which 
are multiplicatively renormalizable in the continuum may lose this property
on the lattice.~\footnote{The fact that additional operators appear in the 
mixings is a feature that occurs not only using the lattice regularization. 
For example, in continuum calculations using dimensional regularization 
in the version known as DRED ``evanescent'' operators, coming from the
additional $-2\epsilon$ dimensions, are generated in the intermediate stages 
of the calculations.}

Thus, the mixing of lattice operators under renormalization presents features 
which are often significantly different from the corresponding continuum 
theory. In general the mixing patterns are more complicated. Moreover,
for Wilson fermions additional mixings (beside those due to the breaking of
Lorentz invariance) can arise because of the breaking of chiral symmetry. 
For staggered fermions, the loss of flavor invariance also opens the door 
for more mixings, although of a different kind. All these additional mixings 
are not physical, are just lattice artifacts which have to be subtracted 
in order to get physical results from the lattice. In practical terms 
the worst situation occurs in the case of mixings with operators of lower 
dimensions. This implies that the corresponding lattice renormalization 
factors contain a power divergent coefficient, proportional to $1/a^n$. 
These are a kind of lattice artifact which is not possible to subtract 
in perturbation theory.

In short, the Lorentz breaking, as well as the breaking of chiral, flavor 
or other symmetries that occur in specific lattice actions, means in general
that the multiplicative renormalizability of continuum operators becomes 
a mixing when the theory is put on a lattice, in some cases even with power 
divergences. The necessary condition for not having any mixing at all is that 
the operator belongs to an irreducible representation of $W_4$, but this is 
sometimes not sufficient, as we will see shortly.

\subsection{Unpolarized structure functions}

Let us discuss some examples involving operators which measure moments of
unpolarized structure functions.~\footnote{It is not possible to compute 
a complete structure function directly on the lattice.
The reason is that the structure functions describe the physics 
close to the light cone, and this region of Minkowski space shrinks to
a point when one goes to Euclidean space, where Monte Carlo simulations
are performed. However, on a Euclidean lattice it is possible to compute
the moments of the structure functions, using an operator product expansion:
\begin{displaymath}
\int_0^1 x^n {\cal F}^{(i)} (x,Q^2) \sim C^{(n,i)} (\frac{Q^2}{\mu^2})
\cdot \langle h | O^{(n,i)} (\mu)  | h \rangle .
\end{displaymath}
The Wilson coefficients contain the short-distance physics, and can be
perturbatively computed in the continuum. The matrix elements contain the
long-distance physics, and can computed using numerical simulations, 
supplemented by a lattice renormalization of the relevant operators.}
These operators appear in the operator product expansion of two 
electromagnetic or weak hadronic currents, and have the form 
\be
O_{\{\mu \mu_1 \cdots \mu_n\}} (x) = 
\overline{\psi} (x) \,\gamma_{\{\mu} \, D_{\mu_1} \cdots D_{\mu_n\}} \,
\psi (x) .
\ee
They are symmetric in all their indices and traceless. The operator
$O_{\{\mu \mu_1 \cdots \mu_n\}}$ measures the $n$-th moment, 
$\langle x^n \rangle$, of the unpolarized structure functions, that is of the 
distribution of the momentum of the quarks inside the hadrons. The 
renormalization of these operators, which presents particular computational 
difficulties due to the presence of the covariant derivatives, has been first 
studied with Wilson fermions in~\cite{Kronfeld:1984zv,Corbo:1989ps,Corbo:1989yj,Caracciolo:pt,Caracciolo:bu} and then 
in~\cite{Capitani:qn,Beccarini:1995iv,Gockeler:1996hg,Brower:1996pe}
and~\cite{Capitani:2000wi,Capitani:2000aq}.
Recent numerical works in full QCD which have made use of these perturbative 
renormalization factors are~\cite{Dolgov:2000ca,Dolgov:2002zm,Dreher:2002qf} 
and~\cite{Gockeler:2002ek}, and recent short reviews of perturbative and 
nonperturbative methods and results can be found 
in~\cite{Capitani:2002wb,Capitani:2002jj,Negele:2002vs}.

Perturbative renormalization factors for all these operators have also been 
computed using overlap fermions, and are reported 
in~\cite{Capitani:2000wi,Capitani:2000aq,Capitani:2001yq}.

We point out that all mixings discussed below, which are artifacts of the 
lattice, are only due to the breaking of Lorentz invariance. They have nothing
to do with the breaking of chiral symmetry for Wilson fermions, and therefore 
they are still present, in exactly the same form, even when one uses 
Ginsparg-Wilson fermions.

In the continuum each of these structure function operators belongs to an 
irreducible representation of the Lorentz group. On the lattice they are 
instead in general reducible, they become linear combinations of irreducible 
representations of the hypercubic group, and this is the reason of the mixings
which appear when radiative corrections are computed. More detailed analyses 
of these mixings can be found in~\cite{Beccarini:1995iv} 
and~\cite{Gockeler:1996mu}.

\subsubsection{First moment} 

The operator is $O_{\mu \nu }  = 
\overline{\psi} \, \gamma_\mu \, D_{\nu} \, \psi $, symmetric and traceless.

An object with a single Lorentz index belongs in the continuum to the  
${\bf (\frac{1}{2},\frac{1}{2})}$ representation of the euclidean Lorentz 
group $O(4)$, while on the lattice it belongs to the ${\bf 4_1}$ 
representation of the hypercubic group $W_4$.

The general decomposition of the 16 (nonsymmetrized) tensor components is
in the continuum:
\be {\bf (\frac{1}{2},\frac{1}{2})} 
                    \otimes {\bf (\frac{1}{2},\frac{1}{2})}
              = {\bf (0,0)} \oplus {\bf (1,0)} 
               \oplus {\bf (0,1)}  \oplus {\bf (1,1)}, 
\ee
while on the lattice is: 
\be {\bf 4_1} \otimes  {\bf 4_1} = {\bf 1_1}  \oplus {\bf 3_1} 
             \oplus {\bf 6_1}  \oplus {\bf 6_3}.
\ee

We have essentially two choices here for the symmetrized operators, that
is the two indices can be different or can be equal. In the latter case,
one has also to subtract the trace component.

The first case can be exemplified by considering the operator $O_{\{01\}}$,
which belongs to the ${\bf 6_1}$ and is multiplicatively renormalizable. 
We will compute its renormalization constant in detail in 
Section~\ref{sec:examplewilson}. A representative of the second case is 
$O_{\{00\}} -\frac{1}{3} ( O_{\{11\}} + O_{\{22\}} + O_{\{33\}} )$, which
belongs to the ${\bf 3_1}$ and is also multiplicatively renormalizable.
The subtracted trace part belongs to the ${\bf 1_1}$. Finally, the 
antisymmetric components (which do not enter in the operator product expansion
for the moments), for example the operator $O_{[01]}$, belong to the 
remaining representation in the expansion, the ${\bf 6_3}$.

Since they belong to different representations of $W_4$, the lattice 
renormalization factors of the operators $O_{\{01\}}$ and 
$O_{\{00\}} -\frac{1}{3} ( O_{\{11\}} + O_{\{22\}} + O_{\{33\}} )$ are 
different, as has been verified by explicit calculations; in the continuum 
however they are the same, as both operators belong to the ${\bf (1,1)}$.

We mention here that from the point of view of Monte Carlo simulations
the choice of two different indices is worse, because in this case one 
has to choose one component of the hadron momentum to be different from zero, 
and this leads to larger systematic effects due to the granularity of the 
lattice.

\subsubsection{Second moment} 

The operator is $O_{\mu \nu \sigma }  = 
\overline{\psi} \, \gamma_\mu \, D_{\nu} D_{\sigma} \, \psi $, symmetric
and traceless.

The general decomposition of the 64 (nonsymmetrized) tensor components 
of this rank-three operator is in the continuum: 
\be 
{\bf (\frac{1}{2},\frac{1}{2})} 
                    \otimes {\bf (\frac{1}{2},\frac{1}{2})}
                    \otimes {\bf (\frac{1}{2},\frac{1}{2})}
             = 4 \cdot {\bf (\frac{1}{2},\frac{1}{2})}
               \oplus 2 \cdot {\bf (\frac{3}{2},\frac{1}{2})}
               \oplus 2 \cdot {\bf (\frac{1}{2},\frac{3}{2})}
                       \oplus {\bf (\frac{3}{2},\frac{3}{2})},
\ee 
while on the lattice is: 
\be 
{\bf 4_1} \otimes {\bf 4_1} \otimes {\bf 4_1}
             = 4 \cdot {\bf 4_1} 
                      \oplus {\bf 4_2} 
                      \oplus {\bf 4_4} 
              \oplus 3 \cdot {\bf 8_1} 
              \oplus 2 \cdot {\bf 8_2} .
\ee

We have essentially three choices here for the symmetrized components.
One is represented by the operator $O_{\{123\}}$, which belongs to the 
${\bf 4_2}$ and is multiplicatively renormalizable. This choice however
is quite unsatisfactory from the point of view of simulations, because 
two components of the hadron momentum have to be different from zero, 
leading to rather large systematic errors. One should minimize these 
systematic errors by including as few nonzero components of the hadron 
momentum as possible. From this point of view, the optimal choice is the 
operator $O_{\{111\}}$, which belongs to the ${\bf 4_1}$. Unfortunately 
this operator mixes with $\overline{\psi} \, \gamma_1 \, \psi $, which is 
a ${\bf 4_1}$ as well. Moreover, the coefficient of this mixing can be seen 
from dimensional arguments to be power divergent, $1/a^2$, and thus this 
mixing cannot be studied in perturbation theory.

There is an intermediate choice between having the indices all different 
or all equal, and is given by the operator 
$O_{\{011\}} -\frac{1}{2} ( O_{\{022\}} + O_{\{033\}} )$, which has no
power divergences due to the particular combination chosen. This operator 
belongs to an irreducible representation of $W_4$, but nonetheless is not 
multiplicatively renormalizable, and mixes with other operators. The way 
in which this happens is not trivial, and was first understood 
in~\cite{Beccarini:1995iv}. The point is that this operator belongs to the 
${\bf 8_1}$, but there are three ``copies'' of the ${\bf 8_1}$ in the 
decomposition of $O_{\mu \nu \sigma }$. It turns out that two of these copies 
mix in our case, at least at the 1-loop level. What happens is that 
$O_A = O_{011} -\frac{1}{2} ( O_{022} + O_{033} )$ and 
$O_B = O_{101} + O_{110} -\frac{1}{2} ( O_{202} + O_{220} + O_{303} 
+ O_{330} )$, which have different tree levels 
($\gamma_0 p_1^2 -\frac{1}{2} (\gamma_0 p_2^2 + \gamma_0 p_3^2)$ and 
$2 \gamma_1 p_0 p_1 -(\gamma_2 p_0 p_2 + \gamma_3 p_0 p_3)$ respectively), 
have different 1-loop corrections and renormalize with different numerical 
factors. In other words, 
$O_{\{011\}} -\frac{1}{2} ( O_{\{022\}} + O_{\{033\}} )=\frac{1}{3} (O_A+O_B)$ 
mixes with an operator of mixed symmetry (nonsymmetrized).

We have thus seen that the choice of indices for this rank-three operator 
is very important, and has practical consequences for the Monte Carlo 
simulations as well as for the calculation of renormalization factors.

In the continuum, all the cases discussed above for $O_{\{\mu \nu \sigma \}}$,
including $O_{\{111\}}$, belong to the ${\bf (\frac{3}{2},\frac{3}{2})}$, 
and therefore they have the same renormalization constant, and of course 
no mixing problem.

\subsubsection{Third moment} 

The operator is $O_{\mu \nu \sigma \rho}  = 
\overline{\psi} \, \gamma_\mu \, D_{\nu} D_{\sigma} D_{\rho} \, \psi $, 
symmetric and traceless.

The general decomposition of the 256 (nonsymmetrized) tensor components is
in the continuum: 
\bea 
{\bf (\frac{1}{2},\frac{1}{2})} 
                    \otimes {\bf (\frac{1}{2},\frac{1}{2})}
                    \otimes {\bf (\frac{1}{2},\frac{1}{2})}
                    \otimes {\bf (\frac{1}{2},\frac{1}{2})} 
&=& 4 \cdot {\bf (0,0)}
               \oplus 6 \cdot {\bf (1,0)}
               \oplus 6 \cdot {\bf (0,1)}
               \oplus 2 \cdot {\bf (2,0)}
               \oplus 2 \cdot {\bf (0,2)}  \nonumber \\
&& \oplus 9 \cdot {\bf (1,1)}
               \oplus 3 \cdot {\bf (2,1)}
               \oplus 3 \cdot {\bf (1,2)}
               \oplus 2 \cdot {\bf (2,2)}, 
\eea 
while on the lattice is: 
\bea
{\bf 4_1} \otimes {\bf 4_1} \otimes {\bf 4_1} \otimes {\bf 4_1} 
&=& 4 \cdot {\bf 1_1} 
                      \oplus {\bf 1_2} 
                      \oplus {\bf 1_4} 
              \oplus 3 \cdot {\bf 2_1} 
              \oplus 2 \cdot {\bf 2_2} 
              \oplus 7 \cdot {\bf 3_1} 
              \oplus 3 \cdot {\bf 3_2} 
              \oplus 3 \cdot {\bf 3_3} 
              \oplus 3 \cdot {\bf 3_4}  \nonumber \\
&&            \oplus 10 \cdot {\bf 6_1} 
              \oplus 6 \cdot {\bf 6_2} 
              \oplus 10 \cdot {\bf 6_3} 
              \oplus 6 \cdot {\bf 6_4}. 
\eea

Without entering a detailed discussion, here we only say that it turns out 
that among the symmetrized operators only $O_{\{0123\}}$, which belongs to the 
${\bf 1_2}$ representation, is multiplicatively renormalizable. 
Any other choice leads to some mixings, in some cases with power-divergent 
coefficients.

A special case is given by the operator 
$O_{\{0011\}} + O_{\{3322\}} - O_{\{0022\}} - O_{\{3311\}}$, which belongs 
to the ${\bf 2_1}$ and in principle mixes with two other operators 
of mixed symmetry which have very complicated expressions, as shown 
in~\cite{Gockeler:1996mu}. However, at one loop this mixing is not seen and we
can consider this operator to be multiplicatively renormalizable. The tadpole 
coming from this operator (as well as the one corresponding to $O_{\{0123\}}$)
will be computed in detail in Section~\ref{sec:operatortadpoles}.

\subsubsection{Higher moments} 

The operator for the fourth moment is $O_{\mu \nu \sigma \rho \tau}  = 
\overline{\psi} \,\gamma_\mu \, D_{\nu} D_{\sigma} D_{\rho} D_{\tau} \,\psi $,
symmetric and traceless.

We have seen that going from the first to the second and then to the third 
moment the mixing structure becomes more and more complicated. For the fourth 
moment and higher, that is for operators at least of rank five, a new feature 
occur: mixings which imply power-divergent coefficients becomes unavoidable, 
because at least two of the indices are bound to be equal. One can always 
find lower-dimensional operators with the same transformation properties, 
and it is not possible to avoid a power-divergent renormalization factor.

Furthermore, even the finite mixings become much more complicated and quite 
entangled, since there happen to be a lot of ``copies'' of the same 
representations around. For example, the rank-five operator has
the decomposition
\be
{\bf 4_1} \otimes {\bf 4_1} \otimes {\bf 4_1} \otimes {\bf 4_1} 
   \otimes {\bf 4_1} = 
                     31 \cdot {\bf 4_1} 
              \oplus 20 \cdot {\bf 4_2} 
              \oplus 15 \cdot {\bf 4_3} 
              \oplus 20 \cdot {\bf 4_4} 
              \oplus 45 \cdot {\bf 8_1} 
              \oplus 40 \cdot {\bf 8_2} . 
\ee
For the higher moments it seems then rather unlikely to find an operator which 
does not mix and does not have power divergences.

\subsection{A mixing due to breaking of chiral symmetry: $\Delta I = 1/2$ 
operators}
\label{sec:amixing}

We now discuss a case in which some of the operator mixings that take place
are entirely due to the breaking of chiral symmetry by Wilson fermions. We 
show then that the calculations on the lattice become much simpler and more 
manageable when overlap fermions are instead used. The physics is the one of 
strangeness-changing weak decays, and the operators that we consider appear in 
the part of the $\Delta S=1$ effective weak nonleptonic Hamiltonian which 
is relevant for $\Delta I=1/2$ transitions~\cite{Gaillard:1974nj,Altarelli:1974ex,Altarelli:1980fi,Buras:1989xd,Buras:1991jm}. The $\Delta I=1/2$ amplitudes,
as is well known, are experimentally much greater than the $\Delta I=3/2$ 
amplitudes. This phenomenon is called ``octet enhancement'' or ``
$\Delta I=1/2$ rule'', and to this day has not been theoretically understood.

The $\Delta I=1/2$ bare operators that we consider are, for scales above the 
charm mass and below the bottom mass,
\be
O_\pm = (O_1-O_1^c) \pm (O_2-O_2^c) ,
\ee
with~\footnote{When color indices are not shown they are trivially
contracted, i.e., the operators are color singlets.} 
\bea
O_1 &=& ( \overline{s}^a \gamma_L^\mu u^b ) \,
        ( \overline{u}^b \gamma_L^\mu d^a ) \\
O_2 &=& ( \overline{s} \gamma_L^\mu u ) \, 
        ( \overline{u} \gamma_L^\mu d ) \\
O_1^c &=& ( \overline{s}^a \gamma_L^\mu c^b ) \,
        ( \overline{c}^b \gamma_L^\mu d^a ) \\
O_2^c &=& ( \overline{s} \gamma_L^\mu c ) \, 
        ( \overline{c} \gamma_L^\mu d ) .
\eea
Without entering into many details, we sketch the structure of their mixing
on the lattice.

When Wilson fermions are used, these operators need in order to be 
renormalized the subtraction of several other operators of the same and of 
lower dimensionality. The pattern of mixing is as follows:
\bea
\widehat{O}_\pm &=& Z_\pm^{\mathrm W} \Bigg[ O_\pm 
+ \sum_i C_\pm^{6,i} \cdot O_\pm^{6,i} \label{eq:zwfive} \\
&& + (m_c-m_u) \, C_\pm^5 \cdot \ii \overline{s} \sigma_{\mu\nu} F^{\mu\nu} d 
+ a (m_c-m_u) (m_d-m_s) \, \widetilde{C}_\pm^5 \cdot 
\ii \overline{s} \sigma_{\mu\nu} \widetilde{F}^{\mu\nu} d \nonumber \\
&& +\frac{1}{a^2} (m_c-m_u) \, C_\pm^3 \cdot \overline{s} d 
   +\frac{1}{a} (m_c-m_u) (m_d-m_s) \, \widetilde{C}_\pm^3 \cdot 
\overline{s} \gamma_5 d \Bigg] +O(a^2) . \nonumber 
\eea
All dimension-6 operators $O_\pm^{6,i}$ have opposite chirality with 
respect to $O_\pm$. The precise form of them, which is given 
in~\cite{Martinelli:1983ac}, will not interest us here.~\footnote{The study 
of weak operators on the lattice has a long history, and several perturbative 
calculations have been done using Wilson
fermions~\cite{Cabibbo:1983xa,Martinelli:1983ac,Maiani:1986db,Bernard:1987rw}.
Recent results for four-fermion operators can be found in~\cite{Gupta:1996yt}.
Four-fermion operators are also useful for other problems, like the 
renormalization of higher-twist operators in deep inelastic scattering, and 
in this case they have a different color, spin and flavor 
structure~\cite{Capitani:1998kj,Capitani:1999ai,Capitani:1999rv,Capitani:2000je}.}

A remarkable thing is that the coefficients of the mixings with the 
dimension-5 operators, which could in principle diverge like $1/a$, are 
instead, thanks to the Glashow-Iliopoulos-Maiani (GIM) mechanism, finite. 
This happens because the GIM mechanism says that this mixing should be zero
for $m_c=m_u$, and then a mass factor $m_c-m_u$ takes the place in which a 
factor $1/a$ would otherwise be.

The coefficients of the mixings with the dimension-3 operators could in 
principle diverge like $1/a^3$. The GIM mechanism and, for the 
parity-violating operator, also a factor $m_d-m_s$ due to the $CPS$ symmetry
(which combines $C$, $P$ and the exchange of the $s$ and $d$ 
quarks~\cite{Bernard:wf}), renders these power divergences less severe.
Still, the fact that these mixings remain power divergent makes impossible 
to calculate the full renormalization of the $\Delta I=1/2$ operators using 
perturbation theory. What can be computed in perturbation theory is only the 
overall renormalization $Z^W$, which is logarithmically divergent, the
coefficients $C_\pm^{6,i}$, and the coefficients $C_\pm^5$ and 
$\widetilde{C}_\pm^5$, which however require a two-loop calculation, which has
been done in~\cite{Curci:1987hu} for $C_\pm^5$.

Let us now see what happens when fermions which respect chiral symmetry are 
used. The renormalization is in this case given by:~\footnote{We should 
mention that to construct overlap operators which have the right chiral 
properties $\gamma_L^\mu$ has to be replaced by 
$\gamma_L^\mu (1 - \frac{a}{2\rho} D)$, and similarly for the other bilinears,
so that for example 
$\overline{s} d$ becomes $\overline{s} (1 - \frac{a}{2\rho} D) d$.}
\be
\widehat{O}_\pm = Z_\pm^{\mathrm ov} \Bigg[ O_\pm 
+ (m_c^2-m_u^2) \, C_\pm^m \cdot \Big( (m_d+m_s) \, \overline{s} d
+ (m_d-m_s) \, \overline{s} \gamma_5 d \Big) \Bigg] +O(a^2) .
\ee
We can see that chiral symmetry has brought a big change in the pattern
of subtractions. 

First of all, chiral symmetry forbids in a direct way any mixings with the 
other dimension-6 operators, $O_\pm^{6,i}$, which are of opposite chirality.
Furthermore, the GIM mechanism from linear becomes now, when combined with 
chiral symmetry, quadratic, and thus it gives coefficients proportional to 
$m_c^2-m_u^2$, like in the continuum. This mass factor counts for {\em two}
powers of $1/a$. Finally, the mixing coefficients with parity-conserving 
and parity-violating operators are now the same. The parity-conserving 
operators then acquire an additional factor $(m_d+m_s)$ which mirrors the 
factor $(m_d-m_s)$ coming from the $CPS$ symmetry for the parity-violating 
operators. This eats away another factor of $a$.

As a result, the mixings with the dimension-5 magnetic operators, which were
finite in Wilson, become now of order $a^2$, and hence one does not have to 
take into account these mixings at all, even in the improved theory. 
What is remarkable is that after all these new mass factors are inserted
the mixings with the dimension-3 operators $\overline{s} d$ and 
$\overline{s} \gamma_5 d$ (which were power divergent in Wilson) become 
finite. Thus, the renormalization of the $\Delta I=1/2$ 
matrix elements can now be carried out entirely with perturbative methods, 
because there are no power-divergent coefficients when one uses overlap 
fermions~\cite{Capitani:2000bm}.

\section{Analytic computations}
\label{sec:analyticcomputations}

Analytic computations of Feynman diagrams in lattice QCD present quite 
a few new and interesting features with respect to the continuum. Of course 
standard rules like a minus sign for each fermionic loop, which derive from 
the general properties of the path integral and the Wick theorem, continue
to be valid on the lattice. The combinatorial rules are also similar to the 
continuum. But there are a few technicalities, many of them connected to the 
breaking of Lorentz invariance, which the reader should be aware of.
We will discuss many of them in this Section. We will first introduce 
the power counting theorem on the lattice and see how divergent integrals
can be treated. We will then show in detail the calculation of a matrix 
element at one loop, using Wilson fermions. A few comments about calculations 
with overlap fermions and with fat links will be also made.

\subsection{The power counting theorem of Reisz}

On the lattice the functions to be integrated are periodic (with period
$2\pi/a$), and a power counting theorem which is appropriate for this kind
of integrals, and which accounts for their properties in the the continuum
limit, has been established by Reisz (1988a; 1988b; 1988c; 1988d). This power 
counting theorem, like the one in the continuum~\cite{HahnZimmermann}, is very
useful for the treatment of divergent integrals, and is fundamental for 
proving the renormalizability of lattice gauge theories.

We will follow the presentation of~\cite{Luscher:1988sd}, where somewhat 
milder conditions are required than in the original papers.
Let us then consider a generic lattice integral at $L$ loops, which will have
the general form
\be
I = \int^{\frac{\pi}{a}}_{-\frac{\pi}{a}} \frac{d^4k_1}{(2\pi)^4} 
    \cdots \int^{\frac{\pi}{a}}_{-\frac{\pi}{a}} \frac{d^4k_L}{(2\pi)^4} \, \, 
    \frac{V(k,q;m,a)}{C(k,q;m,a)} ,
\label{eq:iniint}
\ee
where $q_i~(i=1,\ldots,E)$ are the external momenta and $m$ stands for the 
masses of the theory. The numerator $V$ contains all vertices and the 
numerators of the various propagators, while the denominator $C$ is the 
product of the denominators of these propagators. This overall denominator 
is assumed to have the structure
\be
C(k,q;m,a) = \prod_{i=1}^I C_i(l_i;m,a) ,
\ee
where $I$ is the number of internal lines of the diagram, and the line momenta
$l_i(k,q)$ carried by them are linear combinations of the integration 
variables $k_j$ and the external momenta $q_j$. For the power counting theorem
to be valid, a few conditions have to be satisfied by the numerator $V$, the 
denominators $C_i$ and the line momenta $l_i$. These conditions can be 
stated as follows. 
\begin{description}
\item{(V1)} There exists an integer $\omega$ and a smooth function $F$ such 
that
\be
V(k,q;m,a) = a^{-\omega} F(ak,aq;am) ,
\ee
and F is periodic in $ak_i$ and a polynomial in $am$.
\item{(V2)} The continuum limit of the numerator, 
\be 
P(k,q;m) = \lim_{a \to 0} V(k,q;m,a),
\ee
exists.
\item{(C1)} There exist smooth functions $G_i$ satisfying
\be
C_i(l_i;m,a) = a^{-2} G_i(al_i;am) ,
\ee
and the $G_i$'s are periodic in $al_i$ and polynomials in $am$.
\item{(C2)} The continuum limit of all $C_i$'s exists, and is given by
\be 
\lim_{a \to 0} C_i(l_i;m,a) = l_i^2+m_i^2 ,
\ee
where the positive masses $m_i$ are combinations of the original masses $m$.
\item{(C3)} There exist positive constants $a_0$ and $A$ such that
\be
|C_i(l_i;m,a)| \ge A (\hat{l}_i^2+m_i^2) 
\ee
for all $a \le a_0$ and all $l_i$'s.
\item{(L1)} All line momenta satisfy
\be
l_i(k,q) = \sum_{j=1}^L a_{ij}k_j  + \sum_{l=1}^E b_{il}q_l , 
\ee
for $a_{ij}$ integer and $b_{il}$ real.
\item{(L2)} Given the linear combinations
\be
p_i(k) = \sum_{j=1}^L a_{ij} k_j
\ee
and the associated set
\be
{\cal L} = \{k_1, \dots, k_L, p_1, \dots, p_I\},
\ee
and considering $u_1, \dots, u_L$ linearly independent elements of ${\cal L}$,
then
\be
k_i = \sum_{j=1}^L c_{ij} u_j
\ee
holds, with $c_{ij}$ integer. 
\end{description}
The conditions L1 and L2 define a ``natural'' choice of line momenta. It is 
important that the coefficients $a_{ij}$ in L1 and $c_{ij}$ in L2 are integers.
These conditions guarantee that shifting integration variables by $2\pi/a$ 
and choosing some of the line momenta as new integration variables still gives
a periodic integrand and does not change the domain of integration.

The condition C3 is one of the most significant. While the other conditions
are rather weak, and are fulfilled by any reasonable theory, this one is
strongly discriminating against certain type of integrands.
Condition C3 is in fact satisfied by scalars as well as by Wilson and overlap 
fermions, but not by naive fermions and staggered fermions.
Essentially this condition asks that the denominators $C_i$ diverge 
like $1/a^2$ when the momenta $l_i$ are at the edges of the Brillouin zone,
which is sufficient to forbid any doublers in that region.

We need now a definition of the degree of divergence of an integrand. The 
degree of divergence of the numerator is defined from its asymptotic behavior,
\be
V(\lambda k,q;m,a\lambda) \stackrel{\lambda \to \infty}{=}
K \lambda^{\deg V} + O(\lambda^{\deg V -1}), 
\ee
where $K \neq 0$, and similarly for the degree of divergence of the 
denominator, $\deg C$. The lattice degree of divergence takes into account
the behavior of the integrand functions not only for small lattice spacing,
but also for large loop momenta $k \sim 1/a$. The degree of divergence of the 
integral $I$ is then given (in four dimensions) by
\be
\deg I = 4 + \deg V - \deg C .
\ee
Finally, for integrals beyond one loop we need to introduce the notion of 
Zimmermann subspaces, which are linear subspaces of the momenta.
Let us consider $L$ linear independent elements of the set of momenta 
${\cal L}$ defined in condition L2,
\be
u_1, \dots, u_d, v_1, \dots, v_{L-d}, \qquad (d \ge 1),
\ee
and take them as new integration variables. If we now fix $v_1, \dots, v_{L-d}$
to some value, we obtain a 4d-dimensional Zimmermann subspace, spanned by 
$u_1, \dots, u_d$.~\footnote{One does not distinguish between subspaces 
corresponding to different values of the fixed momenta $v_1, \dots, v_{L-d}$.}
The degree of divergence for $V$ in this Zimmermann subspace is then defined 
as
\be
V(k(\lambda u,v),q;m,a\lambda) \stackrel{\lambda \to \infty}{=}
K \lambda^{\deg_Z V} + O(\lambda^{\deg_Z V -1}), 
\ee
where $K \neq 0$, and similarly for $C_i$. This allows to study the
behavior of the integrand when only some of the momenta are large.

The theorem of Reisz says that the continuum limit of the integral $I$ 
in Eq.~(\ref{eq:iniint}) exists if $\deg_Z I < 0$ for all its possible 
Zimmermann subspaces $Z$, and in this case is given by integrating the naive 
continuum limit of the integrand (as in conditions V2 and C2):
\be
\lim_{a \to 0} I = \int^\infty_{-\infty} \frac{d^4k_1}{(2\pi)^4} \cdots 
\int^\infty_{-\infty} \frac{d^4k_L}{(2\pi)^4} \, \, 
\frac{P(k,q;m)}{\prod_{i=1}^I (l_i^2 +m_i^2)} .
\ee
Thus, in this case we have reduced the initial problem to the computation 
of a simpler continuum integral, which is absolutely convergent. The theorem 
can also be formulated in the case in which massless propagators are present, 
but then one has also to introduce infrared degrees of divergence 
(Reisz, 1988b; 1988d).

The proof of the power counting theorem is rather complicated and goes beyond 
the scope of this review. We refer the interested reader to the original
papers (Reisz, 1988a; 1988b; 1988c; 1988d).

We now discuss a couple of examples which illustrate the meaning of the 
theorem of Reisz. Let us consider the one-dimensional integral
\be
I(N) = \int^{\frac{\pi}{a}}_{-\frac{\pi}{a}} \frac{dk}{2\pi} \, 
\frac{1}{\displaystyle \frac{N^2}{a^2} \sin^2 \frac{ak}{N} + m^2},
\ee
which when $N=1$ describes naive fermions and when $N=2$ describes scalar 
particles. Doing {\em naively} the limit $a \to 0$ of the integrand and of 
the integration region gives
\be
\int^\infty_{-\infty} \frac{dk}{2\pi} \, \frac{1}{k^2+ m^2} = \frac{1}{2m} ,
\label{eq:naivecl}
\ee
which is independent of $N$. The {\em true} value of the integral at finite 
$a$ can be computed using the Schwinger representation
\be
\frac{1}{x^2+m^2} = \int_0^\infty d\alpha \, \e^{-\alpha (x^2+m^2)} ,
\ee
so that it becomes
\be
I(N) = \frac{2 a}{N^2} \int_0^\infty dy \, 
\e^{-\Big(1+\frac{2a^2m^2}{N^2}\Big) y} \, I_0(y),
\ee
where $I_0$ is a modified Bessel function, which for large $y$ behaves as
\be
I_0 (y) \stackrel{y \to \infty}{\longrightarrow} 
\frac{1}{\sqrt{2\pi y}} \, \e^y.
\ee
We can now compute the limit $a \to 0$ of $I(N)$ by replacing in the integrand 
the modified Bessel function with its asymptotic expression, and using 
$\int_0^\infty \e^{-bx} dx / \sqrt{x} = \sqrt{\pi/a}$ one obtains
\be
I(N)  \stackrel{a \to 0}{\longrightarrow} \frac{1}{Nm} ,
\label{eq:truevalue}
\ee
which is the correct continuum limit of $I(N)$. We can see that it depends on 
$N$. For $N=2$ this result is the same as the naive continuum limit 
(\ref{eq:naivecl}), whereas for $N=1$ the naive continuum limit gives only 
half of the true value (\ref{eq:truevalue}). This mismatch corresponds
to a case in which the Reisz theorem cannot be applied, because the propagator
of naive fermions ($N=1$) does not satisfy condition C3. In fact, this 
propagator has a doubler, and the true value of the integral in the continuum 
limit is then precisely twice the result which one would obtain just doing 
the naive continuum limit.

In the above example all integrals have a negative degree of divergence.
The integral
\be
\int^{\frac{\pi}{a}}_{-\frac{\pi}{a}} \frac{d^4k}{(2\pi)^4} \,
\frac{1 -\cos ak_\mu}{{\displaystyle \frac{4}{a^2}} 
\sum_\lambda \sin^2 {\displaystyle \frac{a k_\lambda}{2}} +m^2}  ,
\ee
instead, is divergent like $1/a^2$ in the continuum limit, as can be seen 
by dimensional counting using the rescaled variable $k'=ka$.
Therefore, the theorem of Reisz cannot be used, and in fact a naive continuum 
limit of the integrand gives a zero result, which is incorrect.

The Reisz power counting theorem is quite useful for the calculation of
lattice integrals, especially when they are divergent, as we will see in the
next Section. It is also of great help for the proof of the renormalizability 
of lattice theories at all orders of perturbation theory, which has been given
for pure nonabelian gauge theories by (Reisz, 1989).

\subsection{Divergent integrals}
\label{sec:divergentintegrals}

For the treatment of divergent integrals on the lattice it is convenient 
to use a method which was introduced in~\cite{Kawai:1980ja}. It consists in 
making an expansion of these integrals in powers of the external momenta, 
and computing on the lattice only the integrals with vanishing momentum, 
which are technically much simpler. As an example we consider the case of a 
quadratically divergent integral depending on two external momenta $p$ and $q$,
\be
I = \int {\mathrm d}k \, \, {\cal I}(k,p,q) .
\ee 
This integral can be split as
\be
I = J + (I-J) ,
\label{eq:divint}
\ee
where
\bea
J &&= \int \, {\mathrm d}k \, \, {\cal I}(k,0,0) + \sum_{\rho,\sigma} 
\Bigg[ p_\rho q_\sigma \int \, {\mathrm d}k \, \, 
\frac{\partial^2 {\cal I}(k,p,q)}{\partial p_\rho \partial q_\sigma} 
\Bigg|_{p=q=0} \\ 
&& + \frac{p_\rho p_\sigma}{2}  \int \, {\mathrm d}k \, \,  
\frac{\partial^2 {\cal I}(k,p,0)}{\partial p_\rho \partial p_\sigma}
\Bigg|_{p=0}
+ \frac{q_\rho q_\sigma}{2} \int \, {\mathrm d}k \, \, 
\frac{\partial^2 {\cal I}(k,0,q)}{\partial q_\rho \partial q_\sigma} 
\Bigg|_{q=0} \Bigg] \nonumber
\eea
is the Taylor expansion of the original integral to second order. 
The integrals appearing in $J$ do not depend on the external momenta and are 
thus much easier to calculate on the lattice. The whole dependence on
the external momenta remains in $I-J$ which, because of the subtraction,  
is ultraviolet-finite for $a \rightarrow 0$ and can thus be computed, 
according to the theorem of Reisz, just by taking the naive continuum limit. 
Thanks to this fact, only zero-momentum integrals have to be evaluated on 
the lattice.

Notice that for $p,q \neq 0$ and finite lattice spacing $I$ is well defined, 
but $J$ and $I-J$ are infrared divergent. To compute $J$ and $I-J$ separately,
one must then introduce an intermediate regularization. 
The associated divergences will at the end cancel out in the sum $J + (I-J)$. 
This intermediate regularization is completely independent from the main 
regularization used in the lattice theory, and in particular can be different 
from it. It just comes out because the splitting is somewhat unnatural.

To give an explicit illustration of this method, let us take the 
logarithmically divergent integral
\be
I =  \int^{\frac{\pi}{a}}_{-\frac{\pi}{a}} \frac{d^4k}{(2\pi)^4} \, 
\frac{1}{\displaystyle \Bigg(\frac{4}{a^2} \sum_\mu \sin^2 \frac{a(k+p)_\mu}{2}
\Bigg)^2} . 
\ee 
The splitting is then made as follows:
\bea
 J &=& I(p=0) = \int^{\frac{\pi}{a}}_{-\frac{\pi}{a}} \frac{d^4k}{(2\pi)^4} \, 
\frac{1}{\displaystyle \Bigg(\frac{4}{a^2} \sum_\mu \sin^2 \frac{ak_\mu}{2} 
\Bigg)^2} , \\
I-J &=& \lim_{a \to 0} \,  \int^{\frac{\pi}{a}}_{-\frac{\pi}{a}}
   \frac{d^4k}{(2\pi)^4} \, \Bigg\{ 
   \frac{1}{\displaystyle\Bigg(\frac{4}{a^2} 
          \sum_\mu \sin^2 \frac{a(k+p)_\mu}{2} \Bigg)^2} -
   \frac{1}{\displaystyle\Bigg(\frac{4}{a^2} \sum_\mu \sin^2 \frac{ak_\mu}{2} 
           \Bigg)^2} 
   \Bigg\} \nonumber \\
&=&  \int_{-\infty}^\infty \frac{d^4k}{(2\pi)^4} \, 
\Bigg\{ \frac{1}{((k+p)^2)^2} - \frac{1}{(k^2)^2} \Bigg\} .
\eea
Taking common denominators, it is easy to see that the degree of divergence 
of the above integral is negative, and therefore it can be safely computed 
in the continuum.
 
If we use dimensional regularization we have the result~\footnote{See 
Eq.~(\ref{eq:divintdimreg}) later. Notice that the integral of the second term 
in $I-J$ is zero in this regularization.}
\bea
  J &=&  \frac{1}{16\pi^2} \Big( \frac{2}{d-4} -\log 4\pi +F_0 \Big) , \\
I-J &=&  \frac{1}{16\pi^2} \Big(-\frac{2}{d-4} -\log a^2p^2 +\log 4\pi 
         -\gamma_E \Big) ,
\eea
where $\gamma_E=0.57721566490153286\dots$ and the lattice constant is 
$F_0= 4.369225233874758\dots$ (see Eqs.~(\ref{eq:f0intdr}) 
and~(\ref{eq:f0intm}) and Table~\ref{tab:z0z1f0} later), while if we 
regularize adding a small mass term $m^2$ to $k^2$ in the denominators we 
obtain~\footnote{See Eq.~(\ref{eq:divintmass}) later.}
\bea
  J_m &=& \frac{1}{16\pi^2} \Big( -\log a^2m^2 -\gamma_E +F_0 \Big) , \\
(I-J)_m &=& \frac{1}{16\pi^2} \Big( \log a^2m^2 -\log a^2p^2 \Big) .
\eea
In both cases, adding up $J$ and $I-J$ we obtain for the original integral 
the result
\be
I = -\log a^2p^2 -\gamma_E +F_0.
\ee

To summarize, for the computation of any divergent integral which depends on 
external momenta it is sufficient to compute some lattice integrals at zero
momenta and some continuum integrals.

In computer programs, a convenient way to deal with a generic divergent 
integral (which has to be processed in an automated way) is to subtract to it 
a simple integral with the same divergent behavior for which the numerical
value is exactly known. The difference is then finite and can be computed
with reasonable precision using simple integration routines. This is extremely
convenient in the case of actions which give rise to complicated denominators,
like for example overlap fermions. In this case, Wilson integrals with the 
same divergence are subtracted to the original overlap integral, and then 
overlap denominators, which are much more complicated, appear only in the 
numerical calculation of finite integrals. The calculation of divergent 
integrals is made only using Wilson fermions.

Of course, this is not the only available method for computing divergent
integrals. In Section~\ref{sec:bosoniccase} we will show another technique 
based on the coordinate space method.

\subsection{General aspects of the calculations}

We have seen that the Feynman rules on the lattice are rather different
from the continuum ones. The structure of lattice integrals is also completely
different. The integrands are periodic in the momenta, and the basic objects 
are trigonometric functions and not simple polynomials of the momenta.
Many standard methods which are very useful in continuum perturbation theory,
like Feynman parameterization and partial integration, are then not of much 
relevance for perturbative lattice calculations.~\footnote{We mention that 
recently a computational method has been presented~\cite{Becher:2002if} that 
through a change of variables ($t = \tan k/2$) transforms lattice integrals 
in continuum-like integrals and employs known techniques of continuum 
calculations which use asymptotic expansions.}

We have thought of doing a useful thing by showing a complete lattice 
calculation which illustrates the peculiar aspects of lattice perturbation 
theory.

\subsection{Example (Wilson): 
the first moment of the quark momentum distribution}
\label{sec:examplewilson}

We describe in this Section, as a pedagogical example, a typical lattice 
perturbative calculation. We explain in detail the main steps 
of the computation, with the Wilson action, of the renormalization constant 
of the forward matrix element $\langle q | O_{\{\mu \nu\}} | q \rangle$ 
on one-quark states of the operator 
\be
O_{\{\mu \nu\}} = \overline{\psi} \gamma_{\{\mu} D_{\nu\}} 
\frac{\lambda^f}{2} \psi , \qquad \mu \neq \nu .
\ee
This operator, which is symmetrized in the indices $\mu$ and $\nu$, 
measures the first moment of the fraction of the momentum of the proton 
carried by the quarks. The $\lambda$'s are flavor matrices, which means that 
we are considering a flavor nonsinglet operator. The corresponding singlet 
operator (proportional to the identity flavor matrix) mixes, when radiative 
corrections are included, with the gluon operator 
$\sum_\rho \Tr \, (F_{\mu\rho} F_{\rho\nu})$, 
which measures the first moment of the momentum distribution of the gluon, 
and to make things not too complicated we will not consider it here. 

This example is rather simple (compared to other operators) and contains
all the main interesting features one can think of: a logarithmic divergence, 
a covariant derivative, symmetrized indices and of course the special use 
of Kronecker deltas in lattice perturbative calculations. 
Moreover, it is an example of a calculation which needs an expansion of the 
various propagators and vertices in the lattice spacing $a$ (in this case,
at first order).~\footnote{A simpler calculation with no covariant derivatives
and no need of expansions in $a$, that the reader could try as a warm-up, 
is given by the renormalization of quark currents. This was first computed 
in~\cite{Martinelli:1982mw}, and for the case of extended currents, which are 
defined on more than one lattice site like the conserved vector current 
in Eq.~(\ref{eq:extcurr}), in~\cite{Martinelli:1983be}.}

For what concerns the value of the renormalization constant, the flavor 
matrices are not important (that is, except for the fact that they forbid 
the mixing with gluonic operators), and we will then carry out the explicit 
computations using the operator
\be
\frac{1}{2} \Big( \overline{\psi} \gamma_\mu D_\nu \psi
+ \overline{\psi} \gamma_\nu D_\mu \psi \Big) , 
\ee
where $\mu \neq \nu$ is chosen, so that this operator belongs to the
representation ${\bf 6_1}$ of the hypercubic group (see 
Section~\ref{sec:hypercubic}). 
The choice $\mu = \nu$ would compel us to consider the operator 
$O_{\{00\}} -\frac{1}{3} ( O_{\{11\}} + O_{\{22\}} + O_{\{33\}} )$,
which belongs to the representation ${\bf 3_1}$ and has a different lattice
renormalization constant. This would render the calculations a bit more 
cumbersome, without teaching us much new.

The covariant derivative is defined as follows:
\be
D = \stackrel{\leftrightarrow}{D} = \frac{1}{2} \Big(
    \stackrel{\rightarrow}{D} -  \stackrel{\leftarrow}{D} \Big),
\ee
and the discretization of it that we use for this calculation is
\bea
\stackrel{\rightarrow}{D}_{\mu} \psi(x) &=&  \frac{1}{2a}
\Big[ U_{\mu} (x) \psi(x+a\hat{\mu}) - U^\dagger_{\mu} (x-a\hat{\mu})
\psi(x-a\hat{\mu})\Big] , \\
\overline{\psi}(x)\stackrel{\leftarrow}{D}_{\mu} &=&  \frac{1}{2a}
\Big[ \overline{\psi}(x+a\hat{\mu})U^\dagger_{\mu} (x) - \overline{\psi}
(x-a\hat{\mu}) U_{\mu} (x-a\hat{\mu}) \Big] . \nonumber
\eea

\begin{figure}[t]
\begin{center}
\begin{picture}(500,160)(-20,-20)
\ArrowLine(0,0)(50,80)
\ArrowLine(50,80)(100,0)
\GBoxc(50,80)(6,6){0}
\Gluon(20,32)(80,32){-4}{8}
\ArrowLine(120,0)(170,80)
\ArrowLine(170,80)(220,0)
\GBoxc(170,80)(6,6){0}
\GlueArc(155,56)(28.3,58,238){4}{10}
\ArrowLine(240,0)(290,80)
\ArrowLine(290,80)(340,0)
\GBoxc(290,80)(6,6){0}
\GlueArc(305,56)(28.3,-58,122){4}{10}
\ArrowLine(360,0)(410,80)
\ArrowLine(410,80)(460,0)
\GBoxc(410,80)(6,6){0}
\GlueArc(410,100)(20,-90,270){4}{15}
\Text(50,-15)[t]{(Vertex)}
\Text(170,-15)[t]{(Left Sail)}
\Text(290,-15)[t]{(Right Sail)}
\Text(410,-15)[t]{(Operator Tadpole)}
\Text(5,20)[b]{$p$}
\Text(30,60)[b]{$k$}
\Text(50,10)[b]{$p-k$}
\Text(95,20)[b]{$p$}
\Text(70,60)[b]{$k$}
\Text(125,20)[b]{$p$}
\Text(150,60)[b]{$k$}
\Text(150,90)[b]{$p-k$}
\Text(215,20)[b]{$p$}
\Text(245,20)[b]{$p$}
\Text(310,90)[b]{$p-k$}
\Text(335,20)[b]{$p$}
\Text(310,60)[b]{$k$}
\Text(365,20)[b]{$p$}
\Text(455,20)[b]{$p$}
\Text(410,130)[b]{$k$}
\end{picture}
\end{center}
\caption{\small ``Proper'' diagrams for the 1-loop correction of the 
matrix element
$\langle q | \overline{\psi} \gamma_{\{\mu} D_{\nu\}} \psi | q \rangle $. 
The black squares indicates the insertion of the operator. Shown is also 
the choice of momenta used in the calculations.}
\label{fig:proper}
\end{figure}
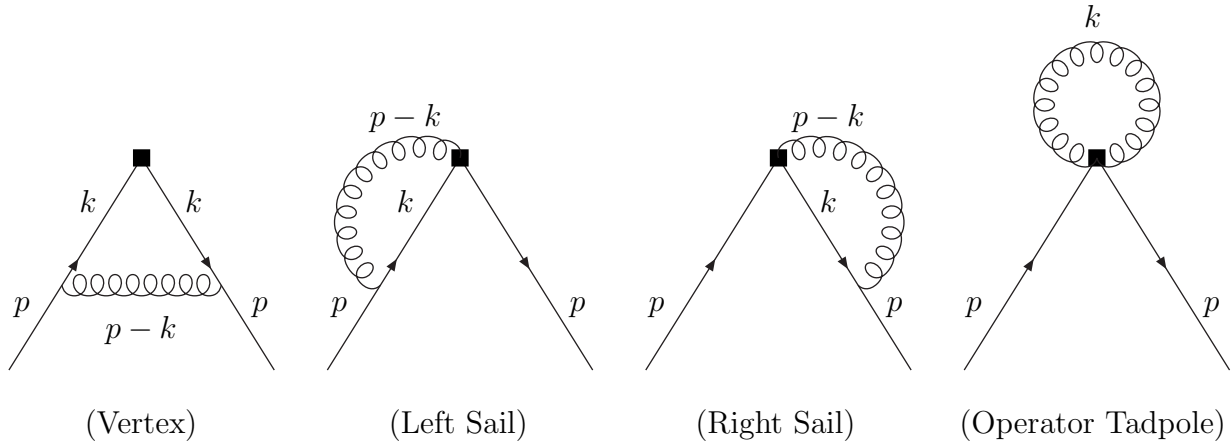

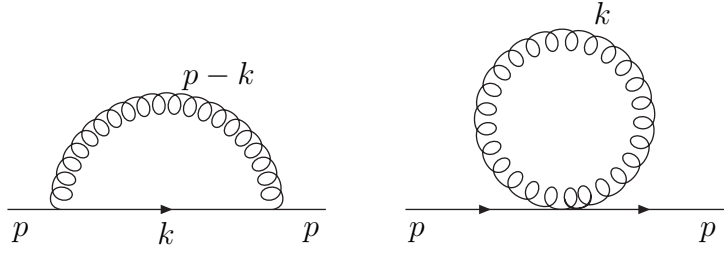
\begin{figure}[t]
\begin{center}
\begin{picture}(300,100)(-10,-10)
\ArrowLine(0,0)(120,0)
\GlueArc(60,0)(40,0,180){4}{20}
\ArrowLine(150,0)(210,0)
\ArrowLine(210,0)(270,0)
\GlueArc(210,34)(30,-90,289){4}{25}
\Text(5,-5)[t]{$p$}
\Text(60,-5)[t]{$k$}
\Text(115,-5)[t]{$p$}
\Text(155,-5)[t]{$p$}
\Text(265,-5)[t]{$p$}
\Text(80,45)[b]{$p-k$}
\Text(225,70)[b]{$k$}
\end{picture}
\end{center}
\caption{\small Diagrams for the quark self-energy. On the left the 
sunset diagram, on the right the tadpole diagram.}
\label{fig:quark_self}
\end{figure}

We consider amputated Green's functions, that is the external propagators 
are removed. The tree level of the amputated forward quark matrix element 
of the operator above is easily seen to be
\be
\langle q | O_{\{\mu \nu\}} | q \rangle \Big|_{\mathrm tree}
= \frac{1}{2} \ii ( \gamma_\mu p_\nu + \gamma_\nu p_\mu ) , 
\ee
and the 1-loop QCD result has, as we will see from the calculation, the form
\be
\langle q | O_{\{\mu \nu\}} | q \rangle \Big|_{\mathrm 1~loop}
= \frac{1}{2} \ii ( \gamma_\mu p_\nu + \gamma_\nu p_\mu ) 
\cdot \frac{g_0^2}{16\pi^2}C_F \Big( c_1 \log a^2p^2 + c_2) ,
\ee
i.e., it is proportional to the tree level and this operator is thus 
multiplicatively renormalized. The renormalization constant for the matching
to the $\ms$ scheme can then be read off from the above 1-loop result plus 
the corresponding continuum calculations made in the $\ms$ scheme (see 
Eq.~(\ref{eq:1loopcontlat}) and Section~\ref{sec:renormalizationoperators}). 
For the computation of the lattice part it is necessary to evaluate six 
Feynman diagrams, which are given in Figs.~\ref{fig:proper} 
and~\ref{fig:quark_self}. The two diagrams in Fig.~\ref{fig:quark_self} 
compute the quark self-energy, and give the renormalization of the wave 
function. The remaining four diagrams, in Fig.~\ref{fig:proper}, are specific 
to the operator considered, and we will call them ``proper'' diagrams.

\subsubsection{Preliminaries}

We work in Feynman gauge ($\alpha=1$), where the form of the gluon propagator 
is simpler, and we set $r=1$. We perform the calculations using massless 
fermions.~\footnote{Calculations in which the quark propagator is massive 
are more complicated. A few examples of these calculations, which use simpler 
operators, can be found in~\cite{Kronfeld:dd,El-Khadra:1996mp,Mertens:1997wx} 
and~\cite{Kuramashi:1997tt}.}
This is the simplest situation one can think of, although it already leads to 
complicated manipulations, as we will shortly see.
We carry out these manipulations starting from the operator
\be
O_{\mu\nu} = \overline{\psi} \gamma_\mu D_\nu \psi ,
\ee
and implement the symmetrization in $\mu$ and $\nu$ at a later stage.
Due to the presence of the link variable $U$ in the covariant derivative, 
this operator has an expansion in the coupling, 
\be
O_{\mu\nu} = O_{\mu\nu}^{(0)} + g_0 O_{\mu\nu}^{(1)} 
+ g_0^2 O_{\mu\nu}^{(2)} + O(g_0^3) .
\ee
To evaluate the one-loop Feynman diagrams in momentum space one has to compute 
the Fourier transforms of the operators in this expansion including the term 
of $O(g_0^2)$. It turns out that to work out these momentum-space insertions 
for our forward matrix element we can use the operator defined with the right 
derivative only, instead of the one involving the difference between the right
and the left derivative (which would lead to more complicated manipulations). 
We have then that the expansion of $a^4 \sum_x \Big( \overline{\psi} 
\gamma_\mu \stackrel{\rightarrow}{D}_\nu \psi \Big) (x)$ is
\bea 
&& a^4 \, \frac{1}{2a} \sum_x \Big( 
\overline{\psi} (x) \gamma_\mu U_\nu (x) \psi (x+a\hat{\nu}) - 
\overline{\psi} (x) \gamma_\mu U_\nu^\dagger (x-a\hat{\nu}) \psi (x-a\hat{\nu})
\Big) \nonumber \\
&& = a^4 \, \Bigg\{ \frac{1}{2a} \sum_x \Big( 
\overline{\psi} (x) \gamma_\mu \psi (x+a\hat{\nu}) 
- \overline{\psi} (x) \gamma_\mu \psi (x-a\hat{\nu})
\Big) \\
&& \quad +\frac{1}{2} \ii g_0 T^a \sum_x \Big( 
\overline{\psi} (x) \gamma_\mu A_\nu^a (x) \psi (x+a\hat{\nu}) 
+ \overline{\psi} (x) \gamma_\mu A_\nu^a (x-a\hat{\nu}) \psi (x-a\hat{\nu})
\Big)  \nonumber \\
&& \quad -\frac{1}{4} a g_0^2 T^a T^b \sum_x \Big( 
\overline{\psi} (x) \gamma_\mu A_\nu^a (x) A_\nu^b (x) \psi (x+a\hat{\nu}) 
- \overline{\psi} (x) \gamma_\mu A_\nu^a (x-a\hat{\nu}) A_\nu^b (x-a\hat{\nu}) 
\psi (x-a\hat{\nu}) \Big) \nonumber \\
&& \quad + O(a^2 g_0^3) \Bigg\} . \nonumber 
\eea
The Fourier transform of the lowest order is 
\bea
&& a^4 \, \frac{1}{2a} \sum_x 
\int_{-\frac{\pi}{a}}^{\frac{\pi}{a}} \frac{d^4k}{(2\pi)^4} \, 
\int_{-\frac{\pi}{a}}^{\frac{\pi}{a}} \frac{d^4k'}{(2\pi)^4} \, 
\overline{\psi} (k') \gamma_\mu \psi (k) \e^{-\ii k'x} 
\Big( \e^{\ii k (x+a\hat{\nu})} - \e^{\ii k (x-a\hat{\nu})} \Big) \\
&& = \frac{1}{2a} 
\int_{-\frac{\pi}{a}}^{\frac{\pi}{a}} \frac{d^4k}{(2\pi)^4} \, 
\overline{\psi} (k) \gamma_\mu \psi (k)
\Big( \e^{\ii ak_\nu} - \e^{-\ii ak_\nu} \Big) \\
&& = \frac{\ii}{a}
\int_{-\frac{\pi}{a}}^{\frac{\pi}{a}} \frac{d^4k}{(2\pi)^4} \, 
\overline{\psi} (k) \gamma_\mu \psi (k)
\sin ak_\nu ,
\eea
where we have used 
$a^4 \sum_x \e^{-\ii k'x} \e^{\ii kx} = (2\pi)^4 \delta^{(4)} (k-k')$.
Similar delta functions arise at each order of the expansion, expressing 
the conservation of momentum at the various vertices. The first order term 
in $g_0$ is
\bea
&& a^4 \, \frac{1}{2} \ii g_0 T^a \sum_x 
\int_{-\frac{\pi}{a}}^{\frac{\pi}{a}} \frac{d^4k}{(2\pi)^4} \, 
\int_{-\frac{\pi}{a}}^{\frac{\pi}{a}} \frac{d^4p}{(2\pi)^4} \, 
\int_{-\frac{\pi}{a}}^{\frac{\pi}{a}} \frac{d^4q}{(2\pi)^4} \, 
\overline{\psi} (p) \gamma_\mu \psi (k) A_\nu^a (q) \nonumber \\
&& \qquad \qquad \qquad \times \e^{-\ii px} \e^{\ii qx} \e^{\ii kx} 
\Big( \e^{\ii q_\nu a/2} \e^{\ii k_\nu a}
     +\e^{-\ii q_\nu a/2}  \e^{-\ii k_\nu a} \Big) \nonumber \\
&& = \frac{1}{2} \ii g_0 T^a
\int_{-\frac{\pi}{a}}^{\frac{\pi}{a}} \frac{d^4k}{(2\pi)^4} \, 
\int_{-\frac{\pi}{a}}^{\frac{\pi}{a}} \frac{d^4p}{(2\pi)^4} \, 
\overline{\psi} (p) \gamma_\mu \psi (k) A_\nu^a (p-k) 
\Big( \e^{\ii (p-k)_\nu a/2} \e^{\ii k_\nu a}
     +\e^{-\ii (p-k)_\nu a/2}  \e^{-\ii k_\nu a} \Big) \nonumber \\
&& = \ii g_0 T^a 
\int_{-\frac{\pi}{a}}^{\frac{\pi}{a}} \frac{d^4k}{(2\pi)^4} \, 
\int_{-\frac{\pi}{a}}^{\frac{\pi}{a}} \frac{d^4p}{(2\pi)^4} \, 
\overline{\psi} (p) \gamma_\mu \psi (k) A_\nu^a (p-k) 
\cos \frac{a(k+p)_\nu}{2} .
\eea
With our convention for the Fourier transform of the gauge fields the gluons 
are always entering the vertices. The calculation above corresponds to the 
left sail in Fig.~\ref{fig:proper}, and the reader can check that the 
calculation for the right sail gives the same function:
\be
\ii g_0 T^a 
\int_{-\frac{\pi}{a}}^{\frac{\pi}{a}} \frac{d^4k}{(2\pi)^4} \, 
\int_{-\frac{\pi}{a}}^{\frac{\pi}{a}} \frac{d^4p}{(2\pi)^4} \, 
\overline{\psi} (k) \gamma_\mu \psi (p) A_\nu^a (k-p) 
\cos \frac{a(k+p)_\nu}{2} .
\ee
Finally, since the only second-order contribution in which we are interested 
is the operator tadpole, we exploit the simplifications which this situation 
brings. In particular, since the gluon is emitted and reabsorbed at the same
vertex, there is a Kronecker delta in color space coming from the gluon
propagator and the color factor becomes 
$\sum_a (T^a)^2_{bb} = (N_c^2-1)/(2N_c) = C_F$, the quadratic Casimir 
invariant of $SU(N_c)$. We have then that the insertion of the operator 
tadpole is 
\bea
&& -\frac{1}{4} a g_0^2 C_F  
\int_{-\frac{\pi}{a}}^{\frac{\pi}{a}} \frac{d^4k}{(2\pi)^4} \, 
\overline{\psi} (p) \gamma_\mu \psi (p) A_\nu^a (k) A_\nu^a (k) 
\Big( \e^{\ii p_\nu a} - \e^{-\ii p_\nu a} \Big) \nonumber \\
&& \qquad \qquad = - \frac{1}{2} \ii a g_0^2 C_F  
\int_{-\frac{\pi}{a}}^{\frac{\pi}{a}} \frac{d^4k}{(2\pi)^4} \, 
\overline{\psi} (p) \gamma_\mu \psi (p) A_\nu^a (k) A_\nu^a (k) 
\sin ap_\nu ,
\eea
where now the color index $a$ is not summed. Note that the factors 
$\exp (\pm \ii ak_\nu/2)$ coming from the gluons have canceled, again because 
it is the same gluon that is emitted and absorbed at the vertex. 
This insertion does not depend on the momentum of the gluon.

\begin{figure}[t]
\begin{center}
\begin{picture}(470,150)(0,-20)
\ArrowLine(10,20)(50,80)
\ArrowLine(50,80)(90,20)
\GBoxc(50,80)(6,6){0}
\ArrowLine(160,20)(200,80)
\ArrowLine(200,80)(240,20)
\GBoxc(200,80)(6,6){0}
\Gluon(200,80)(180,120){-4}{4}
\ArrowLine(310,20)(350,80)
\ArrowLine(350,80)(390,20)
\GBoxc(350,80)(6,6){0}
\Gluon(350,80)(310,120){4}{6}
\Gluon(350,80)(390,120){-4}{6}
\Text(50,-15)[b]{\large $\frac{1}{a} \, \ii \gamma_\mu \sin ak_\nu $}
\Text(200,-15)[b]{\large $T^a \, \ii \gamma_\mu \cos \frac{a(k+p)_\nu}{2}$}
\Text(350,-15)[b]{\large $-\frac{a}{2} \, C_F \, \ii \gamma_\mu \sin ap_\nu$}
\Text(110,80)[l]{\Large $+g_0 \cdot$}
\Text(260,80)[l]{\Large $+g_0^2 \cdot$}
\Text(410,80)[l]{\Large $+ O(g_0^3)$}
\Text(15,50)[b]{$k$}
\Text(86,50)[b]{$k$}
\Text(165,50)[b]{$k$}
\Text(235,50)[b]{$p$}
\Text(215,100)[b]{$p-k$}
\Text(315,50)[b]{$p$}
\Text(385,50)[b]{$p$}
\Text(330,115)[b]{$k$}
\Text(370,115)[b]{$k$}
\Text(170,125)[b]{$\nu$}
\Text(300,125)[b]{$\nu$}
\Text(400,125)[b]{$\nu$}
\end{picture}
\end{center}
\caption{\small Operator insertions for 
$\overline{\psi} \gamma_{\{\mu} D_{\nu\}} \psi$. Please note that in the 
second-order term the color factors have been already worked out as 
they occur in the tadpole.}
\label{fig:opins}
\end{figure}
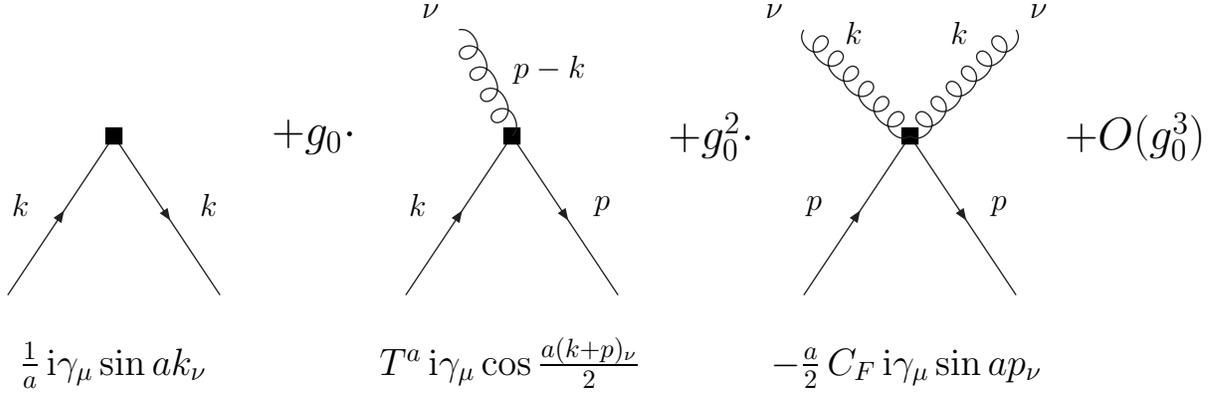

We have thus obtained, in the momentum choice of Fig.~\ref{fig:proper}, 
the operator insertions
\bea
O_{\mu\nu}^{(0)} (k) &=& \frac{1}{a} \, \ii \gamma_\mu \sin ak_\nu , \\
O_{\mu\nu}^{(1)} (k,p) &=& T^a \, \ii \gamma_\mu \cos \frac{a(k+p)_\nu}{2} , \\
O_{\mu\nu}^{(2)} (p) &=& -\frac{a}{2} \, C_F \, \ii \gamma_\mu \sin ap_\nu .
\eea
These operator insertions are shown in Fig.~\ref{fig:opins}. They generate the
vertex, sails and operator tadpole respectively (see Fig.~\ref{fig:proper}). 
The gluons have the same index as the trigonometric functions.

In the intermediate stages of the calculations many terms are present. 
A lot of them originate from the Taylor expansions of the denominators of the 
propagators. With our choice of momenta~\footnote{We always have that the 
external momentum is $p$, the quark propagator has momentum $k$ and the gluon 
propagator has momentum $p-k$ (except in the tadpoles).} only the gluon 
propagator has to be expanded in the lattice spacing, whereas the quark 
propagator does not contain any external momenta, and does not need to be 
expanded. We have found that this leads to less complicated manipulations than
the other choice in which the quark propagators have momentum $p-k$ and the 
gluon propagator has momentum $k$.

\subsubsection{Vertex}

The first diagram that we consider, and for which we will give a rather
detailed explanation of how is computed, is the vertex function in 
Fig.~\ref{fig:proper}. Since this Feynman diagram is divergent, it is 
convenient to split its computation into two parts, as explained in 
Section~\ref{sec:divergentintegrals}: the integral at zero momentum, $J$, 
and the rest of the integral, $I-J$, which can be computed in the continuum 
thanks to the theorem of Reisz. We use dimensional regularization as regulator
of the intermediate divergences, with $d=4-2\epsilon$.

We have then~\footnote{We will assume the implicit summation convention
for the indices $\rho$ and $\lambda$ in the continuum, while on the lattice 
even repeated indices, unless explicitly stated, are not summed.}
\be
J = \int_{-\frac{\pi}{a}}^{\frac{\pi}{a}} \frac{d^d k}{(2\pi)^d} \, 
\sum_{\rho,\lambda} G_{\rho\lambda} (p-k) \cdot V_\rho (k,p) \cdot S(k) 
\cdot O_{\mu\nu}^{(0)} (k) \cdot S(k) \cdot V_\lambda (p,k) \Bigg|_{ap=0} ,
\ee
where (in Feynman gauge)
\bea
S^{ab} (k) &=& a \, \delta^{ab} \cdot 
\frac{\displaystyle -\ii \sum_\lambda \gamma_\lambda 
\sin ak_\lambda + 2 \sum_\lambda \sin^2 \frac{a k_\lambda}{2}}{\displaystyle 
\sum_\lambda \sin^2 ak_\lambda 
+ 4 \Bigg(\sum_\lambda \sin^2 \frac{a k_\lambda}{2}\Bigg)^2},\\
(V^a)^{bc}_\rho(k,p) = (V^a)^{bc}_\rho (p,k) &=&  
- g_0 (T^a)^{bc} \Bigg( \sin \frac{a(k+p)_\rho}{2} 
+ \ii \gamma_\rho \cos \frac{a(k+p)_\rho}{2} \Bigg) \\
&=& -g_0 (T^a)^{bc} \Bigg( \sin \frac{ak_\rho}{2} 
          + \ii \gamma_\rho \cos \frac{ak_\rho}{2} 
          +\frac{ap_\rho}{2} \Bigg( \cos \frac{ak_\rho}{2}
   - \ii \gamma_\rho \sin \frac{ak_\rho}{2} \Bigg) \Bigg) \nonumber \\
&& +O(a^2), \nonumber \\
G^{ab}_{\rho\lambda} (p-k) &=& \delta^{ab} \delta_{\rho\lambda} \, \cdot
\frac{1}{\displaystyle \frac{4}{a^2} 
\sum_\lambda \sin^2 \frac{a(p-k)_\lambda}{2}} \\
&=& \delta^{ab} \delta_{\rho\lambda} \, \cdot a^2 
\left\{ \frac{1}{4 \displaystyle \sum_\lambda \sin^2 \frac{ak_\lambda}{2}}
+a \, \frac{\displaystyle 2\sum_\lambda p_\lambda \sin ak_\lambda}{
\Bigg(\displaystyle 4 \sum_\lambda \sin^2 \frac{ak_\lambda}{2}\Bigg)^2}
+ O(a^2) \right\} . \nonumber 
\eea
Putting everything together we have
\bea
J &=& a^4 \cdot \frac{g_0^2}{a} C_F 
\int_{-\frac{\pi}{a}}^{\frac{\pi}{a}} \frac{d^d k}{(2\pi)^d} \,
\cdot \sum_\rho 
\left\{ 
\frac{1}{4 \sum_\lambda \sin^2 {\displaystyle \frac{ak_\lambda}{2}}}
+\frac{2a \sum_\lambda p_\lambda \sin ak_\lambda}{
\Bigg(4 \sum_\lambda \sin^2 {\displaystyle \frac{ak_\lambda}{2}}\Bigg)^2}
\right\} \label{eq:putalltog} \\
&& \times \Bigg( 
\sin \frac{ak_\rho}{2} + \ii \gamma_\rho \cos \frac{ak_\rho}{2} 
+\frac{ap_\rho}{2} \Bigg( \cos \frac{ak_\rho}{2}
- \ii \gamma_\rho \sin \frac{ak_\rho}{2} \Bigg) \Bigg) \nonumber \\
&& \times \left\{ 
\frac{-\ii \sum_\lambda \gamma_\lambda \sin ak_\lambda
+ 2 \sum_\lambda \sin^2 {\displaystyle \frac{a k_\lambda}{2}}}{
\sum_\lambda \sin^2 ak_\lambda 
+ 4 \Bigg(\sum_\lambda \sin^2 {\displaystyle \frac{a k_\lambda}{2}\Bigg)^2}}
\right\}
\cdot \gamma_\mu \sin ak_\nu \cdot
\left\{
\frac{-\ii \sum_\lambda \gamma_\lambda \sin ak_\lambda
+ 2 \sum_\lambda \sin^2 {\displaystyle \frac{a k_\lambda}{2}}}{
\sum_\lambda \sin^2 ak_\lambda 
+ 4 \Bigg(\sum_\lambda \sin^2 {\displaystyle \frac{a k_\lambda}{2}\Bigg)^2}}
\right\} \nonumber \\
&& \times \Bigg( 
\sin \frac{ak_\rho}{2} + \ii \gamma_\rho \cos \frac{ak_\rho}{2} 
+\frac{ap_\rho}{2} \Bigg( \cos \frac{ak_\rho}{2}
- \ii \gamma_\rho \sin \frac{ak_\rho}{2} \Bigg) \Bigg) +O(a^5) . \nonumber
\eea
The color factors are, for this diagram, the same as in the continuum, 
\be
\sum_a \sum_b (T^a)_{cb} (T^a)_{bc} 
= \sum_a (T^a)_{cc}^2 = \frac{N_c^2-1}{2N_c} = C_F ,
\ee
and $C_F$ becomes an overall factor in front of the expressions.  
We now rescale the integration variable:
\be
k \rightarrow k' = ak ,
\ee
so that~\footnote{We have introduced noninteger dimensions because we are 
using dimensional regularization to compute the finite part of the divergent 
integrals. While in principle there should be a factor $1/a^d$ in the 
right-hand side of this equation, the difference between $d$ and 4 is not 
relevant for the expansion in $a$ and we can set from the beginning 
this factor to be equal to $1/a^4$.}
\be
\int_{-\frac{\pi}{a}}^{\frac{\pi}{a}} \frac{d^dk}{(2\pi)^d} \, f(ak,ap)
= \int_{-\pi}^\pi \frac{d^dk'}{(2\pi)^d} \, \frac{1}{a^4} \, f(k',ap) .
\label{eq:rescaling}
\ee
Note that the domain of integration after the rescaling becomes independent 
of $a$. The factor $a^4$ in front of Eq.~(\ref{eq:putalltog}) cancels the 
factor $1/a^4$ coming from the rescaling in Eq.~(\ref{eq:rescaling}), and thus 
we have an overall factor $1/a$ left. This means that in order to take the 
continuum limit of this lattice integral we have to expand the whole integrand
including factors of order $ap$. This is not surprising, as it goes in the 
right direction of recovering the tree level $\gamma_\mu p_\nu$. 
We continue to call $k$ the integration variable, with the understanding 
that is the rescaled one. We use from now on the shorthand notation
\bea
\Gamma_\lambda &=& \sin k_\lambda , \\
W &=& 2 \sum_\lambda \sin^2 \frac{k_\lambda}{2} , \\
N_\rho &=& \sin \frac{k_\rho}{2} , \\
M_\rho &=& \cos \frac{k_\rho}{2} ,
\eea
so that the calculations will appear somewhat easier to follow visually.
Of course we also have
\be
\Gs = \sum_\lambda \gamma_\lambda \sin k_\lambda 
\ee
and
\be
\Gamma^2 = \sum_\lambda \sin^2 k_\lambda .
\ee
It is worthwhile to keep in mind during the calculations that 
$\Gamma$ and $N$ are odd in $k$, while $M$ and $W$ are even.
Expanding everything to order $a$ (which compensates the remaining $1/a$ factor
in Eq.~(\ref{eq:putalltog})) we have:
\bea
J &=& \ii g_0^2 C_F \int_{-\pi}^\pi \frac{d^d k}{(2\pi)^d} \, \sum_\rho 
\Bigg\{ \frac{2\sum_\lambda p_\lambda \Gamma_\lambda}{(2W)^2} 
(N_\rho + \ii \gamma_\rho M_\rho) \frac{-\ii \Gs + W}{\Gamma^2+W^2}
\cdot \gamma_\mu \Gamma_\nu \cdot \frac{-\ii \Gs + W}{\Gamma^2+W^2}
(N_\rho + \ii \gamma_\rho M_\rho) \nonumber \\
&& + \frac{1}{2W} \frac{p_\rho}{2} 
(M_\rho - \ii \gamma_\rho N_\rho) \frac{-\ii \Gs + W}{\Gamma^2+W^2}
\cdot \gamma_\mu \Gamma_\nu \cdot \frac{-\ii \Gs + W}{\Gamma^2+W^2}
(N_\rho + \ii \gamma_\rho M_\rho) \nonumber \\
&& + \frac{1}{2W} \frac{p_\rho}{2} 
(N_\rho + \ii \gamma_\rho M_\rho) \frac{-\ii \Gs + W}{\Gamma^2+W^2}
\cdot \gamma_\mu \Gamma_\nu \cdot \frac{-\ii \Gs + W}{\Gamma^2+W^2}
(M_\rho - \ii \gamma_\rho N_\rho) \Bigg\} .
\eea
We emphasize at this point that it has been necessary to perform the Taylor 
expansion before doing the gamma algebra, because the Dirac structure is 
``hidden'' inside the unexpanded propagators and vertices.
After a few manipulations we get
\bea
J &=& \ii g_0^2 C_F \int_{-\pi}^\pi \frac{d^d k}{(2\pi)^d}\, \sum_\rho \Bigg\{ 
\frac{2\sum_\lambda p_\lambda \Gamma_\lambda}{(2W)^2(\Gamma^2+W^2)^2}
\, \Gamma_\nu \, \Bigg[ \gamma_\mu N_\rho^2 W^2 
-\gamma_\rho \gamma_\mu \gamma_\rho M_\rho^2 W^2 -\Gs \gamma_\mu \Gs N_\rho^2
\nonumber \\
&& +\gamma_\rho \Gs \gamma_\mu \Gs \gamma_\rho M_\rho^2 
   +(\gamma_\rho \Gs \gamma_\mu + \gamma_\mu \Gs \gamma_\rho
   + \gamma_\rho \gamma_\mu \Gs + \Gs \gamma_\mu \gamma_\rho ) 
   N_\rho M_\rho W + \dots \Bigg] \nonumber \\
&& + \frac{1}{2W(\Gamma^2+W^2)^2} \, \frac{p_\rho}{2} \, \Gamma_\nu \, \Bigg[
2 \gamma_\mu N_\rho M_\rho W^2 - 2 \Gs \gamma_\mu \Gs N_\rho M_\rho
+ 2 \gamma_\rho \gamma_\mu \gamma_\rho N_\rho M_\rho W^2 \\
&& - 2 \gamma_\rho \Gs \gamma_\mu \Gs \gamma_\rho N_\rho M_\rho 
+ (\gamma_\rho \Gs \gamma_\mu + \gamma_\mu \Gs \gamma_\rho
   + \gamma_\rho \gamma_\mu \Gs + \Gs \gamma_\mu \gamma_\rho ) 
   (M_\rho^2 - N_\rho^2) W + \dots \Bigg] \Bigg\} , \nonumber
\eea
where the dots denote terms which are odd in $k$ and therefore 
do not contribute to the final result because their integral is zero
by parity. Notice that the terms which are even in $k$ possess an 
odd number of Dirac matrices, and this again goes in the right direction of 
recovering the tree-level expression.

At this stage we can finally perform the gamma algebra. However, this is one
of the most delicate points of the whole calculation, and one must be careful 
here. What happens is that the expressions are not in general tensors in the 
usual sense. Summed indices can appear more than twice in each monomial, due 
to the breaking of Lorentz invariance. Thus, the well-known formulae for doing
the gamma algebra in the continuum cannot be used straightforwardly on the 
lattice.

The lattice reduction formulae which we need in order to proceed with 
the calculation of the vertex are:
\be
\Gs \gamma_\mu \Gs = -\gamma_\mu \Gamma^2 + 2 \Gs \Gamma_\mu ,
\ee
which is just as it would be in the continuum,
\be
\sum_\rho \gamma_\rho \gamma_\mu \gamma_\rho 
= \sum_\rho \gamma_\rho (-\gamma_\rho \gamma_\mu + 2\delta_{\rho\mu} ) 
= \gamma_\mu \sum_\rho (-\gamma_\rho^2 + 2 \delta_{\rho\mu} ) ,
\ee
which is instead different from the continuum result, and
\bea
\sum_\rho \gamma_\rho \Gs \gamma_\mu \Gs \gamma_\rho 
&=& \sum_\rho \gamma_\rho (-\gamma_\mu
\Gamma^2 + 2 \Gs \Gamma_\mu) \gamma_\rho \label{eq:examplegammas} \\
&=& \sum_\rho ( \gamma_\mu \gamma_\rho^2 \Gamma^2 
- 2 \delta_{\rho\mu} \gamma_\mu \Gamma^2 -2 \Gs \gamma_\rho^2 \Gamma_\mu 
+ 4 \gamma_\rho \Gamma_\rho \Gamma_\mu ) \nonumber .
\eea
The Kronecker deltas in the formulae above, as well as the various 
$\gamma_\rho^2$ factors, are very important and for the moment they have 
to be kept, because their outcome depends on what is present in the rest 
of each monomial. This can be seen in the following examples:
\bea
\sum_\rho \gamma_\rho \gamma_\mu \gamma_\rho \cos k_\mu \cos k_\rho 
&=& \sum_\rho (-\gamma_\mu \gamma_\rho^2 + 2\delta_{\rho\mu} \gamma_\mu ) 
    \cos k_\mu \cos k_\rho \\
&=& -\gamma_\mu \cos k_\mu \sum_\rho \cos k_\rho 
+ 2\gamma_\mu \cos^2 k_\mu \nonumber \\
\sum_\rho \gamma_\rho \gamma_\mu \gamma_\rho \cos^2 k_\mu
&=& \sum_\rho (-\gamma_\mu \gamma_\rho^2 + 2\delta_{\rho\mu} \gamma_\mu ) 
   \cos^2 k_\mu\\
&=& -(4-2\epsilon) \gamma_\mu \cos^2 k_\mu + 2\gamma_\mu \cos^2 k_\mu 
 = -(2-2\epsilon) \gamma_\mu \cos^2 k_\mu . \nonumber
\eea
Similarly, a term like $\gamma_\rho \Gamma_\rho$ in the $J$ expression  
above can be contracted only if in the rest of the monomial there are no other 
functions with index $\rho$. Another gamma algebra relation that we need is
\be
 \gamma_\alpha \gamma_\beta \gamma_\gamma   
+\gamma_\gamma \gamma_\beta \gamma_\alpha 
= 2 ( \delta_{\alpha\beta} \gamma_\gamma 
     +\delta_{\beta\gamma} \gamma_\alpha
     -\delta_{\alpha\gamma} \gamma_\beta ),
\ee
so that
\be
\gamma_\rho \Gs \gamma_\mu + \gamma_\mu \Gs \gamma_\rho
   + \gamma_\rho \gamma_\mu \Gs + \Gs \gamma_\mu \gamma_\rho 
= 4 \gamma_\rho \Gamma_\mu .
\ee
Using these formulae for the reduction of $\gamma$ matrices we then obtain:
\bea
J &=& \ii g_0^2 C_F \int_{-\pi}^\pi \frac{d^d k}{(2\pi)^d} \, \Bigg\{ 
\frac{2\sum_\lambda p_\lambda \Gamma_\lambda}{(2W)^2(\Gamma^2+W^2)^2}
\, \Gamma_\nu \, \Bigg[ \sum_\rho \Bigg( \gamma_\mu N_\rho^2 W^2 
+ \gamma_\mu \Gamma^2 (N_\rho^2+M_\rho^2)
\nonumber \\
&& \qquad -2 \Gs \Gamma_\mu (N_\rho^2 +M_\rho^2) +\gamma_\mu M_\rho^2 W^2 
+4 \gamma_\rho \Gamma_\rho \Gamma_\mu M_\rho^2 
+4 \gamma_\rho \Gamma_\mu N_\rho M_\rho W \Bigg)
\nonumber \\
&& \qquad \qquad \qquad 
-2\gamma_\mu \Gamma^2 M_\mu^2 -2\gamma_\mu M_\mu^2 W^2 \Bigg] 
\nonumber \\
&& + \frac{1}{2W(\Gamma^2+W^2)^2} \, \sum_\rho \frac{p_\rho}{2} \, 
\Gamma_\nu \, \Bigg[ \Bigg( 4 \gamma_\mu \Gamma^2 N_\mu M_\mu
- 8 \gamma_\rho \Gamma_\rho \Gamma_\mu N_\rho M_\rho \\
&& \qquad  
+ 4 \gamma_\rho \Gamma_\mu (M_\rho^2 - N_\rho^2) W \Bigg) 
+ 4 \gamma_\mu N_\mu M_\mu W^2 
 \Bigg] \Bigg\} . \nonumber
\eea
We now exploit the symmetry $k \rightarrow -k$ of the integration region one 
more time: odd powers of sine functions of the various Lorentz components of 
the momentum must be combined together in such a way that only even powers, 
which have a nonzero final integral, are left:
\bea
\sum_{\lambda\rho} \sin k_\rho \sin k_\lambda &=& 
\sum_{\lambda\rho} \sin^2 k_\rho \delta_{\lambda\rho} \\
\sum_{\lambda\rho} \sin k_\rho \sin k_\lambda \sin_\mu \sin_\nu &=& 
\sum_{\lambda\rho} \sin^2 k_\rho \sin^2 k_\lambda 
(\delta_{\rho\mu} \delta_{\lambda\nu} + \delta_{\rho\nu} \delta_{\lambda\mu}) 
\qquad (\mu \neq \nu) .
\eea
Of course, since in our case we have $\mu \neq \nu$, terms of the kind 
$\sin k_\mu \sin k_\nu$ are zero when integrated and can be safely
dropped (unless they appear together with sines of other indices 
like in $\sin k_\mu \sin k_\nu \sin k_\alpha \sin k_\beta$, as in the
formula above). Examples which are relevant for the $J$ expression that
we are computing are:
\bea
\sum_{\lambda\rho} p_\lambda \Gamma_\lambda \Gamma_\nu \gamma_\mu N_\rho^2 W
&=& \sum_\rho \gamma_\mu p_\nu \Gamma_\nu^2 N_\rho^2 W , \\
\sum_\rho \gamma_\mu p_\rho \Gamma_\nu N_\mu M_\mu W^2 &=& 0 , \\
\sum_{\lambda\rho} p_\lambda \Gamma_\lambda \Gamma_\nu \Gs \Gamma_\mu
(N_\rho^2 + M_\rho^2)  
&=& \sum_\rho (p_\nu \Gamma_\nu^2 \gamma_\mu \Gamma_\mu^2
  + p_\mu \Gamma_\mu^2 \gamma_\nu \Gamma_\nu^2) (N_\rho^2 + M_\rho^2) ,
\eea
and another interesting example is
\be
\sum_\rho p_\rho \Gs \Gamma_\mu \Gamma_\rho N_\rho M_\rho 
=  \gamma_\mu p_\mu \Gamma^2_\mu \Gamma_\nu N_\nu M_\nu
 + \gamma_\nu p_\nu \Gamma^2_\nu \Gamma_\mu N_\mu M_\mu   .
\ee
It is easy to see that after this step all that is left in $J$ is multiplied
for either $\gamma_\mu p_\nu$ or $\gamma_\nu p_\mu$, while $p_\rho$ 
has disappeared altogether:
\bea
J &=& \ii g_0^2 C_F \int_{-\pi}^\pi \frac{d^d k}{(2\pi)^d} \, \Bigg\{ 
\frac{2 \gamma_\mu p_\nu \Gamma_\nu^2}{(2W)^2(\Gamma^2+W^2)^2}
\Bigg[ \sum_\rho \Bigg( (N_\rho^2+M_\rho^2) W^2 
+ \Gamma^2 (N_\rho^2+M_\rho^2) 
\nonumber \\
&& \qquad -2 \Gamma_\mu^2 (N_\rho^2 +M_\rho^2) \Bigg)
-2 \Gamma^2 M_\mu^2 -2 M_\mu^2 W^2 +4 \Gamma_\mu^2 M_\mu^2 
+4 \Gamma_\mu N_\mu M_\mu W \Bigg] \label{eq:1exp} \\
&& \qquad + \frac{2 \gamma_\nu p_\mu \Gamma_\mu^2}{(2W)^2(\Gamma^2+W^2)^2}
\Bigg[ -2 \Gamma_\nu^2 \sum_\rho (N_\rho^2 +M_\rho^2)
+4 \Gamma_\nu^2 M_\nu^2 
+4 \Gamma_\nu N_\nu M_\nu W \Bigg] \Bigg\} . \nonumber 
\eea
Notice also that the whole part proportional to $1/(2W(\Gamma^2+W^2)^2)$ has 
disappeared, because all terms were of the kind $\sin k_\mu \sin k_\nu$.

At this point we remark that we started to do our calculation only using
the operator $\gamma_\mu D_\nu$, and what we have just obtained is 
a 1-loop expression of the form 
\be
\gamma_\mu D_\nu \stackrel{\mathrm{1~loop}}{\longrightarrow} 
\frac{g_0^2}{16\pi^2} \, \ii \, 
\Big( \gamma_\mu p_\nu A + \gamma_\nu p_\mu B \Big).
\ee
It seems that we have not obtained a 1-loop expression which is proportional 
to the tree level, which is the necessary condition in order for the operator 
to be multiplicatively renormalized. However, if we now take the 
symmetrization in $\mu$ and $\nu$ into account everything is fine, because 
the operator $O_{\nu\mu}$ gives
\be
\gamma_\nu D_\mu \stackrel{\mathrm{1~loop}}{\longrightarrow} 
\frac{g_0^2}{16\pi^2} \, \ii \, 
\Big( \gamma_\mu p_\nu B + \gamma_\nu p_\mu A \Big),
\ee
with the same $A$ and $B$ of $\gamma_\mu D_\nu$, and this means that 
\be
O_{\{\mu\nu\}} = \frac{1}{2} \Big( \gamma_\mu D_\nu + \gamma_\nu D_\mu \Big)
\stackrel{\mathrm{1~loop}}{\longrightarrow} \frac{g_0^2}{16\pi^2} \, 
\frac{1}{2} \, \ii \, (\gamma_\mu p_\nu + \gamma_\nu p_\mu) (A+B) .
\ee
The symmetrized operator is thus multiplicatively renormalized (at least, 
for now, its vertex contribution), and looking at Eq.~(\ref{eq:1looplat})
we see that the lattice part of the 1-loop renormalization constant that 
matches to the continuum is given by
\be
-\gamma_{\mathrm vert}^{(0)} \cdot \log a^2p^2 
+ R^{\mathrm lat}_{\mathrm vert} = A + B . 
\ee
In practical terms, in order to get the 1-loop expression for the symmetrized 
operator, we need to exchange, in the expression in Eq.~(\ref{eq:1exp}) 
for $J$, the indices $\mu$ and $\nu$ in the terms proportional to 
$\gamma_\nu p_\mu$. We have then the result
\bea
J &=& \ii g_0^2 C_F \int_{-\pi}^\pi \frac{d^d k}{(2\pi)^d} \, \Bigg\{ 
\frac{\gamma_\mu p_\nu \Gamma_\nu^2}{2(\Gamma^2+W^2)^2}
\Bigg[ \sum_\rho (N_\rho^2+M_\rho^2) -2 M_\mu^2 \Bigg] 
+ \frac{2 \gamma_\mu p_\nu \Gamma_\mu^2 \Gamma_\nu^2}{W(\Gamma^2+W^2)^2}
\nonumber \\
&& + \frac{2 \gamma_\mu p_\nu \Gamma_\nu^2}{(2W)^2(\Gamma^2+W^2)^2}
\Bigg( \sum_\rho \Bigg( \Gamma^2 (N_\rho^2+M_\rho^2) 
-4 \Gamma_\mu^2 (N_\rho^2 +M_\rho^2) \Bigg) -2 \Gamma^2 M_\mu^2 
+8 \Gamma_\mu^2 M_\mu^2 \Bigg) \Bigg\} , \nonumber
\eea
in which we have also made some simplifications and used the trigonometric 
identity
\be
N_\rho M_\rho = \frac{1}{2} \Gamma_\rho .
\ee
We can now write the final result in a more explicit way, using also
\be
N_\rho^2+M_\rho^2 = 1 ,
\ee
and we obtain our final expression:
\bea
J &=& \ii g_0^2 C_F \gamma_\mu p_\nu \int_{-\pi}^\pi \frac{d^d k}{(2\pi)^d} \,
\Bigg\{ \frac{\sin^2 k_\nu \Big( 4 -2\epsilon -2\cos^2 \frac{k_\mu}{2} \Big)}{
2 \Big( \sum_\lambda \sin^2 k_\lambda 
+ \Big(2 \sum_\lambda \sin^2 \frac{k_\lambda}{2} \Big)^2 \Big)^2}
\nonumber \\
&&+ \frac{\sin^2 k_\mu \sin^2 k_\nu}{
\Big(\sum_\lambda \sin^2 \frac{k_\lambda}{2} \Big)
\Big( \sum_\lambda \sin^2 k_\lambda 
+ \Big(2 \sum_\lambda \sin^2 \frac{k_\lambda}{2} \Big)^2 \Big)^2}
\nonumber \\
&&+ \frac{\sin^2 k_\nu \Big( \Big( 4-2\epsilon -2\cos^2 \frac{k_\mu}{2} \Big) 
\Big( \sum_\lambda \sin^2 k_\lambda -4 \sin^2 k_\mu \Big) \Big) }{
8 \Big(\sum_\lambda \sin^2 \frac{k_\lambda}{2} \Big)^2
\Big( \sum_\lambda \sin^2 k_\lambda 
+ \Big(2 \sum_\lambda \sin^2 \frac{k_\lambda}{2} \Big)^2 \Big)^2}
\Bigg\} .
\eea
It should be noticed that the function $\cos (k_\alpha/2)$ is present 
in the final expression of the Feynman diagrams only with an even power 
(even though in the Feynman rules, for example in the quark-gluon vertex, 
Eq.~(\ref{eq:qqg}), it appears also at the first power), while $\cos k_\alpha$
can be present also with an odd power. The same will be true for the sails,
and in general for every Feynman diagram in this kind of theories.
This is due to the fact that $\cos (k_\alpha/2)$ has not a period of $2\pi$, 
and hence is not an admissible function, while $\cos^2 (k_\alpha/2)$ has a 
period of $2\pi$ indeed. Of course the sine functions, due to the symmetry of 
the integration region, appear in any case only with an even power.

The last term in the final result for $J$ is logarithmically divergent. 
The coefficient of the divergence is easily extracted, and the finite rest can
be computed with great precision using the algebraic method that we will 
introduce in Section~\ref{sec:algebraic}. We can however proceed here in a 
much faster (although less precise) way by subtracting a simple lattice 
integral which the same divergence, for example
\be
-\frac{1}{3} \int^\pi_{-\pi} \frac{d^{4-2\epsilon} k}{(2\pi)^{4-2\epsilon}} \,
\frac{1}{\Bigg( 
4 \sum_\lambda \sin^2 {\displaystyle \frac{k_\lambda}{2}}\Bigg)^2 } =
\frac{1}{16\pi^2} \, \Bigg[ -\frac{1}{3} \Bigg( -\frac{1}{\epsilon} 
- \log 4\pi + F_0 \Bigg) \Bigg] ,
\ee
which we have taken from Section~\ref{sec:algebraic}. The finite difference 
can then be computed numerically with a simple integration routine.
In this way the value of the zero-momentum integral can be obtained, and the
result is
\be
J = \ii \gamma_\mu p_\nu \,
       \frac{g_0^2}{16\pi^2} C_F \Bigg[ -\frac{1}{3} \Bigg( \frac{1}{\epsilon} 
       -F_0 +\log 4\pi \Bigg) +0.473493 \Bigg]
   = \ii \gamma_\mu p_\nu \,
       \frac{g_0^2}{16\pi^2} C_F \Bigg( -\frac{1}{3\epsilon} 
       +1.086227 \Bigg) .
\ee
The continuum part, $I-J$, is just
\bea
I - J &=& -g_0^2 C_F \frac{1}{(p-k)^2} \,
    \gamma_\rho \, \frac{-\ii \slash{k} }{k^2} \,
    \ii \gamma_\mu k_\nu \,
    \frac{-\ii \slash{k} }{k^2} \, \gamma_\rho  \\
  &=& \ii \gamma_\mu p_\nu \,
      \frac{g_0^2}{16\pi^2} C_F \Bigg[ -\frac{1}{3} \Bigg( -\frac{1}{\epsilon} 
      + \log p^2 +\gamma_E -\log 4\pi \Bigg) + \frac{5}{9} \Bigg] \nonumber \\
  &=& \ii \gamma_\mu p_\nu \,
      \frac{g_0^2}{16\pi^2} C_F \Bigg( \frac{1}{3\epsilon} 
      -\frac{1}{3} \log p^2 +1.206825 \Bigg) . \nonumber 
\eea
In principle in $I-J$ also terms of the form $p_\mu p_\nu \slash{p} /p^2$ 
appear, but they belong to the matrix element and not to the renormalization 
constant, and in any case they cancel when the difference between continuum 
and lattice renormalization factors is taken, because they have the same 
coefficient in both cases.

The final result for the Feynman diagram is~\cite{Capitani:2000wi}:
\be
I = \ii \gamma_\mu p_\nu \, \frac{g_0^2}{16\pi^2} C_F 
    \Bigg( -\frac{1}{3} \log p^2 +2.293052 \Bigg) . 
\label{eq:finalresultvertex}
\ee
We could have chosen a mass regularization instead of dimensional 
regularization (adding a mass $m^2$ to the gluon propagator). 
In this case we would have obtained
\be
J_m = \ii \gamma_\mu p_\nu \,
       \frac{g_0^2}{16\pi^2} C_F \Bigg[ -\frac{1}{3} \Bigg( 
       \log m^2 +\gamma_E -F_0 \Bigg) -0.193173 \Bigg]
   = \ii \gamma_\mu p_\nu \,
       \frac{g_0^2}{16\pi^2} C_F \Bigg( -\frac{1}{3} \log m^2 
       +1.070830 \Bigg) 
\ee
and
\be
(I - J)_m = \ii \gamma_\mu p_\nu \,
      \frac{g_0^2}{16\pi^2} C_F \Bigg[ -\frac{1}{3} \Bigg( 
      \log p^2 - \log m^2 \Bigg) + \frac{11}{9} \Bigg] .
\ee
Of course the sum of $J_m$ and $(I - J)_m$ is still given by 
Eq.~(\ref{eq:finalresultvertex}).

Looking at the expression in Eq.~(\ref{eq:putalltog}), it can be seen that 
after the $a^4$ rescaling it contains also $1/a$ terms. One could ask where 
have the $1/a$ terms gone. The fact is that they are zero for this diagram. 
They are zero also for the sails, while in the case of the self-energy the 
$1/a$ terms are finite and give an important contribution, which is 
linked to the breaking of chiral symmetry.

\subsubsection{Sails}

We now turn to the sails in Fig.~\ref{fig:proper}.
In these diagrams there is only one interaction vertex coming from QCD, 
and the gluon is the contraction of the $A_\nu$ coming from the covariant 
derivative in the operator and of the $A_\rho$ in the QCD vertex $V_\rho$, 
which therefore (in Feynman gauge) becomes $V_\nu$.

The first order expansion of the covariant derivative in the operator gives, 
as we have seen,
\be
O_{\mu\nu}^{(1)} (k,p) 
= \ii  \gamma_\mu \cos \frac{a(k+p)_\nu}{2} = \ii \gamma_\mu 
\Bigg( \cos \frac{ak_\nu}{2} -\frac{ap_\nu}{2} \sin \frac{ak_\nu}{2} \Bigg) ,
\ee
and this expression is valid for both sails. We evaluate the two sails 
together, since they are similar and some simplifications will take place 
at intermediate stages of the calculations when taking their sum.
Again, we first compute the $J$ part at zero-momentum,
\bea
J &=& \int_{-\frac{\pi}{a}}^{\frac{\pi}{a}} \frac{d^d k}{(2\pi)^d} \,
 G_{\nu\nu} (p-k) \cdot g_0  
 \Bigg[  V_\nu (k,p) \cdot S(k) \cdot O_{\mu\nu}^{(1)} (k,p)  
      + O_{\mu\nu}^{(1)} (k,p) \cdot S(k) \cdot V_\nu (p,k)  \Bigg] 
  \Bigg|_{ap=0}  \nonumber \\
&=& -\frac{\ii g_0^2}{a} C_F 
\int_{-\pi}^\pi \frac{d^d k}{(2\pi)^d} \,
\Bigg(\frac{1}{2W} + \frac{2a\sum_\lambda
p_\lambda \Gamma_\lambda}{(2W)^2} \Bigg) \Bigg[ 
\Big(N_\nu + \ii \gamma_\nu M_\nu\Big) \frac{-\ii \Gs + W}{\Gamma^2+W^2} 
\gamma_\mu \Big(M_\nu -\frac{ap_\nu}{2} N_\nu\Big) \nonumber \\  
&& \qquad \qquad \qquad \qquad 
+ \gamma_\mu \Big(M_\nu -\frac{ap_\nu}{2} N_\nu\Big)
\frac{-\ii \Gs + W}{\Gamma^2+W^2} \Big(N_\nu + \ii \gamma_\nu M_\nu\Big) 
\Bigg] ,
\eea
where we have already rescaled the integration variable as we did in the 
vertex. It is easy to see that the integral in the first line contains
an overall factor $a^3$ ($a$ from $S(k)$ and $a^2$ from $G_{\nu\nu}(p-k)$),
and after rescaling (which gives a factor $1/a^4$) one is left with an overall
$1/a$ factor. We have then, in the limit $a \to 0$:
\bea
J &=& -\ii g_0^2 C_F \int_{-\pi}^\pi \frac{d^d k}{(2\pi)^d} \, \nonumber \\ 
&& \times \Bigg\{ \frac{2\sum_\lambda p_\lambda \Gamma_\lambda}{(2W)^2} \Bigg(
(N_\nu + \ii \gamma_\nu M_\nu) \frac{-\ii \Gs + W}{\Gamma^2+W^2}
\cdot \gamma_\mu M_\nu + \gamma_\mu M_\nu \cdot
\frac{-\ii \Gs + W}{\Gamma^2+W^2} (N_\nu + \ii \gamma_\nu M_\nu) \Bigg)
\nonumber \\
&& + \frac{1}{2W} \frac{p_\nu}{2} \Bigg(
(M_\nu - \ii \gamma_\nu N_\nu) \frac{-\ii \Gs + W}{\Gamma^2+W^2}
\cdot \gamma_\mu M_\nu + \gamma_\mu M_\nu \cdot
\frac{-\ii \Gs + W}{\Gamma^2+W^2} (M_\nu - \ii \gamma_\nu N_\nu) \\
&& \qquad + 
(N_\nu + \ii \gamma_\nu M_\nu) \frac{-\ii \Gs + W}{\Gamma^2+W^2}
\cdot \gamma_\mu (-N_\nu) + \gamma_\mu (-N_\nu) \cdot    
\frac{-\ii \Gs + W}{\Gamma^2+W^2} (N_\nu + \ii \gamma_\nu M_\nu) 
\Bigg) \Bigg\} \nonumber \\
&=& - \ii g_0^2 C_F \int_{-\pi}^\pi \frac{d^d k}{(2\pi)^d} \,
\Bigg\{ \frac{2\sum_\lambda p_\lambda \Gamma_\lambda}{
(2W)^2(\Gamma^2+W^2)} 
\Bigg( 2 \gamma_\mu N_\nu M_\nu W +(\gamma_\nu \Gs \gamma_\mu + 
\gamma_\mu \Gs \gamma_\nu ) M_\nu^2 + \dots \Bigg) \nonumber \\
&& + \frac{1}{2W(\Gamma^2+W^2)} \, \frac{p_\nu}{2} \, \Bigg(
2\gamma_\mu (M_\nu^2-N_\nu^2) W -2 (\gamma_\nu \Gs \gamma_\mu 
+ \gamma_\mu \Gs \gamma_\nu ) N_\nu M_\nu
+ \dots \Bigg) \Bigg\} . \nonumber
\eea
Again we drop terms odd in $k$ and perform the gamma algebra, using 
\be
\gamma_\nu \Gs \gamma_\mu + \gamma_\mu \Gs \gamma_\nu
= 2(\gamma_\mu \Gamma_\nu + \gamma_\nu \Gamma_\mu),
\ee
which is valid for $\mu \neq \nu$. After combining the sine functions and
exchanging the indices $\mu$ and $\nu$ in the $\gamma_\nu p_\mu$ terms 
we arrive at
\bea
J &=& - \ii g_0^2 C_F \int_{-\pi}^\pi \frac{d^d k}{(2\pi)^d} \, 
\Bigg\{ \frac{1}{ 2W^2(\Gamma^2+W^2)} \gamma_\mu p_\nu
\Bigg( \Gamma_\nu^2 W + 2\Gamma_\nu^2 (M_\nu^2+M_\mu^2) \Bigg) 
\nonumber \\
&& + \frac{1}{2W(\Gamma^2+W^2)} \Bigg( \gamma_\mu p_\nu (M_\nu^2-N_\nu^2) W 
-\Gamma_\nu^2 \Bigg) \Bigg\} , 
\eea
where we have also replaced the $N_\nu M_\nu$ factor with $\Gamma_\nu/2$. 
At the end, after some further simplifications (which include the replacement
of $M_\rho^2-N_\rho^2$ with $\cos k_\rho$), we obtain the explicit form
\bea
J &=& -\ii g_0^2 C_F \gamma_\mu p_\nu \int_{-\pi}^\pi \frac{d^d k}{(2\pi)^d} \,
\Bigg\{ \frac{\cos k_\nu}{2\Big( \sum_\lambda \sin^2 k_\lambda 
+ \Big(2 \sum_\lambda \sin^2 \frac{k_\lambda}{2} \Big)^2 \Big)^2}
\nonumber \\
&&+ \frac{\sin^2 k_\nu \Big( \cos^2 \frac{k_\mu}{2} + \cos^2 \frac{k_\nu}{2}
\Big)}{
4 \Big(\sum_\lambda \sin^2 \frac{k_\lambda}{2} \Big)^2
\Big( \sum_\lambda \sin^2 k_\lambda 
+ \Big(2 \sum_\lambda \sin^2 \frac{k_\lambda}{2} \Big)^2 \Big)^2}
\Bigg\} .
\eea
This final expression for the sails gives, when integrated,
\be
J = \ii \gamma_\mu p_\nu \,
       \frac{g_0^2}{16\pi^2} C_F \Bigg[ 2 \Bigg( \frac{1}{\epsilon} 
       -F_0 +\log 4\pi \Bigg) +6.506752 \Bigg]
   = \ii \gamma_\mu p_\nu \,
       \frac{g_0^2}{16\pi^2} C_F \Bigg( \frac{2}{\epsilon} 
       +2.830350 \Bigg) ,
\ee
and
\bea
I - J &=& g_0^2 C_F \frac{1}{(p-k)^2} \,
 \Bigg[ \gamma_\nu \, \frac{-\ii \slash{k} }{k^2} \, \gamma_\mu
  +\gamma_\nu \, \frac{-\ii \slash{k} }{k^2} \, \gamma_\mu \Bigg] \\
  &=& \ii \gamma_\mu p_\nu \, 
       \frac{g_0^2}{16\pi^2} C_F \Bigg[ 2 \Bigg( -\frac{1}{\epsilon} + \log p^2
       +\gamma_E -\log 4\pi \Bigg) -4 \Bigg] \nonumber \\
  &=& \ii \gamma_\mu p_\nu \,
       \frac{g_0^2}{16\pi^2} C_F \Bigg( -\frac{2}{\epsilon} + 2 \log p^2
       -7.907617 \Bigg) . \nonumber 
\eea
The final result for the Feynman diagram is~\cite{Capitani:2000wi}:
\be
I = \ii \gamma_\mu p_\nu \, \frac{g_0^2}{16\pi^2} C_F  
    \Bigg( 2 \log p^2 -5.077267 \Bigg) . 
\ee
With a mass regularization the intermediate results would be:
\be
J_m = \ii \gamma_\mu p_\nu \,
       \frac{g_0^2}{16\pi^2} C_F \Bigg[ 2 \Bigg( 
       \log m^2 +\gamma_E -F_0 \Bigg) +6.506752 \Bigg]
   = \ii \gamma_\mu p_\nu \,
       \frac{g_0^2}{16\pi^2} C_F \Bigg( 2 \log m^2 -1.077267 \Bigg) ,
\ee
and
\be
(I - J)_m = \ii \gamma_\mu p_\nu \,
      \frac{g_0^2}{16\pi^2} C_F \Bigg[ 2 \Bigg( 
      \log p^2 - \log m^2 \Bigg) - 4 \Bigg] .
\ee

Also in this diagram the $1/a$ terms are zero.

\subsubsection{Operator tadpole}

We now consider the diagram arising from the second order expansion of the 
covariant derivative in the operator, whose Fourier transform is, in the 
special case in which the two gluons are contracted to make a tadpole,
\be
O_{\mu\nu}^{(2)} (p) = -\frac{a}{2} \, C_F \, \ii \gamma_\mu \sin ap_\nu .
\ee
The operator that enters into the tadpole depends only on the external 
quark momentum, but not on the integration variable, which is carried 
by the gluon, and so it has a simple continuum limit. 
The integral is rather easy to compute, and the result is 
\bea
I &=& \int_{-\frac{\pi}{a}}^{\frac{\pi}{a}} \frac{d^d k}{(2\pi)^d} \, 
G_{\nu\nu} (k) \cdot g_0^2 O_{\mu\nu}^{(2)} (p)  \nonumber \\
&=& g_0^2 C_F \Big( -\frac{a}{2} \ii \gamma_\mu \sin ap_\nu \Big) 
\int_{-\frac{\pi}{a}}^{\frac{\pi}{a}} \frac{d^d k}{(2\pi)^d} \,
a^2 \frac{1}{4 \displaystyle \sum_\lambda 
\sin^2 \frac{ak_\lambda}{2}}  
\nonumber \\
&=& g_0^2 C_F \Big( -\frac{1}{2} \ii \gamma_\mu \frac{\sin ap_\nu}{a} \Big)
\int_{-\pi}^\pi \frac{d^d k}{(2\pi)^d} \,
\frac{1}{4 \displaystyle \sum_\lambda \sin^2 \frac{k_\lambda}{2}} 
 \nonumber \\
&=& -\frac{1}{2} g_0^2 C_F \, \ii \gamma_\mu p_\nu
\int_{-\pi}^\pi \frac{d^d k}{(2\pi)^d} \,
\frac{1}{4 \displaystyle \sum_\lambda \sin^2 \frac{k_\lambda}{2}} .
\eea
Although this diagram comes from $O(a^2A^2)$ terms in the action which are 
naively zero in the continuum limit, the gluon loop gives an extra factor 
of $1/a^2$ and so these contributions do not disappear in the limit $a=0$ 
of the diagram.

So, this diagram is not zero, and we can see that it is also not divergent.
This finite integral can be encountered very frequently in lattice 
calculations, is in fact one of the most basic quantities of perturbation 
theory and is taken as a fundamental constant, called $Z_0$, in the algebraic
method (see Section~\ref{sec:algebraic}). This constant can in principle can 
be computed with arbitrary precision. It is now known with a very great 
precision, about 400 significant decimal places, as we will see in 
Section~\ref{sec:coordinatespacemethods}. For our calculation is enough to 
know that 
\be
\int_{-\pi}^\pi \frac{d^d k}{(2\pi)^d} \,
\frac{1}{4 \displaystyle \sum_\lambda \sin^2 \frac{k_\lambda}{2}} 
= Z_0 = 0.15493339,
\ee
so that the value of the operator tadpole is
\be
I = - \frac{g_0^2}{16\pi^2} C_F \ii \gamma_\mu p_\nu \cdot 12.233050 . 
\ee
Of course the symmetrization of this result is trivial.

The computation of operator tadpoles for more complicated operators, which can
be done in an exact way (that is, expressing the results in terms of only
two integrals known with arbitrary precision) using the algebraic method, 
is discussed in Section~\ref{sec:algebraic}.

\subsubsection{Quark self-energy (sunset diagram)}

The zero-momentum part for the sunset diagram of the quark self-energy 
(that is, without considering the tadpole) is:
\bea
J &=& \int_{-\frac{\pi}{a}}^{\frac{\pi}{a}} \frac{d^d k}{(2\pi)^d} \, 
\sum_\rho G_{\rho\rho} (p-k) \cdot 
 \Big[  V_\rho (k,p) \cdot S(k) \cdot V_\rho (p,k)  \Big] \Bigg|_{ap=0} 
\nonumber \\
&=& \frac{g_0^2}{a} C_F \int_{-\pi}^\pi \frac{d^d k}{(2\pi)^d} \,
  \sum_\rho \Bigg( \frac{1}{2W} 
  + \frac{2a\sum_\lambda p_\lambda \Gamma_\lambda}{(2W)^2} \Bigg) 
 \Bigg(N_\rho + \ii \gamma_\rho M_\rho 
    +\frac{ap_\rho}{2} (M_\rho - \ii \gamma_\rho N_\rho) \Bigg) \nonumber \\
&& \times
   \frac{-\ii \Gs + W}{\Gamma^2+W^2} \Bigg(N_\rho + \ii \gamma_\rho M_\rho 
   +\frac{ap_\rho}{2} (M_\rho - \ii \gamma_\rho N_\rho) \Bigg) ,
\eea
where we have already rescaled the integration variable. After combining the 
various factors $a$ coming from the propagator and the vertices as well as 
from the rescaling of $k$, what remains in front of the expression is an 
overall factor $1/a$. 
This means that we have to keep all terms of order $ap$. We have then
\bea
J &=& \frac{g_0^2}{a} C_F \int_{-\pi}^\pi \frac{d^d k}{(2\pi)^d} \,
   \sum_\rho \frac{1}{2W} 
   (N_\rho + \ii \gamma_\rho M_\rho) \frac{-\ii \Gs + W}{\Gamma^2+W^2} 
   (N_\rho + \ii \gamma_\rho M_\rho) \\
&+& g_0^2 C_F \int_{-\pi}^\pi \frac{d^d k}{(2\pi)^d} \, \sum_\rho \Bigg\{ 
\frac{2\sum_\lambda p_\lambda \Gamma_\lambda}{(2W)^2}
  (N_\rho + \ii \gamma_\rho M_\rho) \frac{-\ii \Gs + W}{\Gamma^2+W^2}
  (N_\rho + \ii \gamma_\rho M_\rho) \nonumber \\
&&+ \frac{1}{2W} \frac{p_\rho}{2}
\Bigg[ (M_\rho - \ii \gamma_\rho N_\rho) \frac{-\ii \Gs + W}{\Gamma^2+W^2}
(N_\rho + \ii \gamma_\rho M_\rho) \nonumber \\
&& + (N_\rho + \ii \gamma_\rho M_\rho) 
\frac{-\ii \Gs + W}{\Gamma^2+W^2} (M_\rho - \ii \gamma_\rho N_\rho) \Bigg] 
\Bigg\} , \nonumber 
\eea
which gives
\bea
J &=& \frac{g_0^2}{a} C_F \int_{-\pi}^\pi \frac{d^d k}{(2\pi)^d} \,
  \sum_\rho \frac{1}{2W(\Gamma^2+W^2)} 
\Bigg( N_\rho^2 W -\gamma_\rho^2 M_\rho^2 W 
      + (\gamma_\rho \Gs + \Gs \gamma_\rho) N_\rho M_\rho \Bigg) \nonumber \\
&+& g_0^2 C_F \int_{-\pi}^\pi \frac{d^d k}{(2\pi)^d} \, \sum_\rho \Bigg\{ 
\frac{2\sum_\lambda p_\lambda \Gamma_\lambda}{(2W)^2(\Gamma^2+W^2)}
\Bigg( -\ii \Gs N_\rho^2 + \ii \gamma_\rho \Gs \gamma_\rho M_\rho^2
+ 2 \ii \gamma_\rho N_\rho M_\rho W \Bigg) \nonumber \\
&& +\frac{1}{2W(\Gamma^2+W^2)} \frac{p_\rho}{2} 
\Bigg( 2 \ii \gamma_\rho (M_\rho^2 -N_\rho^2) W - 2 \ii \Gs N_\rho M_\rho
 - 2 \ii \gamma_\rho \Gs \gamma_\rho N_\rho M_\rho \Bigg) \Bigg\} ,
\label{eq:intmse} 
\eea
where again we have not considered terms which are odd in $k$. 
Here we have also kept the contribution proportional to $1/a$ because,
contrary to vertex and sails, for this diagram it turns out not to be zero. 
In fact, this is a very important quantity for Wilson fermions. 
It contributes to the linearly divergent $\Sigma_0/a$ term in the 1-loop 
self-energy~\footnote{Of course, since in our example we are doing the 
computations using a massless fermion propagator, we do not obtain the 
$\Sigma_2$ factor.}
\be
\frac{g_0^2}{16\pi^2} \Bigg(
\frac{\Sigma_0}{a} + \ii \slash{p} \, \Sigma_1 + m_0 \, \Sigma_2
\Bigg) ,
\ee
which is due to the breaking of chiral symmetry for Wilson fermions
(in fact it is proportional to $r$, and vanishes for naive fermions), 
and gives the critical mass to one loop:
\be
m_c = \Sigma_0 .
\ee
We can see that after reduction there are no Dirac matrices in the 
contribution to $m_c$ coming from this diagram. The corresponding integral 
is finite and given by
\bea
m_c^{(a)} &=& 
g_0^2 C_F \int_{-\pi}^\pi \frac{d^d k}{(2\pi)^d} \,
\sum_\rho \frac{1}{2W(\Gamma^2+W^2)} 
\Bigg( (N_\rho^2 -M_\rho^2 )W + \Gamma_\rho^2 \Bigg) \\
&=& g_0^2 C_F \int_{-\pi}^\pi \frac{d^d k}{(2\pi)^d} \,
\Bigg\{ \frac{\sum_\rho \cos k_\rho}{2\Big( \sum_\lambda \sin^2 k_\lambda 
+ \Big(2 \sum_\lambda \sin^2 \frac{k_\lambda}{2} \Big)^2 \Big)^2}
\nonumber \\
&& \qquad \qquad + \frac{\sum_\rho \sin^2 k_\rho}{
4 \Big(\sum_\lambda \sin^2 \frac{k_\lambda}{2} \Big)
\Big( \sum_\lambda \sin^2 k_\lambda 
+ \Big(2 \sum_\lambda \sin^2 \frac{k_\lambda}{2} \Big)^2 \Big)^2}
\Bigg\} \nonumber \\
&=& - \frac{g_0^2}{16\pi^2} C_F \cdot 2.502511
\nonumber .
\eea
This is the contribution to the critical mass coming from the sunset diagram 
of the self-energy. Another contribution to it, $m_c^{(b)}$, comes from the 
tadpole and we will compute it soon. Let us now turn to the rest of the 
expression (\ref{eq:intmse}), which contributes to the renormalization of the 
operator that we are studying. From now on $J$ will only denote the $a=0$ 
part of the zero-momentum expression, in which after reduction there is one 
Dirac matrix, and which gives $\Sigma_1$. We have then
\bea
J &=& g_0^2 C_F \int_{-\pi}^\pi \frac{d^d k}{(2\pi)^d} \, \sum_\rho \Bigg\{ 
\frac{2\sum_\lambda p_\lambda \Gamma_\lambda}{(2W)^2(\Gamma^2+W^2)}
\Bigg( -\ii \Gs (N_\rho^2+M_\rho^2) + 2 \ii \gamma_\rho \Gamma_\rho M_\rho^2
+ \ii \gamma_\rho \Gamma_\rho W \Bigg) \nonumber \\
&& +\frac{1}{2W(\Gamma^2+W^2)} p_\rho 
\Bigg( \ii \gamma_\rho (M_\rho^2 -N_\rho^2) W 
- \ii \gamma_\rho \Gamma^2_\rho \Bigg) \Bigg\} \\
&=& g_0^2 C_F \int_{-\pi}^\pi \frac{d^d k}{(2\pi)^d} \, \Bigg\{ 
\frac{\ii \slash{p} }{(2W)^2(\Gamma^2+W^2)}
\Bigg( -2 \Gamma_\nu^2 \sum_\rho (N_\rho^2+M_\rho^2) 
+ 4 \sum_\rho \Gamma_\rho^2 M_\rho^2 + 2 \Gamma_\nu^2 W \Bigg) \nonumber \\
&& +\frac{\ii \slash{p} }{2W(\Gamma^2+W^2)} 
\Bigg( (M_\nu^2 -N_\nu^2) W - \Gamma^2_\nu \Bigg) \Bigg\} .
\eea
In the last passage we have used the substitution
\be
\sum_\lambda \gamma_\lambda p_\lambda \int f_\lambda (k) =
\slash{p} \int f_\mu (k) ,
\ee
since this kind of integrals does not depend on the direction, with the 
understanding that the index $\mu$ is fixed and must not appear in the rest of
the monomial. This reconstructs the $\ii \slash{p}$ factor and allows the 
extraction of the value of $\Sigma_1$. We thus have:
\bea
J &=& g_0^2 C_F \ii \slash{p} 
\int_{-\pi}^\pi \frac{d^d k}{(2\pi)^d} \,
\Bigg\{ \frac{\cos k_\nu}{2\Big( \sum_\lambda \sin^2 k_\lambda 
+ \Big(2 \sum_\lambda \sin^2 \frac{k_\lambda}{2} \Big)^2 \Big)^2}
\nonumber \\
&&+ \frac{-(4-2\epsilon)\sin^2 k_\nu +2 \sum_\rho \sin^2 k_\rho
\cos^2 \frac{k_\rho}{2}}{
8 \Big(\sum_\lambda \sin^2 \frac{k_\lambda}{2} \Big)^2
\Big( \sum_\lambda \sin^2 k_\lambda 
+ \Big(2 \sum_\lambda \sin^2 \frac{k_\lambda}{2} \Big)^2 \Big)^2}
\Bigg\} .
\eea
The lattice result at zero momentum is then
\be
J = \ii \slash{p} \,
       \frac{g_0^2}{16\pi^2} C_F \Bigg( \frac{1}{\epsilon} 
       -F_0 +\log 4\pi +4.411364 \Bigg)
   = \ii \slash{p} \,
       \frac{g_0^2}{16\pi^2} C_F \Bigg( \frac{1}{\epsilon} 
       +2.573163 \Bigg) .
\ee
We have still to consider the continuum ($I-J$):
\bea
I - J &=& -g_0^2 C_F \frac{1}{(p-k)^2} \,
   \gamma_\rho \, \frac{-\ii \slash{k} }{k^2} \, \gamma_\rho \\
     &=& \ii \slash{p} \,
       \frac{g_0^2}{16\pi^2} C_F \Bigg( -\frac{1}{\epsilon} 
       + \log p^2 +\gamma_E -\log 4\pi -1 \Bigg) \nonumber \\
     &=& \ii \slash{p} \,
       \frac{g_0^2}{16\pi^2} C_F \Bigg( -\frac{1}{\epsilon}
       + \log p^2 -2.953809 \Bigg) . \nonumber 
\eea
The final result for the $\ii \slash{p}$ part of the sunset diagram 
of the lattice self-energy is then:
\be
\Sigma_1^{(a)} = \frac{g_0^2}{16\pi^2} C_F 
    \Bigg( \log p^2 -0.380646 \Bigg) . 
\ee
This term, together with the analog term coming from the self-energy tadpole
which we will compute soon, needs to be added to the results of the proper 
diagrams (vertex, sails and operator tadpole). In fact this is the wave 
function renormalization, and inserted into the tree-level diagram for the 
operator gives a leg correction:
\be
\ii \slash{p} \, \Sigma_1 \cdot S(p) \cdot \ii \gamma_\mu p_\nu
= \ii \slash{p} \, \Sigma_1 \cdot
\frac{-\ii \slash{p} }{p^2} \cdot \ii \gamma_\mu p_\nu
= \ii \gamma_\mu p_\nu \, \Sigma_1 
\ee

If a mass regularization is used one gets
\be
J_m = \ii \slash{p} \,
       \frac{g_0^2}{16\pi^2} C_F \Bigg( 
       \log m^2 +\gamma_E -F_0 +5.411364 \Bigg)
    = \ii \slash{p} \,
       \frac{g_0^2}{16\pi^2} C_F \Bigg( \log m^2 +1.619354 \Bigg) ,
\ee
and
\be
(I - J)_m = \ii \slash{p} \,
      \frac{g_0^2}{16\pi^2} C_F \Bigg( \log p^2 - \log m^2 - 2 \Bigg) .
\ee

\subsubsection{Quark self-energy (tadpole diagram)}

The last diagram, which completes the lattice self-energy and has no
analog in the continuum, is the self-energy tadpole, which originates from
the irrelevant vertex in Eq.~(\ref{eq:qqgg}). This vertex, when put in the 
tadpole diagram, depends only on the external momentum.
The diagram is then~\footnote{Note that we have used 
$\lim_{a \to 0} \sum_\rho \cos ap_\rho = 4$.} 
\bea
I &=& \frac{1}{2} 
\int_{-\frac{\pi}{a}}^{\frac{\pi}{a}} \frac{d^d k}{(2\pi)^d} \, 
\sum_\rho G_{\rho\rho} (k) \cdot (V^{aa}_2)_{\rho\rho} (p,p) \nonumber \\
&=& \frac{1}{2} \int_{-\frac{\pi}{a}}^{\frac{\pi}{a}} \frac{d^d k}{(2\pi)^d} \,
a^2 \frac{1}{4 \displaystyle \sum_\lambda \sin^2 \frac{ak_\lambda}{2}} 
\Big( - \frac{1}{2} a g_0^2 \, \sum_a \{T^a,T^a\}_{cc} \Big)   
\sum_\rho \Bigg( -\ii \gamma_\rho \sin ap_\rho + \cos ap_\rho \Bigg)
\nonumber \\
&=& - \frac{1}{2} g_0^2 C_F \int_{-\pi}^\pi \frac{d^d k}{(2\pi)^d} \,
\frac{1}{4 \displaystyle \sum_\lambda \sin^2 \frac{k_\lambda}{2}}  
\, \Bigg( -\ii \slash{p} +\frac{4}{a} \Bigg)
\nonumber \\
&=&
- \frac{1}{2} g_0^2 C_F Z_0 \, \Bigg( -\ii \slash{p} +\frac{4}{a} \Bigg) .
\eea
The last term diverges like $1/a$, and therefore is part of $\Sigma_0$.
In fact it gives a substantial contribution to the critical mass:
\be
m_c^{(b)} = - g_0^2 C_F \cdot 2 Z_0 
  = - \frac{g_0^2}{16\pi^2} C_F \cdot 48.932201 .
\ee
The 1-loop critical mass for Wilson fermions is then:~\footnote{We will later 
see that the 1-loop critical mass can be expressed in terms of only two 
constants (Eq.~(\ref{eq:mcconst})), which are calculable with very high 
precision. Its 2-loop value is given at the end of 
Section~\ref{sec:coordinatespacemethods}.}
\be
m_c = m_c^{(a)} + m_c^{(b)} = - \frac{g_0^2}{16\pi^2} C_F \cdot 51.434712
= - g_0^2 C_F \cdot 0.325714 .
\ee

The term proportional to $\ii \slash{p}$ gives the contribution 
of the self-energy tadpole to the renormalization of the first moment 
operator:
\be
\Sigma_1^{(b)} = \frac{g_0^2}{16\pi^2} C_F \cdot 12.233050 .
\label{eq:valuetadpole}
\ee

The total result for the 1-loop self-energy on the lattice in the massless
case is then
\be
\Sigma_1 = \Sigma_1^{(a)} + \Sigma_1^{(b)} = \frac{g_0^2}{16\pi^2} C_F 
\Bigg( \log p^2 +11.852404 + (1-\alpha) \Big( -\log p^2 +4.792010 \Big) \Bigg),
\label{eq:quarkself}
\ee
where we have also included the part which one would obtain if the calculation
were carried out in a general covariant gauge.

\subsubsection{Concluding remarks}

We have thus computed all diagrams which are necessary for the 1-loop 
renormalization of the operator $O_{\{\mu\nu\}}=\overline{\psi} \gamma_{\{\mu}
D_{\nu\}} \frac{\lambda^\alpha}{2} \psi$, with different indices. Collecting 
the results of the various diagrams together, we have the final 
result~\cite{Capitani:2000wi}
\be
\langle q | O_{\{0 1\}} | q \rangle \Big|_{\mathrm 1~loop}
= \frac{1}{2} \ii ( \gamma_0 p_1 + \gamma_1 p_0 ) \cdot
\frac{g_0^2}{16\pi^2} C_F \Bigg( \frac{8}{3} \log a^2p^2 -3.16486 \Bigg) .
\ee
This allows us to specify Eq.~(\ref{eq:1looplat}) for this operator, with 
$R_{ij}^{\mathrm lat}=-3.16486 \cdot C_F$. For the 1-loop matching to the 
$\ms$ scheme one needs also to know the $R_{ij}^{\ms}$ factor in 
Eq.~(\ref{eq:1loopcont}). Its value is $-40/9 \cdot C_F$, and can be inferred 
from the calculations of the various $I-J$ integrals that we have just done, 
by summing the contributions of vertex, sails and the sunset diagram of the 
self-energy (which are $5/9$, $-4$ and $-1$ respectively). We then obtain for 
this operator 
\bea
\langle q | O_{\{0 1\}}^{\mathrm lat} | q \rangle &=& \Bigg( 1
+\frac{g_0^2}{16 \pi^2} C_F \Big( \frac{8}{3} \log a^2p^2 -3.16486 
\Big) \Bigg) \cdot \langle q | O_{\{0 1\}}^{\mathrm tree} | q \rangle  \\
\langle q | O_{\{0 1\}}^{\ms} | q \rangle &=& \Bigg( 1 
+\frac{g_{\ms}^2}{16 \pi^2} C_F \Big( \frac{8}{3} \log \frac{p^2}{\mu^2} 
-\frac{40}{9} \Big) \Bigg) \cdot \langle q | O_{\{0 1\}}^{\mathrm tree} 
| q \rangle ,
\eea
so that the renormalization factor that converts the raw lattice results 
(in the Wilson formulation) to the $\ms$ scheme is (for $\mu=1/a$) 
\be
\langle q | O_{\{0 1\}}^{\ms} | q \rangle = \Big( 1 -\frac{g_0^2}{16 \pi^2} 
C_F \cdot 1.27958 \Big) \cdot \langle q | O_{\{0 1\}}^{\mathrm lat} 
| q \rangle .
\ee
For the typical value $g_0=1$ (which corresponds to a scale $1/a$ of about 
2 GeV in the quenched approximation) we then obtain
\be
\langle q | O_{\{0 1\}}^{\ms} | q \rangle = 0.98920 \cdot 
\langle q | O_{\{0 1\}}^{\mathrm lat} 
| q \rangle .
\ee

We would now like to make some comments on these lattice calculations.
First we notice that the magnitude of the self-energy tadpole is much larger 
than the result for the vertex and the sails, and this is a general feature of 
lattice calculations, which is called ``tadpole dominance'' of perturbation 
theory. It has been in some cases used to estimate the value of the radiative 
corrections of matrix elements, by neglecting all diagrams other than the 
tadpole.

However, even if the tadpole of the self-energy is large, sometimes this is 
not the end of the story. An uncritical application of tadpole dominance can 
be at times misleading. For example, in the calculations that we have just 
done we have encountered a situation in which the operator tadpole, which also
gives a large number, exactly cancels the self-energy tadpole, so that the 
final result for the matrix element is given only from the contributions of 
the vertex and the sails (plus the sunset diagram of the self-energy). 
This cancellation only happens for the first moment. For higher moments, 
that is the case in which we have an operator with $n$ covariant derivatives 
with indices all different from each other, the result of the operator tadpole
is $n Z_0/2$ (see also Section~\ref{sec:operatortadpoles}),~\footnote{This 
result is only valid for $n$ between 2 and 4, as for higher $n$ is impossible 
to have all indices distinct from each other, and the contractions of the 
$A_\mu$'s become more complicated and do not give just $n Z_0/2$. The numerical
result for $n \ge 5$ is however not far from this number, and one has again a 
final positive result when all diagrams are summed, with dominance of the 
operator tadpole and not of the self-energy tadpole.}
so the final result has even a different sign from the one that would be 
inferred from computing the self-energy tadpole alone. 

As discussed in Section~\ref{sec:operatormixing}, there exists also another 
class of operators which measure the first moment of the quark momentum 
distribution and which belong to another representation of the hypercubic 
group:
\be
O_{00} -\frac{1}{3} (O_{11}+O_{22}+O_{33}).
\ee
Due to the breaking of the Lorentz invariance, on the lattice this operator 
has a renormalization constant different from $O_{\{0 1\}}$. We leave as an 
exercise for the interested reader to reproduce the final 1-loop 
result~\footnote{It might turn useful to know that the numerical results for 
all individual diagrams are the same as in the calculation with different 
indices, except for the vertex diagram, in which the number for the finite 
part is now $3.575320$.}
\be
\langle q | O_{00} -\frac{1}{3} \sum_{i=1}^3 O_{ii} | q \rangle 
\Big|_{\mathrm 1~loop} = \frac{1}{2} \ii 
\Big( \gamma_0 p_0 -\frac{1}{3} \sum_{i=1}^3 \gamma_i p_i \Big) \cdot
\frac{g_0^2}{16\pi^2} C_F \Bigg( \frac{8}{3} \log a^2p^2 -1.88259 \Bigg) ,
\ee
which gives the 1-loop matching factor to the $\ms$ scheme
\be
\langle q | (O_{00} -1/3 \sum_{i=1}^3 O_{ii})^{\ms} | q \rangle = 
0.97837 \cdot 
\langle q | (O_{00} -1/3 \sum_{i=1}^3 O_{ii})^{\mathrm lat} 
| q \rangle .
\ee

We want to conclude this Section by mentioning that in the case of a 
calculation with the improved action we have to add the 
Sheikholeslami-Wohlert improved vertex of Eq.~(\ref{eq:impvertex}), 
and also the operators have to be improved. 
This renders the calculations much more cumbersome. 
Since the Sheikholeslami-Wohlert vertex includes a $\sigma$ matrix, more 
Dirac matrices appear in the manipulations. In the case in which both vertices
in the vertex function are taken to be the improved ones, the chains of Dirac 
matrices can become quite long.

The improvement of the operators, in the case that we have just calculated, 
means that we have to consider also the contribution to the renormalization
constants coming from the operators in Eq.~(\ref{eq:imprfirstmom}).
These calculations are quite complicated, and the results are given 
in~\cite{Capitani:2000xi}.

\subsection{Example of overlap results}
\label{sec:exoverlap}

To follow an overlap calculation step by step in the same way as we did for the
Wilson case would be quite cumbersome. We give here, as an example of results, 
the analytic expressions for the tadpole of the self-energy of the quark,
$\Sigma_1^{(b)}$, and for the vertex of the scalar current 
$\overline{\psi} \psi$ (in the Feynman gauge, for $r=1$). In order to be able 
to write them in a compact form it is convenient to introduce further 
abbreviations:
\bea
B &=& b(k) = 2 \sum_\lambda \sin^2 \frac{k_\lambda}{2} -\rho, \\
D &=& 2 \rho  \Big( \omega (k) + b(k) \Big) , \\
A &=& \frac{\omega^2(k)}{\rho^2}= 1-\frac{4}{\rho} \sum_\lambda \sin^2 
\frac{k_\lambda}{2} +\frac{1}{\rho^2} \Bigg( \sum_\lambda \sin^2 k_\lambda 
+ \left( 2 \sum_\lambda \sin^2 \frac{k_\lambda}{2} \right)^2 \Bigg)  ,
\eea
with $\omega(k)$ given in Eq.~(\ref{eq:omega}). The result for the 1-loop 
tadpole of the quark self-energy for overlap fermions is then given by
\bea
&& \frac{1}{2} g_0^2 \int \frac{d^4k}{2\pi^4} G_{\mu\mu}(k) 
\Bigg( 1 - \frac{4}{\rho} \Bigg) + g_0^2 \int \frac{d^4k}{2\pi^4} 
G_{\mu\mu}(k) \frac{1}{\rho^2 (1+\sqrt{A})^2} 
\label{eq:tadpoverlap} \\
&& \quad \times \sum_\lambda \Bigg[ M_\lambda^2+N_\lambda^2 
+ \Bigg(1+\frac{1}{\sqrt{A}} \Bigg) \Bigg( - \Gamma_\mu^2 
+ B (M_\mu^2-N_\mu^2) \Bigg) + \frac{2+\sqrt{A}}{\rho\sqrt{A}} \Bigg(  
- B (M_\lambda^2-N_\lambda^2)  + \Gamma^2 \Bigg) \Bigg]  \nonumber \\
&& + g_0^2 \int \frac{d^4k}{2\pi^4} G_{\mu\mu}^2(k) \frac{1}{\rho^2 
(1+\sqrt{A})^2} \sum_\lambda \Bigg[ -2 \Gamma_\mu^2 N_\lambda^2  
+ \Bigg(1+\frac{1}{\sqrt{A}} \Bigg) \Bigg( 2 \Gamma_\mu^2 
( B + 2 M_\mu^2 - M_\lambda^2 ) \Bigg) \Bigg] .  \nonumber 
\eea
The first term comes from the part of the overlap vertex $V_2$ in 
Eq.~(\ref{eq:overlapqqgg}) containing $W_2$ and $W_2^\dagger$, 
and its value for $\rho=1$ is quite large,
\be
\frac{1}{2} g_0^2 \cdot Z_0 \Bigg( 1 - \frac{4}{\rho} \Bigg) \Bigg|_{\rho=1}
= -\frac{g_0^2}{16\pi^2} \, 36.69915 ,
\ee
while the 1-loop result for the whole self-energy tadpole in 
Eq.~(\ref{eq:tadpoverlap}) is slightly smaller, but still large:
\be
-\frac{g_0^2}{16\pi^2} \, 23.35975 .
\ee
Adding now the value $-14.27088 \, g_0^2/(16\pi^2)$ for the sunset 
diagram of the overlap self-energy of the quark, which is much harder to 
compute by hand and would produce a very lengthy analytic expression, gives 
the result $-37.63063 \, g_0^2/(16\pi^2)$ for the complete self-energy in the 
Feynman gauge, for $\rho=1$~\cite{Alexandrou:2000kj,Capitani:2000wi}. 
The values of the overlap self-energy for various choices of $\rho$ and in
a general covariant gauge are given in Table~\ref{tab:overlapself}.
One advantage of using overlap fermions is that $\Sigma_0$, the
power divergent part of the self-energy that for Wilson fermions gives
a nonzero additive mass renormalization, vanishes here.

\begin{table}[t]
\begin{center}
\begin{tabular}{|r|r|r|r|} 
\hline
$\rho$ & self-energy (sunset) & self-energy (tadpole) & total self-energy
\\ 
\hline
0.2 &  -27.511695 +11.911596 $\xi$  &  -213.087934 -7.119586 $\xi$  & 
      -240.599629 +4.792010 $\xi$  \\
0.3 &  -23.687573 +11.098129 $\xi$  &  -131.723110 -6.306119 $\xi$  &  
      -155.410693 +4.792010 $\xi$  \\
0.4 &  -21.172454 +10.520210 $\xi$  &   -91.817537 -5.728200 $\xi$  &  
      -112.989991 +4.792010 $\xi$  \\
0.5 &  -19.337313 +10.071356 $\xi$  &   -68.315503 -5.279346 $\xi$  &   
       -87.652816 +4.792010 $\xi$  \\
0.6 &  -17.912921  +9.704142 $\xi$  &   -52.931363 -4.912132 $\xi$  &   
       -70.844284 +4.792010 $\xi$  \\
0.7 &  -16.760616  +9.393275 $\xi$  &   -42.140608 -4.601265 $\xi$  &   
       -58.901224 +4.792010 $\xi$  \\
0.8 &  -15.800204  +9.123666 $\xi$  &   -34.193597 -4.331656 $\xi$  &   
       -49.993801 +4.792010 $\xi$  \\
0.9 &  -14.981431  +8.885590 $\xi$  &   -28.125054 -4.093580 $\xi$  &   
       -43.106485 +4.792010 $\xi$  \\
1.0 &  -14.270881  +8.672419 $\xi$  &   -23.359746 -3.880409 $\xi$  &   
       -37.630627 +4.792010 $\xi$  \\
1.1 &  -13.645294  +8.479438 $\xi$  &   -19.534056 -3.687428 $\xi$  &   
       -33.179350 +4.792010 $\xi$  \\
1.2 &  -13.087876  +8.303183 $\xi$  &   -16.407174 -3.511173 $\xi$  &   
       -29.495050 +4.792010 $\xi$  \\
1.3 &  -12.586126  +8.141044 $\xi$  &   -13.813486 -3.349034 $\xi$  &   
       -26.399612 +4.792010 $\xi$  \\
1.4 &  -12.130497  +7.991018 $\xi$  &   -11.635482 -3.199008 $\xi$  &    
       -23.765979 +4.792010 $\xi$  \\
1.5 &  -11.713524  +7.851554 $\xi$  &    -9.787582 -3.059544 $\xi$  &   
       -21.501106 +4.792010 $\xi$  \\
1.6 &  -11.329238  +7.721442 $\xi$  &    -8.206069 -2.929432 $\xi$  &   
       -19.535307 +4.792010 $\xi$  \\
1.7 &  -10.972744  +7.599750 $\xi$  &    -6.842630 -2.807740 $\xi$  &   
       -17.815374 +4.792010 $\xi$  \\
1.8 &  -10.639905  +7.485778 $\xi$  &    -5.660084 -2.693768 $\xi$  &   
       -16.299989 +4.792010 $\xi$  \\
1.9 &  -10.327042  +7.379023 $\xi$  &    -4.629539 -2.587013 $\xi$  &   
       -14.956581 +4.792010 $\xi$  \\
\hline
\end{tabular} 
\caption{Results for the finite constant of the quark self-energy with 
overlap fermions in a general covariant gauge, from~(Capitani, 2001a).
We have used the abbreviation $\xi=1-\alpha$. The first column (sunset) 
refers to the diagram on the left in Fig.~\ref{fig:quark_self}, and the second 
column (tadpole) to the diagram on the right.}
\label{tab:overlapself}
\end{center}
\end{table}

The results for the 1-loop vertex diagram of the scalar operator~\footnote{We
note that the sails are not present in this case, as there are no covariant 
derivatives.} can be written in the form
\be
g_0^2 \int \frac{d^4k}{2\pi^4} G(p-k) \frac{1}{(1+\sqrt{A})^2}
\Bigg[ \Bigg( -\frac{\Gamma^2}{D^2} + \frac{1}{4 \rho^2} \Bigg) \cdot X 
+ \frac{1}{\rho D} \cdot Y \Bigg] 
\ee
for small $p$, where
\bea
X &=& \sum_\lambda \Bigg[ - (M_\lambda^2 - N_\lambda^2) 
+ \frac{1}{\rho \sqrt{A}} 2  (M_\lambda^2 + N_\lambda^2) 
+ \frac{1}{\rho^2 A} \Bigg( (\Gamma^2 -B^2) (M_\lambda^2 -N_\lambda^2) 
+2 B \Gamma^2 \Bigg) \Bigg] \nonumber \\
Y &=&  \sum_\lambda \Bigg[ \Gamma^2 + \frac{1}{\rho \sqrt{A}} 2 \Gamma^2 
(M_\lambda^2 + N_\lambda^2) 
+ \frac{1}{\rho^2 A} \Gamma^2 \Bigg( -2 B (M_\lambda^2 -N_\lambda^2) 
+ \Gamma^2 -B^2 \Bigg) \Bigg] .
\eea

These are the only diagrams for which the author has had enough patience 
to perform their calculation by hand using the overlap 
action~\cite{Capitani:2000wi}. Other overlap calculations of renormalization
factors made using FORM codes are reported there and 
in~\cite{Capitani:2000aq,Capitani:2000da,Capitani:2000bm}.

The renormalization constants of operators computed using overlap fermions 
are sometimes large, and in general larger than the corresponding Wilson 
results~\cite{Capitani:2000wi,Capitani:2000aq}. For example for the first 
moment of the unpolarized quark distribution (operator $O_{\{0 1\}}$) the 
constant is $-53.25571$, while (as we have just seen) in the Wilson case is 
$-3.16486$. However, for the proper diagrams the overlap results show only
relatively small differences from the Wilson numbers. The biggest contribution
to the renormalization constant comes from the operator tadpole, and it is 
exactly the same for overlap and Wilson fermions. The difference between 
overlap and Wilson results is then almost entirely due to the quark 
self-energy. In the Feynman gauge, the constant of the finite part of 
the self-energy in the overlap (for $\rho=1$) is $-37.63063$, while for Wilson
fermions is $+11.85240$; their difference is a large number, $-49.48303$, 
and quite close to the difference of the total renormalization constants for
the two kinds of fermions, $-50.09085$. We can notice from 
Table~\ref{tab:overlapself} that the value of the overlap self-energy 
decreases when $\rho$ increases, and if one would consider the overlap 
fermions for $\rho=1.9$, the difference between the self-energies would go 
from $-49.48303$ down to $-26.80898$. 
However, the quark propagator becomes singular for $\rho=2$, so simulations 
would likely be more expensive when approaching this value of $\rho$.

We note that, contrary to the Wilson case, where the tadpole gives a 
much larger contribution than the sunset, in the overlap things are more 
entangled, and for certain values of $\rho$ it is the sunset that gives
a larger result than the tadpole.

\subsection{Tadpole improvement}
\label{sec:tadpoleimprovement}

We have seen that gluon tadpoles give large numerical results compared to other
diagrams (although sometimes they happen to cancel with each other). Quite 
often the large corrections which occur in lattice perturbation theory are 
caused by these tadpole diagrams. Since they are an artifact of the lattice
(the corresponding interaction vertex is zero in the naive continuum limit),
in order to attempt to make the lattice perturbative expansions closer to the 
continuum ones a tadpole resummation method has been proposed 
in~\cite{Parisi:1980pe} and~\cite{Lepage:1992xa}. The lectures 
of~\cite{Mackenzie:wr} contain a pedagogical introduction to these ideas.

This resummation of tadpoles amounts to a mean-field improvement, in which 
one makes a redefinition of the link, separating the contributions of their 
infrared modes:
\be
U_\mu (x) = \e^{\ii g_0 a A_\mu (x)} = 
u_0 \, \e^{\ii g_0 a A_\mu^{IR} (x)} = u_0 \, \widetilde{U}_\mu (x) .
\ee
The rescaling factor $u_0$ (which is a number between 0 and 1) contains
the high-energy part of the link variables. One makes in the gluon action 
the substitution
\be
U_\mu (x) = u_0 \, \widetilde{U}_\mu (x) ,
\ee
and takes $\widetilde{U}_\mu (x)$ as the new link variable. This implies
$\widetilde{\beta} = u_0^4 \beta$. The effective coupling becomes then
$\widetilde{g}_0^2 = g_0^2/u_0^4$, and is claimed to be closer to the coupling 
defined in the $\ms$ scheme than the original lattice coupling. In fact, the 
perturbative expansions in terms of $\widetilde{g}_0$ have often smaller 
coefficients and are better behaved than the standard perturbative expansions.

The value of the rescaling factor $u_0$ is taken from the mean value of the 
link in the Landau gauge,
\be
u_0 = \Bigg\langle \frac{1}{N_c} \, \mathrm{Re} \, \Tr U_\mu (x) \Bigg\rangle,
\ee
because in this gauge its value is higher (in other gauges $u_0$ can assume 
very small values). However, the fourth root of the mean value of the 
plaquette,
\be
u_P = \Bigg\langle \frac{1}{N_c} \, \mathrm{Re} \, \Tr P_{\mu\nu} (x) 
\Bigg\rangle^{1/4},
\ee
is easier to compute and is then taken in place of $u_0$ in most applications,
although one should always keep in mind that this is a good approximation only
for small lattice spacings. When $a \sim 0.4$ fm the two definitions differ 
already by about 10\%. In this case, if one sticks to the plaquette 
definition, every perturbative quantity should be computed at least to 2 loop. 
There is then a small ambiguity in this choice.

In the resummed theory one works with tadpole-improved actions and operators. 
Of course all quantities that contain a field $U$ have to be rescaled 
accordingly, and will be multiplied by some positive or negative power of 
$u_0$. This then applies also to the covariant derivatives appearing in 
operators measuring moments of structure functions. When all variables are 
properly rescaled the large tadpole contributions can be reabsorbed, and after
this tadpole improvement has been implemented the coefficients in the results 
of perturbative lattice calculations get in general smaller.

The above construction can be seen as a different choice of the strong 
coupling (and of other parameters), and is equivalent to a reorganization of 
lattice perturbation theory. At the end of the day, this is a theory in which 
a redefinition of the coupling is taking place, and then it has a different 
value of the $\Lambda$ parameter.~\footnote{This is similar to what happens 
in continuum QCD renormalized in the minimal subtraction ($\mathrm{MS}$) 
scheme, where some second-order perturbative corrections can sometimes be 
large. If however one systematically drops the factors $\gamma_E$ and 
$\log 4\pi$ these corrections become much smaller, and this defines the 
$\ms$ scheme, where one has a different renormalized coupling and a different
value of the $\Lambda$ parameter.} 
So, one should be careful and do things consistently, because one is 
effectively changing scheme. This would not matter if one knew the complete  
series, but at the lowest orders it makes a difference. It is however
difficult to estimate the error that results from this. Moreover,
not always this different definition of the coupling gives a better
perturbative expansions. Not in all cases the contributions get smaller.

It is possible to set up tadpole improvement also on Symanzik-improved 
theories, rescaling the affected quantities with appropriate powers of $u_0$. 
For example, the improvement coefficient $c_{sw}$ undergoes the rescaling
$\widetilde{c}_{sw} = u_0^3 \, c_{sw}$. The coefficients of the 
L\"uscher-Weisz improved gauge action are modified as follows: 
\be
c_0 = \frac{5}{3u_0^4}, \qquad c_1 = - \frac{1}{12u_0^6} .
\ee

A different kind of improvement of lattice perturbation theory has been 
proposed and used in~\cite{Panagopoulos:1998xf,Panagopoulos:1998vc}.
It works through the resummation of ``cactus'' diagrams, i.e., tadpoles 
which become disconnected if any one of their vertices is removed. These 
diagrams are gauge invariant and it seems that in some cases, for example
the renormalization of the topological charge, this kind of resummation 
achieves better results than standard tadpole improvement.

We conclude mentioning that for overlap fermions simple tadpole improvements
like the ones discussed here seem not to be of much help, because also the 
sunset diagram of the self-energy gives large results, and this is a diagram 
that is present also in the continuum. Probably a different kind of diagram 
resummation has to be devised in this case.

\subsection{Perturbation theory for fat links}
\label{sec:fatlinks}

Another way to reduce the magnitude of the large renormalization factors 
seems to be given, at least in some cases, by fat link 
actions~\cite{DeGrand:1998mn}. Under the name of fat links are meant actions 
in which the quarks couple to gauge links which are smeared (see 
Fig.~\ref{fig:fatlink}). They present some advantages, like the facts that 
exceptional configurations are suppressed and the additive mass 
renormalization is small. Furthermore, these actions exhibit better chiral 
properties. From the point of view of perturbation theory the interesting 
feature is that renormalization factors often have values quite close to 
unity, and hence this could be worth of some consideration in the cases in 
which the 1-loop corrections of matrix elements on the lattice are large 
compared to their tree-level values. Damping the large perturbative 
corrections can be particularly useful in the case of overlap fermions. 
Simulations using overlap-like fermions with fat links have been reported
in~\cite{DeGrand:2000tf}, and the improvement of the locality and topological 
properties using overlap and overlap-like fermions with fat links 
has been studied in~\cite{DeGrand:2002vu} and~\cite{Kovacs:2002nz}.

\begin{figure}[t]
\begin{center}
\begin{picture}(300,160)(0,0)
\Line(145,30)(50,30)
\Line(50,30)(50,130)
\LongArrow(50,130)(145,130)
\Line(155,30)(250,30)
\Line(250,30)(250,130)
\LongArrow(250,130)(155,130)
\Text(150,10)[b]{$x$}
\Text(50,10)[b]{$x-\hat{\nu}$}
\Text(250,10)[b]{$x+\hat{\nu}$}
\Text(150,140)[b]{$x+\hat{\mu}$}
\Text(50,140)[b]{$x-\hat{\nu}+\hat{\mu}$}
\Text(250,140)[b]{$x+\hat{\nu}+\hat{\mu}$}
\Text(55,80)[l]{$\displaystyle \frac{c}{6}$}
\Text(155,80)[l]{$c$}
\Text(245,80)[r]{$\displaystyle \frac{c}{6}$}
\SetWidth{1}
\LongArrow(150,30)(150,130)
\end{picture}
\end{center}
\caption{\small The APE smearing of the thin link $U_\mu (x)$ which produces 
the fat link.}  
\label{fig:fatlink}
\end{figure}

We follow~\cite{Bernard:1999kc}, and we consider the particular construction
known as APE blocking~\cite{Albanese:ds}, where the smearing of a link is done 
as follows. Starting with the original link, also known as thin link,
\be
V_\mu^{(0)} (x) = U_\mu (x) ,
\ee
the first smearing step is done like in Fig.~\ref{fig:fatlink}.
The general smearing step constructs the fat link recursively as
\bea
V_\mu^{(m+1)} (x) &=& {\cal P} \, \Big( (1-c) V_\mu^{(m)} (x) + \frac{c}{6}
\sum_{\nu \neq \mu} \Big[
   V_\nu^{(m)} (x) V_\mu^{(m)} (x + \hat{\nu}) 
     V_\nu^{\dagger (m)} (x + \hat{\mu}) \\
&& \qquad \qquad \qquad \qquad
 + V_\nu^{\dagger (m)} (x - \hat{\nu}) V_\mu^{(m)} (x - \hat{\nu})
     V_\nu^{(m)} (x - \hat{\nu} + \hat{\mu}) \Big] \Big) , \nonumber 
\eea
where ${\cal P}$ projects back into $SU(3)$ matrices. A typical value that 
is used is $c=0.45$, and a typical value for the total number of iterations is 
$N=10$ (see for example~\cite{Bernard:1999ic}).

In perturbation theory the basic variables are the $A_\mu$'s, and for 1-loop 
computations of quark operators only the linear part of the relation between 
them,
\be
A_\mu^{(1)} (x) = \sum_y \sum_\nu h_{\mu\nu} (y) A_\nu^{(0)} (x+y) ,
\ee
is relevant, because the quadratic part, being antisymmetric, gives 
no contributions to the tadpoles (which at one loop are the only diagrams
that can be constructed from two gluons stemming from the same point). 
In momentum space this convolution becomes a form factor,
\be
A_\mu^{(1)} (q) = \sum_\nu \widetilde{h}_{\mu\nu} (q) A_\nu^{(0)} (q) ,
\ee
where
\be
\widetilde{h}_{\mu\nu} (q) = \Big( 1 - \frac{c}{6} \widehat{q}^2 \Big) 
\Big( \delta_{\mu\nu} - \frac{\widehat{q}_\mu\widehat{q}_\nu}{\widehat{q}^2}
\Big) + \frac{\widehat{q}_\mu\widehat{q}_\nu}{\widehat{q}^2} .
\ee 
We notice that the longitudinal part $\widetilde{h}$ is not affected by the 
smearing, because this part is unphysical and does not depend on which path 
of links between two fixed sites is chosen. After $N$ smearings one gets
\be
A_\mu^{(N)} (q) = \sum_\nu \widetilde{h}_{\mu\nu}^{(N)} (q) A_\nu^{(0)} (q) ,
\ee
with
\be
\widetilde{h}_{\mu\nu}^{(N)} (q) = \Big( 1 - \frac{c}{6} \widehat{q}^2 \Big)^N 
\Big( \delta_{\mu\nu} - \frac{\widehat{q}_\mu\widehat{q}_\nu}{\widehat{q}^2}
\Big) + \frac{\widehat{q}_\mu\widehat{q}_\nu}{\widehat{q}^2} .
\ee 
Since this is a correction to the gauge interaction of the quarks, the effect
is that each quark-gluon vertex is multiplied by a form factor  
$\widetilde{h}_{\mu\nu}^{(N)} (q)$, where $q$ is the gluon momentum. 
Now we consider the case, which often happens in practice, in which all gluon 
lines of a diagram start and end on quark lines. If this is true one can 
imagine these form factors as attached to the gluon propagator instead than
to the vertices. The effect of smearing can then be summarized in a simple 
modification of the gluon propagator~\footnote{Here $G_{\mu\nu} (q)$ is the 
standard Wilson gluon propagator, because the fat link action changes the 
quark-gluon interaction but leaves the plaquette action unaltered, and in 
particular the gluon propagator. This is in some sense complementary to the 
case of improved gluons (Section~\ref{sec:improvedgluons}), where the pure 
gauge action is modified but not the quark-gluon interaction. Recently actions
which contain both fat links and an improvement of the pure gauge part have 
also been considered in perturbation theory~\cite{DeGrand:2002vu}.} 
\be
G_{\mu\nu} (q)  \longrightarrow  \widetilde{h}_{\mu\lambda}^{(N)} (q) 
G_{\lambda\rho} (q) \widetilde{h}_{\rho\nu}^{(N)} (q) .
\ee
It is now clear that the Landau gauge is the most natural setup for fat link 
calculations. The Landau gauge propagator kills the longitudinal components of 
$\widetilde{h}$, and it is then quite easy to convert a thin link calculation
made in Landau gauge to a fat link result, provided the loop integration 
momentum was chosen to be the same as the gluon momentum $q$ (which we have 
{\em not} done in Section \ref{sec:examplewilson}):
\be
\int \frac{d^4q}{(2\pi)^4} \, {\cal I}(q) \rightarrow 
\int \frac{d^4q}{(2\pi)^4} \, 
\Big( 1 - \frac{c}{6} \widehat{q}^2 \Big)^{2N} {\cal I}(q) . 
\label{eq:thinfat}
\ee
Thus, no matter how many iterations $N$ one makes, the final result is simply 
given by the multiplication of the old thin link integrand with a form factor.
In the case in which the thin link calculation was instead done in Feynman
gauge, one can observe that doing the same calculation using fat links 
in Landau gauge will generate new terms, coming from the 
$\Big( 1 - \frac{c}{6} \widehat{q}^2 \Big)^N 
\frac{\widehat{q}_\mu\widehat{q}_\nu}{\widehat{q}^2}$ part of 
$\widetilde{h}_{\mu\nu}^{(N)} (q)$. However, when 
$ \Big( 1 - \frac{c}{6} \widehat{q}^2 \Big) = 1$, the total contribution of 
these terms must vanish by gauge invariance. If this cancellation already 
occurs at the level of the integrands, the further multiplication by 
$\Big( 1 - \frac{c}{6} \widehat{q}^2 \Big)^N$ will not spoil it, and then one 
can use again Eq.~(\ref{eq:thinfat}) for passing from thin to fat links, even
if the thin link calculation was done in Feynman gauge. This has been verified
for the additive and multiplicative renormalization of the 
mass~\cite{Bernard:1999kc}. However, if the cancellation is more subtle, for 
example involving integration by parts, then Eq.~(\ref{eq:thinfat}) is not
valid anymore and these new fat link terms have to be computed from scratch.

As we mentioned at the beginning, one of the attractive features of fat links
is that renormalization factors seem to be much closer to their tree level 
values. The Wilson tadpole diagram $12.233050\, g_0^2/(16\pi^2) C_F$ 
(see Eq.~(\ref{eq:valuetadpole})), which is responsible for many of the large 
corrections in lattice perturbation theory, becomes in the fat link case
$0.346274\, g_0^2/(16\pi^2) C_F$ (when $c=0.45$ and $N=10$ are used). 
Thus tadpole improvement is not necessary in this kind of calculations. 
Other quantities, like the 1-loop correction to the renormalization of the 
various currents, have been verified to be smaller of nearly two orders 
of magnitudes with respect to the Wilson results.

The reason of these small factors is that if $c$ is not too large, $c < 0.75$,
the absolute value of $\Big( 1 - \frac{c}{6} \widehat{q}^2 \Big)$ is less than
one, and so $\Big( 1 - \frac{c}{6} \widehat{q}^2 \Big)^N$ is a very small 
factor, except perhaps for the region of very small momenta. Thus, the region 
of large momenta is completely suppressed, and this is precisely the dominant 
contribution to the tadpole diagrams. We can then understand why these 
fat link integrals are strongly suppressed.~\footnote{We should however point 
out that a too strong suppression could render the integrals infrared 
sensitive.} This damping however does not necessarily take place when 
divergent diagrams are considered. In this case one has to perform appropriate
subtractions, and not all terms in the original integral remain proportional 
to $\Big( 1 - \frac{c}{6} \widehat{q}^2 \Big)^{2N}$. The finite part of a 
divergent fat link integral is then not necessarily small, and more 
investigations are needed to understand the general situation of divergent 
integrals.

Recently a new bunch of perturbative calculations with fat link actions has 
been reported~\cite{DeGrand:2002va,DeGrand:2002vu}. They mainly use another 
smearing choice known as hypercubic 
blocking~\cite{Hasenfratz:2001hp,Hasenfratz:2001tw,Hasenfratz:2002jn}, which
is more complicated and involves three free parameters which need to be fixed. 
Using this version of fat links the renormalization constants of the quark 
currents as well as of weak four-fermion operators have been computed, with
the fattening of the links applied to the Wilson, the improved (for the quark 
as well as the gluon part) and the overlap actions. The case of fat links 
applied to staggered fermions has been extensively studied 
in~\cite{Lee:2002ui,Lee:2002fj,Lee:2002bf}.

\section{Computer codes}
\label{sec:computercodes}

The calculation of the 1-loop renormalization constant of the unimproved 
matrix element $\langle q | O_{\{\mu \nu\}} | q \rangle$, that we have 
described in detail in the previous Section, can be done entirely by hand. 
However, even in this relatively simple case a computer program (written 
in the FORM language) has also been used to have some cross-checks on the 
correctness of all results. It is in general useful to have this kind of 
mutual checks between calculations made by hand and calculations made using 
a computer.

If one wants to compute the matrix element 
$\langle q | O_{\{\mu \nu\}} | q \rangle$ in the improved theory, adding
the vertex in Eq.~(\ref{eq:impvertex}) to the usual Wilson vertex, or even 
including the improvement of the operator as well, which means computing the 
renormalization of the operators in Eq.~(\ref{eq:imprfirstmom}), or even worse
if one considers more complicated operators, containing perhaps more covariant
derivatives, computer programs become necessary, as the huge number of 
manipulations and the size of the typical monomials makes the completion 
of these calculations entirely by hand almost impossible. Of course these 
codes turn out at the end to be necessary also because they can provide an 
output file of the result of the analytic manipulations which is already 
formatted (for example in Fortran) as an input file for the numerical 
integration.

One of the main reasons for the increasing difficulty in the manipulations 
for the moments of unpolarized structure functions is that the covariant 
derivative is proportional to the inverse of the lattice spacing, 
$D \sim 1/a$, so that one has
\be
\langle x^n \rangle \quad \sim \quad \langle \overline{\psi} \, \gamma_\mu \, 
D_{\mu_1} \cdots \, D_{\mu_n} \, \psi \rangle \quad \sim \quad \frac{1}{a^n}.
\ee
This means that to compute the $n$-th moment one needs to perform a Taylor 
expansion in $a$ to order $n$, and then every single object (propagators, 
vertices, operator insertions) has to be expanded to this order. 
One does not need too much imagination to see what happens. It is sufficient
to have a look to the Wilson quark-quark-gluon vertex to order $a^2$,
\bea
(V^a)^{bc}_\mu (k,ap) &=& -g_0 \, (T^a)^{bc} 
\cdot \Bigg\{ \ii \gamma_\mu 
\Big[ \cos \frac{k_\mu}{2} -\frac{1}{2} a p_\mu \, \sin \frac{k_\mu}{2}
-\frac{1}{8} a^2 p_\mu^2 \, \cos \frac{k_\mu}{2} \Big] \\
&& \qquad \qquad \qquad + r \Big[ \sin \frac{k_\mu}{2} 
+\frac{1}{2} a p_\mu \, \cos \frac{k_\mu}{2} 
-\frac{1}{8} a^2 p_\mu^2 \, \sin \frac{k_\mu}{2} \Big] \Bigg\} , \nonumber 
\eea
or to the expansion of the Wilson quark propagator even only to order $a$,
\bea
&& S^{ab} (k+aq,am_0 ) = \delta^{ab} \cdot \Bigg\{
\frac{- \ii  \sum_\mu \gamma_\mu \sin k_\mu + 2 r 
\sum_\mu \sin^2 {\displaystyle \frac{k_\mu}{2}} }{ \sum_\mu \sin^2 k_\mu +
\Big[ 2 r \sum_\mu \sin^2 {\displaystyle \frac{k_\mu}{2}} \Big]^2} \\
&& \qquad \qquad
+ a \cdot \Bigg[ 
\frac{- \ii  \sum_\mu \gamma_\mu q_\mu \cos k_\mu 
+ r \sum_\mu q_\mu \sin k_\mu + m_0}{ \sum_\mu \sin^2 k_\mu 
+ \Big[ 2 r \sum_\mu \sin^2 {\displaystyle \frac{k_\mu}{2}} \Big]^2} 
\nonumber \\
&&
- \Big(- \ii  \sum_\rho \gamma_\rho \sin k_\rho 
+ 2 r \sum_\rho \sin^2 {\displaystyle \frac{k_\rho}{2}} \Big) 
 \frac{\sum_\mu q_\mu \sin 2 k_\mu 
+ 4 r \sum_\mu \sin^2 {\displaystyle \frac{k_\mu}{2}} \,
\Big( r \sum_\nu q_\nu \sin k_\nu + m_0 \Big) }{ 
\Big\{ \sum_\mu \sin^2 k_\mu 
+ \Big[ 2 r \sum_\mu \sin^2 {\displaystyle \frac{k_\mu}{2}} \Big]^2 \Big\}^2 }
\Bigg] \Bigg\} \nonumber .
\eea
The algebraic manipulations become thus quite complex. The main consequence of
all this is the generation of a huge number of terms, at least in the initial 
stages of the manipulations, even in the case of matrix elements where all 
Lorentz indices are contracted. The multiplication of two vertices and two 
quark propagators which are expanded to order $a$ can be seen from the 
formulae above to give rise to about $4^2 \cdot 11^2 \sim 2000$ monomial 
terms. Initial expansions of Feynman diagrams containing operators which 
measure the second and third moment of structure functions can easily reach 
the order of $10^6$ terms. This slows down the execution of the codes 
considerably. Most of these terms become zero after doing the Dirac algebra, 
or do not contribute to the sought Dirac structure, or are zero after 
integration. The terms which do not contribute to the final expression have to
be killed as early as possible to speed up the computations. Of course this is
not easy; for example terms like the ones proportional to 
$1/(2W(\Gamma^2+W^2)^2)$ which we have mentioned after Eq.~(\ref{eq:1exp}) are
zero because $\mu \neq \nu$, but this can be seen only after the Dirac algebra
has been solved.

Thus, the fact that an operator with $n$ covariant derivatives requires Taylor
expansions in $a$ to order $n$ also implies a limitation on the number of 
moments of structure functions that one can practically compute on the 
lattice. This is something different from the limitation coming from operator 
mixings, seen in Section~\ref{sec:operatormixing}, and the combination of 
these two computational challenges renders in practice the computation of the 
renormalization of the fourth moment or higher very difficult.

Also the gamma algebra becomes cumbersome to do by hand. This is even worse 
when one adds the improvement, because, as noted in the previous Section,
adding a $\sigma$ matrix for each improved vertex at the end builds up long 
chains of Dirac matrices.

At the end of the day, perturbation theory, even only at 1-loop level, is quite
cumbersome on the lattice, and due to the complexity of the calculations, to 
the great number of diagrams, and to the huge amount of terms for each 
diagram, computer codes have to be used. To evaluate the Feynman diagrams and 
obtain the algebraic expressions for the renormalization factors, the author 
has developed sets of computer codes written in the symbolic manipulation 
language FORM (for recent developments see~\cite{Vermaseren:2000nd}).
These codes are able to take as input the Feynman rules for the particular 
combination of operators, propagators (Wilson or overlap)
and vertices (Wilson, and Sheikholeslami-Wohlert-improved, or overlap)
appearing in each diagram, to expand them in the lattice spacing $a$ at the 
appropriate order, to evaluate the gamma algebra on the lattice, 
and then to work out everything until the final sought-for expressions 
are obtained. Due to the enormous number of terms in the initial stages
of the manipulations one needs in many cases a large working memory.

To properly deal with the $\gamma_5$ matrices additional computer routines 
have also been written which are able to perform computations in the 
't~Hooft-Veltman scheme, the only scheme proven to have consistent trace
properties when $\gamma_5$ matrices are involved.~\footnote{A definition
of the Dirac matrices in this scheme is given in~\cite{Veltman:1988au}. 
The work of~\cite{Jegerlehner:2000dz} contains an interesting discussion about
dimensional regularization, the use of chiral fields and their relation with 
the properties of $\gamma_5$ in noninteger dimensions.} 
This has been important especially for doing calculations involving weak 
interactions and four-fermion operators~\cite{Capitani:1998kj,Capitani:1999ai,Capitani:1999rv,Capitani:2000je}. There is however an increase of about one 
order of magnitude in computing times when one uses the 't~Hooft-Veltman 
scheme, due to the sum splittings, and also a careful memory management is 
required. These FORM codes are able to use Dimensional Regularization (NDR 
or 't~Hooft-Veltman), and a mass regularization, and some independent checks 
are thus possible. As a further check on the codes, the author has in many 
cases performed the calculations also by hand.

Computer codes are of course often employed nowadays also for continuum 
perturbation theory, where however in general there are more external 
legs and one can reach higher loops, because the building blocks 
(the various Feynman rules, the typical monomials, etc.) are much simpler.
But there are differences also in the codes themselves.

We have already mentioned several times that the Lorentz symmetry is broken 
on the lattice. This gives rise to a whole new series of problems, regarding 
for example the validity of the Einstein summation convention.
One of the biggest challenges of computer codes for lattice perturbation 
theory is to deal with the fact that the summation convention on repeated 
indices is suspended. FORM, and similar programs, have been developed having 
in mind the usual continuum calculations.~\footnote{A recent description of 
this kind of programs can be found in~\cite{Weinzierl:2002cg}.}
There are therefore many useful built-in features that are in principle
somewhat of an hindrance in doing lattice perturbative calculations.
These built-in functions cannot be used straightforwardly on the lattice.
This is for example what FORM would normally do, because two equal
indices are assumed to be contracted:
\bea
\sum_{\lambda} \gamma_{\lambda} p_{\lambda} 
&\longrightarrow& \slash{p} \\
\sum_{\lambda} \gamma_{\lambda} p_{\lambda} \sin k_{\lambda} 
&\longrightarrow& \slash{p} \sin k_{\lambda} \\ 
\sum_{\lambda} \gamma_{\lambda} \sin k_{\lambda} \cos^2 k_{\lambda}
&\longrightarrow& (\gamma \cdot \sin k) \cos^2 k_{\lambda} \\ 
\sum_{\lambda,\rho} \gamma_{\rho} \gamma_{\lambda} \gamma_{\rho}  
\sin k_{\lambda} \cos^2 k_{\rho} &\longrightarrow& 
-2 \sum_{\lambda} \gamma_{\lambda} \sin k_{\lambda} \cos^2 k_{\rho} .
\eea
Here however the typical terms are monomials which contain more than twice
the same index. Only the first case is then correctly handled by FORM. 
For example, in the last case the right answer is instead
\be
- \sum_{\lambda, \rho} \gamma_{\lambda} \sin k_{\lambda} \cos^2 k_{\rho}
+ 2 \sum_{\rho} \gamma_{\rho} \sin k_{\rho} \cos^2 k_{\rho} .
\ee
For this reason one needs the development of special routines to deal with
the gamma algebra on the lattice (for more details see 
also~\cite{Capitani:1995nk,Capitani:qn}).
One solution is to introduce generalized Kronecker delta 
symbols~\cite{Luscher:1995np}
\be
\delta_{\mu_1 \mu_2 \dots \mu_n} 
\ee
which are equal to one only if all indices are equal, 
$\mu_1 = \mu_2 = \dots = \mu_n$, and are zero otherwise. In general one needs 
special routines, with appropriate modifications to the usual commands, to 
properly treat Dirac matrices and handle terms like 
in Eq.~(\ref{eq:examplegammas}).

In Feynman diagrams that involve a few gluons the color structure can become 
quite involved, especially if the 4-gluon vertex is present. A program for 
the automatic generation of gluon vertices and the reduction of the color 
structure has been described in~\cite{Luscher:1985wf}. Expression containing 
color tensors can be computed by repeatedly using the 
identities~\cite{Luscher:1995np}
\bea
\Tr (T^a X T^a Y) &=& \frac{1}{2N_c} \Tr (X Y) -\frac{1}{2} \Tr (X) \,\Tr (Y)\\
\Tr (T^a X) \, \Tr (T^a Y) &=& \frac{1}{2N_c} \Tr (X) \, \Tr (Y) 
   - \frac{1}{2} \Tr (XY) ,
\eea
where $X$ and $Y$ stand for general complex $N_c \times N_c$ matrices.

Of course when the computations become so complicated that it is not possible 
to carry them out by hand, a number of additional checks on the codes
is desirable. One can use different regularizations, like a mass regularization
and dimensional regulation (in its various forms), and one can develop
various routines which use different methods.

The case of overlap fermions is one in which manual checks are much more 
difficult, because only a few computations can be performed by hand. 
When computing renormalization factors it is useful then to exploit 
the cancellation of the gauge-dependent part between the continuum and the 
lattice results (as we noted in Section~\ref{sec:renormalizationoperators}).
In particular, in the calculations made in covariant gauge the contributions 
proportional to $(1-\alpha)$ must be independent of the parameter $\rho$, and 
this can only be seen after the numerical integration, since in the thousands 
of terms which have to be integrated the dependence on $\rho$ cannot be 
factored out. The dependence on $\rho$ is in fact highly nontrivial, and a 
look at the quark propagator shows that the monomials in the integrand are not
even rational functions of $\rho$. Thus, this is really a nontrivial check. 
Of course calculations made in a general covariant gauge are much more 
expensive than the restriction to the Feynman gauge, due to the more 
complicated form of the gluon propagator, but it is worthwhile to do them. 
They can give a strong check of the behavior of the FORM codes in the 
case of overlap fermions, as well as of the integration routines.

The contributions proportional to $(1-\alpha)$, besides being independent of 
$\rho$, seem to a certain extent also to be independent of the fermion action 
used. In particular, for overlap fermions they have the same value as for the 
Wilson case.~\footnote{In the case of overlap fermions however, due do the 
greater number of terms and the more complicated functional forms, one gets 
less precise numbers for these contributions (for a given computing time).} 
These terms are in general equal to their Wilson counterparts at the level of 
the single diagrams. An exception is given by the self-energy, where only the 
sum of the sunset and the tadpole diagrams is independent of $\rho$, as can be 
seen in Table~\ref{tab:overlapself}, and has the same value as in the Wilson 
action, as can be seen looking at Eq.~(\ref{eq:quarkself}).~\footnote{This is 
probably connected to the fact that the 1-loop self-energy for Wilson fermions
has a nonvanishing $\Sigma_0$ contribution, which gives the additive quark 
mass renormalization due to the breaking of chiral symmetry, while for overlap
fermions this term is zero.} The component of the total self-energy 
proportional to $(1-\alpha)$ is proportional only to the combination 
$F_0-\gamma_E+1=4.792009568973\cdots$ (see Eqs.~(\ref{eq:f0intdr}) 
and~(\ref{eq:f0intm}) and Table~\ref{tab:z0z1f0} later), and therefore is the 
same for all plaquette actions, irrespective of the fermionic part. Another 
case in which we have found that individual overlap diagrams do not correspond
to their Wilson result is given by operators measuring the moment of the 
structure function $g_2$~\cite{Capitani:2000aq}. These operators are of the 
form
\be
\overline{\psi} \gamma_{[\mu} \gamma_5 D_{\{\nu]} D_\rho \cdots D_{\lambda\}}
\psi ,
\ee
that is they involve both a symmetrization and an antisymmetrization of 
indices, and in the Wilson case they mix with other operators of lower 
dimension, with a $1/a$ divergent coefficient, while in the overlap case they 
are multiplicatively renormalizable, because these mixings are forbidden 
by chiral symmetry. For this operator only the sum of vertex and sails is 
independent of the fermion action used.

Lattice perturbative calculations generally involve the manipulation of a 
huge number of terms, but often a large number of terms remain also in the 
final analytic expressions which have to be numerically integrated. 
The integrations then require a lot of computer time, which in some cases can 
be of the order of hundreds of hours.

\section{Lattice integrals}
\label{sec:latticeintegrals}

After the analytical manipulations in Feynman diagrams have been carried out, 
it still remains to integrate the resulting expressions. Lattice integrals are 
quite complicated rational functions of trigonometric expressions, and so far
it has not been possible to compute them analytically. The only way to obtain 
a number from these expressions is to use numerical integration methods. 
The results of the analytic calculations obtained with computer programs can 
be stored as a sum of terms suitable to be integrated, which can be formatted 
in an appropriate way and passed on to become the input of a, say, Fortran 
program.

Integrals can be approximated by discrete sums. At the basic level, one 
computes the function to be integrated in a number of points, and uses the sum 
over these points (sometimes with appropriate weights) as an approximation to 
the true value of the integral. Since we are working in four dimensions 
(usually in momentum space), even choosing only 100 points in each direction 
means that the function has to be evaluated in $10^8$ points, and the 
computational requirements grow then quite rapidly. This becomes much worse 
for 2-loop integrals, where one has to evaluate sums in eight dimensions. 
It is therefore invaluable, in order to compute 2-loop integrals, to use the 
coordinate space methods that we will introduce in 
Section~\ref{sec:coordinatespacemethods}, with which these 8-dimensional 
integrals can be expressed in terms of only 4-dimensional sums.

Using simple integration routines it is possible to evaluate numerically 
lattice 1-loop integrals with reasonable precision, and one can without much 
effort obtain results with five or six significant decimal places. There are 
also some more refined methods which have been devised and used to obtain 
faster and cheaper evaluations of these integrals and to improve on the 
precision of the integrations. We are going to explain some of them in the 
rest of this Section.

One expects~\cite{Symanzik:1983gh} that a lattice Feynman diagram $D$ with 
$l$ loops will have near the continuum limit the asymptotic behavior 
\be
D(a) \stackrel{a \rightarrow 0}{\sim}
a^{-\omega} \sum_{n=0}^\infty \sum_{m=0}^l c_{nm} \, a^n (\log a)^m ,
\label{eq:symasex}
\ee
where the nonnegative exponent $\omega$ is related to the convergence 
properties of $D$ and its subdiagrams. L\"uscher and Weisz have devised a 
recursive blocking method for the calculation of the first coefficients of the
expansion in $a$ of a 1-loop Feynman diagram: 
\be
D(a) \stackrel{a \rightarrow 0}{\sim}
a^{-\omega} \sum_{n=0}^\infty a^n \, ( c_{n0} + c_{n1} \log a ) ,
\ee
so that one does not need to evaluate the diagram for very small
lattice spacings, which would be numerically quite demanding.
To use this method~\cite{Luscher:1985wf}, the diagram has to be known 
numerically (with a good precision) for a sequence of different lattice 
spacings, which for simplicity will be taken to be $a_k = 1 / \mu k$, with $k$
integer and $\mu$ a constant. The continuum limit $a \to 0$ means 
$k \to \infty$.
We consider the case in which only even powers of $a$ appear in the expansion,
which includes diagrams which are $O(a)$ improved (if one is interested 
in the coefficients with $n \le 2$).
It is also assumed that one knows the coefficients of the logarithm terms 
exactly, so that the task is reduced to compute numerically only the 
coefficient $c_{00}$ (which is the finite part of the limit $a=0$ of the 
Feynman diagram), and $c_{20}$. Let us see how one can compute $c_{00}$.

One starts by defining an auxiliary function
\be
f_0 (k) = \{ a^\omega D(a) + (c_{01}+a^2 c_{21} ) \log k \} |_{a=a_k} ,
\ee
which can be seen to have the asymptotic expansion
\be
f_0 (k)  \stackrel{k \rightarrow \infty}{\sim}
A_0 + \frac{A_1}{(\mu k)^2} + \sum_{n=2}^\infty 
\frac{(A_n + B_n \log k)}{(\mu k)^{2n}} ,
\label{eq:a0exp}
\ee
where
\bea
A_0 &=& c_{00} - c_{01} \log \mu, \\
A_1 &=& c_{20} - c_{21} \log \mu, \nonumber 
\eea
and so on. We can then consider taking
\be
A_0 \simeq f_0 (k_{max})
\ee
as a first approximation to compute $c_{00}$ (remember that $c_{01}$ is 
exactly known). It is however easy to do better. In fact, the function
\be
f_1(k) = \frac{(k+\delta_0)^2}{4\delta_0 k} f_0 (k+\delta_0) 
- \frac{(k-\delta_0)^2}{4\delta_0 k} f_0 (k-\delta_0)
\ee
has an expansion similar to Eq.~(\ref{eq:a0exp}) with the same initial term 
$A_0$ but no $1/k^2$ term, so that the first correction becomes of order 
$1/k^4$ only. Therefore the approach to the limiting value for 
$k \rightarrow \infty$ is faster, and
\be
A_0 \simeq f_1 (k_{max}-\delta_0)
\ee
gives a better approximation for the computation of $c_{00}$. Notice that 
$f_1$ is not defined at $k_{max}$, because it is a discrete difference 
involving the nearest point, and thus one has to use its value at 
$k_{max}-\delta_0$. Usually one chooses $\delta_0=1$ or $\delta_0=2$. 

The above blocking transformation can be iterated $i$ times to give closer 
and closer approximations to the right value of $A_0$:
\be
f_i (k) = A_0 + O(1/k^{2i+2}) .
\ee
However, because of the logarithms present in Eq.~(\ref{eq:a0exp}) 
for $n \ge 2$, the transformations become now slightly more complicated:
\bea
f_{i+1}(k) &=& w_1 f_i(k+\delta_i) + w_2 f_i(k) + w_3 f_i(k-\delta_i), \\
w_j &=& v_j / (v_1 + v_2 + v_3), \qquad (j=1,2,3) , \\
v_1 &=& (k+\delta_i)^{2i+2} \log (1-\delta_i / k) , \\
v_2 &=& k^{2i+2} [\log (1+\delta_i / k) -\log (1-\delta_i / k)] , \\
v_3 &=& -(k-\delta_i)^{2i+2} \log (1+\delta_i / k) .
\eea
While each new iteration gives in principle a better approximation to $A_0$, 
it is to be noticed that the domain of the function $f_i$ becomes smaller and 
smaller with every new blocking step, and therefore after a while (depending 
on the total number of different lattice measurements available, $k$) there 
is a natural halt to the blocking steps. Moreover, the numerical precision 
is lost a little after each iteration; L\"uscher and Weisz estimated that 
one loses one or two decimal places with each blocking step.

There is an optimal choice for the number of iterations, and 
in~\cite{Luscher:1985wf} a stopping criterion and an estimate for the total 
error are given. At the end there is an optimal estimate of $c_{00}$ given by
\be
A_0 \simeq f_{i^*} (k^*)
\ee
where $k^*$ and $i^*$ minimize the error.

Another method which is useful for the extraction of the leading behavior 
of lattice integrals has been used in~\cite{Bode:1999dn,Bode:1999sm}. 
One assumes as before to know the integrals $F(k)$ for a sequence of different
lattice spacings $k_i$, $(i=1, \dots, n)$, with the necessary computer 
precision. One wants then to determine the first $n_f$ coefficients of the 
expansion
\be
F(k) = \sum_{i=1}^{n_f} \alpha_i f_i(k) + R(k) 
\label{eq:impfits}
\ee
in terms of general functions $f_i(k) = \log^\mu k /k^\nu$ (which also include
$f_0(k)=1$).

Of course one has always $n_f \le n$, because the parameters to be determined 
cannot exceed the data set. In matrix form the above equation reads
\be
F = f \alpha + R ,
\ee
where $F$ is the $n$-dimensional vector of the data $F(k_i)$, $\alpha$ is
the $n_f$-dimensional vector of the leading coefficients that we want
to determine, and $f$ is an $n \times n_f$ matrix.

The method calls for $\alpha$ to be determined by minimizing the quadratic 
form in the residues
\be
\chi^2 = (F - f \alpha)^T W^2 (F - f \alpha) ,
\ee
where $W^2$ is a matrix of positive weights.~\footnote{These weights could be 
useful in order to give more importance to particular data points, as for 
larger $a$ there is less roundoff error, while for smaller $a$ the asymptotic 
expansions are better satisfied. In~\cite{Bode:1999dn,Bode:1999sm}
it was however reported that uniform weights $W=1$ work just fine as well.}
This leads to 
\be
f^T W^2 f \alpha = f^T W^2 F ,
\ee
and if the columns of the matrix $Wf$ are linearly independent, and $P$ 
is a projector on the corresponding $n_f$-dimensional subspace, then 
the task is reduced to finding a solution of
\be
W f \alpha = P W F .
\ee
This can be done using the singular value decomposition for $Wf$,
\be
Wf = USV^T ,
\ee
where $S$ and $V$ are diagonal and orthonormal $n_f \times n_f$ matrices
respectively, and $U$ is a column-orthonormal $n_f \times n_f$ matrix, i.e.,
$U^TU=1$ and $UU^T=P$. The solution for the $n_f$ leading coefficients 
in Eq.~(\ref{eq:impfits}) is then
\be
\alpha = V S^{-1} U^T W F .
\ee

The blocking procedure of L\"uscher and Weisz discussed previously can be 
considered as a particular case of this method. In fact, if one repeats
these improved fits modifying the component proportional to one of the 
functions $f_j$, then only the corresponding $\alpha_j$ changes. 
Then, if $n=n_f$ (i.e., one has the minimal data set which is able to 
determine the coefficients), this method is equivalent to performing blocking 
transformations which cancel the $n_f-1$ components not related to $\alpha_j$.

We want to conclude this Section by discussing a simple method which is useful
to accelerate the convergence of the numerical evaluation of the integral of 
a periodic analytic function over a compact domain. This kind of integrals 
arises in theories with twisted boundary conditions~\cite{Luscher:1985wf},
\be
U_\mu (x + L \hat{\nu}) = \Omega_\nu U_\mu (x) \Omega^{-1}_\nu ,
\label{eq:twistedbc}
\ee
where the gauge fields cross through the lattice boundaries. At least two 
directions must be twisted, otherwise the twisting is equivalent to 
a field redefinition of a theory with standard boundary conditions.
The twist matrices $\Omega$ are constant and gauge-field independent, and 
are $SU(N_c)$ matrices which satisfy the algebra
\be
\Omega_\mu \Omega_\sigma  = 
    \e^{\frac{2\pi \ii}{N_c}} \, \Omega_\sigma \Omega_\mu .
\ee 
Explicit representations of these matrices are not needed. 
These boundary conditions remove the zero modes and the theory acquires a mass
gap, so that the gluon propagator is not singular. Beside inducing an infrared 
cutoff in Feynman diagrams, they also cause the spectrum of momenta to be
continuous, because the momentum components are not quantized by the boundary 
conditions. Similarly, the ghosts are also twisted periodic fields, and the 
Faddeev-Popov determinant has no zero modes. One then ends up with integrals 
of periodic analytic functions. More technicalities and the Feynman rules can 
be found in~\cite{Luscher:1985wf}.

For simplicity we consider one-dimensional integrals. We want to compute 
\be
I = \int_{-\pi}^\pi \frac{d k}{2\pi} \, f(k), 
\ee
where $f(k)$ is analytic and periodic, $f(k+2\pi) = f(k)$.
A first approximation of this integral is given by the sum
\be
I(N) = \frac{1}{N} \sum_{j=1}^{N} f (\frac{2\pi}{N}j) 
\ee
for $N$ large enough. While in general this is not a very efficient 
approximation, for periodic functions the convergence turns out to be 
exponential,
\be
I(N)-I = O(\e^{-\epsilon N}) ,
\ee
where $\epsilon$ is the absolute value of the imaginary part of the
singularity of the integrand which is closest to the real axis.

One can see that problems can arise when $\epsilon$ happens to be very small, 
because in this case the convergence of $I(N)$ to $I$ becomes quite slow. 
However, things can be improved by making a change of 
variable~\cite{Luscher:1985wf},
\be
k = k' - \alpha \sin k' ,  \qquad 0 \le \alpha (\epsilon) < 1 ,
\ee
where $\alpha$ is chosen to be near to one, so that the singularity 
in the new variables is pushed away from the real axis: $\hat{\epsilon}=O(1)$.
The sums
\be
\hat{I}(N) = \frac{1}{N} \sum_{j=1}^{N} \hat{f} (\frac{2\pi}{N}j),
\ee
calculated using the transformed function 
\be
\hat{f} (k') = (1- \alpha \cos k') f(k(k'))  ,
\ee
become then better convergent to the integral $I$:
\be
\hat{I}(N)-I = O(\e^{-\hat{\epsilon} N}) .
\ee
Thus, much lower values of $N$, and less computing power, are sufficient 
to get the same precision in the numerical evaluation of the integral $I$.

\section{Algebraic method for 1-loop integrals}
\label{sec:algebraic}

If one is happy with computing 1-loop lattice integrals with a precision of
only five or six significant decimal places, their values can be easily 
estimated by using a simple rectangle integration. In order to have some more 
precise results one can implement the methods discussed in the previous 
Section.

The algebraic method for Wilson fermions, that we are going to explain in this
Section, is an enormous improvement on all that. It allows every integral 
coming from 1-loop Feynman diagrams to be computed in a completely symbolic 
way and to be reduced to a linear combination of a few basic constants. 
This means that the generic integral can then be calculated numerically with 
a very large precision with a very small effort. In fact, once determined
these few basic constants with the desired precision once for all, the 
original integral is just an appropriate linear combination of them. It is 
then possible to compute every integral with a very high precision, for 
example with seventy significant decimal places~\cite{Capitani:1997rz}, and in
some cases with even nearly 400 significant decimal places (see Appendix B).
Computing 1-loop integrals with sixty or seventy significant decimal places 
turns out to be absolutely necessary if one wants to evaluate 2-loop integrals 
with a precision of at least ten significant decimal places, which can be
accomplished using the coordinate space method 
(Section~\ref{sec:coordinatespacemethods}).

It is sufficient to apply the algebraic method only to integrals with zero
external momenta, since the momentum-dependent part can be evaluated in the 
continuum, as we have seen in Section~\ref{sec:divergentintegrals}. 
Furthermore, the general zero-momentum integral on the lattice can always be 
written as a linear combination of terms of the form
\be
{\cal F} (p,q;n_x,n_y,n_z,n_t) =\,
 \int_{-\pi}^\pi {d^4 k \over (2\pi)^4} \,
 {\hat{k}_x^{2 n_x} \hat{k}_y^{2 n_y} \hat{k}_z^{2 n_z} \hat{k}_t^{2 n_t}\over
     D_F(k,m_f)^p D_B(k,m_b)^q} .
\ee
The algebraic method allows then to express these terms in terms of a certain 
number of basic integrals. The complete reduction of a generic 
${\cal F} (p,q;n_x,n_y,n_z,n_t)$ is achieved using an iteration procedure 
which makes use of appropriate recursion relations in noninteger dimensions.

At the end of this procedure, using the algebraic method every purely bosonic 
integral can be expressed in terms of 3 basic constants, every purely 
fermionic integral in terms of 9 basic constants, and every general 
fermionic-bosonic integral in terms of 15 basic constants.

\subsection{The bosonic case}

We are now going to explain in detail the recursive algorithm in
the bosonic case, when the gluon action is given by the Wilson 
plaquette~\cite{Caracciolo:1991vc,Caracciolo:1991cp}.

We note first of all that it is enough to consider lattice integrals at zero 
external momenta. We are in fact interested in the continuum limit of
\be
G(p) = \int {d^4 k\over (2 \pi)^4} F(k;p)
\ee
where $p$ is some external momentum. As explained in 
Section~\ref{sec:divergentintegrals}, we can split this integral in 
two parts: a subtracted (ultraviolet-finite) integral, which can be 
evaluated in the continuum, and a certain number of lattice integrals with 
zero external momenta. We have then
\bea
G(p) &=& \int {d^4 k\over (2 \pi)^4} \,  
   \left[ F(k;p) - (T^{n_F} F)(k;p)\right]  \nonumber \\
     && +\int {d^4 k\over (2 \pi)^4} (T^{n_F} F)(k;p) ,
\eea
where $n_F$ is the degree of the divergence of the integral, and 
\be
(T^{n_F} F)(k;p) = \sum_{n=0}^{n_F} {1\over n!} \, p_{\mu_1} \ldots p_{\mu_n} 
\left[ {\partial\over \partial p_{\mu_1} } \ldots
{\partial\over \partial p_{\mu_n}} F(k;p) \right]_{p=0} .
\ee
If the propagators are massless, we need to at this point to introduce an 
intermediate regularization for $k=0$. It is convenient to choose an infrared 
mass cutoff $m$ or to use dimensional regularization. The singularities will 
cancel when the contribution of the momentum-dependent part is added at the
end.

Any zero-momentum integral coming from the calculation of lattice Feynman 
diagrams in the pure gauge Wilson theory can be expressed as a linear
combination of terms of the form~\footnote{It is always possible
to reduce any numerator to contain only factors of $\sin^2 k_\mu/2$,
using $\sin^2 k_\mu = 4\sin^2 k_\mu/2 - 4 \sin^4 k_\mu/2$ and similar
formulae for the cosine functions.}
\be
{\cal B}(p;n_x,n_y,n_z,n_t) = 
\int_{-\pi}^\pi {d^4 k\over (2 \pi)^4}  \, {
\hat{k}^{2 n_x}_x \hat{k}^{2 n_y}_y \hat{k}^{2 n_z}_z \hat{k}^{2 n_t}_t\over
D_B(k,m)^p } ,
\ee
where $p$ and $n_i$ are positive integers, and the inverse bosonic propagator,
which we take in general to be massive in order to regularize the divergences
coming from the separation in $J$ and $I-J$, is
\be
D_B(k,m) = \hat{k}^2 + m^2 .
\ee
Actually, due to the appearance of other kind of singularities at some 
intermediate stages of the reductions, we have to consider the more general 
integrals
\bd
{\cal B}_\delta(p;n_x,n_y,n_z,n_t) = 
\int_{-\pi}^\pi {d^4k\over (2 \pi)^4}  \, {
\hat{k}^{2 n_x}_x \hat{k}^{2 n_y}_y  \hat{k}^{2 n_z}_z \hat{k}^{2 n_t}_t\over
D_B(k,m)^{p+\delta} }  ,
\ed
where $p$ is an arbitrary integer (not necessarily positive) 
and $\delta$ is a real number which will be set to zero at the end
of the calculations.

To begin with, each integral ${\cal B}_\delta(p;n_x,n_y,n_z,n_t)$ 
can be reduced through purely algebraic manipulations to a sum of integrals 
of the same type with $n_x=n_y=n_z=n_t=0$ (i.e., pure denominators).
This is done by using the recursion relations~\footnote{In the following 
when one of the arguments $n_i$ is zero it will be omitted.}
\bea
{\cal B}_\delta(p;1) & = & 
{1\over 4} \,[{\cal B}_\delta(p-1) - m^2 {\cal B}_\delta(p)] \\
{\cal B}_\delta(p;x,1) & = & {1\over 3}\,[{\cal B}_\delta(p-1;x) 
- {\cal B}_\delta(p;x+1) - m^2 {\cal B}_\delta(p;x) ] \\
{\cal B}_\delta(p;x,y,1) & = & {1\over 2}\,[{\cal B}_\delta(p-1;x,y) 
- {\cal B}_\delta(p;x+1,y) - {\cal B}_\delta(p;x,y+1) \nonumber \\ 
       && - m^2 {\cal B}_\delta(p;x,y) ] \\
{\cal B}_\delta(p;x,y,z,1) & = & {\cal B}_\delta(p-1;x,y,z) 
- {\cal B}_\delta(p;x+1,y,z) \nonumber \\ 
       && - {\cal B}_\delta(p;x,y+1,z)-\!\!{\cal B}_\delta(p;x,y,z+1) 
      - m^2 {\cal B}_\delta(p;x,y,z) ,
\eea
which can be obtained from the trivial identity
\be
D_B(k,m)\, = \sum_{i=1}^4 \hat{k}^2_i +\, m^2 .
\ee

With these recursion relations one can eliminate each numerator argument, 
$n_i$, of the ${\cal B}_\delta$ function, provided that it has the value 1.
When it is greater than 1, one has to lower its value until it reaches 1,
so that it is then possible to use the above set of recursion relations. 
The lowering of $n_i$ is done by using another recursion relation,
\be
{\cal B}_\delta(p;\ldots,r) = {r-1\over p+\delta-1} 
{\cal B}_\delta(p-1;\ldots,r-1) 
    - {4r-6\over p+\delta-1} {\cal B}_\delta(p-1;\ldots,r-2)
    + 4 {\cal B}_\delta(p;\ldots,r-1) ,
\ee
which is obtained integrating by parts the equation (for $r>1$)
\be
{({\hat k}^2_w )^r \over D_B(k,m)^{p+\delta}} = 
4 {({\hat k}^2_w )^{r-1} \over D_B(k,m)^{p+\delta}} 
+ \!\!2 {({\hat k}^2_w )^{r-2} \over p + \delta - 1} \sin k_w 
{\partial \over \partial k_w} { 1 \over D_B(k,m)^{p+\delta-1}} .
\ee
Notice that for $p=1$ some coefficients in this recursion relation
diverge as $1/\delta$, and therefore in order to compute 
${\cal B}_\delta(1;\ldots)$ for $\delta=0$ we need to compute 
${\cal B}_\delta(0;\ldots)$ including terms of order $\delta$.
In general one needs to compute the intermediate expressions for the 
integrals ${\cal B}_\delta(p;n_x,n_y,n_z,n_t)$ with $p\le 0$ keeping all 
terms of order $\delta$.

Using the recursion relations introduced so far, every integral 
${\cal B}_\delta(p;n_x,n_y,n_z,n_t)$ can thus be reduced to a sum of the form 
\be
{\cal B}_\delta(p;n_x,n_y,n_z,n_t) = \sum_{r=p-n_x-n_y-n_z-n_t}^{p} 
a_r (m,\delta) {\cal B}_\delta(r) ,
\ee
where $a_r (m,\delta)$ are polynomials in $m^2$, which may 
diverge as $1/\delta$ for $p > 0$ and $r \le 0$ .

At this point, it only remains to reexpress all ${\cal B}_\delta(p)$'s 
appearing in the above formula in terms of a small finite number of them. 
To accomplish this we need some other recursion relations, which can be 
obtained considering the trivial identity
\be
{\cal B}_\delta(p;1,1,1,1) - 4{\cal B}_\delta(p+1;2,1,1,1)
- m^2\,{\cal B}_\delta(p+1;1,1,1,1) \: =  \: 0 ,
\label{eq:idp1111}
\ee
and applying to it the previous procedure until it is reduced to a relation 
between the ${\cal B}_\delta(r)$'s only.
One then arrives to a nontrivial relation of the form
\be
\sum_{r=p-4}^p b_r(p;\delta) {\cal B}_\delta(r) +\, {\cal S}(p;m,\delta) = 0 ,
\ee
where ${\cal S}(p;m,\delta)=O(m^2)$ for $p\le 2$, while for $p>2$ 
is a polynomial in $1/m^2$ (which is finite for $\delta \to 0$).

We can now use the last relation to express all ${\cal B}_\delta(p)$'s 
in terms of ${\cal B}_\delta(r)$'s which are only in the range $0\le r\le 3$. 
To do this, when $p \ge 4$ we just write ${\cal B}_\delta(p)$ in terms of 
${\cal B}_\delta(p-1),\ldots,{\cal B}_\delta(p-4)$ and iterate until needed.
When $p\le -1$, we solve the relation in terms of ${\cal B}_\delta(p-4)$, 
make the shift $p\to p+4$, and then use it to write ${\cal B}_\delta(p)$ 
in terms of ${\cal B}_\delta(p+1),\ldots,{\cal B}_\delta(p+4)$.
Again we iterate until needed.
Applying recursively these two relations we get then, for $p \neq 0,1,2,3$:
\be
{\cal B}_\delta(p) \, =\, \sum_{r=0}^3 c_r(p;\delta) {\cal B}_\delta(r) 
+\, {\cal T}(p;m,\delta) ,
\ee
where ${\cal T}(p;m,\delta)$ is a polynomial in $1/m^2$.

The above procedure allows the general bosonic integral to be written,
after a finite number of steps, as
\be 
{\cal B}_\delta(p;n_x,n_y,n_z,n_t)=A(\delta) {\cal B}_\delta(0) 
+ B(\delta) {\cal B}_\delta(1)
+ C(\delta) {\cal B}_\delta(2) + D(\delta) {\cal B}_\delta(3) 
+ E(m,\delta) ,
\ee
where $E(m,\delta)$ is a polynomial in $1/m^2$.
It can be shown that the limit $\delta\to 0$ is safe at this stage, and 
one finally obtains 
\be
{\cal B}(p;n_x,n_y,n_z,n_t) =  A(0) + B(0) {\cal B}(1)
+ C(0) {\cal B}(2) + D(0) {\cal B}(3) + E(m,0) ,
\ee
in terms of three basic constants, ${\cal B}(1)$, ${\cal B}(2)$ and 
${\cal B}(3)$. This is a minimal set, i.e, no further reduction can be done.

It is common practice to write the bosonic results in terms of the three
constants $Z_0$, $Z_1$ and $F_0$, which are defined by
\bea
Z_0 &=& \left. {\cal B}(1) \right|_{m = 0} \label{eq:z0} \\
Z_1 &=& {1\over 4} \left. {\cal B}(1;1,1) \right|_{m = 0} \label{eq:z1} \\
F_0 &=& \lim_{m\to 0} \, \Big(16\pi^2 {\cal B}(2) + \log m^2 + \gamma_E \Big) .
\eea
Explicitly,
\bea
Z_0 &=& \int^\pi_{-\pi} \frac{d^4 k}{(2\pi)^4} \, 
   \frac{1}{4 \sum_\lambda \sin^2 {\displaystyle \frac{k_\lambda}{2}}}, \\
Z_1 &=&
\int^\pi_{-\pi} \frac{d^4 k}{(2\pi)^4} \, \frac{\sin^2 {\displaystyle 
\frac{k_1}{2}} \, \sin^2 {\displaystyle \frac{k_2}{2}} }{\sum_\lambda \sin^2 
{\displaystyle \frac{k_\lambda}{2}}} ,
\eea
and~\footnote{In case dimensional regularization is used, $F_0$ is defined by
\be
\int^\pi_{-\pi} \frac{d^{4-2\epsilon} k}{(2\pi)^{4-2\epsilon}} \, \frac{1}{
\Big( 4 \sum_\lambda \sin^2 {\displaystyle \frac{k_\lambda}{2}}\Big)^2 } =
\frac{1}{16\pi^2} \, \Big( -\frac{1}{\epsilon} - \log 4\pi + F_0 \Big) .
\label{eq:f0intdr}
\ee
}
\be
\int^\pi_{-\pi} \frac{d^4k}{(2\pi)^4} \, \frac{1}{
\Big( 4 \sum_\lambda \sin^2 {\displaystyle \frac{k_\lambda}{2}}\Big)^2 +m^2} =
\frac{1}{16\pi^2} \, \Big( - \log m^2 - \gamma_E + F_0 \Big) .
\label{eq:f0intm}
\ee
Their values are given in Table~\ref{tab:z0z1f0}. We remind that 
$\gamma_E=0.57721566490153286\dots$ is the well-known Euler's constant 
appearing in continuum integrals.

\begin{table}[t]
\begin{center}
\begin{tabular}{|c|l|}
\hline
$Z_0$ & 0.154933390231060214084837208 \\
$Z_1$ & 0.107781313539874001343391550 \\
$F_0$ & 4.369225233874758 \\
\hline
\end{tabular}
\caption{Numerical values of the basic bosonic constants.}
\label{tab:z0z1f0}
\end{center}
\end{table}

Rewriting ${\cal B}(1)$ and ${\cal B}(2)$ in terms of $F_0$ and $Z_0$ 
is rather trivial. For ${\cal B}(3)$ one has 
\be
{\cal B}(3) =\, {1\over 32 \pi^2 m^2} - {1\over 128 \pi^2} 
    \left(\log m^2 + \gamma_E - F_0\right)
- {1\over 1024}- {13 \over 1536 \pi^2} + {Z_1\over 256} ,
\ee
which is a special case of Eq.~(\ref{eq:divintmass}). 

In general $d-1$ basic constants are enough for all bosonic integrals in $d$ 
dimensions, and $d-2$ if one only considers finite integrals. This means that 
in two spacetime dimensions any finite bosonic integral can be written in 
terms of rational numbers and $1/\pi^2$ factors only, and one constant has 
to be introduced for divergent integrals.

This concludes the illustration of the algebraic method. However, if one wants 
to use it, it comes handy to know some formulae regarding divergent integrals,
that is the ${\cal B}_\delta(p)$'s for $p \ge 2$. These formulae can be 
derived using their expression in terms of the modified Bessel function 
$I_0 (x)$, 
\be
{\cal B}_\delta(p) \, =\, \frac{1}{2^{p+\delta}\Gamma(p+\delta)} 
\int_0^\infty dx \, x^{p+\delta-1} \e^{-m^2x/(2-4x)}
I_0^4 (x) ,
\ee
obtained using the Schwinger representation. 
The basic formula for the divergent integrals is then
\be
{\cal B}(r) = \frac{1}{\Gamma(r)} \sum_{i=2}^{r-1} \frac{b_{i-2} \Gamma(r-i)
}{2^i(m^2)^{r-i}} + 
\frac{b_{r-2}}{2^r\Gamma(r)} \Big( - \log m^2 - \gamma_E 
+ F_{r-2} \Big) + H_{r-2} .
\label{eq:divintmass}
\ee
If dimensional regularization is instead used one has
\be
{\cal B}_d(r) = \frac{b_{r-2}}{2^r\Gamma(r)} \Big( \frac{2}{d-4} - \log 4\pi
+ F_{r-2} \Big) + G_{r-2} .
\label{eq:divintdimreg}
\ee
In the latter case, the recursion relations have to be extended to noninteger
dimensions, which can be done without too much effort. These formulae 
can be found in~\cite{Caracciolo:1991cp}. 

The constants $b_i$ appearing in the formulae for divergent integrals are 
defined by the asymptotic expansion of the modified Bessel function $I_0$,
\be
I_0^d (x) = \frac{\e^{dx}}{(2\pi x)^{d/2-2}} \sum_{i=0}^\infty
\frac{b_i - (d-4) c_i +O((d-4)^2)}{x^{i+2}} + O(\e^{-x}), 
\ee
and the constants $F_i$ are also related to this modified Bessel function:
~\footnote{These and other constants to them related were introduced 
in \cite{Gonzalez-Arroyo:1981ce}. Note that $F_0$ was called $F_{0000}$ there.
Many properties of these functions are discussed there and 
in~\cite{Ellis:1983af}. We also note that the finite constant $Z_0$ 
can also be expressed in terms of modified Bessel functions:
\be
Z_0 \, =\, \frac{1}{2} \int_0^\infty dx \, \e^{-4x} I_0^4 (x) .
\ee
}
\be
F_p = \frac{1}{b_p} \Bigg\{ \int_0^2 dx \, x^{p+1} \e^{-4x} I_0^4 (x) 
 + \int_2^\infty dx \, x^{p+1} \Big[ \e^{-4x} I_0^4 (x)
 - \sum_{i=0}^p \frac{b_i}{x^{i+2}}
\Big] \Bigg\} .
\ee
The first $F_i$'s are given explicitly by~\footnote{These relations can be 
obtained by applying the recursion relations to an appropriate identity 
like Eq.~(\ref{eq:idp1111}).}
\bea
F_1 &=& F_0-\frac{1}{8}\pi^2+\frac{35}{12}+\frac{1}{2}\pi^2 \, Z_1 \\
F_2 &=& F_0-\frac{31}{144}\pi^2+\frac{1349}{216}+\frac{1}{9}\pi^2 \, Z_0
           +\frac{31}{36}\pi^2 \, Z_1 \\
F_3 &=& F_0-\frac{523}{1872}\pi^2+\frac{24257}{2808}+\frac{25}{117}\pi^2 \, Z_0
           +\frac{523}{468}\pi^2 \, Z_1 \\
F_4 &=& F_0 -\frac{7145}{22176}\pi^2  +\frac{294919}{33264}  
          +\frac{401}{1386}\pi^2 \, Z_0 +\frac{7145}{5544}\pi^2 \, Z_1 \\
F_5 &=& F_0 -\frac{27971}{80190}\pi^2  +\frac{3347101}{481140}  
          +\frac{13582}{40095}\pi^2 \, Z_0
          +\frac{55942}{40095}\pi^2 \, Z_1 \\
F_6 &=& F_0 -\frac{27039607}{74221920}\pi^2 +\frac{448657133}{111332880}
          +\frac{1708783}{4638870}\pi^2 \, Z_0
          +\frac{27039607}{18555480}\pi^2 \, Z_1 \\
F_7 &=& F_0 -\frac{751956319}{2016614880}\pi^2 +\frac{3823946741}{3024922320}
          +\frac{48529351}{126038430}\pi^2 \, Z_0
          +\frac{751956319}{504153720}\pi^2 \, Z_1 ,
\eea
while the first $b_i$'s, together with the first $H_i$'s and $G_i$'s, which 
are defined by
\bea
H_p &=& \frac{1}{\Gamma (p+2)} \sum_{i=0}^{p-1} \frac{b_i}{2^{i+2}(i-p)} , \\
G_p &=& H_p - \frac{1}{\Gamma (p+2)}\frac{c_p}{2^{p+1}} ,
\eea
are given in Table~\ref{tab:bosonicconstants}.

\begin{table}[t]
\begin{center}
\begin{tabular}{|c|c|c|c|c|}
\hline
$r$  &  $b_r$  &  $c_r$  &  $G_r$  & $H_r$ \\
\hline 
& & & & \\
$0$  &  $ \displaystyle \frac{1}{4\pi^2}$   
     &  $ \displaystyle 0$   
     &  $ \displaystyle 0$   
     &  $ \displaystyle 0$  \\ [0.5cm]
$1$  &  $ \displaystyle \frac{1}{8\pi^2}$   
     &  $ \displaystyle -\frac{1}{32\pi^2}$   
     &  $ \displaystyle -\frac{7}{256\pi^2}$   
     &  $ \displaystyle -\frac{1}{32\pi^2}$  \\ [0.5cm]
$2$  &  $ \displaystyle \frac{3}{32\pi^2}$   
     &  $ \displaystyle -\frac{1}{32\pi^2}$   
     &  $ \displaystyle -\frac{11}{1536\pi^2}$   
     &  $ \displaystyle -\frac{1}{128\pi^2}$  \\ [0.5cm]
$3$  &  $ \displaystyle \frac{13}{128\pi^2}$   
     &  $ \displaystyle -\frac{55}{1536\pi^2}$   
     &  $ \displaystyle -\frac{793}{589824\pi^2}$   
     &  $ \displaystyle -\frac{53}{36864\pi^2}$  \\ [0.5cm]
$4$  &  $ \displaystyle \frac{77}{512\pi^2}$   
     &  $ \displaystyle -\frac{5}{96\pi^2}$   
     &  $ \displaystyle -\frac{311}{1474560\pi^2}$   
     &  $ \displaystyle -\frac{331}{1474560\pi^2}$  \\ [0.5cm]
$5$  &  $ \displaystyle \frac{297}{1024\pi^2}$   
     &  $ \displaystyle -\frac{1973}{20480\pi^2}$   
     &  $ \displaystyle -\frac{27251}{943718400\pi^2}$   
     &  $ \displaystyle -\frac{3653}{11796480\pi^2}$  \\ [0.5cm]    
$6$  &  $ \displaystyle \frac{5727}{8192\pi^2}$   
     &  $ \displaystyle -\frac{54583}{245760\pi^2}$   
     &  $ \displaystyle -\frac{559001}{158544691200\pi^2}$   
     &  $ \displaystyle -\frac{4261}{110100480\pi^2}$  \\ [0.5cm]
$7$  &  $ \displaystyle \frac{66687}{32768\pi^2}$   
     &  $ \displaystyle -\frac{8558131}{13762560\pi^2}$   
     &  $ \displaystyle -\frac{7910171}{20293720473600\pi^2}$   
     &  $ \displaystyle -\frac{1331861}{295950090240\pi^2}$  \\ [0.5cm]
\hline
\end{tabular}
\caption{Values of $b_r$, $c_r$, $G_r$ and $H_r$.} 
\label{tab:bosonicconstants}
\end{center}
\end{table}

If the integral to be computed is finite from the start, all $\log m^2$ terms 
must cancel, and this means that the $F_0$ constant is not present. 
In fact $F_0$ appears only in divergent integrals 
(${\cal B}(2)$, ${\cal B}(3)$, $\dots$), and always in the combination 
\be
\log m^2 + \gamma_E - F_0.
\ee
The absence of $\log m^2$ factors implies then the cancellation of the 
$\gamma_E$ and $F_0$ terms, and thus all finite integrals turn out to be 
functions of $Z_0$ and $Z_1$ only.

These two basic constants are now known with an incredible high precision, 
about 400 significant decimal places, as it will be shown in 
Section~\ref{sec:highprecisionintegrals} and Appendix~B, where they will be 
computed using coordinate space methods. Thus, every finite bosonic integral 
can now be calculated with about 400 significant decimal places.

\subsection{Examples of bosonic integrals}

As an illustration of the algebraic method, we show here in detail how to 
compute a few bosonic integrals which will afterwards be used for the 
calculation of the operator tadpoles necessary for the renormalization of the
operators corresponding to the third moment of the quark momentum 
distribution. We will start from the simpler integrals, which in some cases 
will be needed as intermediate results for the calculation of the more 
complicated ones.

Since these integrals all are finite, in general it is sufficient to use
the recursion relations of the previous subsection with $m=0$. Exceptions
will be duly noted. We will not write the subscript $\delta$ explicitly, 
although the integrals are supposed to be computed at $\delta \neq 0$ when 
necessary. We also remind that ${\cal B}(1) = Z_0$, from its definition.

\begin{enumerate}

\item{${\cal B}(1;1)$}

This very simple integral is given, using the first recursion relation, by
\be
{\cal B}(1;1) = \frac{1}{4} {\cal B}(0) = \frac{1}{4} .
\ee

\item{${\cal B}(1;1,1)$}

We have
\be
{\cal B}(1;1,1) = 4Z_1 ,
\ee
from its definition.

\item{${\cal B}(2;1,1)$}

By applying the various recursion relations we get
\bea
{\cal B}(2;1,1) &=& \frac{1}{3} \Bigg( {\cal B}(1;1)-{\cal B}(2;2) \Bigg) 
\nonumber \\ 
&=& \frac{1}{3} \Bigg( \frac{1}{4} - \Bigg( \frac{1}{1+\delta} {\cal B}(1;1) 
- \frac{2}{1+\delta} {\cal B}(1) + 4 {\cal B}(2;1) \Bigg) \Bigg) \nonumber \\
&=& \frac{1}{3} \Bigg( \frac{1}{4} - \Bigg( \frac{1}{4} -2 Z_0 
+4 \cdot \frac{1}{4} {\cal B}(1) \Bigg) \Bigg) \nonumber \\
&=& \frac{1}{3} Z_0 .
\eea
We have taken the limit $\delta = 0$, because it is safe to do so here.

\item{${\cal B}(2;2,1)$}

We have to compute this integral because it is necessary for 
${\cal B}(2;1,1,1)$ (the last example). The manipulations are as follows:
\bea
{\cal B}(2;2,1) &=& \frac{1}{3} \Bigg( {\cal B}(1;2) - {\cal B}(2;3) \Bigg) 
\nonumber  \\
&=& \frac{1}{3} \Bigg( \Bigg(\frac{1}{\delta} {\cal B}(0;1) - \frac{2}{\delta}
{\cal B}(0) + 4 {\cal B}(1;1)  \Bigg) - \Bigg( 2{\cal B}(1;2) 
- 6{\cal B}(1;1) +4{\cal B}(2;1)  \Bigg) \Bigg) \nonumber \\
&=& \frac{1}{3} \Bigg( \frac{1}{\delta} {\cal B}(0;1) - \frac{2}{\delta}
{\cal B}(0) + 4 {\cal B}(1;1)  \nonumber \\
&& \quad -\frac{2}{\delta} {\cal B}(0;1) + \frac{4}{\delta} {\cal B}(0) 
- 8 {\cal B}(1;1) + 6{\cal B}(1;1) -4{\cal B}(2;1) \Bigg) \nonumber \\
&=& \frac{1}{3} \Bigg( - \frac{1}{\delta} {\cal B}(0;1) + \frac{2}{\delta}
{\cal B}(0) +2 {\cal B}(1;1) -4{\cal B}(2;1)  \Bigg) ,
\eea
where we have taken the limit $\delta=0$ when safe. Now
\be
{\cal B}(0;1) = \frac{1}{4} {\cal B}(-1) ,
\ee
and we need the expression of ${\cal B}(-1)$ including terms of order $\delta$.
Applying the recursion relations to the identity 
(\ref{eq:idp1111}),~\footnote{Now we have to keep the $m^2$ term in this 
identity, because some intermediate integrals will be divergent.}
\be
{\cal B}_\delta(3;1,1,1,1) - 4{\cal B}_\delta(4;2,1,1,1)
- m^2\,{\cal B}_\delta(4;1,1,1,1) \: =  \: 0 ,
\ee
we obtain
\be
{\cal B}(-1) = 8 + \delta \cdot \Big( -20Z_0 -48Z_1 +8 \Big) +O(\delta^2) .
\ee
We have then
\be
-\frac{1}{\delta} {\cal B}(0;1) + \frac{2}{\delta} {\cal B}(0) = 
-\frac{1}{4\delta} \Bigg( 8 +\delta \Big( -20Z_0 -48Z_1 +8 \Big) \Bigg) \
+ \frac{2}{\delta} = 5Z_0 +12Z_1 -2 ,
\ee
which is finite, as it should be.
The final result is
\be
{\cal B}(2;2,1) = \frac{4}{3} Z_0 +4 Z_1 -\frac{1}{2} .  
\ee

\item{${\cal B}(2;1,2)$}

Also this integral appears in the computation of ${\cal B}(2;1,1,1)$. 
Of course it has to be equal to ${\cal B}(2;2,1)$ by symmetry, 
but the manipulations are slightly different. Actually, they are
much simpler than the previous one:
\bea
{\cal B}(2;1,2) &=& {\cal B}(1;1,1) - 2 {\cal B}(1;1) + 4 {\cal B}(2;1,1) 
\nonumber \\
&=& 4 Z_1 -2 \cdot \frac{1}{4} +4 \cdot \frac{1}{3} Z_0 \nonumber \\
&=& \frac{4}{3} Z_0 +4 Z_1 -\frac{1}{2} .
\eea
This example shows that by a judicious choice of the order of the indices
it is possible to obtain the same result with less effort.

\item{${\cal B}(2;1,1,1)$}

The first decomposition of the integral gives
\be
{\cal B}(2;1,1,1) = \frac{1}{2} \Bigg( {\cal B}(1;1,1) - {\cal B}(2;2,1) 
- {\cal B}(2;1,2) \Bigg) .
\ee
Taking the results of the two integrals that we have just computed, we have 
then
\bea
{\cal B}(2;1,1,1) &=& \frac{1}{2} \Bigg( 4Z_1 
 - 2 \cdot \Bigg( \frac{4}{3} Z_0 +4 Z_1 -\frac{1}{2} \Bigg) \Bigg)
 \nonumber \\
&=& -\frac{4}{3}Z_0 -2 Z_1 +\frac{1}{2} .
\eea

\end{enumerate}

These are all the integrals which will be needed in the next Section. 
In~\cite{Panagopoulos:1989zn} the exact expressions of various other bosonic 
integrals (useful for the renormalization of the trilinear gluon condensate)
in terms of the constants $Z_0$ and $Z_1$ can also be found.

\subsection{Operator tadpoles}
\label{sec:operatortadpoles}

A very important class of diagrams on the lattice is given by the operator 
tadpoles, which are generated when an operator in a matrix element contains 
in its definition $U$ fields. In the case in which the pure gauge action 
is given by the plaquette, and this includes Wilson and Ginsparg-Wilson 
fermions, these tadpoles can be computed exactly, using the algebraic methods 
that we have just discussed. 

We give here the results of the operator tadpoles in a general covariant 
gauge for operators which contain one, two and three covariant derivatives. 
These operators are useful for example for the calculations of the 
renormalization constants of operators measuring the lowest moments of the 
structure functions. Dirac matrices can then be added at will, since they do 
not influence the calculation of the tadpoles. What is important instead is 
the choice of the indices of the derivatives, and this is not surprising 
because these operators fall in different representations of the hypercubic 
group depending on this choice, and the results will differ accordingly. 
The operator tadpoles for the various operators in terms of $Z_0$ and $Z_1$ are
\bea
{\cal T}_{\overline{\psi} D_\mu \psi} & = & 
\frac{1}{2} Z_0 +(1-\alpha) \frac{1}{8} Z_0  , \\
{\cal T}_{\overline{\psi} D_\mu D_\nu \psi} & = & 
-Z_0 + (1-\alpha) \frac{1}{6} Z_0  , \\
{\cal T}_{\overline{\psi} D_\mu D_\mu \psi} & = & 
-2 Z_0 +\frac{1}{8} + (1-\alpha) \Bigg( Z_0 - \frac{1}{8} \Bigg)  , \\
{\cal T}_{\overline{\psi} D_\mu D_\nu D_\sigma \psi} & = & 
-\frac{3}{2} Z_0 + (1-\alpha) \Bigg( -\frac{1}{24} Z_0 
-\frac{1}{4} Z_1 +\frac{1}{16} \Bigg)  , \label{eq:optad123} \\
{\cal T}_{\overline{\psi} D_\mu D_\mu D_\nu \psi} & = & 
-\frac{5}{2} Z_0 +\frac{1}{8}
+ (1-\alpha) \Bigg(\frac{9}{8} Z_0 +Z_1 -\frac{1}{4} \Bigg)  , 
\label{eq:tadex1} \\
{\cal T}_{\overline{\psi} D_\nu D_\mu D_\mu \psi} & = & 
-\frac{5}{2} Z_0 +\frac{1}{8}
+ (1-\alpha) \Bigg(\frac{9}{8} Z_0 +Z_1 -\frac{1}{4} \Bigg)  , 
\label{eq:tadex2} \\
{\cal T}_{\overline{\psi} D_\mu D_\nu D_\mu \psi} & = & 
-\frac{5}{2} Z_0 -Z_1 +\frac{1}{4}
+ (1-\alpha) \Bigg(\frac{9}{8} Z_0 +Z_1 -\frac{1}{4} \Bigg)  ,
\label{eq:tadex3}
\eea
where $\mu \neq \nu \neq \sigma$, and repeated indices are not summed.

These results, as we mentioned, are valid also for Wilson and overlap fermions
(among others), because they depend only on the structure of the gluon 
propagator, whereas this is not true for the self-energy tadpole, which 
for overlap fermions is different from the Wilson result (as we have seen 
in Section~\ref{sec:exoverlap}), because there the interaction vertex 
is different.

We now show explicitly how these computations are done, deriving the results
of Eqs.~(\ref{eq:optad123}), (\ref{eq:tadex1}), (\ref{eq:tadex2}) and 
(\ref{eq:tadex3}), and taking as final task the calculation of the operator 
tadpoles of the operators
\be
O_{\{0123\}} = \overline{\psi} \gamma_{\{0} D_1 D_2 D_{3\}} \psi
\ee
and
\bea
&&O_{\{0011\}} + O_{\{3322\}} - O_{\{0022\}} -O_{\{3311\}} \label{eq:op0011}\\
&& \qquad = \overline{\psi} \gamma_{\{0} D_0 D_1 D_{1\}} \psi
+ \overline{\psi} \gamma_{\{3} D_3 D_2 D_{2\}} \psi
- \overline{\psi} \gamma_{\{0} D_0 D_2 D_{2\}} \psi
- \overline{\psi} \gamma_{\{3} D_3 D_1 D_{1\}} \psi
\nonumber ,
\eea
which was carried out in~\cite{Capitani:2000aq}. These operators are 
multiplicatively renormalizable at one loop, and measure the third moment 
of the unpolarized quark distribution.

Each of the four terms in Eq.~(\ref{eq:op0011}) gives the same value for the 
tadpole. Furthermore 
\be
O_{\{0011\}} = \frac{1}{6} \Bigg( 
O_{0011} + O_{0101} + O_{0110} + O_{1001} + O_{1010} + O_{1100} \Bigg) ,
\ee
and since for the tadpole what is important is the position of
the covariant derivatives, and not of the Dirac matrices, we have
\be
{\cal T}_{O_{\{0011\}}} = \frac{1}{3} \Bigg( {\cal T}_{O_{0011}} 
+ {\cal T}_{O_{0101}} + {\cal T}_{O_{0110}} \Bigg) 
\ee
(the remaining terms have the indices 0 and 1 exchanged, and therefore they 
give the same result). The results corresponding to these three terms are 
given in Eqs.~(\ref{eq:tadex2}), (\ref{eq:tadex3}) and (\ref{eq:tadex1}) 
respectively.

Let us begin by computing the tadpoles for operators with three covariant 
derivatives which have general indices $\mu$, $\nu$ and $\rho$ (that is, 
for the moment they are left open to be equal or different).
We have again that one can consider, for the sake of the computation of the
tadpole, 
$\overline{\psi} \stackrel{\rightarrow}{D} \stackrel{\rightarrow}{D}
\stackrel{\rightarrow}{D}\psi$ instead of 
$1/8 \, \overline{\psi} \stackrel{\leftrightarrow}{D}
\stackrel{\leftrightarrow}{D} \stackrel{\leftrightarrow}{D}\psi$,
which would be a lot more tedious to calculate.
Applying the three covariant derivatives in cascade gives
\bea
\stackrel{\rightarrow}{D}_\rho \stackrel{\rightarrow}{D}_\nu 
\stackrel{\rightarrow}{D}_\mu \psi(x) &=& \frac{1}{8a^3} \Bigg[ 
     U_\rho (x)  U_\nu (x+a\hat{\rho})  U_\mu (x+a\hat{\nu}+a\hat{\rho})  
     \psi (x+a\hat{\mu}+a\hat{\nu}+a\hat{\rho})  \label{eq:dddu} \\ 
&& - U_\rho (x)  U_\nu (x+a\hat{\rho})  
     U_\mu^\dagger (x-a\hat{\mu}+a\hat{\nu}+a\hat{\rho})  
     \psi (x-a\hat{\mu}+a\hat{\nu}+a\hat{\rho})  \nonumber \\
&& - U_\rho (x)  U_\nu^\dagger (x-a\hat{\nu}+a\hat{\rho})  
     U_\mu (x-a\hat{\nu}+a\hat{\rho})  
     \psi (x+a\hat{\mu}-a\hat{\nu}+a\hat{\rho})  \nonumber \\ 
&& + U_\rho (x)  U_\nu^\dagger (x-a\hat{\nu}+a\hat{\rho})  
     U_\mu^\dagger (x-a\hat{\mu}-a\hat{\nu}+a\hat{\rho})  
     \psi (x-a\hat{\mu}-a\hat{\nu}+a\hat{\rho})  \nonumber \\
&& - U_\rho^\dagger (x-a\hat{\rho})  U_\nu (x-a\hat{\rho})  
     U_\mu (x+a\hat{\nu}-a\hat{\rho})  
     \psi (x+a\hat{\mu}+a\hat{\nu}-a\hat{\rho})  \nonumber \\ 
&& + U_\rho^\dagger (x-a\hat{\rho})  U_\nu (x-a\hat{\rho})  
     U_\mu^\dagger (x-a\hat{\mu}+a\hat{\nu}-a\hat{\rho})  
     \psi (x-a\hat{\mu}+a\hat{\nu}-a\hat{\rho})  \nonumber \\
&& + U_\rho^\dagger (x-a\hat{\rho})  U_\nu^\dagger (x-a\hat{\nu}-a\hat{\rho})  
     U_\mu (x-a\hat{\nu}-a\hat{\rho})  
     \psi (x+a\hat{\mu}-a\hat{\nu}-a\hat{\rho})  \nonumber \\ 
&& - U_\rho^\dagger (x-a\hat{\rho})  U_\nu^\dagger (x-a\hat{\nu}-a\hat{\rho})  
     U_\mu^\dagger (x-a\hat{\mu}-a\hat{\nu}-a\hat{\rho})  
     \psi (x-a\hat{\mu}-a\hat{\nu}-a\hat{\rho}) \Bigg] \nonumber . 
\eea
For the calculation of the tadpoles we have now to expand the $U$'s in the 
above expression to second order, and keep the $A^2$ terms coming from the
same $U$ as well as the products $A\cdot A$ coming from two different $U$'s. 
Regarding the latter, if we were doing the calculations in Feynman gauge only 
the pairs of $A$ with the same index would give nonzero contributions. Here 
however we work in general covariant gauge, so we must consider all terms. 
In addition, we have to expand the $\psi$ terms to order $a^3p^3$, which 
reconstructs the tree level of the operator,
\be
-\ii p_\mu p_\nu p_\rho .
\ee
For example, one of such terms gives
\bea
\frac{1}{a^3} \psi (x+a\hat{\mu}+a\hat{\nu}+a\hat{\rho})
&\longrightarrow & 
\frac{1}{a^3} \e^{\ii ap_\mu}  \e^{\ii ap_\nu}  \e^{\ii ap_\rho} \\
&=& -\ii p_\mu p_\nu p_\rho - \delta_{\mu\nu} \ii p_\mu^2 p_\rho
- \delta_{\mu\rho} \ii p_\mu^2 p_\nu- \delta_{\nu\rho} \ii p_\nu^2 p_\mu
- \frac{\ii}{2} \delta_{\mu\nu} \delta_{\nu\rho} p_\mu^3 . \nonumber 
\eea
This factor $a^3$ compensates the factor $1/a^3$ coming from the covariant
derivatives, while the factor $a^2$ coming from the $U$ expansions multiplied 
with the factor $a^2$ present in the gluon propagator cancels the rescaling
factor of the integration variable, $a^4$. The tadpoles are then finite in 
the limit $a \to 0$.

Let us first consider the terms where both $A$'s come from the expansion
of the same $U$. This is a simple calculation which gives
\bea
&& C_F \, \frac{1}{8a^3} \Bigg( -\frac{g_0^2a^2}{2} \Bigg)
\int_{-\frac{\pi}{a}}^{\frac{\pi}{a}} \frac{d^4k}{(2\pi)^4} \, 
\Big( A_\mu^a (k) A_\mu^a (-k) + A_\nu^a (k) A_\nu^a (-k) 
    + A_\rho^a (k) A_\rho^a (-k) \Big)
\cdot 8 a^3 \ii p_\mu \ii p_\nu \ii p_\rho \nonumber \\
&& = -\frac{3g_0^2a^2}{2}
\int_{-\frac{\pi}{a}}^{\frac{\pi}{a}} \frac{d^4k}{(2\pi)^4} \, 
G_{\mu\mu}^{aa} (k) 
\cdot ( - \ii p_\mu p_\nu p_\rho ) ,
\eea
where we have contracted the $A$'s to form the propagator (note that now the
color index $a$ is not summed). Inserting the covariant gauge propagator 
(Eq.~(\ref{eq:gg})) and dividing for the tree level we have that the 
above expression is reduced to $g_0^2 C_F$ multiplied by~\footnote{It is easy 
to see that ${\cal B} (2;1) = 1/4 \cdot {\cal B} (1) = 1/4 \cdot Z_0$.}
\be
{\cal T}_1 = -\frac{3}{2} \Big( {\cal B} (1) - (1-\alpha) {\cal B} (2;1) \Big)
= -\frac{3}{2} Z_0 +(1-\alpha) \frac{3}{8} Z_0  .
\ee

The remaining part of the calculation of the tadpoles involves the terms
in which the $A$'s that have to be contracted come from the expansion
of different $U$'s, and is much more complicated. In the case in which all 
indices all different, which corresponds to the operator $O_{\{0123\}}$, 
the relevant part of the expansion of Eq.~(\ref{eq:dddu}) is
\bea
&&\frac{1}{8a^3} (-g_0^2 a^2) 
\int_{-\frac{\pi}{a}}^{\frac{\pi}{a}} \frac{d^4k}{(2\pi)^4} \,   \\
&& \times \Bigg[ 
  \Big(  A_\rho (-k) A_\nu (k) \e^{\ii a k_\rho/2} \e^{\ii a k_\nu/2}  
       + A_\nu (-k) A_\mu (k) \e^{\ii a k_\nu/2} \e^{\ii a k_\mu/2} 
\nonumber \\ && \qquad \qquad   
       + A_\rho (-k) A_\mu (k) \e^{\ii a k_\rho/2} \e^{\ii a k_\mu/2} 
         \e^{\ii a k_\nu} \Big) \Big( -\ii p_\mu p_\nu p_\rho 
 - \ii \delta_{\nu\rho} p_\nu^2 p_\mu - \ii \delta_{\mu\nu} p_\mu^2 p_\rho 
 - \ii \delta_{\mu\rho} p_\mu^2 p_\nu \Big) \nonumber  \\
&& + \Big( A_\rho (-k) A_\nu (k) \e^{\ii a k_\rho/2} \e^{\ii a k_\nu/2}  
       - A_\nu (-k) A_\mu (k) \e^{\ii a k_\nu/2} \e^{-\ii a k_\mu/2}  
\nonumber \\ && \qquad \qquad   
       - A_\rho (-k) A_\mu (k) \e^{\ii a k_\rho/2} \e^{-\ii a k_\mu/2} 
         \e^{\ii a k_\nu} \Big) \Big( -\ii p_\mu p_\nu p_\rho 
 + \ii \delta_{\nu\rho} p_\nu^2 p_\mu - \ii \delta_{\mu\nu} p_\mu^2 p_\rho 
 + \ii \delta_{\mu\rho} p_\mu^2 p_\nu \Big) \nonumber  \\
&& + \Big( - A_\rho (-k) A_\nu (k) \e^{\ii a k_\rho/2} \e^{-\ii a k_\nu/2}  
       - A_\nu (-k) A_\mu (k) \e^{-\ii a k_\nu/2} \e^{\ii a k_\mu/2}  
\nonumber \\ && \qquad \qquad   
       + A_\rho (-k) A_\mu (k) \e^{\ii a k_\rho/2} \e^{\ii a k_\mu/2} 
         \e^{-\ii a k_\nu} \Big) \Big( -\ii p_\mu p_\nu p_\rho 
 - \ii \delta_{\nu\rho} p_\nu^2 p_\mu + \ii \delta_{\mu\nu} p_\mu^2 p_\rho 
 + \ii \delta_{\mu\rho} p_\mu^2 p_\nu \Big) \nonumber  \\
&& + \Big( - A_\rho (-k) A_\nu (k) \e^{\ii a k_\rho/2} \e^{-\ii a k_\nu/2}  
       + A_\nu (-k) A_\mu (k) \e^{-\ii a k_\nu/2} \e^{-\ii a k_\mu/2}  
\nonumber \\ && \qquad \qquad   
       - A_\rho (-k) A_\mu (k) \e^{\ii a k_\rho/2} \e^{-\ii a k_\mu/2} 
         \e^{-\ii a k_\nu} \Big) \Big( -\ii p_\mu p_\nu p_\rho 
 + \ii \delta_{\nu\rho} p_\nu^2 p_\mu + \ii \delta_{\mu\nu} p_\mu^2 p_\rho 
 - \ii \delta_{\mu\rho} p_\mu^2 p_\nu \Big) \nonumber  \\
&& + \Big( - A_\rho (-k) A_\nu (k) \e^{-\ii a k_\rho/2} \e^{\ii a k_\nu/2}  
       + A_\nu (-k) A_\mu (k) \e^{\ii a k_\nu/2} \e^{\ii a k_\mu/2}  
\nonumber \\ && \qquad \qquad   
       - A_\rho (-k) A_\mu (k) \e^{-\ii a k_\rho/2} \e^{\ii a k_\mu/2} 
         \e^{\ii a k_\nu} \Big) \Big( -\ii p_\mu p_\nu p_\rho 
 + \ii \delta_{\nu\rho} p_\nu^2 p_\mu + \ii \delta_{\mu\nu} p_\mu^2 p_\rho 
 - \ii \delta_{\mu\rho} p_\mu^2 p_\nu \Big) \nonumber  \\
&& + \Big( - A_\rho (-k) A_\nu (k) \e^{-\ii a k_\rho/2} \e^{\ii a k_\nu/2}  
       - A_\nu (-k) A_\mu (k) \e^{\ii a k_\nu/2} \e^{-\ii a k_\mu/2}  
\nonumber \\ && \qquad \qquad   
       + A_\rho (-k) A_\mu (k) \e^{-\ii a k_\rho/2} \e^{-\ii a k_\mu/2} 
         \e^{\ii a k_\nu} \Big) \Big( -\ii p_\mu p_\nu p_\rho 
 - \ii \delta_{\nu\rho} p_\nu^2 p_\mu + \ii \delta_{\mu\nu} p_\mu^2 p_\rho 
 + \ii \delta_{\mu\rho} p_\mu^2 p_\nu \Big) \nonumber  \\
&& + \Big(   A_\rho (-k) A_\nu (k) \e^{-\ii a k_\rho/2} \e^{-\ii a k_\nu/2}  
       - A_\nu (-k) A_\mu (k) \e^{-\ii a k_\nu/2} \e^{\ii a k_\mu/2}  
\nonumber \\ && \qquad \qquad   
       - A_\rho (-k) A_\mu (k) \e^{-\ii a k_\rho/2} \e^{\ii a k_\mu/2} 
         \e^{-\ii a k_\nu} \Big) \Big( -\ii p_\mu p_\nu p_\rho 
 + \ii \delta_{\nu\rho} p_\nu^2 p_\mu - \ii \delta_{\mu\nu} p_\mu^2 p_\rho 
 + \ii \delta_{\mu\rho} p_\mu^2 p_\nu \Big) \nonumber  \\
&& + \Big(   A_\rho (-k) A_\nu (k) \e^{-\ii a k_\rho/2} \e^{-\ii a k_\nu/2}  
       + A_\nu (-k) A_\mu (k) \e^{-\ii a k_\nu/2} \e^{-\ii a k_\mu/2}  
\nonumber \\ && \qquad \qquad   
       + A_\rho (-k) A_\mu (k) \e^{-\ii a k_\rho/2} \e^{-\ii a k_\mu/2} 
         \e^{-\ii a k_\nu} \Big) \Big( -\ii p_\mu p_\nu p_\rho 
 - \ii \delta_{\nu\rho} p_\nu^2 p_\mu - \ii \delta_{\mu\nu} p_\mu^2 p_\rho 
 - \ii \delta_{\mu\rho} p_\mu^2 p_\nu \Big) \Bigg] \nonumber ,
\eea
which gives
\bea
&& ( - \ii p_\mu p_\nu p_\rho ) g_0^2 a^2
\int_{-\frac{\pi}{a}}^{\frac{\pi}{a}} \frac{d^4k}{(2\pi)^4} \, 
\Bigg( G_{\rho\nu} (k) \sin \frac{ak_\rho}{2} \sin \frac{ak_\nu}{2} \\
&& \qquad + G_{\nu\mu} (k) \sin \frac{ak_\nu}{2} \sin \frac{ak_\mu}{2} 
+ G_{\rho\mu} (k) \cos ak_\nu \sin \frac{ak_\rho}{2} \sin \frac{ak_\mu}{2} 
\Bigg) \nonumber .
\eea
The result is
\bea
{\cal T}_2^{(\mu\nu\sigma)} &=& -(1-\alpha) 
\int_{-\pi}^\pi \frac{d^4k}{(2\pi)^4} \, 
\Bigg( \frac{ 4 \sin^2 \frac{k_\rho}{2} \sin^2 \frac{k_\nu}{2} 
+ 4 \sin^2 \frac{k_\nu}{2} \sin^2 \frac{k_\mu}{2} 
+ \Big( 1 - 2 \sin^2 \frac{k_\nu}{2} \Big) 
\sin^2 \frac{k_\rho}{2} \sin^2 \frac{k_\mu}{2} }{
\Big( 4 \sum_\lambda \sin^2 \frac{k_\lambda}{2} \Big)^2 } \Bigg) \nonumber \\
&=&  (1-\alpha) \Bigg( -\frac{3}{4} {\cal B} (2;1,1)
 +\frac{1}{8} {\cal B} (2;1,1,1) \Bigg) \nonumber \\ 
&=&  (1-\alpha) \Bigg( -\frac{5}{12} Z_0
 -\frac{1}{4} Z_1  + \frac{1}{16} \Bigg) .
\eea
In the last line we have used the results which we have obtained in the 
previous subsection applying the recursion relations of the algebraic method.
This result, added to ${\cal T}_1$, gives the operator tadpole for 
$O_{\{0123\}}$, Eq.~(\ref{eq:optad123}):
\be
{\cal T}^{(\mu\nu\sigma)} = -\frac{3}{2} Z_0 
+ (1-\alpha) \Bigg( -\frac{1}{24} Z_0 -\frac{1}{4} Z_1  + \frac{1}{16} \Bigg) 
= 36.69915049 +(1-\alpha) \, 4.59514785 .
\ee

We now consider the cases in which two of the indices are equal, which are
necessary for the computation of the operator tadpole of $O_{\{0011\}}$.
We have
\bea
{\cal T}_2^{(\nu\nu\mu)} &=& a^2
\int_{-\frac{\pi}{a}}^{\frac{\pi}{a}} \frac{d^4k}{(2\pi)^4} \, 
\Bigg( G_{\nu\nu} (k) \Bigg( \sin^2 \frac{ak_\nu}{2} - \cos^2 \frac{ak_\nu}{2} 
\Bigg) \nonumber \\ &&
+ G_{\mu\nu} (k) \Bigg( \sin \frac{ak_\mu}{2} \sin \frac{ak_\nu}{2} 
+ \cos ak_\nu \sin \frac{ak_\mu}{2} \sin \frac{ak_\nu}{2} 
+ \sin ak_\nu \sin \frac{ak_\mu}{2} \sin \frac{ak_\nu}{2} \Bigg) \Bigg) 
\nonumber \\
&=& \int_{-\pi}^\pi \frac{d^4k}{(2\pi)^4} \, \Bigg(
\frac{\sin^2 \frac{k_\nu}{2} - \cos^2 \frac{k_\nu}{2}}{ 
\Big( 4 \sum_\lambda \sin^2 \frac{k_\lambda}{2} \Big)} -(1-\alpha) \times
\nonumber \\
&& \times
\frac{ \sin^4 \frac{k_\nu}{2} -\sin^2 \frac{k_\nu}{2} \cos^2 \frac{k_\nu}{2}
+\sin^2 \frac{k_\mu}{2} \sin^2 \frac{k_\nu}{2} \Big( 1+\cos k_\nu \Big)
+\sin^2 \frac{k_\mu}{2} \sin \frac{k_\nu}{2} \cos \frac{k_\nu}{2} \sin k_\nu
}{ \Big( 4 \sum_\lambda \sin^2 \frac{k_\lambda}{2} \Big)^2 } \Bigg)
\nonumber \\
&=& \frac{1}{2} {\cal B} (1;1) - {\cal B} (1)
-(1-\alpha) \Bigg( \frac{1}{2} {\cal B} (2;2) - {\cal B} (2;1) 
+ {\cal B} (2;1,1) - \frac{1}{4} {\cal B} (2;2,1) \Bigg) \nonumber \\
&=& - Z_0 +\frac{1}{8} - (1-\alpha) \Bigg( \frac{1}{2} \Bigg( \frac{1}{4} 
-Z_0 \Bigg) - \frac{1}{4} Z_0 +\frac{1}{3} Z_0 -\frac{1}{4} 
\Bigg( \frac{4}{3} Z_0 +4 Z_1 -\frac{1}{2} \Bigg) \Bigg) \nonumber \\
&=& - Z_0 +\frac{1}{8} + (1-\alpha) \Bigg( \frac{3}{4} Z_0 +Z_1 
-\frac{1}{4} \Bigg).
\eea
Again, in the last line we have substituted the results obtained in the 
previous subsection using the algebraic method.
The willing reader can check that
\be
{\cal T}_2^{(\rho\mu\mu)} = {\cal T}_2^{(\nu\nu\mu)} ,
\ee
but the remaining combination gives a different result: 
\bea
{\cal T}_2^{(\mu\nu\mu)} &=& 
\int_{-\pi}^\pi \frac{d^4k}{(2\pi)^4} \, \Bigg(
\frac{\cos k_\nu \Big( \sin^2 \frac{k_\nu}{2} - \cos^2 \frac{k_\nu}{2} \Big)}{
\Big( 4 \sum_\lambda \sin^2 \frac{k_\lambda}{2} \Big) }  \nonumber \\
&& - (1-\alpha)
\frac{ \cos k_\nu \Big( \sin^4 \frac{k_\nu}{2} -\sin^2 \frac{k_\nu}{2} \cos^2
 \frac{k_\nu}{2} \Big) +2 \sin^2 \frac{k_\mu}{2} \sin^2 \frac{k_\nu}{2} 
}{ \Big( 4 \sum_\lambda \sin^2 \frac{k_\lambda}{2} \Big)^2  } \Bigg)
\nonumber \\
&=& - \frac{1}{4} {\cal B} (1;1,1) +{\cal B} (1;1) - {\cal B} (1)
-(1-\alpha) \Bigg( \frac{1}{2} {\cal B} (2;2) - {\cal B} (2;1) 
+ {\cal B} (2;1,1) - \frac{1}{4} {\cal B} (2;2,1) \Bigg) \nonumber \\
&=& -Z_0 -Z_1 +\frac{1}{4} + (1-\alpha) \Bigg( \frac{3}{4} Z_0 +Z_1 
-\frac{1}{4} \Bigg) .
\eea

Adding the term ${\cal T}_1$ to each of the above cases we have
\bea
{\cal T}^{(\rho\mu\mu)}  = {\cal T}^{(\nu\nu\mu)} &=& 
- \frac{5}{2} Z_0 + \frac{1}{8} + (1-\alpha) \Bigg( \frac{9}{8} Z_0 +Z_1 
-\frac{1}{4} \Bigg) \\ 
{\cal T}^{(\mu\nu\mu)}  &=& 
- \frac{5}{2} Z_0 - Z_1 + \frac{1}{4} + (1-\alpha) \Bigg( \frac{9}{8} Z_0 +Z_1 
-\frac{1}{4}\Bigg) ,
\eea
so that the operator tadpole for the operator $O_{\{0011\}}$, and hence
for the operator $O_{\{0011\}} + O_{\{3322\}} - O_{\{0022\}} -O_{\{3311\}}$,
is finally
\bea
{\cal T}_{O_{\{0011\}}} 
&=& \frac{1}{3} \Bigg({\cal T}^{(\mu\nu\mu)} +2 {\cal T}^{(\nu\nu\mu)}\Bigg) \\
&=& - \frac{5}{2} Z_0 - \frac{1}{3} Z_1 + \frac{1}{6} 
+ (1-\alpha) \Bigg( \frac{9}{8} Z_0 +Z_1 -\frac{1}{4} \Bigg) \\
&=& -40.5196866756  + (1-\alpha) \, 5.0660880895 .
\eea
All these results can be given with much more precision using the numbers
reported in Appendix B.

The tadpoles of operators with one or two covariant derivatives, which are 
given at the beginning of this Section, are much simpler to calculate and are 
left as an exercise for the reader.

\subsection{The first moment of the gluon momentum distribution}

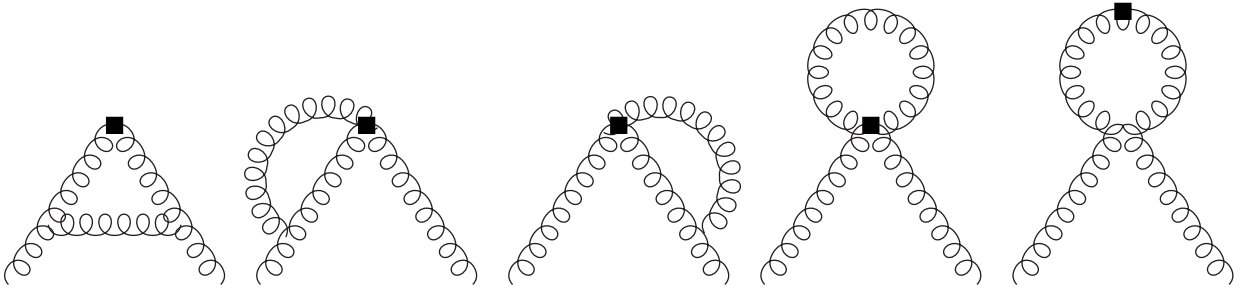
\begin{figure}[t]
\begin{center}
\begin{picture}(500,150)(-5,0)
\Gluon(0,0)(40,60){4}{8}
\Gluon(40,60)(80,0){4}{8}
\Gluon(15,22.5)(65,22.5){-4}{6}
\GBoxc(40,60)(6,6){0}
\Gluon(95,0)(135,60){4}{8}
\Gluon(135,60)(175,0){4}{8}
\GlueArc(120,40)(27,45,235){-4}{12}
\GBoxc(135,60)(6,6){0}
\Gluon(190,0)(230,60){4}{8}
\Gluon(230,60)(270,0){4}{8}
\GlueArc(245,40)(27,-50,140){-4}{12}
\GBoxc(230,60)(6,6){0}
\Gluon(285,0)(325,60){4}{8}
\Gluon(325,60)(365,0){4}{8}
\GlueArc(325,80)(20,-90,270){4}{15}
\GBoxc(325,60)(6,6){0}
\Gluon(380,0)(420,60){4}{8}
\Gluon(420,60)(460,0){4}{8}
\GlueArc(420,80)(20,-90,270){4}{15}
\GBoxc(420,103)(6,6){0}
\end{picture}
\end{center}
\caption{\small ``Proper'' diagrams for the 1-loop correction to the 
matrix element
$\langle g | \sum_\rho \Tr \, (F_{\mu\rho} F_{\rho\nu}) | g \rangle $. 
The black squares indicate the insertion of the operator. Notice that the 
last two diagrams are quite different: the last diagram contains a 4-gluon 
vertex (and vanishes for this matrix element), while the previous one is
the tadpole coming from the second-order expansion of the operator.}
\label{fig:properff}
\end{figure}

We want now to show the 1-loop result of a gluonic matrix element, which is 
probably the most complicated case in which the resulting integrals have been 
reduced to an expression containing only the two bosonic constants $Z_0$ and 
$Z_1$. The calculation of the operator measuring the first moment of the gluon 
momentum distribution, which also corresponds to the gluonic contribution to 
the energy-momentum tensor, has been done analytically and then the integrals 
reduced using the algebraic method~\cite{Caracciolo:1991cp,Capitani:qn}. 
The renormalization of this operator can be obtained by computing the radiative
corrections to the gluonic matrix element 
\be
\langle g | \sum_\rho \Tr \, (F_{\mu\rho} F_{\rho\nu}) | g \rangle .
\ee
The relevant diagrams are shown in Fig.~\ref{fig:properff}.

The vertex function gives
\be
N_c \cdot \Bigg( -\frac{13}{192} +\frac{23}{48\pi^2} -\frac{53}{144}Z_0 
+\frac{1}{3}Z_1 -\frac{3}{16\pi^2} \Big( \log p^2 + \gamma_E - F_0 \Big)\Bigg),
\ee
the result for the sails is 
\be
N_c \cdot \Bigg( \frac{1}{192} -\frac{7}{9\pi^2} +\frac{31}{24}Z_0 
-\frac{19}{48}Z_1 +\frac{7}{24\pi^2} \Big( \log p^2 + \gamma_E - F_0 \Big) 
\Bigg) ,
\ee
and the operator tadpole gives
\be
N_c \cdot \Bigg( -\frac{3}{64} -\frac{4}{3}Z_0 \Bigg) +\frac{1}{4N_c}.
\ee
The diagram containing the 4-gluon vertex (the rightmost in 
Fig.~\ref{fig:properff}) is zero for this operator.

One has still to add the gluon self-energy at one 
loop~\cite{Caracciolo:1991cp,Capitani:qn},
\be
N_c \cdot \Bigg( \frac{1}{16} +\frac{7}{36\pi^2} +\frac{7}{72}Z_0 
-\frac{5}{48\pi^2} \Big( \log p^2 + \gamma_E - F_0 \Big) \Bigg) 
-\frac{1}{8N_c} .
\ee
This is the case $N_f=0$, that is the quenched approximation. The diagrams 
are as in Fig.~\ref{fig:gluon_self}, without the quark loops. 
The numerical result for the gluon self-energy is
\be
\frac{1}{16\pi^2} \Bigg( 21.679380\, N_c - 19.739209\, \frac{1}{N_c} \Bigg) .
\label{eq:numgluonself}
\ee

Summing everything together we have that the complete renormalization of the 
operator $\sum_\rho \Tr \, (F_{\mu\rho} F_{\rho\nu})$ at 1-loop is given by 
\be
N_c \cdot \Bigg( -\frac{3}{64} -\frac{5}{48\pi^2} -\frac{5}{16}Z_0 
-\frac{1}{16}Z_1 \Bigg) +\frac{1}{8N_c} .
\ee
The divergences of the individual diagrams have canceled, and so the 
energy-momentum tensor has zero anomalous dimensions, as it should be. 
Numerically one has
\be
\frac{1}{16\pi^2} \Bigg( -17.778285\, N_c + 19.739209\, \frac{1}{N_c} \Bigg) .
\ee
Using the values of $Z_0$ and $Z_1$ reported in Appendix B this result 
(as well as the gluon self-energy Eq.~(\ref{eq:numgluonself}) can be 
stated with almost 400 significant decimal places.

We want to point out that working out the color structure for the diagrams 
considered here is more complicated than the case of the quark matrix elements
measuring the moments of the unpolarized quark distributions. The reduction 
of the color factors, the $f_{abc}$ tensors etc. can become quite cumbersome. 
On the other hand, the momentum integrals are much simpler in these gluonic 
matrix elements than in the quark case, because the fermion propagator 
is missing. Numerically evaluating integrals that have only gluon 
propagators poses much less of a computational challenge.

\subsection{The general fermionic case}
\label{sec:thegeneralfermioniccase}

Similarly to the bosonic case, one only needs to compute lattice integrals with
vanishing external momenta. Any lattice zero-momentum integral coming from the
calculation of lattice Feynman diagrams in the general Wilson case can be 
written as a linear combination of terms of the form
\be
{\cal F} (p,q;n_x,n_y,n_z,n_t) =\,
 \int_{-\pi}^\pi {d^4 k \over (2\pi)^4} \,
 {\hat{k}_x^{2 n_x} \hat{k}_y^{2 n_y} \hat{k}_z^{2 n_z} \hat{k}_t^{2 n_t}\over
     D_F(k,m_f)^p D_B(k,m_b)^q} ,
\ee
where $p$, $q$ and $n_i$ are positive integers, the inverse bosonic propagator
is
\be
D_B(k,m_b) = \hat{k}^2 + m_b^2 ,
\ee
and the denominator appearing in the propagator of Wilson fermions is taken
to be
\be
D_F(k,m_f) = \sum_i \sin^2 k_i + {r^2\over 4} (\hat{k}^2)^2 + m_f^2 .
\label{eq:falseprop}
\ee
Actually, the correct Wilson denominator would be
\be
\hat{D}_F(k,m_f) = \sum_i \sin^2 k_i + \left({r\over 2} \hat{k}^2 + m_f
    \right)^2 ,
\ee
however in this algorithm $m_f$ only play the role of an infrared regulator 
and thus it does not need to be the true fermion mass. This form of the 
fermion propagator turns out to be much easier to handle than the true one.
Moreover, for integrals which are only logarithmically divergent,
the two forms give exactly the same result in the limit of small quark masses.

As we did in the fermionic case, we first generalize the integrals to 
noninteger dimensions introducing
\be
{\cal F}_\delta (p,q;n_x,n_y,n_z,n_t) =
 \int_{-\pi}^\pi {d^4 k\over (2\pi)^4} \,
 {\hat{k}_x^{2 n_x} \hat{k}_y^{2 n_y} \hat{k}_z^{2 n_z} \hat{k}_t^{2 n_t}\over
     D_F(k,m_f)^{p+\delta} D_B(k,m_b)^q} ,
\ee
where $p$ and $q$ are arbitrary integers (not necessarily positive) 
and $\delta$ is used in the intermediate steps of the calculations and then
safely set to zero at the end of them.

It turns out that every ${\cal F}_\delta(p,q;n_x,n_y,n_z,n_t)$ with $q\le 0$ 
(i.e., a purely fermionic integral) can be expressed iteratively in terms 
of nine purely fermionic integrals, in our case 
${\cal F} (1,0)$, ${\cal F} (1,-1)$, ${\cal F} (1,-2)$, ${\cal F} (2,0)$, 
${\cal F} (2,-1)$, ${\cal F} (2,-2)$, ${\cal F} (3,-2)$, ${\cal F} (3,-3)$ 
and ${\cal F} (3,-4)$.
Purely fermionic integrals can always be expressed in terms of integrals of 
the same type. This is a general property of all recursion relations.
The integral ${\cal F} (2,0)$ appears only in the case of divergent integrals.
Only eight constants are then needed if the original purely fermionic integral 
is finite (i.e., $q\le 0$ and $p\le 1$). 

In the general case in which $q$ can be positive one needs three additional 
constants, called $Y_1$, $Y_2$, and $Y_3$, to describe the mixed 
fermionic-bosonic integrals, plus the constants $Z_0$, $Z_1$ and $F_0$
which already appeared in the purely bosonic case.

While for the bosonic case we could give a complete treatment of the reduction
steps, for fermions, due to the complexity of the procedure, we can only 
sketch them. For the readers interested to apply the method, 
in~\cite{Burgio:1996ji} all details can be found.

There are four steps in this general fermionic method.

The first step consists in expressing each integral 
${\cal F}_\delta(p,q;n_x,n_y,n_z,n_t)$ in terms of ${\cal F}_\delta(p,q)$ 
only (that is, pure denominators). For this we need three sets of recursion 
relations. From the trivial identity $D_B(k,m_b) = \sum_i \hat{k}_i^2 + m_b^2$
we obtain the first set of recursion relations:
\bea
{\cal F}_\delta (p,q;1) &=& {1\over4} \left[ {\cal F}_\delta (p,q-1) 
     - m_b^2 {\cal F}_\delta (p,q)\right] , \\ 
{\cal F}_\delta (p,q;x,1) &=& {1\over3} \left[ {\cal F}_\delta (p,q-1;x) 
     - m_b^2 {\cal F}_\delta (p,q;x) \right. \\
     && - \left. {\cal F}_\delta (p,q;x+1) \right] , \nonumber \\ 
{\cal F}_\delta (p,q;x,y,1) &=& 
     {1\over2} \left[ {\cal F}_\delta (p,q-1;x,y) 
     - m_b^2 {\cal F}_\delta (p,q;x,y) \right. \\
     && \left. - {\cal F}_\delta (p,q;x+1,y) 
     - {\cal F}_\delta (p,q;x,y+1) \right] , \nonumber \\ 
{\cal F}_\delta (p,q;x,y,z,1) &=&  {\cal F}_\delta (p,q-1;x,y,z) 
     - m_b^2 {\cal F}_\delta (p,q;x,y,z) \\
     && - {\cal F}_\delta (p,q;x+1,y,z) 
     - {\cal F}_\delta (p,q;x,y+1,z) 
     - {\cal F}_\delta (p,q;x,y,z+1) . \nonumber 
\eea
From the identity
\be
\sum_i \hat{k}^4_i = 4 (D_B (k,m_b) - D_F(k,m_f) - m_b^2 + m_f^2)
+ r^2 (D_B(k,m_b) - m_b^2)^2
\ee
we get a second set of recursion relations, of which
we give here only some examples: 
\bea
{\cal F}_\delta (p,q;x,y,2) &=& 
2 \Big[\vphantom{{1\over4}}{\cal F}_\delta(p,q-1;x,y) 
-{\cal F}_\delta(p-1,q;x,y) \\ 
&&
+ (m_f^2 - m_b^2) {\cal F}_\delta (p,q;x,y) 
- {1\over 4} {\cal F}_\delta (p,q;x+2,y) 
- {1\over 4} {\cal F}_\delta (p,q;x,y+2) \Big] \nonumber \\
&&
+ {r^2\over2} \Big[ {\cal F}_\delta (p,q-2;x,y) 
-2 m_b^2 {\cal F}_\delta (p,q-1;x,y) +m_b^4 {\cal F}_\delta (p,q;x,y) \Big] ,
\nonumber \\
{\cal F}_\delta (p,q;x,y,z,2) &=& 4 \Big[\vphantom{{1\over4}} 
{\cal F}_\delta (p,q-1;x,y,z) - {\cal F}_\delta(p-1,q;x,y,z)  \\
&&+ (m_f^2 - m_b^2) {\cal F}_\delta (p,q;x,y,z) 
- {1\over4} {\cal F}_\delta (p,q;x+2,y,z) \nonumber \\
&&- {1\over4} {\cal F}_\delta (p,q;x,y+2,z)
- {1\over4} {\cal F}_\delta (p,q;x,y;z+2)\Big] \nonumber \\
&& + r^2 \Big[ {\cal F}_\delta (p,q-2;x,y,z)
- 2 m_b^2 {\cal F}_\delta (p,q-1;x,y,z) + m_b^4 {\cal F}_\delta (p,q;x,y,z) 
\Big] . \nonumber
\eea
Integrating by parts, for $s\ge3$, the equation 
\bea
{(\hat{k}^2_w)^s\over D_F(k,m_f)^{p+\delta}} &=&
{4 (\hat{k}^2_w)^{s-1} - 4 (2 + r^2 \hat{k}^2) (\hat{k}^2_w)^{s-3}
     \sin^2 k_w\over
      D_F(k,m_f)^{p+\delta} } \nonumber \\ 
&& -{4 (\hat{k}^2_w)^{s-3}\over p + \delta -1} 
\sin k_w {\partial\over\partial k_w} {1\over D_F(k,m_f)^{p-1+\delta}} ,
\eea
we obtain the third set of recursion relations:
\bea
{\cal F}_\delta (p,q;\ldots,s) &=&
6 {\cal F}_\delta (p,q;\ldots,s-1) - 8 {\cal F}_\delta (p,q;\ldots,s-2)
\nonumber \\
&& - 4 r^2 {\cal F}_\delta (p,q-1;\ldots,s-2) 
+ 4 r^2 m_b^2 {\cal F}_\delta (p,q;\ldots,s-2) 
\nonumber \\
&& + r^2 {\cal F}_\delta (p,q-1;\ldots,s-1) 
- r^2 m_b^2 {\cal F}_\delta (p,q;\ldots,s-1)
\nonumber \\
&& +{4\over p+\delta -1} \Big[-2 q {\cal F}_\delta (p-1,q+1,\ldots,s-2)
\nonumber \\
&& + {q\over2} {\cal F}_\delta (p-1,q+1;\ldots,s-1) 
+ (2 s-5) {\cal F}_\delta (p-1,q;\dots,s-3) 
\nonumber \\
&& - {1\over2} (s-2) {\cal F}_\delta (p-1,q;\ldots,s-2) \Big] .
\eea
As before, by looking at this relation for $p=1$ we see that in 
general we have to keep contributions of order $\delta$ in the 
intermediate stages of the algebraic reductions.

Using the recursion relations introduced up to now each integral 
${\cal F}_\delta (p,q;n_x,n_y,n_z,n_t)$ can be reduced at the end of 
the first step to a sum of pure denominator integrals of the form
\be
{\cal F}_\delta (p,q;n_x,n_y,n_z,n_t) \,=\,
  \sum_{r=p-k+1}^p \sum_{s=q-k}^{q+k} a_{rs}(m,\delta) {\cal F}_\delta(r,s) ,
\ee
where $k= (n_x + n_y + n_z + n_t)$, and $m=m_b=m_f$.
In the limit $m\to 0$ this becomes
\be
{\cal F}_\delta (p,q;n_x,n_y,n_z,n_t) \,=\,
  \sum_{r=p-k+1}^p \sum_{s=q-k}^{q+k} a_{rs}(0,\delta) {\cal F}_\delta(r,s) 
+\, {\cal R}(m,\delta)+\, O(m^2) ,
\ee
where ${\cal R}(m,\delta)$ depends on the values of $p$ and $q$ as follows:
\begin{enumerate}
\item $p>0$, $q\le 0$:
${\cal R}(m,\delta)$ is a polynomial in $1/m^2$, finite for $\delta\to 0$;
\item $p>0$, $q>0$:
${\cal R}(m,\delta) = {1\over\delta} \left(1 - \delta\log m^2\right) 
   {\cal R}^{(1)}(m)+\, {\cal R}^{(2)}(m) + O(\delta)$,
where ${\cal R}^{(i)}(m)$ are polynomials in $1/m^2$;
\item $p\le 0$,$q\le 0$: 
${\cal R}(m,\delta) = 0$;
\item $p\le 0$, $q>0$:
${\cal R}(m,\delta) = \left(1 - \delta\log m^2\right) 
   {\cal R}^{(1)}(m)+\,\delta\, {\cal R}^{(2)}(m) + O(\delta^2)$.
\end{enumerate}

In the second step, with the systematic use of the identity
\be
{\cal F}_\delta (p,q;1,1,1,1) - 4 {\cal F}_\delta (p,q+1;2,1,1,1)
- m^2 {\cal F}_\delta (p,q+1;1,1,1,1) =\, 0 ,   
\label{idf1}
\ee
one obtains a nontrivial relation of the form
\be
   \sum_{r,s} f_{rs}(p,q;\delta) {\cal F}_\delta (r,s) 
  +{\cal R}_\delta (p,q;m,\delta) = 0 ,
\ee
where $p-4\le r \le p$. 
As in the bosonic case, it can be used to obtain new recursion relations;
in this case, one can exploit it in three different ways.

At the end of the second step one is then able to reduce every 
${\cal F}_\delta(p,q)$ in terms of only the ${\cal F}_\delta(r,s)$'s with
$0\le r \le 3$ and arbitrary $s$, or $r \leq -1$ and $s=1,2,3$, or 
$r\ge 4$ and $s=0,-1,-2$.

In the third step we systematically use the identity 
\bea
&& {\cal F}_\delta (p,q;1,1,1,1) 
 -{\cal F}_\delta (p+1,q-1;1,1,1,1) 
+{\cal F}_\delta (p+1,q;3,1,1,1)  \nonumber \\
&& 
- {1\over4} \Big[ {\cal F}_\delta (p+1,q-2;1,1,1,1)- 2 m^2 
{\cal F}_\delta (p+1,q-1;1,1,1,1)  \nonumber \\
&& 
+ m^4 {\cal F}_\delta (p+1,q;1,1,1,1)\Big] = 0 ,
\label{idf2}
\eea
which is applied in four different ways according to the particular properties
of the four regions $q\le 0$ and $0\le p \le 3$, $q> 0$ and $0\le p \le 3$,
$q=1,2,3$ and $p \le -1$, $q=-2,-1,0$ and $p \ge 4$.

We can at the end of the third step express the remaining 
${\cal F}_\delta(p,q)$ in terms of only the ${\cal F}_\delta(r,s)$'s with 
$r=3$ and $-4\le s \le 0$, or $r=2$ and $-4\le s \le 2$, or
$r=1$ and $-4\le s \le 4$, or $r=0$ and $-4\le s \le 6$, or $r=-1$ and $s=2$.

In the fourth step the identities in Eqs.~(\ref{idf1}) and (\ref{idf2}), 
which so far have not been used for all possible values of $p$ and $q$, 
are further exploited to provide additional relations between the remaining 
integrals. One has to look systematically for further values of $p$ and $q$ 
for which these two identities were not trivially satisfied. 
At the end of this process one can then achieve a further decrease 
of the number of independent constants.
We give an example of these additional relations:
\bea
{\cal F}_\delta(0,4) &=& (1 - \delta \log m^2) \left[
{1\over {96\pi^2 m^4}}
-{{19}\over {4608\pi^2 m^2}}
+ {1\over {9216\pi^2}}\right]
\\ &&
+ {{31}\over {144}} {\cal F}_\delta(0,3)  
-{{13}\over {1152}} {\cal F}_\delta(0,2) 
+ {1\over {9216}} {\cal F}_\delta(0,1)
\nonumber \\ &&
+ \delta \left[
-{5\over {576\pi^2 m^4}}
+ {{61}\over {18432\pi^2 m^2}}
+ {{80989}\over {88473600\pi^2}} \right.
+ {{347}\over {1440}} {\cal F}_\delta(0,3)
\nonumber \\ &&
-{{83}\over {2560}} {\cal F}_\delta(0,2)  
+ {{137}\over {184320}} {\cal F}_\delta(0,1)
-{{689}\over {2880}} {\cal F}_\delta(1,2) 
+ {{1139}\over {23040}} {\cal F}_\delta(1,1)
\nonumber \\ &&
-{{415}\over {147456}} {\cal F}_\delta(1,0)
+ {{23}\over {4423680}} {\cal F}_\delta(1,-1)  
-{{329}\over {11520}} {\cal F}_\delta(2,0)  
+ {{13283}\over {1105920}} {\cal F}_\delta(2,-1)
\nonumber \\ &&
-{{391}\over {221184}} {\cal F}_\delta(2,-2)  
-{{30479}\over {2211840}} {\cal F}_\delta(3,-2)
\left. + {{437}\over {245760}} {\cal F}_\delta(3,-3) 
+ {{161}\over {589824}} {\cal F}_\delta(3,-4) \right] . \nonumber 
\eea

At the end of these complicated reductions, one is able to write a 
generic purely fermionic integral ($p>0$ and $q\le 0$) in terms of 
eight infrared-finite integrals:
${\cal F}(1,0)$, ${\cal F}(1,-1)$, ${\cal F}(1,-2)$, 
${\cal F}(2,-1)$, ${\cal F}(2,-2)$, ${\cal F}(3,-2)$, 
${\cal F}(3,-3)$ and ${\cal F}(3,-4)$,
plus a constant, $Y_0$, which appears in the logarithmic divergent integral
\be
{\cal F}(2,0) =\, - {1\over 16 \pi^2} \left( \log m^2 + \gamma_E - F_0\right) 
+ Y_0 .
\ee
For the mixed fermionic-bosonic integrals ($p>0$ and $q> 0$)
one must introduce three additional constants, which have been chosen as
\bea
Y_1 &=& {1\over 8}\, {\cal F} (1,1;1,1,1) , \\
Y_2 &=& {1\over 16}\, {\cal F} (1,1;1,1,1,1) , \\
Y_3 &=& {1\over 16}\, {\cal F} (1,2;1,1,1) .
\eea
In Table~\ref{tab:fermionicconstants} we report the values of the new basic 
constants introduced in the general case.

\begin{table}[t]
\begin{center}
\begin{tabular}{|c|c|}
\hline
${\cal F}(1,0)$  & \hphantom{$-$}  0.08539036359532067914  \\
${\cal F}(1,-1)$ & \hphantom{$-$}  0.46936331002699614475  \\
${\cal F}(1,-2)$ & \hphantom{$-$}  3.39456907367713000586  \\
${\cal F}(2,-1)$ & \hphantom{$-$}  0.05188019503901136636  \\
${\cal F}(2,-2)$ & \hphantom{$-$}  0.23874773756341478520  \\
${\cal F}(3,-2)$ & \hphantom{$-$}  0.03447644143803223145  \\
${\cal F}(3,-3)$ & \hphantom{$-$}  0.13202727122781293085  \\
${\cal F}(3,-4)$ & \hphantom{$-$}  0.75167199030295682254  \\
$Y_0$            & $-$             0.01849765846791657356  \\
$Y_1$            & \hphantom{$-$}  0.00376636333661866811  \\
$Y_2$            & \hphantom{$-$}  0.00265395729487879354  \\
$Y_3$            & \hphantom{$-$}  0.00022751540615147107  \\
\hline
\end{tabular}
\caption{New constants appearing in the general fermionic case.}
\label{tab:fermionicconstants}
\end{center}
\end{table}

Most of the basic recursion relations for fermions presented above can be 
easily generalized to integrals which are dimensionally regularized, however 
some of them are instead intrinsically four-dimensional identities. For this 
reason it is more convenient to apply the above reductions which use a mass 
as a regulator. We don't give here the complicated relation between the two 
schemes, but only say that fortunately the case of logarithmically divergent 
integrals can still be dealt with rather easily, because it is enough to make 
the substitution $2/(d-4) + \log 4 \pi$ for $\log m^2 + \gamma_E$. 
The relation between integrals computed using the true fermion propagator and 
integrals computed using the propagator (\ref{eq:falseprop}) is also rather 
complicated. However, if the integrals are only logarithmically divergent, for 
$m\to 0$ the two results coincide. For power-divergent expressions instead, if
one uses the true fermion propagator the divergent part is a polynomial in 
$1/m$ instead of $1/m^2$, and the expressions are in general more cumbersome 
than those involving the propagator in Eq.~(\ref{eq:falseprop}).

The method discussed here depends on the form of the quark propagator,
but not on the vertices. It can thus be applied to $O(a)$ improved fermions
as well. A version for, say, overlap fermions, which involves more complicated 
denominators, has not yet been developed. In this case one has to find a 
generalization of the method, but it can be expected that the corresponding 
recursion relations will turn out to be much more complicated, since already 
Wilson fermions bring rather cumbersome recursion relations.
In these cases, a convenient method to reduce all integrals could be the one 
of~\cite{Laporta:2001dd}, which has been used in~\cite{Becher:2002if}. It uses
a brute force approach which reduces integrals to simpler ones integrals 
(with lower values of the indices) by means of a classification which uses 
a lexicographic order, without the need of finding a complicated system 
of recurrence relations like the one exposed above.

\subsection{The quark self-energy}

Using the algebraic methods explained above, it is possible to give a purely 
algebraic result for the 1-loop quark self-energy (for $r=1$, in the Feynman 
gauge). This was first computed 
in~\cite{GonzalezArroyo:1982ts,Hamber:1983qa,Groot:1983ng}.
The result for the 1-loop quark self-energy
\be
\Sigma(p^2,m^2) = g_0^2 C_F 
\left( \widetilde{m}_c + \ii \slash{p} \, 
\widetilde{\Sigma}_1 (p^2,m^2) + m \, \widetilde{\Sigma}_2 (p^2,m^2)\right) ,\
\ee
in terms of the basic constants is~\cite{Burgio:1996ji}
\bea
\widetilde{m}_c &=& -Z_0 -2{\cal F}(1,0) \approx 
-0.32571411742170157236 ,
\label{eq:mcconst} \\
\widetilde{\Sigma}_1 (p^2,m^2) &=& 
     \frac{1}{16\pi^2} (2 G(p^2a^2,m^2 a^2) + \gamma_E - F_0)
   + \frac{1}{8} Z_0 + {1\over 192} 
\nonumber \\ &&
   - \frac{1}{32 \pi^2} - Y_0 + \frac{1}{4} Y_1 
   - \frac{1}{16}Y_2 + 12\, Y_3 - \frac{1}{768} {\cal F}(1,-2)
\nonumber \\ &&
   - \frac{1}{192} {\cal F}(1,-1) + \frac{109}{192} {\cal F}(1,0) 
   - \frac{1}{768} {\cal F}(2,-2) + \frac{25}{48} {\cal F}(2,-1) 
\nonumber \\ &&
\approx \frac{1}{8\pi^2} G(p^2a^2,m^2 a^2) + 0.0877213749 , \\
\widetilde{\Sigma}_2 (p^2,m^2) &=&
     \frac{1}{4\pi^2} (F(p^2a^2,m^2 a^2) + \gamma_E - F_0) + 
     \frac{1}{48} - \frac{1}{4\pi^2}
\nonumber \\ &&
   - 4\, Y_0 +\, Y_1 - \frac{1}{4} Y_2 
   - \frac{1}{192} {\cal F}(1,-2) - \frac{1}{48} {\cal F}(1,-1)
\nonumber \\ &&
   - \frac{83}{48} {\cal F}(1,0) - \frac{1}{192} {\cal F}(2,-2) 
   + \frac{49}{12} {\cal F}(2,-1)
\nonumber \\ &&
\approx \frac{1}{4\pi^2} F(p^2 a^2,m^2 a^2) + 0.0120318529 ,
\eea
where 
\bea
F(p^2a^2,m^2 a^2) &=& \int_0^1 dx \log[(1-x) (p^2 x + m^2) a^2] , \\
G(p^2a^2,m^2 a^2) &=& \int_0^1 dx \, x \log[(1-x) (p^2 x + m^2) a^2] .
\eea
The importance of having explicit expressions like these cannot be 
underestimated. Thanks to them, the 1-loop self-energy $\Sigma(p^2,m^2)$ 
can now be computed with many significant decimal places, provided 
the basic constants are determined with sufficient accuracy.

\section{Coordinate space methods}
\label{sec:coordinatespacemethods}

It is very useful to consider propagators in coordinate space. In this Section
we are going to illustrate the coordinate space method, developed 
by~\cite{Luscher:1995zz}, which is very powerful for a variety of reasons. 
In particular it turns out to be very important for the calculation of 1-loop 
integrals with very high precision. Having 1-loop integrals determined with 
such precision is necessary for the implementation of the only known method 
(which will be also presented here) with which the computation of 2-loop 
integrals with good precision can be carried out.

A fundamental object for the coordinate space method is the mixed 
fermionic-bosonic propagator in position space, which will be denoted by
\be
{\cal G}_F (p,q;x) =\, \int {d^4k\over (2\pi)^4} \, 
{\e^{\ii kx}\over D_F(k,m)^p D_B(k,m)^q} .
\ee
Any of these propagators ${\cal G}_F(p,q;x)$ can be expressed as a linear
combination of the integrals ${\cal F}$ introduced in 
Section~\ref{sec:thegeneralfermioniccase}, and consequently as a linear 
combination of the fifteen basic constants 
$Z_0, Z_1, F_0, \dots, Y_0, Y_1, Y_2, Y_3$.
Of course the general position space propagator ${\cal G}_F(p,q;x)$ can also 
be expressed in terms of a different set of fifteen basic constants, and
general recursion relations between the ${\cal G}_F$'s can be derived.
We can always consider the ${\cal G}_F$'s instead of the ${\cal F}$'s 
as an intermediate representation of the general integral coming from 
the calculations of Feynman diagrams, and then express every ${\cal G}_F$ 
in terms of a chosen set of fifteen constants.

To begin with we consider the finite bosonic case, in which, as we know, 
only two basic constants are needed. In the bosonic case a simple reduction 
algorithm was developed by~\cite{Luscher:1995zz}, following ideas by 
Vohwinkel. For $(p,q)=(0,1)$ (i.e., the standard boson propagator), a 
recursion relation which involves only terms with the same $(p,q)=(0,1)$ 
was obtained.
This simple recursion relation avoids the introduction of noninteger 
dimensions and nonpositive values of $q$, as is instead the case for 
algorithms for general $p$ and $q$. The free lattice gluon propagator 
in position space can then be evaluated recursively, and is a linear function 
of its values near the origin. L\"uscher and Weisz chose as basic constants 
two values of the propagator close to the origin:
\bea
{\cal G}_F(0,1;(0,0,0,0)) &=& Z_0 , \\
{\cal G}_F(0,1;(1,1,0,0)) &=& Z_0 + Z_1 - {1\over4} \nonumber ,
\eea
where the relation between these two constants and the constants
of Section~\ref{sec:algebraic} is also shown.

We denote the gluon propagator by $G(x)={\cal G}_F(0,1;x)$.
The key observation in the L\"uscher-Weisz algorithm is that
\be
(\nabla_\mu^\star +\nabla_\mu) \, G(x) = x_\mu H(x) ,
\ee
where the function
\be
H(x)=\int_{-\pi}^\pi {d^4 p\over (2\pi)^4} \, \e^{\ii px} \, \log \hat{p}^2
\label{eq:hfunction}
\ee
is independent of $\mu$. Summing now this key formula over $\mu$ and using 
$-\Delta G(x) = \delta_{x,0}$ (where 
$\Delta = \sum_{\mu=0}^3 \nabla_\mu^\star \nabla_\mu$) we get
\be
H(x)={2\over \rho} \sum_{\mu=0}^3 
[G(x)-G(x-\hat{\mu})], \qquad \rho=\sum_{\mu=0}^3 x_\mu ,
\ee
which can be used back to eliminate $H(x)$ from the key formula. We then 
obtain the fundamental recursion relation for the gluon propagator in 
position space:
\be
G(x+\hat{\mu})\
=G(x-\hat{\mu})+{2x_\mu\over \rho}\sum_{\nu=0}^3[G(x)-G(x-\hat{\nu})] .
\label{eq:recrelvlw}
\ee
This formula is to be used for $\rho \neq 0$.
Since the propagator is independent of the sign and order of the four 
coordinates, we can restrict ourselves to $x_0 \ge x_1 \ge x_2 \ge x_3 \ge 0$.
In this sector, the recursion relation allows to express $G(x)$
in terms of its values in five points: $G(0,0,0,0)$, $G(1,0,0,0)$, 
$G(1,1,0,0)$, $G(1,1,1,0)$, $G(1,1,1,1)$. Now, using the properties of the 
propagator three more relations between these five constants can be found,
\bea
&& G(0,0,0,0) -G(1,0,0,0) = 1/8 \\
&& G(0,0,0,0) -3 \, G(1,1,0,0) -2 \, G(1,1,1,0) = 1/\pi^2 \nonumber \\
&& G(0,0,0,0) -6 \, G(1,1,0,0) -8 \, G(1,1,1,0) -3 \, G(1,1,1,1) = 0 
\nonumber , 
\eea
and thus we can finally write the generic bosonic propagator $G(x)$ 
in terms of only two basis constants:
\be
G(x) = r_1(x) \, G(0,0,0,0)+r_2(x) \, G(1,1,0,0) 
+{r_3(x)\over \pi^2} +r_4(x) .
\label{eq:expgx}
\ee
The coefficients $r_k(x)$ are rational numbers which can be computed in a 
recursive manner. We have thus expressed the free gluon propagator in terms 
of two basic constants, which can be reinterpreted as the values of the 
propagator near the origin.

The generalization of the coordinate method by L\"uscher and Weisz to fermions
is more complicated, and we will not discuss it here. For fermions a recursion 
relation analog to Eq.~(\ref{eq:recrelvlw}) which only involves the fermion 
propagator itself (i.e., without excursions to other values of $p$ and $q$) 
has not yet been found, and the reduction to the basic constants has to be 
carried out along more complicated procedures. The free fermion propagator 
can be written at the end in terms of eight basic constants, which can be 
reinterpreted as some values of the quark propagator near the origin.

\subsection{High-precision integrals}
\label{sec:highprecisionintegrals}

The reduction of the coordinate space propagators to a small set of basic 
constants, which we have just described, has interesting properties that 
turn out to be very useful for the high-precision computation of these basic 
constants. One can show that the coefficients $r_k$ in Eq.~(\ref{eq:expgx})
increase exponentially with the distance $x$, while the propagator $G(x)$ 
remains of course bounded. There are therefore huge cancellations and
significance losses when $x$ is large.

One can look at this apparent nuisance with other eyes, and exploit this 
numerical instability the other way around to compute the basic constants 
with very high precision~\cite{Luscher:1995zz}.
Let us consider the boson propagator at the points
\bea
x_1 &=& (n,0,0,0) \\
x_2 &=& (n,1,0,0) \nonumber
\eea
for large $n$. 
The associated sets of coefficients $r_k(x_1)$ and $r_k(x_2)$ in
Eq.~(\ref{eq:expgx}),
\bea
G(x_1) &=& r_1(x_1) \, G(0,0,0,0)+r_2(x_1) \, G(1,1,0,0) 
+{r_3(x_1)\over \pi^2} +r_4(x_1) \\
G(x_2) &=& r_1(x_2) \, G(0,0,0,0)+r_2(x_2) \, G(1,1,0,0) 
+{r_3(x_2)\over \pi^2} +r_4(x_2) \nonumber ,
\eea
are of order $10^n$. If this $2\times 2$ algebraic system is now inverted 
in terms of the unknowns $G(0,0,0,0)$ and $G(1,1,0,0)$, the coefficients 
multiplying $G(x_1)$ and $G(x_2)$ will be of order $10^{-n}$. Therefore the 
constants $G(0,0,0,0)$ and $G(1,1,0,0)$ can be determined to this level of 
precision, if one just neglects $G(x_1)$ and $G(x_2)$ in the $2\times 2$
inverse system. $G(0,0,0,0)$ and $G(1,1,0,0)$ then remain functions only 
of the coefficients $r_3$ and $r_4$.

It costs very little to go to very high $n$ and obtain the values of these 
two constants with very high precision. One can then systematically improve 
on the accuracy until the desired level of precision is reached. The method 
is exponentially convergent, and the error can also be estimated with good 
accuracy. On the contrary, a direct evaluation of the integrals defining 
the basic constants can never provide such accurate results.

Using this method we have calculated $Z_0$ and $Z_1$, the two constants which 
are the basis of finite bosonic integrals, with almost 400 significant decimal
places. These new results are given in Appendix B.

In the general mixed fermionic case, the expressions for a generic 
${\cal G}_F(p,q;x)$ in terms of fifteen values of it near the origin are also 
numerically unstable for $|x|\to\infty$. If we consider for example the 
fermion propagator, we can choose eight points with $|x|\approx n$ 
(say $y_1,\ldots,y_8$) and then we can express the propagator for $|x|<n$ 
in terms of ${\cal G}_F(1,0;y_i)$. These expressions are numerically stable: 
considering for example the set of eight points 
$X^{(n)}\equiv \{(n,[0-3],0,0),(n+1,[0-3],0,0)\}$ for high $n$ one can obtain 
the values of the propagator near the origin with great precision. It is also 
possible to apply the same procedure to ${\cal G}_F(1,q;x)$ with $q<0$.
The main advantage is that, using larger negative values of $q$, one can 
obtain more precise estimates of ${\cal G}_F(1,q;x)$ at this set of points.

Once one has extracted the fifteen values for the mixed propagators near the 
origin, it is straightforward to change basis and compute the fifteen basic 
constants for 1-loop integrals, $Z_0, Z_1, \dots, Y_2, Y_3$.
All constants (except $F_0$) are now known with a precision of sixty 
decimal places~\cite{Caracciolo:2001ki}.

\subsection{Coordinate space methods for 2-loop computations}

The determination of 1-loop integrals with high precision is an essential 
component of refined 2-loop calculations which are carried out using 
coordinate space methods. They allow to reach a precision unmatched by more 
conventional methods.

A few 2-loop calculations have also been performed by means of sophisticated 
extrapolations to infinite $L$ of momentum sums, using fitting functions and 
blocking transformations of the kind that we have discussed in 
Section~\ref{sec:latticeintegrals}.~\footnote{In a couple of cases even
three-loop calculations have been carried 
out~\cite{Alles:1992yh,Alles:1993dn,Alles:1998is}.}
The precision reached with these conventional methods can be good in some
cases, but the precision which can be achieved using the coordinate method
is already higher at present, and can be easily increased (by computing the 
1-loop ``building blocks'' with higher precision).
Sometimes the absolute values of the integrals to be computed are rather 
small, less than $10^{-6}$,~\footnote{An example of this is given by the 
calculation of the coefficients $C_\pm^5$ of Eq.~(\ref{eq:zwfive}), which has 
been carried out using conventional 2-loop integration 
methods~\cite{Curci:1987hu}.} and in order to get a good relative error the 
coordinate method seems to be the most suitable.

The 2-loop calculations completed so far, using either conventional or 
coordinate space methods, only concern quantities that are finite, like
the coefficient $b_2$ of the $\beta$ function, of which we have spoken in 
Section~\ref{sec:approach}. No divergent matrix elements have yet been 
computed. Although some of the individual diagrams for finite matrix elements 
can contain subdivergences, the situation in which the matrix element are
themselves divergent is certainly more challenging.

\subsubsection{Bosonic case}
\label{sec:bosoniccase}

We will now illustrate the calculation of two loop integrals using the 
coordinate space method by~\cite{Luscher:1995zz}. At first we discuss the case
of integrals with zero external momenta, which is simpler. Let us then 
consider the 2-loop integral in momentum space
\be
I_1 = \int_{-\pi}^\pi \frac{d^4k}{(2\pi)^4} \, 
    \int_{-\pi}^\pi \frac{d^4q}{(2\pi)^4} \,
    \frac{1}{\widehat{k}^2 \widehat{q}^2 (\widehat{k+q})^2} .
\label{cintconst}
\ee
This integral can be reexpressed as follows:
\bea
I_1 &=& \int_{-\pi}^\pi \frac{d^4k}{(2\pi)^4} \,
      \int_{-\pi}^\pi \frac{d^4q}{(2\pi)^4} \,
      \int_{-\pi}^\pi \frac{d^4r}{(2\pi)^4} \, 
      (2\pi)^4 \delta^{(4)} (k+q+r) \, 
      \frac{1}{\widehat{k}^2 \widehat{q}^2 \widehat{r}^2} \nonumber \\
  &=& \sum_x 
      \int_{-\pi}^\pi \frac{d^4k}{(2\pi)^4} \, 
      \int_{-\pi}^\pi \frac{d^4q}{(2\pi)^4} \, 
      \int_{-\pi}^\pi \frac{d^4r}{(2\pi)^4} \, 
      \e^{\ii kx} \frac{1}{\widehat{k}^2} \, \e^{\ii qx}
      \frac{1}{\widehat{q}^2} \, \e^{\ii rx} \frac{1}{\widehat{r}^2} 
   \nonumber \\
  &=& \sum_x \Bigg(\int_{-\pi}^\pi \frac{d^4k}{(2\pi)^4} \,
      \e^{\ii kx} \, \frac{1}{\widehat{k}^2} \Bigg)^3 \nonumber \\
  &=& \sum_x G^3(x) . 
\eea
The 2-loop integral in momentum space has thus been written in terms of a sum 
of a simple function of the position-space propagator $G(x)$ over the lattice 
sites. We have changed from an eight-dimensional sum in momentum space to a 
four-dimensional sum in position space. Since $G(x) \sim 1/x^2$, this sum is 
absolutely convergent. An evaluation of the sum over a finite domain 
containing the origin, say the region $|x| < 20$, can then be taken as a first
approximation of the integral $I_1$. 
Of course one must also know the propagator $G(x)$ with very good accuracy, 
because some precision will be lost when computing the sums. It is here that
the high-precision technique which we have just discussed turns out to be vary 
handy. In particular, the recursion relations in Eq.~(\ref{eq:recrelvlw})
allow to compute $G(x)$ in terms of two constants (which can be 
chosen to be $Z_0$ and $Z_1$) which can be computed with arbitrary precision,
and thus $G(x)$ can be computed with that precision.

If $G(x)$ can be determined (for any $x$ in the domain) with arbitrary
precision, the only remaining challenge is to evaluate the sums over a 
reasonable domain of sites in the most effective way. Of course the domain 
cannot be too large because of computational limitations. The sums can then be 
better evaluated by exploiting the knowledge of the asymptotic expansion of 
the propagator for large $x$, which is given by~\footnote{For the derivation 
of this and similar asymptotic expansions, see~\cite{Luscher:1995zz}.}
\be
G(x) \stackrel{x \to \infty}{\longrightarrow}
\frac{1}{4\pi^2 x^2} \Bigg\{ 1 - \frac{1}{x^2} + 2 \frac{x^4}{(x^2)^3}
- 4 \frac{1}{(x^2)^2} + 16 \frac{x^4}{(x^2)^4} 
- 48 \frac{x^6}{(x^2)^5} + 40 \frac{(x^4)^2}{(x^2)^6} + \dots 
\Bigg\} .
\ee
Here and in the following the notation
\be
x^n = \sum_{\mu=0}^3 (x_\mu)^n 
\ee
is used. We have then
\bea
G^3(x) &\stackrel{x \to \infty}{\longrightarrow}& 
\frac{1}{(4\pi^2)^3} \frac{1}{(x^2)^3} \Bigg\{ 1 - \frac{3}{x^2} 
+ 6 \frac{x^4}{(x^2)^3} - 9 \frac{1}{(x^2)^2} + 36 \frac{x^4}{(x^2)^4}  
\nonumber \\ 
&& \qquad \qquad \qquad
- 144 \frac{x^6}{(x^2)^5} + 132 \frac{(x^4)^2}{(x^2)^6} \Bigg\} 
+ O(|x|^{-12}) \\
&=& \frac{1}{(4\pi^2)^3} \Bigg\{ 
\Bigg[ \frac{1}{(x^2)^3} + \frac{3}{10(x^2)^5} \Bigg] h_0 (x)
+ \Bigg[ \frac{3}{(x^2)^6} + \frac{24}{7(x^2)^7} \Bigg] h_1 (x) \nonumber \\
&& \qquad \qquad 
- \frac{3}{4(x^2)^8} h_2 (x) + \frac{33}{140(x^2)^9} h_3 (x) \Bigg\} 
+ O(|x|^{-12}) ,
\eea
where in the last line we have rewritten $G^3(x)$ in terms of the homogeneous 
harmonic polynomials
\bea
h_0 (x) & = & 1                                                     , \\
h_1 (x) & = & 2x^4 - (x^2)^2                                        , \\
h_2 (x) & = & 16x^6 - 20x^2x^4 + 5(x^2)^3                           , \\
h_3 (x) & = & 560 (x^4)^2 - 560 x^2x^6 + 60 (x^2)^2x^4 - 9(x^2)^4   .
\eea
It is useful at this point to define a generalized zeta function
\be
Z(s,h) = \sum_{x:\, x\neq 0} h(x) (x^2)^{-s} , 
\ee
where the site $x=0$ is not to be included in the sum. 
These zeta functions can be calculated with the necessary accuracy. 
To this end one introduces the heat kernel
\be
k(t,h) = \sum_{x:\, x\neq 0} h(x) \e^{-\pi t x^2} ,
\ee
so that the generalized zeta function can be reexpressed as
\bea
Z(s,h) &=& \frac{\pi^s}{\Gamma (s)} \int_0^\infty dt \, t^{s-1}
[k(t,h) - h(0)]   \\
&=& \frac{\pi^s}{\Gamma (s)} \Bigg\{ \frac{2h(0)}{s(s-2)} 
+\int_1^\infty dt \, [ t^{s-1} + (-1)^{d/2} t^{d-s+1} ]
[k(t,h) - h(0)]  \Bigg\} ,
\eea
where in the last line the integration has been split into two parts, and the 
formula
\be
k(t,h) = (-1)^{d/2} \, t^{-d-2} \, k(1/t,h)
\ee
has been used.~\footnote{This formula derives from the fact that $h(x)$ is 
harmonic and from the Poisson summation formula
\be
\sum_x \e^{-\ii qx} \e^{-\pi t x^2}
= t^{-2} \sum_x \e^{-(q+2\pi x)^2/(4\pi t)}.
\ee
}
We have then
\bea
\sum_x G^3(x) &=& \sum_{x:\, x\neq 0} \Big[ G^3(x) - G_{\mathrm as}^3(x) \Big] 
 \nonumber\\
 && + \frac{1}{(4\pi^2)^3} \Bigg\{ Z(3,h_0) + \frac{3}{10} Z(5,h_0)  
    + 3 Z(6,h_1) + \frac{24}{7} Z(7,h_1) \nonumber \\ 
 && \qquad \qquad - \frac{3}{4} Z(8,h_2) + \frac{33}{140} Z(9,h_3) \Bigg\} 
 \nonumber \\
 && + G^3(0) .
\eea
We can see that once the values of the zeta function are known, it only 
remains to compute 
$\sum_{x:\, x\neq 0} \Big[ G^3(x) - G_{\mathrm as}^3(x) \Big]$, 
which is rapidly convergent since each term goes at least like $|x|^{-12}$.
This sum can then be evaluated with a reasonable approximation using domains 
which are not too large, and in this way one can obtain
\be
I_1 = 0.0040430548122(3) .
\ee

The coordinate space method can be used to compute more complicated integrals
in which nontrivial numerators are present, like
\bea
I_2 &=& \int_{-\pi}^\pi \frac{d^4k}{(2\pi)^4} \,
      \int_{-\pi}^\pi \frac{d^4q}{(2\pi)^4} \,
      \sum_{\mu=0}^3 \frac{\widehat{k}_\mu^2 \widehat{q}_\mu^2}{\widehat{k}^2
            \widehat{q}^2 (\widehat{k+q})^2} \nonumber \\
  &=& \int_{-\pi}^\pi \frac{d^4k}{(2\pi)^4} \, 
      \int_{-\pi}^\pi \frac{d^4q}{(2\pi)^4} \, 
      \int_{-\pi}^\pi \frac{d^4r}{(2\pi)^4} \, 
      (2\pi)^4 \delta^{(4)} (k+q+r) 
      \sum_{\mu=0}^3 \frac{\widehat{k}_\mu^2 \widehat{q}_\mu^2}{\widehat{k}^2 
            \widehat{q}^2 \widehat{r}^2} \nonumber \\
  &=& \sum_x 
      \int_{-\pi}^\pi \frac{d^4k}{(2\pi)^4} \,
      \int_{-\pi}^\pi \frac{d^4q}{(2\pi)^4} \,
      \int_{-\pi}^\pi \frac{d^4r}{(2\pi)^4} \, 
      \e^{\ii kx} \e^{\ii qx} \e^{\ii rx} 
      \sum_{\mu=0}^3 \frac{\widehat{k}_\mu^2 \widehat{q}_\mu^2}{\widehat{k}^2 
             \widehat{q}^2 \widehat{r}^2} 
   \nonumber \\
  &=& \sum_x \sum_{\mu=0}^3 \Bigg(\int_{-\pi}^\pi \frac{d^4k}{(2\pi)^4} \,
      \e^{\ii kx} \, \frac{\widehat{k}_\mu^2}{\widehat{k}^2} \Bigg)^2
             \Bigg(\int_{-\pi}^\pi \frac{d^4r}{(2\pi)^4} \,
      \e^{\ii rx} \, \frac{1}{\widehat{r}^2} \Bigg)
  \nonumber \\
  &=& \sum_x \sum_{\mu=0}^3 \Big(-\nabla_\mu^* \nabla_\mu G(x)\Big)^2 G(x) 
      \nonumber \\
  &=& 0.0423063684(1) .
\eea
Here the derivatives of the propagator in coordinate space generate the 
$\widehat{k}_\mu^2$ and $\widehat{q}_\mu^2$ factors in the numerators 
of the momentum-space integrals. Similarly,
\bea
I_3 &=& \int_{-\pi}^\pi \frac{d^4k}{(2\pi)^4} \, 
      \int_{-\pi}^\pi \frac{d^4q}{(2\pi)^4} \,
      \sum_{\mu=0}^3 \frac{\widehat{k}_\mu^2 \widehat{q}_\mu^2
           (\widehat{k+q})_\mu^2}{\widehat{k}^2
            \widehat{q}^2 (\widehat{k+q})^2} \nonumber \\
  &=& \int_{-\pi}^\pi \frac{d^4k}{(2\pi)^4} \,
      \int_{-\pi}^\pi \frac{d^4q}{(2\pi)^4} \,
      \int_{-\pi}^\pi \frac{d^4r}{(2\pi)^4} \, 
      (2\pi)^4 \delta^{(4)} (k+q+r) 
      \sum_{\mu=0}^3 \frac{\widehat{k}_\mu^2 \widehat{q}_\mu^2
           \widehat{r}_\mu^2}{\widehat{k}^2 
            \widehat{q}^2 \widehat{r}^2} \nonumber \\
  &=& \sum_x 
      \int_{-\pi}^\pi \frac{d^4k}{(2\pi)^4} \,
      \int_{-\pi}^\pi \frac{d^4q}{(2\pi)^4} \,
      \int_{-\pi}^\pi \frac{d^4r}{(2\pi)^4} \, 
      \e^{\ii kx} \e^{\ii qx} \e^{\ii rx} 
      \sum_{\mu=0}^3 \frac{\widehat{k}_\mu^2 \widehat{q}_\mu^2
            \widehat{r}_\mu^2}{\widehat{k}^2 
             \widehat{q}^2 \widehat{r}^2} 
   \nonumber \\
  &=& \sum_x \sum_{\mu=0}^3 \Bigg(\int_{-\pi}^\pi \frac{d^4k}{(2\pi)^4} \,
      \e^{\ii kx} \, \frac{\widehat{k}_\mu^2}{\widehat{k}^2} \Bigg)^3
         \nonumber \\
  &=& \sum_x \sum_{\mu=0}^3 \Big(-\nabla_\mu^* \nabla_\mu G(x)\Big)^3 
      \nonumber \\
  &=& 0.054623978180(1) .
\eea
In general, integrals where the numerator is a polynomial in sines and cosines
can be computed using these techniques. However, in more complicated cases,
where for example the denominator has a higher power than the integrals 
discussed so far, it is necessary to introduce auxiliary functions.

One of such cases is given by the calculation of 
\be
I_4 = \int_{-\pi}^\pi \frac{d^4k}{(2\pi)^4} \,
    \int_{-\pi}^\pi \frac{d^4q}{(2\pi)^4} \,
    \sum_{\mu=0}^3 \frac{\widehat{k}_\mu^2 \widehat{q}_\mu^2}{(\widehat{k}^2)^2
            \widehat{q}^2 (\widehat{k+q})^2} .
\ee
The problem here is that the factor $1/(\widehat{k}^2)^2$ cannot be related in
a simple way to $G(x)$. In this case, the auxiliary function
\be
K(x) = \int_{-\pi}^\pi \frac{d^4p}{(2\pi)^4} \, \frac{(\e^{\ii px}-1)}{
      (\widehat{p}^2)^2} 
\ee
is just what we need. In fact,
\be
\nabla_\mu^* \nabla_\mu K(x) = - \int_{-\pi}^\pi \frac{d^4p}{(2\pi)^4} \,
\widehat{p}_\mu^2 \, \frac{\e^{\ii px}}{(\widehat{p}^2)^2} ,
\ee
so that
\bea
I_4 &=& \sum_x \sum_{\mu=0}^3 \Big(-\nabla_\mu^* \nabla_\mu K(x)\Big) 
\Big(-\nabla_\mu^* \nabla_\mu G(x)\Big) G(x)
\nonumber \\
&=& 0.006603075727(1) .
\eea
The function $K(x)$ is related to $G(x)$ by
\bea
-\Delta K(x) &=& G(x) , \\
(\nabla_\mu^* + \nabla_\mu) K(x) &=& -x_\mu G(x) , 
\eea 
so that it can be recursively computed in terms of $G(x)$ and of the
values of $K(x)$ at the corner of the unit hypercube.

Finally, the integral
\be
I_5 = \int_{-\pi}^\pi \frac{d^4k}{(2\pi)^4} \, 
          \int_{-\pi}^\pi \frac{d^4q}{(2\pi)^4} \,
          \sum_{\mu=0}^3 \frac{\widehat{k}_\mu^4}{(\widehat{k}^2)^3
          \widehat{q}^2 (\widehat{k+q})^2} 
\ee
can be computed introducing the auxiliary function  
\be
L(x) = \int_{-\pi}^\pi \frac{d^4p}{(2\pi)^4} \, \frac{\Big(\e^{\ii px}-1
   -\ii  \sum_\mu x \sin p_\mu +\frac{1}{2} (\sum_\mu x \sin p_\mu)^2\Big)}{
   (\widehat{p}^2)^3} ,
\ee
which has the properties 
\bea
-\Delta L(x) &=& K(x) + \frac{1}{8} G(0) - \frac{1}{32\pi^2} , \\
(\nabla_\mu^* + \nabla_\mu) L(x) &=& -\frac{1}{2} x_\mu 
\Big[ K(x) + \frac{1}{8} G(0) \Big] ,
\eea
and can be computed recursively. The integral is then given by 
\bea
I_5 &=& \sum_x \sum_{\mu=0}^3 
\Big(\nabla_\mu^* \nabla_\mu \nabla_\mu^* \nabla_\mu L(x)\Big) G^2(x)
\nonumber \\
&=& 0.00173459425(1) .
\eea

We now turn to 2-loop integrals which also depend on an external momentum. Of 
course these integrals can also be computed using the decomposition explained 
in Section~\ref{sec:divergentintegrals} and the theorem of Reisz, thanks to 
which only zero-momentum integrals have at the end to be really computed on 
the lattice. However, it is also possible to compute these 2-loop integrals 
using the coordinate space methods directly, as we are going to show.

To understand how external momenta are incorporated into the method, 
let us first discuss the simpler case of a 1-loop integral depending
on an external momentum $p$. The logarithmically divergent integral
\be
I_6 = \int_{-\pi}^\pi \frac{d^4k}{(2\pi)^4} \,  
\frac{1}{\widehat{k}^2 (\widehat{k-p})^2} 
\ee
can be rewritten in position space as follows:
\bea
I_6 &=& \int_{-\pi}^\pi \frac{d^4k}{(2\pi)^4} \,
      \int_{-\pi}^\pi \frac{d^4q}{(2\pi)^4} \, 
      (2\pi)^4 \delta^{(4)} (q+k-p) \, 
      \frac{1}{\widehat{k}^2 \widehat{q}^2} \nonumber \\
  &=& \sum_x \e^{-\ii px} 
      \int_{-\pi}^\pi \frac{d^4k}{(2\pi)^4} \, 
      \int_{-\pi}^\pi \frac{d^4q}{(2\pi)^4} \, 
      \e^{\ii kx} \frac{1}{\widehat{k}^2} \, \e^{\ii qx} 
      \frac{1}{\widehat{q}^2} \nonumber \\
  &=& \sum_x \e^{-\ii px} G^2(x) . 
\eea
The integral can then be computed by evaluating the sums
\be
I_6 = \lim_{\epsilon \to 0} \,\sum_x \e^{-\epsilon x^2} \e^{-\ii px}\, G^2(x) .
\ee
It is now useful to employ the function $H(x)$ which we have introduced 
in Eq.~(\ref{eq:hfunction}) when deriving the recursion relations for $G(x)$ 
in coordinate space:
\be
H(x)=\int_{-\pi}^\pi {d^4 p\over (2\pi)^4} \, \e^{\ii px} \log \hat{p}^2 .
\ee
Its asymptotic expansion for large $x$ is
\bea
H(x) &\stackrel{x \to \infty}{\longrightarrow}& 
-\frac{1}{\pi^2} \frac{1}{(x^2)^2} \Bigg\{ 1 - \frac{4}{x^2} 
+ 8 \frac{x^4}{(x^2)^3} - 7 \frac{1}{(x^2)^2} + 40 \frac{x^4}{(x^2)^4}  
\nonumber \\ 
&& \qquad \qquad \qquad
- 288 \frac{x^6}{(x^2)^5} + 280 \frac{(x^4)^2}{(x^2)^6} + \dots \Bigg\} ,
\eea
and thus its leading order term has (up to a constant) the same asymptotic
behavior of $G^2(x)$. We now subtract and add to the original integral 
an appropriate expression containing $H(x)$:
\be
I_6 = -\frac{1}{16\pi^2} \lim_{\epsilon \to 0} \, \sum_x \e^{-\epsilon x^2} 
\e^{-\ii px} H(x) 
+ \sum_x \e^{-\ii px} \Big[ G^2(x) + \frac{1}{16\pi^2} H(x) \Big].
\ee
The first part is just the Fourier transform of $H(x)$, which can be read off
from its definition above, and is thus given by
\be
-\frac{1}{16\pi^2} \lim_{\epsilon \to 0} \, \sum_x \e^{-\epsilon x^2} 
\e^{-\ii px} H(x) 
= -\frac{1}{16\pi^2} \log \hat{p}^2 = -\frac{1}{16\pi^2} \log p^2 +O(p^2) .
\ee
The finite constant is obtained from the second part, in which the leading 
$1/|x|^4$ terms of $G^2(x)$ and $H(x)$ cancel in the subtraction, so that this
part goes like the Fourier transform of $1/|x|^6$ (and is therefore finite). 
In the limit $p \to 0$ we have then
\be
I_6 = -\frac{1}{16\pi^2} \log p^2 
   + \sum_x \Big[ G^2(x) + \frac{1}{16\pi^2} H(x) \Big].
\ee
The sum can be computed using generalized zeta functions as explained before.
The result is
\be
I_6 = -\frac{1}{16\pi^2} \log p^2  +\frac{1}{8\pi^2}
   +0.02401318111946489(1).
\ee

Let us now illustrate this powerful method for a 2-loop integral with 
an external momentum $p$. We consider
\be
I_7 = \int_{-\pi}^\pi \frac{d^4k}{(2\pi)^4} \,
    \int_{-\pi}^\pi \frac{d^4q}{(2\pi)^4} \,  
\frac{1}{\widehat{k}^2 \widehat{q}^2 (\widehat{k+q-p})^2} ,
\ee
which by dimensional arguments will give a result of the form 
\be
I_7 = c_1 + c_2 \, p^2 \log p^2 +c_3 \, p^2 +O(p^4). 
\ee
We would like to determine the coefficients $c_2$ and $c_3$. We already know
the constant $c_1$, which can be computed setting $p=0$ and is given by $I_1$, 
Eq.~(\ref{cintconst}). This integral can be written in position space as
\bea
I_7 &=& \int_{-\pi}^\pi \frac{d^4k}{(2\pi)^4} \, 
      \int_{-\pi}^\pi \frac{d^4q}{(2\pi)^4} \, 
      \int_{-\pi}^\pi \frac{d^4s}{(2\pi)^4} \, 
      (2\pi)^4 \delta^{(4)} (s+q+k-p) \, 
      \frac{1}{\widehat{k}^2 \widehat{q}^2 \widehat{s}^2} \nonumber \\
  &=& \sum_x \e^{-\ii px} G^3(x) . 
\eea
A function that has the same leading asymptotic behavior (up to a constant) 
of $G^3(x)$, and for which the Fourier transform is easily calculable
in an exact way, is the 4-dimensional Laplacian of $H(x)$. We thus make 
the decomposition
\be
I_7 = -\frac{1}{2(4\pi)^4} \sum_x \e^{-\ii px} \Delta H(x) 
+ \sum_x \e^{-\ii px} \Big[ G^3(x) + \frac{1}{2(4\pi)^4} \Delta H(x) \Big].
\ee
The first part is the Fourier transform of $\Delta H(x)$, which can 
be easily computed and gives
\be
  -\frac{1}{2(4\pi)^4} \sum_x \e^{-\ii px} \Delta H(x) 
= -\frac{1}{2(4\pi)^4} \, \hat{p}^2 \log \hat{p}^2 
= -\frac{1}{2(4\pi)^4} \, p^2 \log p^2 +O(p^4) .
\ee
We have thus computed the coefficient $c_2$. The coefficient $c_3$ can be 
obtained from the second part, in which the leading $1/|x|^6$ terms of 
$G^3(x)$ and $H(x)$ cancel in the subtraction, which then goes like $1/|x|^8$.
This fact is important, because to compute $c_3$ we have to expand the 
exponential to order $p^2$, and then the second part becomes 
\be
-\frac{1}{8} \sum_x p^2 x^2 
\Big[ G^3(x) + \frac{1}{2(4\pi)^4} \Delta H(x) \Big] ,
\ee
and only in this way the function to be summed goes like $1/|x|^6$ again 
and is then finite and can be computed using generalized zeta functions. 
Putting everything together we have the result
\bea
I_7 &=& I_1 
  -\frac{1}{2(4\pi)^4} \, p^2 \log p^2
  -\frac{1}{8} \, p^2 \sum_x x^2 
   \Big[ G^3(x) + \frac{1}{2(4\pi)^4} \Delta H(x) \Big] +O(p^4) \\
&=& 0.0040430548122(3) 
  -\frac{1}{2(4\pi)^4} \, p^2 \log p^2 
  -p^2 \cdot 0.00007447695(1) +O(p^4) . \nonumber 
\eea

A more complicated example can be found in~\cite{Luscher:1995zz}, to which we 
also refer for further details on the method. These and other integrals have 
been used for the calculation of the coefficient $b_2$ of the $\beta$ function
in the pure Wilson gauge theory~\cite{Luscher:1995nr,Luscher:1995np}.
The complete 2-loop calculation of this coefficient requires a combination 
of momentum space and coordinate space methods .

To summarize, coordinate space is of great help for the calculation of 
important 2-loop momentum space integrals. One can make use of the important 
advantages that only four-dimensional lattice sums must be performed, instead 
of eight-dimensional ones, and one can also exploit the asymptotic expansion 
of the gluon propagator $G(x)$ for large values of $x$ to improve on the 
convergence.

\subsubsection{Fermionic case}
\label{sec:fermioniccase}

We now briefly discuss the computations of 2-loop lattice diagrams with Wilson
fermions based on the coordinate space method by L\"uscher and Weisz, which 
have been presented in~\cite{Capitani:1997rz,Caracciolo:2001ki}.

An essential ingredient of these 2-loop calculations is the high-precision 
determination of 1-loop mixed fermionic-bosonic propagators
(Section~\ref{sec:thegeneralfermioniccase}).
The algebraic method for general Wilson fermions, thanks to which any 1-loop 
lattice integral with Wilson fermions can be written as a combination of 
fifteen basic constants, which can be computed once for all with arbitrarily 
high precision, allows then the implementation of the coordinate space method 
to two-loop integrals with Wilson fermions. We remind that the algebraic 
method depends only on the structure of the Wilson propagators, and not on the
vertices, and thus it can be applied in calculations with the Wilson action as
well as with the improved clover action.

Let us consider a very simple integral
\be
I = \int_{-\pi}^\pi {d^4l\over (2\pi)^4} \,
    \int_{-\pi}^\pi {d^4r\over (2\pi)^4} \, 
    {1\over D_F(l) D_F(r) D_F(l+r)} ,
\ee
where $D_F$ is the denominator of the quark propagator. The standard 
alternative (in momentum space) would consists in replacing each integration 
with a discrete sum over $L$ points and then extrapolate to infinite $L$:
\be
I = {1\over L^8} \sum_{l,r,l+r \neq 0} {1\over D_F(l) D_F(r) D_F(l+r)} .
\ee
Here $l$ and $r$ run over the set 
$(n + 1/2) \cdot 2\pi/L$, 
$n=0,\ldots,L-1$, and one excludes from the sum the points 
$l+r=0 \bmod 2\pi$ (where the third propagator would diverge).
Increasing the values of $L$ one gets the following approximations for 
the integral $I$:
\bea
L=10 &\quad&  0.000 799 652  \\
L=18 &\quad&  0.000 848 862  \\
L=20 &\quad&  0.000 853 822  \\
L=26 &\quad&  0.000 863 064 . 
\eea
Then, using an extrapolation function of the form 
\be
a_0 + {a_1 \log L + a_2\over L^2} +
      {a_3 \log L + a_4\over L^4} 
\ee
and fitting with it the results of the sum for $6\le L \le 26$, one gets 
the estimate $I \approx 0.000 879 776$. Note that the computation at $L=26$
requires the evaluation of the function on an integration grid of 
$26^8 \sim 2 \cdot 10^{11}$ points. With the 2-loop techniques based on the 
coordinate space method one can instead without much effort obtain 
$I \approx 0.000 879 777 918 1 (12)$.

In order to compute 2-loop momentum-space integrals with high precision, 
they are first rewritten in coordinate space along the lines shown by 
L\"uscher and Weisz, trading an additional $x$ integration for the 
$\delta^{(4)} (l+r+s)$ which expresses the vanishing of external 
momenta~\cite{Capitani:1997rz}. For example one can write
\bea
&& I = 
\int_{-\pi}^\pi {d^4l\over (2\pi)^4} \, 
\int_{-\pi}^\pi {d^4r\over (2\pi)^4} \, 
\,{1\over D_F^{p_1}(l) D_B^{q_1}(l)}
\,{1\over D_F^{p_2}(r) D_B^{q_2}(r)} 
\,{1\over D_F^{p_3}(l+r) D_B^{q_3}(l+r)} \nonumber\\
&& = 
\int_{-\pi}^\pi {d^4l\over (2\pi)^4} \,
\int_{-\pi}^\pi {d^4r\over (2\pi)^4} \,
\int_{-\pi}^\pi {d^4s\over (2\pi)^4} \, 
\,{1\over D_F^{p_1}(l) D_B^{q_1}(l)} 
\,{1\over D_F^{p_2}(r) D_B^{q_2}(r)} 
\,{1\over D_F^{p_3}(s) D_B^{q_3}(s)} \, \delta^{(4)} (l+r+s) \nonumber \\
&& = \sum_x \,\, {\cal G}_F(p_1,q_1;x) \,\, {\cal G}_F(p_2,q_2;x)
\,\, {\cal G}_F(p_3,q_3;x) .
\eea
To compute this simple integral we have to evaluate in coordinate space 
the lattice sums in the last line. The price to be paid is to compute all 
necessary ${\cal G}_F(p_j,q_j;x)$ integrals with huge precision. This does not
constitute a particular challenge. These ${\cal G}_F(p_j,q_j;x)$ integrals can
be determined with the desired precision (for a sufficiently large domain of 
values of $x$) by using the 1-loop algebraic algorithm, and one can exploit 
their asymptotic expansions for large values of $x$. After the subtraction 
of the asymptotic behavior, the sums are much better convergent.

The task is then to evaluate with enough accuracy sums of the kind
\be
   \Sigma = \sum_{\Lambda} f(x)
\ee
where $\Lambda$ should in principle be the whole lattice, using only a not 
too big finite lattice domain, and be able to estimate the error. The sums 
are restricted over domains defined as 
\be
D_p = \{x\in \Lambda: |x|_1 \le p\}
\ee
where $|x|_1 = \sum_{\mu} |x_{\mu}|$. In~\cite{Capitani:1997rz} it was found 
that the domain $D_{21}$ is a reasonable choice. For the computation of the 
generic Feynman diagram one then chooses some convenient representation of 
2-loop integrals, and assembles a database of them. To do this, the 1-loop 
integrals necessary for their computations can be recursively decomposed in 
terms of the fifteen basic 1-loop constants and then evaluated with very high 
precision.

Recently with this method the 2-loop critical mass $m_c$, defined by the 
vanishing of the inverse renormalized propagator $S^{-1} (p,m_0)$ at $p=0$,
\be
S^{-1} (0,m_c)=0,
\ee
has been computed in the Wilson case. Using one-dimensional integrals 
calculated with high accuracy (about sixty significant decimal places), 
the 2-loop diagrams relevant for the critical mass have been obtained with a 
precision of about ten significant decimal places. The result for $SU(N_c)$ 
with $N_f$ fermion flavors can be put in the form
\be
m_c = g_0^2 \frac{N_c^2-1}{N_c} c_{1} +g_0^4 (N_c^2-1) \Big( 
c_{2,1} + \frac{1}{N_c^2} c_{2,2} + \frac{N_f}{N_c} c_{2,3} \Big)
\ee
where 
\be
c_1 = -0.16285705871085078618 
\ee
is the 1-loop result, and 
\bea
c_{2,1} &=& -0.0175360218(2) \\
c_{2,2} &=& 0.0165663304(2) \\
c_{2,3} &=& 0.001186203(6)                  
\eea
is the result of the 2-loop computations made with the coordinate space method.
These 2-loop numbers are in agreement with the less precise results given
in~\cite{Follana:2000mn}, obtained using conventional momentum-space 
methods, which have also produced the value of the critical mass for improved 
fermions~\cite{Panagopoulos:2001fn}.

\section{Numerical perturbation theory}

We want to conclude this review mentioning some numerical methods that are 
completely different from all that we have presented so far, but represent 
interesting alternative ideas for the computation of perturbative expansions, 
and in principle can be able to perform high-loop calculations, at least in 
some cases. In the last decade a numerical approach to perturbation theory, in 
which one extracts perturbative coefficients using Monte Carlo simulations 
instead of calculating Feynman diagrams analytically, has emerged and has 
produced a few interesting results.

One of these techniques is given by the so-called numerical stochastic 
perturbation theory~\cite{DiRenzo:1994sy,DiRenzo:1995qc}, which is based on 
the stochastic quantization by Parisi and Wu (1981). It uses numerical 
simulations of the Langevin equation. An additional parameter, the stochastic 
time $\tau$, is introduced in the theory, and the gauge field is taken to be 
a random variable, $U_\mu (x;\tau)$, which evolves according to the Langevin 
equation:
\be
\frac{d}{d\tau} U_\mu (x;\tau) = -\frac{\delta S [U]}{\delta U_\mu} 
+\eta (x;\tau).
\ee
The last term, $\eta (x;\tau)$, is a Gaussian noise matrix, i.e., its 
expectation value is 
\be
\langle \eta (x;\tau) \eta (x';\tau') \rangle = 
2 \delta_{x,x'} \delta_{\tau,\tau'} , 
\ee
and is the stochastic part of the equation. This approach lends itself quite 
naturally to lattice investigations. One has to discretize the stochastic time
$\tau$, introducing a nonzero Langevin time step $\epsilon$. This causes an 
$O(\epsilon)$ systematic error, but for $\epsilon \to 0$ and $\tau \to \infty$ 
the time average is expected to reach asymptotically the expectation values 
corresponding to a path integral with action $S[U]$. At the end of the lattice
calculations one has then to make extrapolations to the limit $\epsilon = 0$.

The numerical solution to the Langevin equation consists in updating the gauge
field according to~\cite{Batrouni:1985jn},
\be
U_\mu (x;\tau+\epsilon) = 
\e^{\displaystyle - F [U(\tau), \eta]} \, U_\mu (x;\tau) ,
\ee
where the driving function is
\be
F [U, \eta] = \sum_i T^i (\epsilon \nabla^i_{x,\mu} S [U] + \sqrt{\epsilon} 
\eta^i ),
\ee
$\nabla$ being the Lie derivative on the group.
For the Wilson plaquette action one has 
\be
\sum_i T^i \nabla^i_{x,\mu} S_G [U] = \frac{\beta}{4N_c} 
\sum_{U_P \ni U_\mu (x)} (U_P - U^\dagger_P)_\Tr ,
\ee
where $\Tr$ stands for the traceless part.

Using stochastic perturbation theory it has been possible for simulations 
involving only gluons to reach much higher orders in $g_0^2$ than in 
conventional perturbative calculations, where only in a few cases one has 
reached to two-loop stage. Simulations done using stochastic perturbation 
theory have instead reached something like the tenth loop order in the case 
of the plaquette~\cite{DiRenzo:1995qc}. Where results are available for 
stochastic and conventional perturbation theory, they agree within errors.

The inclusion of fermions in stochastic perturbation theory has been
accomplished only recently~\cite{DiRenzo:2000qe,DiRenzo:2002vg}. 
Although passing from quenched to unquenched calculations requires little 
computational overhead, one needs to use a fast Fourier transform. 

To perform the field updating in the unquenched theory one has to include
a more complicated derivative term in the driving function,
\be
\nabla S_G \rightarrow \nabla S_G - \nabla (\Tr \log M) = 
                       \nabla S_G - \Tr ((\nabla M)M^{-1}) ,
\ee
where $M$ is the fermionic matrix. We see that the Lie derivative generates 
an inverse fermionic matrix $M^{-1}$, which is nonlocal and quite expensive 
to compute numerically. An ingenious way to simulate this term has been 
proposed long ago~\cite{Batrouni:1985jn}, and consists in taking instead
\be
\nabla S_G - \mathrm{Re} (\xi^\dagger (\nabla M) M^{-1} \xi),
\label{eq:fermstoc}
\ee
where $\xi$ is a Gaussian random variable: 
$\langle \xi_i \xi_j \rangle = \delta_{ij}$,
so that after averaging over this new variable one recovers the fermionic term
\be
\langle \xi (\nabla M) M^{-1} \xi \rangle = \Tr ((\nabla M)M^{-1}).
\ee
This random fermionic term can be computed recursively, expanding the relevant 
quantities order by order in the coupling. In practical terms one generates 
the random variable $\xi_0$ (which does not depend on the coupling and is 
then of order zero) and then recursively computes the variable $\psi$ in
\be
M \psi = \xi .
\ee
This is a lot easier than doing the inversion of the Dirac operator, and 
gives a local evolution, because Eq.~(\ref{eq:fermstoc}) becomes
\be
\nabla S_G - \mathrm{Re} (\xi^\dagger (\nabla M) \psi ) .
\ee
To compute $\psi$ order by order one needs the expansions
\be
M = M^{(0)} + \sum_{k>0} \beta^{-k/2} M^{(k)}
\ee
and its inverse
\be
M^{-1} = {M^{(0)}}^{-1} + \sum_{k>0} \beta^{-k/2} {M^{-1}}^{(k)} .
\ee
Note that inverting the zeroth order of the fermionic matrix is trivial: 
${M^{-1}}^{(0)} = {M^{(0)}}^{-1}$. The non-trivial orders of $M^{-1}$
can be obtained recursively as follows:
\bea
{M^{-1}}^{(1)} &=& - {M^{(0)}}^{-1} M^{(1)} {M^{(0)}}^{-1} \\
{M^{-1}}^{(2)} &=& - {M^{(0)}}^{-1} M^{(2)} {M^{(0)}}^{-1} 
                   - {M^{(0)}}^{-1} M^{(1)} {M^{-1}}^{(1)} \nonumber \\
{M^{-1}}^{(3)} &=& - {M^{(0)}}^{-1} M^{(3)} {M^{(0)}}^{-1} 
                   - {M^{(0)}}^{-1} M^{(2)} {M^{-1}}^{(1)} 
                   - {M^{(0)}}^{-1} M^{(1)} {M^{-1}}^{(2)} , \nonumber
\eea
and so forth. Since ${M^{(0)}}^{-1}$ is diagonal in momentum space, one can 
perform its computation going to momentum space, back and forth, and for 
this reason a fast Fourier transform code is needed. At the end one has
\bea
\psi^{(0)} &=& {M^{(0)}}^{-1} \xi_0 \\
\psi^{(1)} &=& {M^{-1}}^{(1)} \xi_0 = - {M^{(0)}}^{-1} M^{(1)} \psi^{(0)} 
\nonumber \\
\psi^{(2)} &=& {M^{-1}}^{(2)} \xi_0 = - {M^{(0)}}^{-1} M^{(2)} \psi^{(0)} 
                                      - {M^{(0)}}^{-1} M^{(1)} \psi^{(1)} ,
\nonumber 
\eea
and so forth. These are the lowest order terms in the expansion
\be
\psi^{(i)} = {M^{-1}}^{(i)} \xi_0 ,
\ee
with which one can compute recursively Eq.~(\ref{eq:fermstoc}), and hence
simulate the fermion system using the Langevin equation.

Every quantity that one wants to compute with stochastic perturbation theory 
has to be expanded in powers of 
\be
\beta^{-1/2} = \frac{g_0}{\sqrt{6}}. 
\ee
For example,
\be
U_\mu (x;\tau) = 1 + \sum_{k>0} \beta^{-k/2} U_\mu^{(k)} (x;\tau) ,
\qquad A_\mu (x;\tau) = \sum_{k>0} \beta^{-k/2} A_\mu^{(k)} (x;\tau),
\ee
where
\be
U_\mu (x;\tau) = \exp (A_\mu (x;\tau)/\sqrt{\beta}).
\ee
Observables are composite operators, and every observable depending on $U$
can be expanded in the coupling. For example the first order of the plaquette 
(Eq.~(\ref{eq:uuuu})) is given by 
\bea
    P^{(1)}_{\mu \nu}(x;\tau)
&=& \quad U_\mu^{(1)} (x;\tau) U_\nu^{(0)} (x + a\hat{\mu};\tau) 
   (U_\mu^\dagger)^{(0)} (x + a\hat{\nu};\tau) (U_\nu^\dagger)^{(0)} (x;\tau) 
\nonumber \\ && 
+\, U_\mu^{(0)} (x;\tau) U_\nu^{(1)} (x + a\hat{\mu};\tau) 
   (U_\mu^\dagger)^{(0)} (x + a\hat{\nu};\tau) (U_\nu^\dagger)^{(0)} (x;\tau)
\nonumber \\ && 
+\, U_\mu^{(0)} (x;\tau) U_\nu^{(0)} (x + a\hat{\mu};\tau) 
   (U_\mu^\dagger)^{(1)} (x + a\hat{\nu};\tau) (U_\nu^\dagger)^{(0)} (x;\tau)
\nonumber \\ && 
+\, U_\mu^{(0)} (x;\tau) U_\nu^{(0)} (x + a\hat{\mu};\tau) 
   (U_\mu^\dagger)^{(0)} (x + a\hat{\nu};\tau) (U_\nu^\dagger)^{(1)} (x;\tau).
\eea

So far the calculations which use stochastic perturbation theory have been 
limited to finite quantities. The computation of the quark currents, of 
operators measuring structure functions and also weak operators, which in 
general have nonvanishing anomalous dimensions, still seems a long way to go.

There is another way to compute the coefficients of perturbative expansions
numerically. It makes use of Monte Carlo simulations at very weak couplings to 
measure short-distance quantities~\cite{Dimm:1994fy,Trottier:2001vj}. 
The perturbative coefficients are then extracted by making fits to the results 
of these numerical simulations. In this method some input from conventional 
perturbation theory is still required, to define a physical coupling constant. 

Monte Carlo results are thus fitted to truncated polynomial expansions in 
the coupling. One must be careful in doing this, because if the truncation is 
too short a poor fit to the Monte Carlo data will come out, while if it is too
long then the lowest coefficients, which are the dominant contributions, 
become poorly constrained. These fits can be improved by constraining some of 
the parameters by means of the techniques known as ``constrained curve 
fitting''~\cite{Lepage:2001ym}. They are especially useful in order 
to constrain parameters which are poorly determined statistically. 

Couplings and volumes are chosen in such a way that the lattice momenta
are perturbative. The computations are done at very small lattice spacings
and couplings. For example in~\cite{Trottier:2001vj} Wilson fermions have been
used with couplings ranging from $\alpha_{lat}=0.008$ to $\alpha_{lat}=0.053$.
The lattice spacing and the volume must satisfy
\be
q^\star \ll \frac{aL}{2\pi} \ll \frac{1}{\Lambda_{QCD}}
\ee
where $q^\star$ is a typical gluonic momentum scale. This condition also 
insures that the density of the discrete momenta is not small. In recent works
the lattice spacing was able to cover a wide range of values,
\be
10^{-29} < a \Lambda_{QCD} < 10^{-3} .
\ee

Unfortunately when doing these simulations one has to take into account 
the problem of the appearance of zero modes, and these infrared effects 
are potentially dangerous. Twisted boundary conditions, like the ones in  
Eq.~(\ref{eq:twistedbc}), can however eliminate these zero modes, and 
they are also useful to suppress nonperturbative finite-volume effects.

With this method one has studied problems like the mass renormalization, small 
Wilson loops and the static-quark self-energy. In the case of Wilson loops, 
their three-loop coefficients have been extracted and very good agreement 
has been found with existing conventional perturbative results at two 
loops~\cite{Hattori:1981ac,DiGiacomo:vd,Curci:1983cc,Curci:1983wh,Hasenfratz:1984bx,Heller:1984hx,Bali:2002wf}.

Numerical perturbation theory is still in its infancy. Studies have been
mostly limited to gluonic quantities, and moreover finite ones, although some
progress has been seen recently in the fermionic case. It seems that much work
still needs to be done before one can think of reproducing for example the 
results of Section~\ref{sec:examplewilson} for the renormalization of the 
first moment of the unpolarized quark distribution, which involves a fermionic 
operator with a nonzero anomalous dimension.

\section{Conclusions}

We have discussed in this review many different aspects of the perturbative
calculations made with gauge fields and fermions defined on a hypercubic 
lattice. Much progress has been made in the last decade. Perturbative 
calculations have been carried out using a variety of actions and in a variety
of physical situations, and recently they have been of great help in the study
of chiral fermions on the lattice. This long-standing issue has been solved, 
and the construction of chiral gauge theories on the lattice presents features
that are theoretically relevant also for general quantum field theories.

We have seen the consequences of the loss of exact Lorentz invariance, and 
discussed the mixings that derive from this, as well as mixings caused by the 
breaking of chiral symmetry. In this respect Ginsparg-Wilson fermions represent
a great step forward in lattice calculations, and they promise to solve 
long-standing problems like the calculation from first principles of 
$\Delta I = 1/2$ weak amplitudes and of the CP-violating parameter 
$\epsilon'/\epsilon$. 

From the more technical side, new methods for the calculation of one- and 
two-loop integrals, with at times incredible precision, have been invented. 
It is now possible to compute one-loop bosonic integrals with very high 
precision, and also fermionic integrals with rather good precision. More 
challenging have been the calculations of two-loop Feynman diagrams. Here new 
methods based on the calculation of propagators in position space have been 
very useful.
 
These coordinate space methods present many advantages for the computation of 
two-loop integrals, and is likely that more development on this side will make 
two-loop calculations much easier to compute and more precise.

\section*{Acknowledgments}

I would like to thank Martin L\"uscher for having motivated and encouraged me 
to write this article, and Karl Jansen for much useful advice during its 
preparation. I profited from discussions with them and with Rainer Sommer 
and Oleg Tarasov. I finally thank Karl Jansen, Martin L\"uscher and 
Giancarlo Rossi for useful comments on a first version of this article.

\appendix

\newpage

\section{Notation and conventions}

Generally we use Greek letters $\mu, \nu, \lambda, \dots$ for the 
4-dimensional Lorentz (or Euclidean) indices, which run from 0 to 3, 
and Latin letters $i, j, k, \dots$ for the 3-dimensional indices. 
Latin letters $a, b, c, \dots$ are instead used for color indices. 
Matrices in the fundamental representation of $SU(N_c)$ are denoted by $T$, 
while in the adjoint representation they are denoted by $t$. 
The normalization is 
\be
\Tr (T^a T^b) =\frac{1}{2} \delta^{ab} , \qquad [T^a,T^b] = \ii f^{abc} T^c .
\ee

The symbol $\partial_\mu$ is used for continuum derivatives, while
the forward and backward lattice derivatives are
\be
\nabla_\mu f(x) = \frac{f(x+\hat{\mu})-f(x)}{a}, \qquad 
\nabla_\mu^\star f(x) = \frac{f(x)-f(x-\hat{\mu})}{a} .
\ee
The gauge covariant lattice derivatives are denoted by 
$\widetilde{\nabla}_\mu$, $\widetilde{\nabla}_\mu^\star$ and are given in 
the main text.

The shorthand notation
\be
\widehat{a k}_\mu = \frac{2}{a} \sin \frac{a k_\mu}{2}
\ee
is also often used.

The euclidean Dirac matrices in the chiral representation are
\be
\gamma_0 = \Bigg( 
\begin{tabular}{cc} $0$ & $-1$ \\ $-1$ & $0$ \\ \end{tabular} 
\Bigg), \qquad
\gamma_i = \Bigg( 
\begin{tabular}{cc} $0$ & $-i\sigma_i$ \\ $i\sigma_i$ & $0$ \\ \end{tabular} 
\Bigg), \qquad
\gamma_5 = \gamma_0 \gamma_1 \gamma_2 \gamma_3 = \Bigg( 
\begin{tabular}{cc} $1$ & $0$ \\ $0$ & $-1$ \\ \end{tabular} 
\Bigg), 
\ee
where $\sigma_i$ are the Pauli matrices, and $1$ is the $2\times 2$
identity matrix. The chiral projectors are
\be
P_\pm = \frac{1\pm \gamma_5}{2} = \frac{1}{2} \, \Bigg( 
\begin{tabular}{cc} $1$ & $\mp 1$ \\ $\mp 1$ & $1$ \\ \end{tabular} 
\Bigg) .
\ee
Also, 
\be
\sigma_{\mu\nu} = \frac{\ii}{2} \, [ \gamma_\mu, \gamma_\nu ] .
\ee

\newpage

\section{High-precision values of $Z_0$ and $Z_1$}

We give here the new results of a high-precision calculation of the 
fundamental bosonic constants $Z_0$ and $Z_1$, defined in Eqs.~(\ref{eq:z0}) 
and~(\ref{eq:z1}). They have been obtained using the recursion relation 
Eq.~(\ref{eq:recrelvlw}) starting with the gluon propagator at 
$x=396$.~\footnote{This was the maximum value compatible with the computational
resources at my disposal.} 
The resulting values are:
\bea
Z_0 &=& 0.15493339023106021408483720810737508876916113364521\%\\
&& \quad  98321191752313395351673319454163790491630919236741\%\nonumber\\
&& \quad  07489754149497376290387736082594941817577598499678\%\nonumber\\
&& \quad  92951387264251940296570608026229566408322643387967\%\nonumber\\
&& \quad  84914774223913881583813529174816118783903355821052\%\nonumber\\
&& \quad  96552782448948240231078335735055832848473775143559\%\nonumber\\
&& \quad  80401738187671539786446652153505144942596811258480\%\nonumber\\
&& \quad  8043251280463983474068128158341212164145185669 (1) \nonumber ,
\eea
\bea
Z_1 &=& 0.10778131353987400134339155028381651483289553031166\%\\
&& \quad  39233465607465024738935201734450177503973077057462\%\nonumber\\
&& \quad  05621844437484365688328635749227594895147284883092\%\nonumber\\
&& \quad  15513746596880936011669949517608632177321337226921\%\nonumber\\
&& \quad  37793898141534158628242723648006344189984806003623\%\nonumber\\
&& \quad  35938862675314675890326849096822755901010023056671\%\nonumber\\
&& \quad  38098768018557508588203302625651590185630674198459\%\nonumber\\
&& \quad  5509105334593113724314740915504203882005031989 (1) \nonumber .
\eea
These numbers have 396 significant decimal places. Thus, every 1-loop Feynman 
diagram in the pure gauge theory can always be given with such numerical
accuracy (if the plaquette action is used), since it can be expressed as
a linear combination of $Z_0$ and $Z_1$ only.

\end{document}